\documentclass[onecolumn]{aastex631}
\usepackage{natbib}
\usepackage{xspace}
\usepackage{longtable}
\usepackage{placeins}
\usepackage{xcolor}
\usepackage{graphicx}
\usepackage{rotating}
\usepackage{url}

\shorttitle{Binary X-ray Source SEDs}
\shortauthors{Dickey, J.M. et al.}

\begin{document}
\title{Spectral Energy Distributions of Southern Binary X-Ray Sources}
\author[0000-0002-6300-7459]{John M. Dickey}
\affiliation{School of Natural Sciences, Private Bag 37, University of Tasmania, Hobart, TAS, 7001, Australia}
\author[0000-0002-7521-9897]{S. D. Vrtilek}
\affiliation{Harvard-Smithsonian Center for Astrophysics, 60 Garden Street, Cambridge, MA 02138, USA}
\author{Michael McCollough}
\affiliation{Harvard-Smithsonian Center for Astrophysics, 60 Garden Street, Cambridge, MA 02138, USA}
\author{Bram Boroson}
\affiliation{Dept. of Physics and Astronomy, University of South Carolina, 712 Main St., Columbia, SC 29208 USA}
\author{John A. Tomsick}
\affiliation{Space Sciences Laboratory, University of California, Berkeley, 7 Gauss Way, Berkeley, CA 94720-7450, USA}
\author{Charles Bailyn}
\affiliation{Department of Astronomy, Yale University, PO Box 208101, New Haven, CT 06520-8101, USA}
\author[0000-0002-2756-395X]{Jay M. Blanchard}
\affiliation{National Radio Astronomy Observatory, P.O. Box O, 1003 Lopezville Rd., Socorro, NM 87801-0387, USA}
\affiliation{School of Natural Sciences, Private Bag 37, University of Tasmania, Hobart, TAS, 7001, Australia}
\author{Charlotte Johnson}
\affiliation{Dept. of Physics and Astronomy, University of South Carolina, 712 Main St., Columbia, SC 29208 USA}

\begin{abstract}

The rapid variability of X-ray binaries produces a wide range of X-ray states that are linked to activity across the electromagnetic spectrum. It is particularly challenging to study a sample of sources large enough to include all types in their various states, and to cover the full range of frequencies that show flux density variations. Simultaneous observations with many telescopes are necessary. In this project we monitor 48 X-ray binaries with seven telescopes across the electromagnetic spectrum from 5$\times 10^9$ Hz to 10$^{19}$ Hz, including ground-based radio,  IR, and optical observatories and five instruments on two spacecraft over a one-week period.  We construct spectral energy distributions and matching X-ray color-intensity diagrams for 20 sources that have the most extensive detections.  Our observations are consistent with several models of expected behavior proposed for the different classes: we detect no significant radio emission from pulsars or atoll sources, but we do detect radio emission from Z sources in the normal or horizontal branch, and from black holes in the high/soft, low/hard and quiescent states.  The survey data provide useful constraints for more detailed models predicting behavior from the different classes of sources.
\end{abstract}


\section{Introduction}\label{sec:Introduction}
Many of the brightest X-ray sources in the sky are mass-exchange binary stars in which a collapsed object accretes
material from its main sequence or red giant companion.  X-ray binaries (XRBs) generally
show evidence for an accretion disk; if the collapsed object is a neutron star or black hole the inner
region of the disk is hot enough that the black-body peak can be at energies (h$\nu >$ 1 keV),
but other emission processes may dominate.  Often the accretion disk powers a jet of relativistic 
electrons and magnetic field that \color{black}emits \color{black}synchrotron radiation and/or inverse Compton scatters lower
frequency photons into the X-ray and $\gamma$-ray bands. Comptonization can also be due to scattering by electrons in the corona of the accretion disk, a jet is not necessary.  XRBs have been called micro-quasars \citep{Mirabel_1993}
because they are similar to active 
\color{black}galactic \color{black} nuclei in both their structure and emission
processes, but with masses, luminosities, and dynamical time scales 10$^6$ to 10$^9$ times smaller.
XRBs are among the few objects that are bright across the spectrum from cm-wave radio to MeV $\gamma$-rays. 
Rapid variations in their emission spectra show the effects of different states linked to accretion rate
changes.  These can happen much faster in micro-quasars than in active galactic nuclei, typically 
the time scale is proportional to the collapsed object mass.  
Monitoring XRBs with telescopes to cover all wavelengths may reveal how the accretion disk drives the
jet, and the structural connection between them. For example, most jet theories invoke a poloidal magnetic field arising from the accretion disk that collimates the jet flow.  Variations in accretion rate on time scales of days to weeks are common in XRBs
\color{black}; \color{black}
 similar variations take
much longer in active galactic nuclei.

Well-studied XRBs can be classified according to the properties of the collapsed object and its magnetic
field \citep[e.g.][]{Migliari_Fender_2006}.  In black hole (BH) systems the compact object 
is a stellar mass black hole, the companion star can have either high or low mass. Those with low mass companions are sometimes referred to as soft X-ray transients (SXTs). SXTs can increase by factors of 100 to 10,000 in brightness at both X-ray and optical wavelengths. These variations are due to dramatic changes in mass accretion rate from the companion to the BH. 
While few SXT's are persistent, i.e. continuously active, most systems with high mass companions are persistent.

\begin{figure}[ht]
\hspace{-.2in} \includegraphics[width=3.5in]{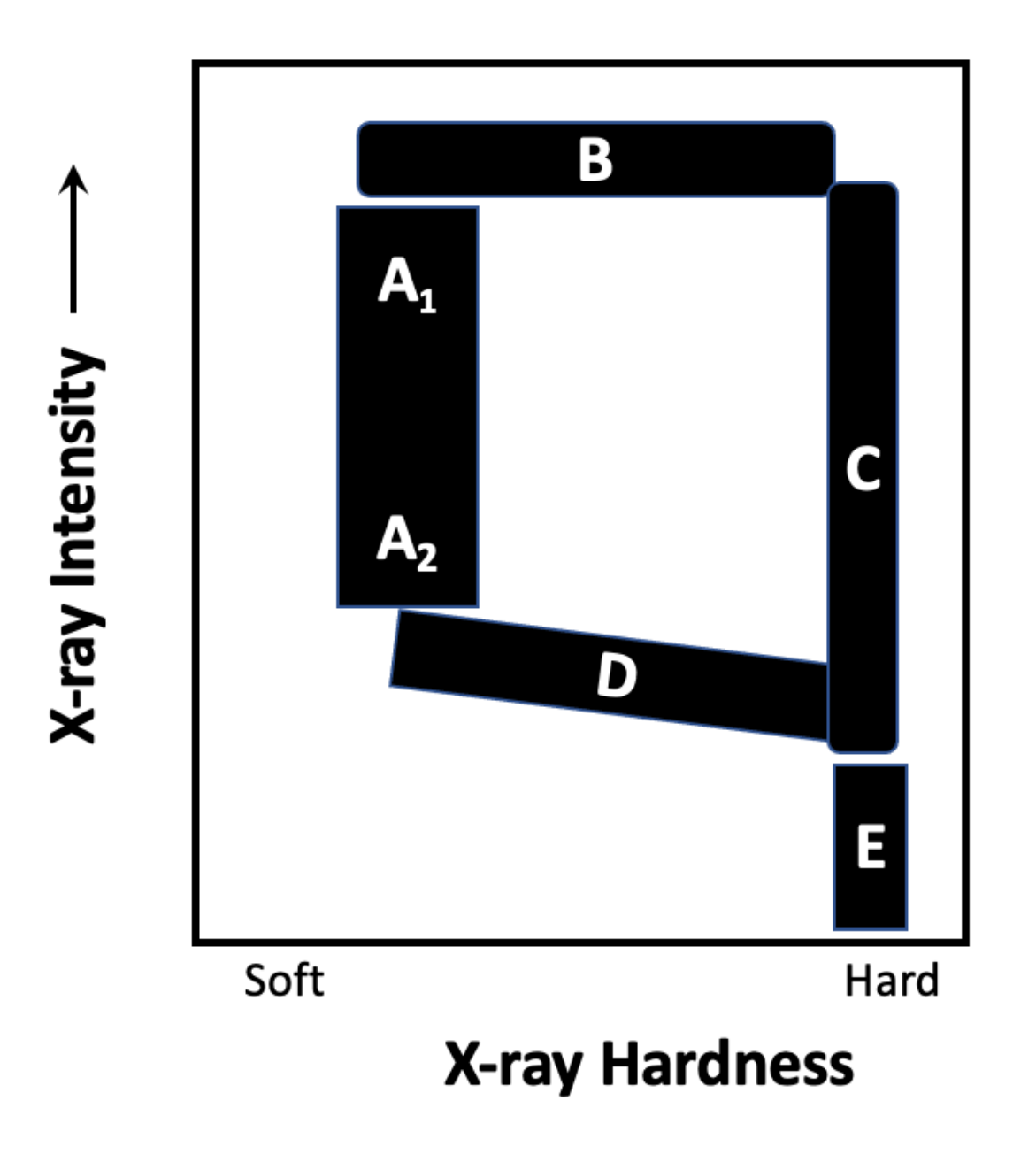}
\hspace{-.1in} \includegraphics[width=3.5in]{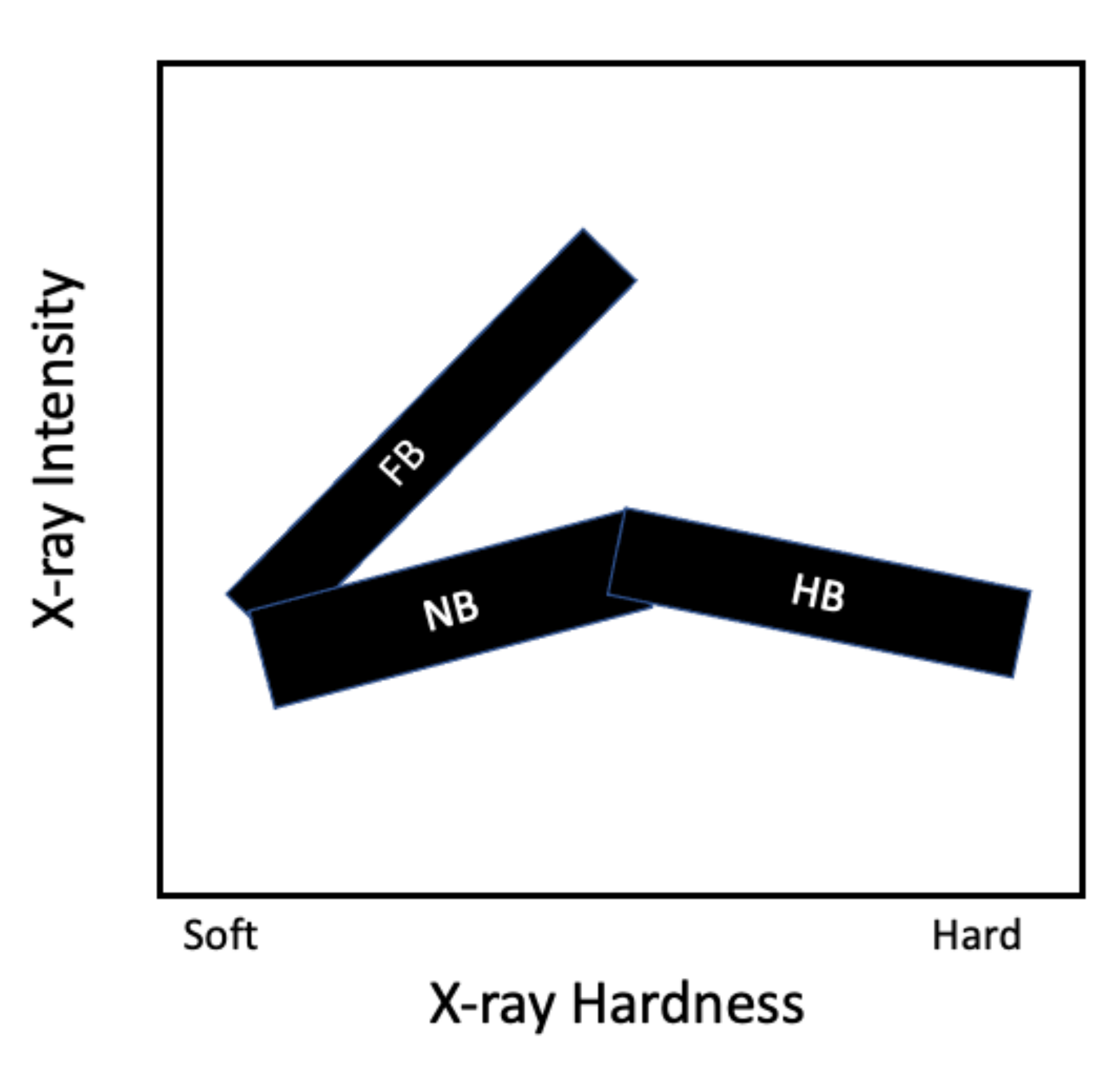}
\caption{Schematics illustrating the characteristic behavior of 
BH (left panel) and Z-type (right-panel) XRBs on plots of X-ray color vs. intensity (CI).  The various states (labeled A-E for clarity on the left panel) are described in 
Section \ref{sec:Introduction}.  \label{fig:BHschematic} }
\end{figure}

Black Holes: BHs have very distinctive patterns in X-ray color-color (CC) and color-intensity (CI) diagrams \citep[e.g.][]{Remillard_McClintock_2006}.  Fig. \ref{fig:BHschematic}, left panel, presents a schematic with black rectangles labeled A-E to indicate the different BH states as defined here:
\begin{itemize}
\item{A1 the high/soft state. In this state the X-rays are dominated by thermal emission. There are relatively large jets, typically hundreds of AU in size, that are optically thin, highly relativistic and transient on short time scales. \citet{Remillard_McClintock_2006} call this the Steep Power Law (SPL), \citet{Belloni_etal_2005} call it the Soft-Intermediate State (SIMS).  }
\item{A2 the soft/low state. In this state the radio jet is quenched and the accretion disk dominates the emission at all wavelengths \citep[e.g.][]{Fender_etal_1999}.}
\item{B the bright, intermediate state. The source shows superluminal ejection and a radio flare. The X-rays indicate a hot corona in addition to the accretion disk. }
\item{C the hard/low state. The X-ray luminosity is high, greater than 10$^{-2}$ times the Eddington luminosity. There is a hot corona, and a persistent, flat-spectrum radio source, typically in the mJy range, whose emission extends to the IR and optical. The radio source is compact, roughly 10 AU in size. The radio, IR, and X-rays show correlation in their variability. The X-ray spectrum indicates a reflection component. }
\item{D the faint, intermediate state. Here there is no radio emission, the X-rays indicate a hot corona in addition to the accretion disk. }
\item{E the quiescent state. The source includes a hot corona plus a black body component that dominates the ultra-violet and optical. There is a persistent, flat-spectrum, compact radio source at the $\mu$Jy level \citep[e.g.][]{ Kalemci_etal_2005,Plotkin_Gallo_Jonker_2013,Gallo_etal_2015, Gallo_etal_2018, Shaw_Plotkin_Miller-Jones_2021}.}
\end{itemize}
In the quiescent state, the luminosity of a BH XRB is lower than that of a neutron star XRB in quiescence, because the BH has no surface \citep{Garcia_etal_2001}

Z-sources: The compact object is a neutron star with a low-mass companion, in which the mass accretion is typically more than 50 percent of the Eddington limit.  
Z-sources 
are named for the shape they form on X-ray CC diagram as a result of changes in both spectral and timing properties.  The schematic in Fig. \ref{fig:BHschematic}, right panel,  shows the three nominal states in a CI \color{black}diagram\color{black}: Horizontal Branch (HB, right side); Normal Branch (NB, middle); Flaring Branch (FB, upper left).  Radio emission is strongest in the HB, decreases during the NB, and is \color{black} usually \color{black} not detectable during the FB.  In the HB state there is a steady radio jet that changes into a transient compact radio jet in the NB \citep[e.g.][]{Migliari_etal_2007}. 
\color{black} The slope of the HB is much steeper in ``Cyg X-2-like'' Z-track sources than in ``Sco X-1-like'' sources \color{black} and the flaring branch in Sco X-1-like sources goes to much higher intensities than in Cyg X-2-like sources. 
Several papers suggest that the flaring branch in 
Sco X-1-like sources is a combination of unstable burning and an increase of $\dot{M}$,
the mass accretion rate, which makes flaring much stronger; whereas the flaring
branch in the Cyg X-2-like sources is associated with a release of energy on the
neutron star consistent with unstable nuclear burning.
\citep{Balucinska-Church_etal_2010,Church_etal_2012}.
Fig. \ref{fig:BHschematic} shows a generic CI diagram for Z-sources
that is \color{black} roughly \color{black} consistent for both types: both NB and FB \color{black} usually \color{black} increase in intensity and
hardness from the node between the FB and the NB.  The HB decreases in intensity but
increases in hardness from the node between the HB and the NB.
\color{black} Z-sources typically have higher accretion rates than Atolls.

\begin{figure}[ht]
\vspace*{-1.5in}
\hspace{-.6in} \includegraphics[width=4in]{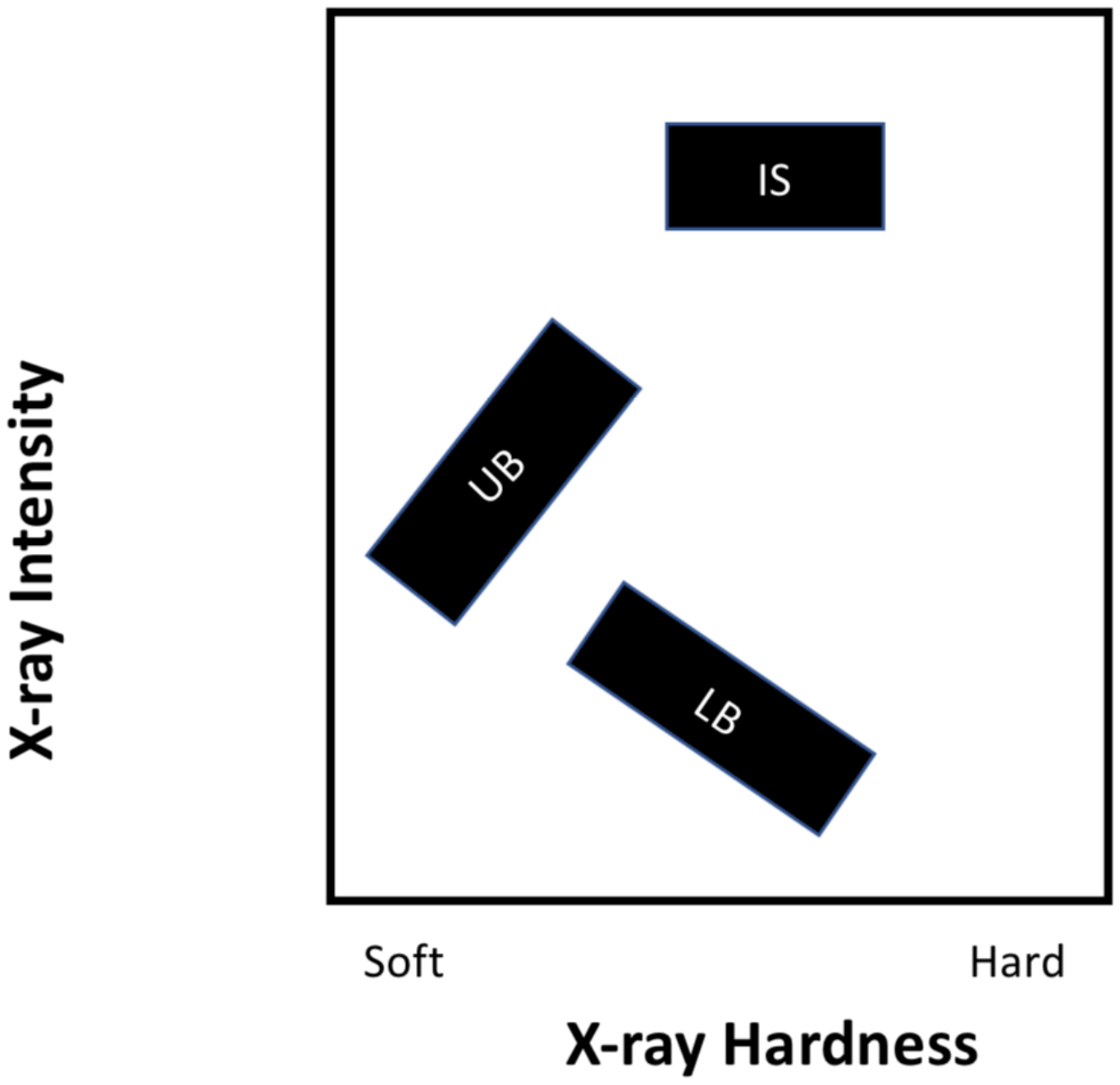}
\hspace{-1.2in} \includegraphics[width=4in]{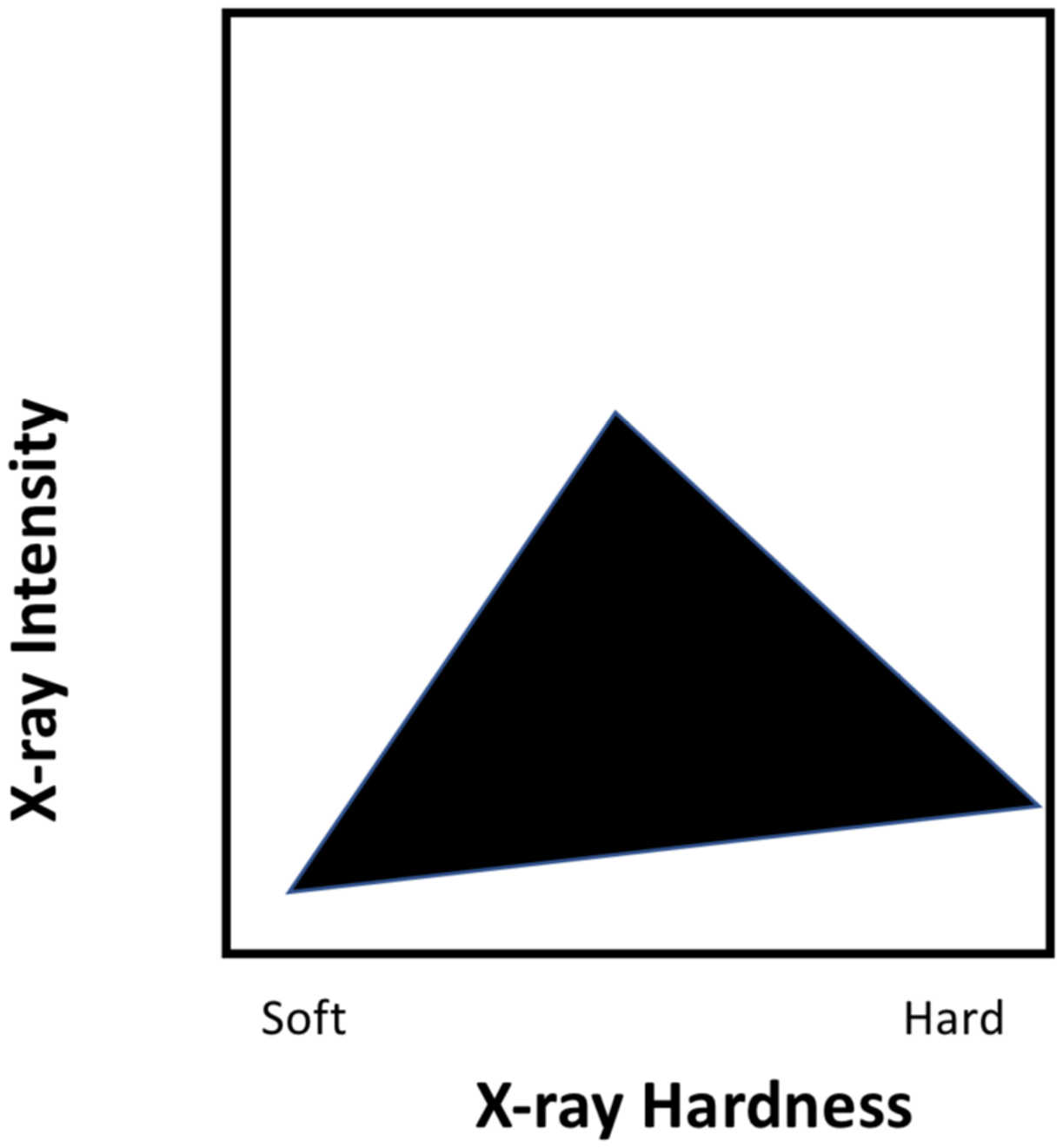}

\vspace*{-1.2in}
\caption{Schematics illustrating the characteristic behavior of
Atoll (left panel) and Pulsar (right panel) XRBs on plots of X-ray color vs. intensity (CI).  The various states are described in  
Section \ref{sec:Introduction}. \label{fig:AtollSchematic}}
\end{figure}

Atolls: The compact object is a neutron star and the companion star is low-mass and the mass accretion is generally less than ten percent of Eddington. 
Atoll sources are named for the shapes they occasionally form in X-ray CC and CI diagrams (Fig.
\ref{fig:AtollSchematic}, left panel). They fall into several spectral states: a hard, high luminosity `island' (IS) or a soft, lower luminosity `banana'; the banana branch is subdivided into upper banana (UB) and lower banana (LB) based on changes in both spectral and timing properties. The transition between these states is rapid and not covered well by observations. A subset of atolls are referred to as bursters: these increase by factors of 10 to 100 in X-ray luminosity due to thermonuclear flashes on the surface of the neutron star. Atolls typically have lower accretion rates than Z-sources. It has been shown that some Z and Atoll sources can morph into each other with major changes in mass accretion rate \citep{Homan_etal_2010}.

Pulsars: The compact object is a neutron star and the companion can be either high or low-mass. The neutron star is highly magnetized ($|B|\ >\ 10^{12}$ G); gas accreted from the companion is guided by the neutron star's magnetic field onto the magnetic poles producing two or more localized X-ray hot spots. As the neutron star rotates the X-rays appear to pulse as the spots rotate in and out of view. This X-ray emission is generally not associated with any pulsed radio emission
\color{black}
but some millisecond pulsars do display radio pulses (rotationally powered) although with a significant phase offset between X-ray and radio pulses
\citep{Papitto_etal_2013, Guillot_etal_2019}.
\color{black}  In X-ray CC and CI diagrams they cluster towards hard colors with little or no dependence on intensity (Fig. \ref{fig:AtollSchematic}, right panel).  No resolved jets have been observed but \citet{vandenEijnden_etal_2018} detect evolving radio emission from several X-ray pulsars.

This project constitutes an extensive program to obtain spectral energy distributions (SEDs) from the radio through the X-rays for a representative selection of accreting binaries.  The impetus for such a survey was the simultaneous availability of multiple space observatories in conjunction with ground-based radio and optical facilities in combinations that allowed us to span the full range of available wavelengths efficiently.  SEDs are often obtained based on target of opportunity telescope observations of a given source in a specific state.  Such campaigns have revealed a lot about the time evolution of emission in individual sources \citep[e.g.][]{McClintock_etal_2001, Vrtilek_etal_2001, Sivakoff_etal_2016}.  Here we obtained observing time on seven telescopes for a large sample of sources over a one week interval: this allowed us to trace patterns of behavior among different categories of XRBs.

The telescopes and observations are described in section \ref{sec:Observations}.
The results for each source are presented in section \ref{sec:SEDs}.  In section \ref{sec:conclusions} we interpret our results in the context of the characteristic patterns of behavior in color-intensity diagrams.

\begin{sidewaystable}[p]
\resizebox{\textwidth}{!}{%
\begin{tabular}{|l|c|c|c|c|c|c|c|c|c|c|c|c|c|c|c|c|c|c|}
\hline
Source & RA & Dec & \multicolumn{2}{|c|}{ATCA}&
\multicolumn{4}{|c|}{SMARTS}&
\multicolumn{6}{|c|}{SWIFT UVOT}&
SWIFT & RXTE & RXTE & SWIFT \\
 & J2000 & J2000 & C & X & K & H & J & V$_g$ & V & B & U & UVW1 & UVM2 & UVW2 & XRT & ASM & PCA & BAT\\
\hline
{\bf Black Holes} (11)&&&&&&&&&&&&&&&&&&  \\
  LMC X-3$^a$            & 05:38:56.63  & -64:05:03.3  &
- & - & \dots & \dots & \dots & \dots & + & + & + &
\dots & \dots & \dots & \dots & - & + & \dots\\
  LMC X-1            & 05:39:38.83  & -69:44:35.5  & 
- & - & \dots & \dots & \dots & \dots & \dots & \dots & \dots &
\dots & \dots & \dots & \dots & + & \dots & \dots\\
  XTE J1550-564      & 15:50:58.65  & -56:28:35.3  & 
- & - & \dots & \dots & - & - & \dots & \dots & \dots &
\dots & \dots & \dots & \dots & - & \dots & \dots\\
  4U 1630-47         & 16:34:01.61  & -47:23:34.8  & 
- & - & \dots & - & \dots & \dots & \dots & \dots & \dots &
\dots & \dots & \dots & \dots & - & \dots & \dots\\
  GRO J1655-40       & 16:54:00.14  & -39:50:44.9  & 
- & - & + & \dots & + & + & \dots & \dots & \dots &
\dots & \dots & \dots & \dots & - & \dots & \dots\\
  GX 339-4$^a$           & 17:02:49.38  & -48:47:23.2  & 
+ & + & \dots & + & + & + & + & + & + &
+ & + & + & + & - & + & +\\ 
  1E 1740.7-2942$^a$     & 17:43:54.83  & -29:44:42.6  & 
- & - & \dots & \dots  & \dots & \dots & \dots & \dots & - &
\dots & \dots & \dots & + & \dots  & \dots & +\\
  GRS 1758-258$^a$       & 18:01:12.40  & -25:44:36.1 & 
- & - & \dots & \dots & \dots & \dots & \dots & \dots & - &
\dots & \dots & \dots & + & + & \dots & +\\ 
  V4641 Sgr          & 18:19:21.63  & -25:24:25.8  & 
- & - & + & + & + & + & \dots & \dots & \dots &
\dots & \dots & \dots & \dots & - & \dots & \dots\\
  GRS1915+105$^a$        & 19:15:11.56  & +10:56:44.9   &
\dots & \dots & + & \dots & \dots & \dots & \dots & \dots & \dots &
\dots & \dots & \dots & \dots & + & + & +\\
  4U 1957+11$^a$        & 19:59:24.01 & +11:42:29.9  & 
- & - & \dots & \dots & \dots & \dots & \dots & \dots & \dots &
\dots & \dots & + & + & + & + & +\\
 \hline
{\bf Z Sources} (8) &&&&&&&&&&&&&&&&&&  \\
  LMC X-2            & 05:20:28.04  & -71:57:53.3  & 
- & - & \dots & \dots & \dots & \dots & \dots & \dots & \dots &
\dots & \dots & \dots & \dots & + & \dots & + \\
  Cir X-1$^a$            & 15:20:40.85  & -57:10:00.1  & 
+ & + & + & \dots & + & + & \dots & \dots & \dots &
\dots & - & - & + & - & \dots & \dots\\
  Sco X-1$^a$            & 16:19:55.07  & -15:38:25.0  & 
+ & + & \dots & \dots & + & + & \dots & \dots & \dots &
\dots & \dots & \dots & \dots & + & + & +\\
  GX 340+0           & 16:45:47.70  & -45:36:40.0  & 
- & - & \dots & \dots & \dots & \dots & \dots & \dots & \dots &
\dots & \dots & \dots & \dots & + & \dots & \dots\\
  XTE J1701-462$^a$      & 17:00:58.46  & -46:11:08.6  & 
- & - & + & + & + & - & \dots & \dots & \dots &
\dots & \dots & \dots & + & + & + & +\\
  GX 349+2$^a$           & 17:05:44.49  & -36:25:23.0  & 
- & - & \dots & \dots & + & + & \dots & \dots & \dots &
\dots & \dots & - & + & + & + & +\\
  GX 5-1             & 18:01:09.73  & -25:04:44.1  & 
+ & + & \dots & \dots & + & + & \dots & \dots & \dots &
\dots & \dots & \dots & \dots & + & \dots & \dots\\
  GX 17+2$^a$            & 18:16:01.39  & -14:02:10.6  & 
+ & + & \dots & \dots & + & + & \dots & \dots & \dots &
\dots & \dots & - & + & + & + & +\\
\hline
\end{tabular}}
\end{sidewaystable}

\begin{sidewaystable}[p]
\resizebox{\textwidth}{!}{%
\begin{tabular}{|l|c|c|c|c|c|c|c|c|c|c|c|c|c|c|c|c|c|c|}
\hline
Source & RA & Dec & \multicolumn{2}{|c|}{ATCA}&
\multicolumn{4}{|c|}{SMARTS}&
\multicolumn{6}{|c|}{SWIFT UVOT}&
SWIFT & RXTE & RXTE & SWIFT \\
 & J2000 & J2000 & C & X & K & H & J & V & V & B & U & UVW1 & UVM2 & UVW2 & XRT & ASM & PCA & BAT\\
\hline
 {\bf Atoll Sources} (18)&&&&&&&&&&&&&&&&&&         \\
  EXO 0748-676$^a$       & 07:48:33.71  & -67:45:07.7  & 
- & - & \dots & \dots & \dots & \dots & \dots & \dots & \dots &
\dots & \dots & \dots & \dots & + & + & +\\
  2S 0921-630        & 09:22:34.68  & -63:17:41.4  & 
- & - & \dots & \dots & \dots & \dots & \dots & \dots & \dots &
\dots & \dots & \dots & \dots & - & \dots & \dots\\
  4U 1254-690$^a$        & 12:57:37.15  & -69:17:19.0  & 
- & - & \dots & \dots & - & + & \dots & \dots & \dots &
\dots & \dots & \dots & \dots & + & + & +\\
  Cen X-4            & 14:58:21.93  & -31:40:07.5  & 
- & - & + & + & \dots & + & \dots & \dots & \dots &
\dots & \dots & \dots & \dots & - & \dots & \dots\\
  4U 1543-624        & 15:47:54.6  & -62:34:06  & 
- & - & \dots & \dots & + & + & \dots & \dots & \dots &
\dots & \dots & \dots & \dots & + & \dots & \dots\\
  4U 1556-60         & 16:01:02.30  & -60:44:18.0  & 
- & - & \dots & \dots & \dots & \dots & \dots & \dots & \dots &
\dots & \dots & \dots & \dots & + & \dots & \dots\\
  4U 1608-52$^a$         & 16:12:43.0  & -52:25:23  & 
- & - & \dots & \dots & - & - & \dots & \dots & + &
- & - & \dots & + & + & + & +\\
  4U 1636-536$^a$        & 16:40:55.60  & -53:45:05.0  & 
- & - & \dots & \dots & + & + & \dots & \dots & \dots &
\dots & \dots & \dots & \dots & + & + & +\\
  MXB 1658-298        & 17:02:06.54  & -29:56:44.1  & 
- & - & \dots & \dots & - & + & \dots & \dots & \dots &
\dots & \dots & \dots & \dots & + & \dots & \dots\\
  4U 1702-429        & 17:06:15.31  & -43:02:08.7  & 
- & - & \dots & \dots & \dots & \dots & \dots & \dots & \dots &
\dots & \dots & \dots & \dots & + & \dots & \dots\\
  4U 1705-44$^a$         & 17:08:54.47  & -44:06:07.4  & 
- & - & \dots & \dots & \dots & \dots & \dots & \dots & \dots &
\dots & \dots & \dots & \dots & + & + & + \\
  GX 9+9$^a$             & 17:31:44.17  & -16:57:40.9  & 
- & - & \dots & \dots & + & + & \dots & \dots & \dots &
\dots & + & + & + & + & + & +\\
  GX 354+0$^a$           & 17:31:57.73  & -33:50:02.5  & 
- & + & \dots & \dots & \dots & \dots & \dots & \dots & \dots &
\dots & - & \dots & + & + & \dots & +\\
  4U 1735-444$^a$        & 17:38:58.3  & -44:27:00  & 
- & - & \dots & \dots & + & + & \dots & \dots & \dots &
\dots & + & \dots & + & + & + & +\\
  GX 3+1             & 17:47:56.00  & -26:33:49.0  & 
- & - & \dots & \dots & \dots & \dots & \dots & \dots & \dots &
\dots & \dots & \dots & \dots & + & \dots & \dots\\
  GX 9+1             & 18:01:32.30  & -20:31:44.0  & 
- & - & \dots & \dots & \dots & \dots & \dots & \dots & \dots &
\dots & \dots & \dots & \dots & + & \dots & \dots\\
  GX 13+1            & 18:14:31.55  & -17:09:26.7  & 
- & - & \dots & \dots & \dots & \dots & \dots & \dots & \dots &
\dots & \dots & \dots & \dots & + & \dots & \dots\\
  Aql X-1$^a$            & 19:11:16.06 & +00:35:05.9   &
\dots & \dots & \dots & \dots & + & + & \dots & \dots & \dots &
\dots & \dots & \dots & + & - & + & \dots\\
\hline
\end{tabular}}
\end{sidewaystable}

\begin{sidewaystable}[p]
\resizebox{\textwidth}{!}{%
\begin{tabular}{|l|c|c|c|c|c|c|c|c|c|c|c|c|c|c|c|c|c|c|}
\hline
Source & RA & Dec & \multicolumn{2}{|c|}{ATCA}&
\multicolumn{4}{|c|}{SMARTS}&
\multicolumn{6}{|c|}{SWIFT UVOT}&
SWIFT & RXTE & RXTE & SWIFT \\
 & J2000 & J2000 & C & X & K & H & J & V & V & B & U & UVW1 & UVM2 & UVW2 & XRT & ASM & PCA & BAT\\
\hline
 {\bf Pulsars} (11)  &&&&&&&&&&&&&&&&&& \\
  SMC X-3            & 00:52:05.63  & -72:26:04.2  & 
- & - & \dots & \dots & + & + & \dots & \dots & \dots &
\dots & \dots & \dots & \dots & - & \dots & \dots\\
  SMC X-1$^a$            & 01:17:05.14  & -73:26:36.0  & 
- & - & \dots & \dots & + & + & \dots & \dots & \dots &
+ & \dots & \dots & + & + & \dots & +\\
  Vela X-1           & 09:02:06.86  & -40:33:16.9  & 
- & - & \dots & \dots & \dots & \dots & \dots & \dots & \dots &
\dots & \dots & \dots & \dots & + & \dots & \dots\\
  Cen X-3            & 11:21:15.09  & -60:37:25.6  & 
- & - & \dots & \dots & + & + & \dots & \dots & \dots &
\dots & \dots & \dots & \dots & + & \dots & \dots\\
  GX 301-2           & 12:26:37.56  & -62:46:13.3  & 
- & - & \dots & \dots & + & + & \dots & \dots & \dots &
\dots & \dots & \dots & \dots & + & \dots & \dots\\
  4U 1538-522        & 15:42:23.36  & -52:23:09.6  & 
- & - & \dots & \dots & + & + & \dots & \dots & \dots &
\dots & \dots & \dots & \dots & + & \dots & \dots\\
  4U 1700-37$^b$        & 17:03:56.77  & -37:50:38.9  & 
- & - & \dots & \dots & \dots & \dots & \dots & \dots & \dots &
\dots & \dots & \dots & \dots & + & \dots & \dots\\
  GX 1+4             & 17:32:02.15  & -24:44:44.1  & 
- & - & \dots & \dots & + & + & \dots & \dots & \dots &
\dots & \dots & \dots & \dots & - & \dots & \dots\\
  Swift J1756.9-2508 & 17:56:57.35  & -25:06:27.8  & 
- & - & \dots & \dots & \dots & \dots & \dots & \dots & - &
- & - & - & \dots & \dots & \dots & \dots\\
  4U 1822-371$^a$        & 18:25:46.82  & -37:06:18.5  & 
- & - & \dots & \dots & + & + & \dots & \dots & + &
\dots & \dots & \dots & \dots & + & \dots & \dots\\
  4U 1907+097$^a$        & 19:09:37.14 & +09:49:55.3  & 
\dots & \dots & \dots & \dots & + & - & \dots & \dots & \dots &
\dots & \dots & \dots & \dots & + & + & +\\
 \hline

\hline
\end{tabular}}
\caption{XRBs observed in this experiment.  Observations are indicated by $+$ for detections,
$-$ for upper limits, \dots indicates the object was not observed by the telescope indicated. 
Positions are provided by Simbad (\url{simbad.cds.unistra.fr}), mostly measured by Gaia \citep[eDR3][]{Bailer-Jones_etal_2021} in ICRS coordinates.\\
Footnote $^a$ indicates sources with sufficient frequency ranges to analyse the spectral
energy distribution (SED) in section \ref{sec:SEDs}.\\
$^b\ $While pulsations have not been detected, the position of 4U 1700-37 in color-color-intensity diagrams locates it as a pulsar \citep{Vrtilek_Boroson_2013, Gopalan_Vrtilek_Bomn_2015, deBeurs_Islam_Gopalan_Vrtilek_2022}.
\label{tab:spreadsheet}}
\end{sidewaystable}

\section{Observations}\label{sec:Observations}

Our coordinated observing program took place from June 29-July 6, 2007 (\color{black} MJD 54280-54287 \color{black}).  Sources were selected from the low-mass X-ray binary (LMXB) and high-mass X-ray binary (HMXB) catalogs of \citet{Liu_etal_2001, Liu_etal_2006}. We include sources that have known optical and/or near-IR counterparts and are at least 5\arcsec \xspace away from objects of similar brightness; we exclude Globular Clusters. Since  the ground-based telescopes are in the Southern Hemisphere, we include only sources south of $\delta = +12^o$.  The resulting sample has 48 sources including 11 black holes (BHs), 8 Z-sources, 19 Atoll sources; and 10 X-ray Pulsars.  Seven telescopes were coordinated to cover the spectrum from radio at 4.8 GHz to X-rays at 50 keV.  These include
the Australia Telescope Compact Array (ATCA) at 4.8 and 8.64 GHz, the Small and Moderate Aperture Research Telescope System (SMARTS) in the near-IR K, H, and J bands and in the optical V band, the UV-Optical Telescope (UVOT) aboard the Neil Gehrels Swift Observatory (Swift) in the optical V and B bands and the ultraviolet U, UVW1, UVM2 and UVW2 bands, the Swift X-ray Telescope (XRT) from 0.5 to 10~keV, the Swift Burst Alert Telescope (BAT) from 5 to 50~keV, the Rossi X-ray Timing Explorer (RXTE) All-sky monitor (ASM) from 1.3 to 12~keV, and the RXTE Proportional Counter Array (PCA) from 2 to 20~keV.   A summary spreadsheet listing the XRB names and positions, the telescopes, and the observations and detections for each band is shown on Table \ref{tab:spreadsheet}.
Another class of 
\color{black} interacting binaries, \color{black}
symbiotic stars, were observed as part of this project, but only at radio frequencies. Results for the symbiotics are described in \citet{Dickey_etal_2021}, 
\color{black} 
but GX 1+4, the prototype symbiotic with a neutron star companion, is included in this paper along with the \color{black} pulsars.
\color{black}

\subsection{Radio Observations}\label{sec:Radio}

The data were taken with the Australia Telescope Compact Array in the 6C
configuration, that has 15 East-West baselines with lengths from 150 m to 6 km,
giving angular resolution of about 2\arcsec\xspace at 4.8 GHz and 1.1\arcsec\xspace
at 8.64 GHz.  These two frequencies were observed simultaneously with total bandwidth
128 MHz for each, of which the central 96 MHz was used to avoid filter edge-effects
(correlator configuration full\_128\_2).

Flux densities were calibrated by daily observations of 1934-638, the absolute flux
standard for southern hemisphere observations at cm wavelengths.  The standard 
calibration procedure using the {\textsc MIRIAD} software package was followed
\citep{Sault_Killeen_2004}.  

As described in \citet{Dickey_etal_2021},
we did not attempt to map every source.
The standard approach of aperture synthesis imaging is not well suited to monitoring a rapidly
varying source, as the aperture or {\it u,v} plane, which is the Fourier Transform of the image or
sky plane, is not well sampled until Earth rotation has caused the baselines between telescopes
to follow tracks that cover a large fraction of the aperture.  This typically takes 12 hours.
For this project we could spend only a few minutes of telescope time on each source each day,
and the sources might change their flux densities from day to day, or even hour to hour.  
Analysis of short scans on each XRB taken on the first three days was done using the 
complex fringe visibility function as measured at each {\it u,v} point (baseline and time).
These visibilities are vector-added by the {\textsc MYRIAD} task \textsc{UVFLUX} to
determine the flux density of
a point source at the phase center (the real part of the vector sum), and the rms
error on the flux density.  
Sources that were not detected on any of the first three days of observations were not
reobserved, upper limit flux densities are in most cases between 2 and 3 mJy at both frequencies.
Starting on \color{black}MJD 54284 \color{black}(3 July 2007) we concentrated on the sources that showed detections or
tentative detections, i.e. with flux density above about 1.5 mJy,
For these detected sources, combining the data from all days
gave sufficient {\it u,v} coverage for mapping and cleaning.  We then fitted two dimensional Gaussians
to each source.
Table \ref{tab:All_Radio} gives the Gaussian parameters for each detected source at 8.64 and 4.80 GHz,
and upper limits for the non-detections.

\begin{figure}[p]
\hspace{.3in} \includegraphics[width=3in]{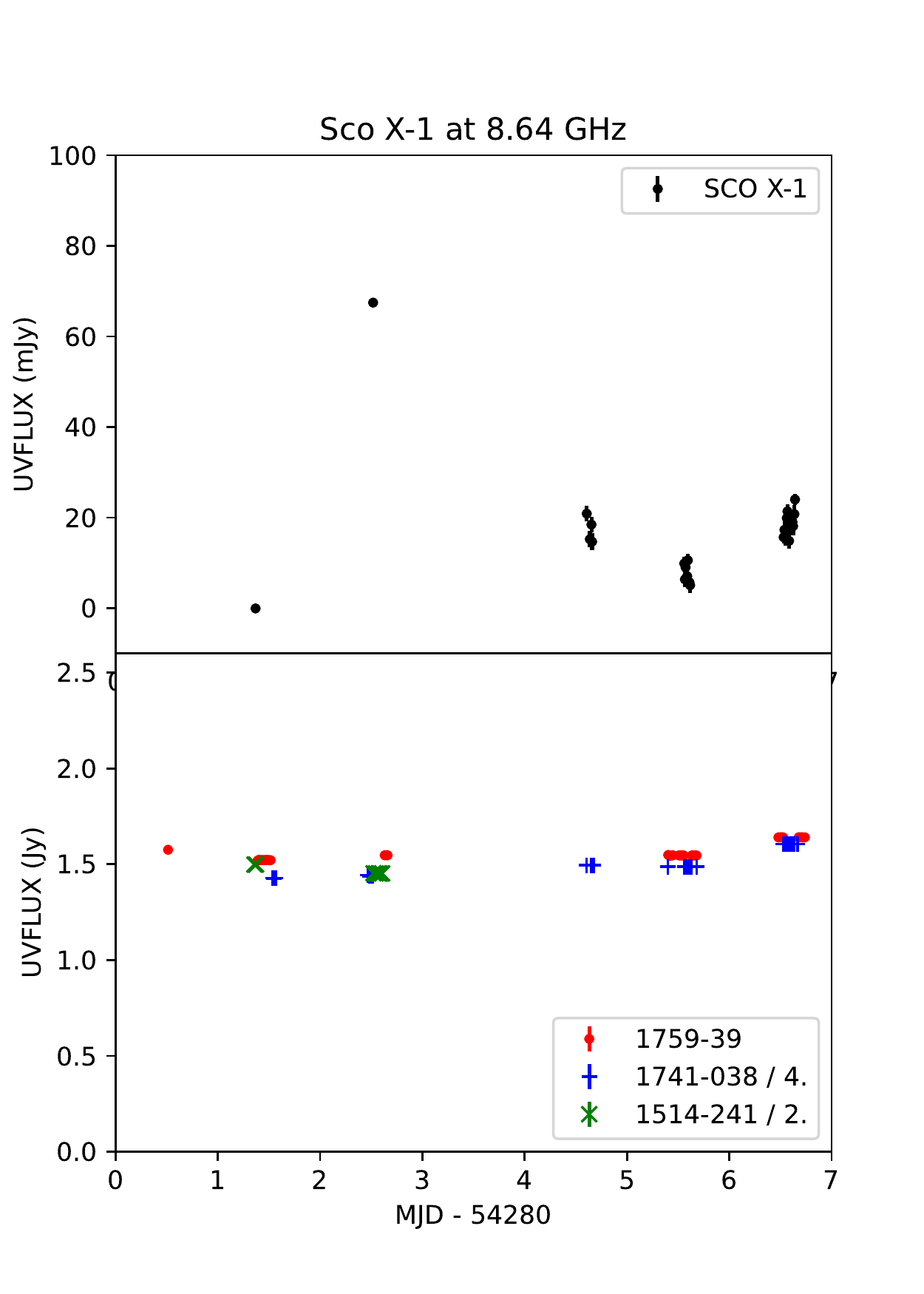} \includegraphics[width=3in]{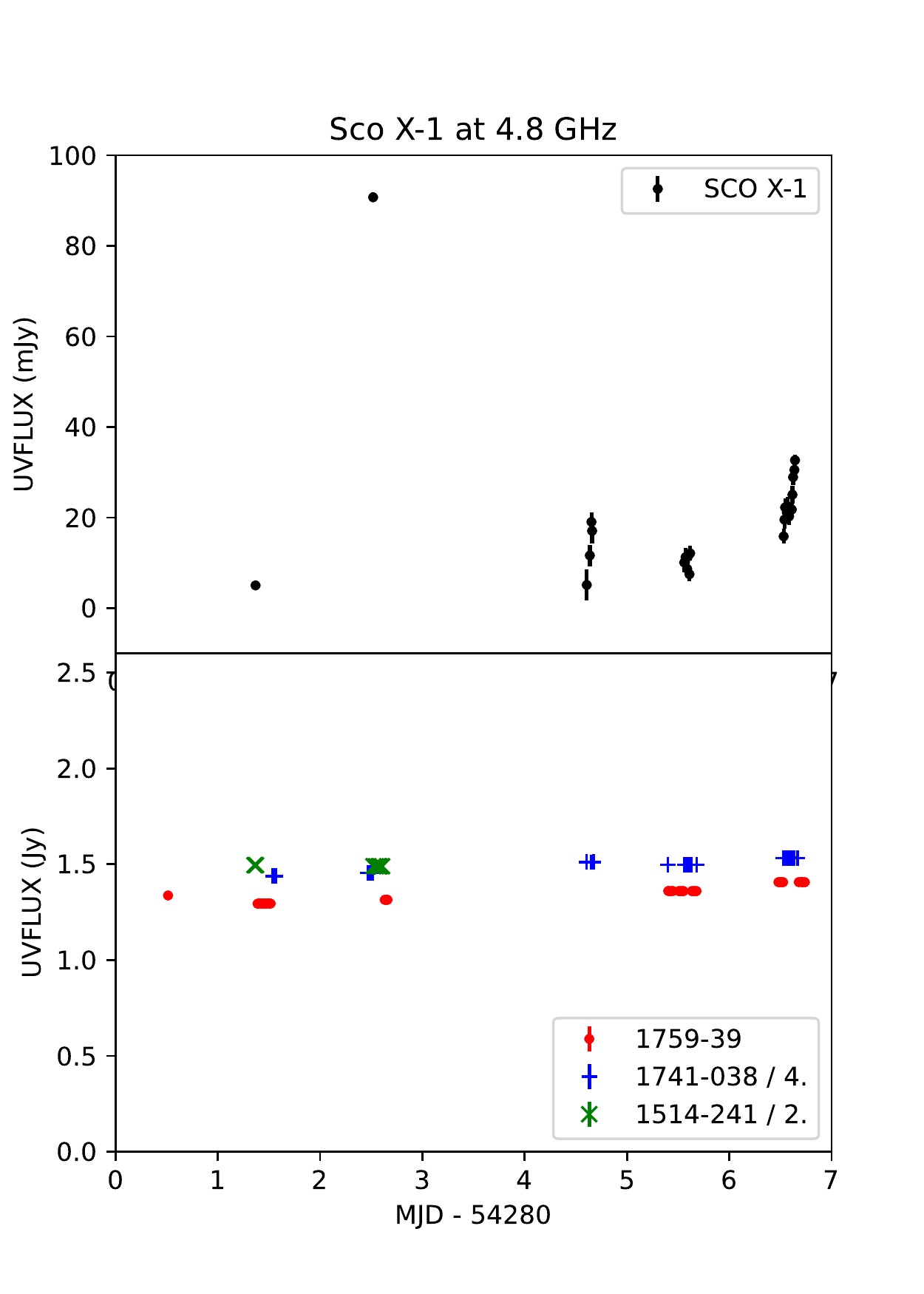}
\caption{The variation in flux density of Sco X-1 measured at 8.64 GHz (left panel) and at 4.8 GHz (right panel).  \color{black}{Error bars represent the errors on the means of the 10 second scans in each observation, they are mostly smaller than the markers.
Measurements of three gain calibrators, 1759-39, 1741-038 and 1514-241, that were observed at nearly the same time as Sco X-1, are shown for comparison in the lower panels.  
Conveniently, their fluxes are very nearly in ratio 1:4:2, so dividing the measured fluxes for 1741-038 by four and those of 1514-241 by two puts them all on the same scale.  The standard deviations of the calibrator fluxes at 8.64 GHz are 2.9\% of the mean
for 1759-39, 4.6\% of the mean for 1741-038, and 1.4\% of the mean for 1514-241.
Thus the overall calibration errors due to gain variations are very likely less than 5\%.}\color{black}
\label{fig:sco_x-1_flux_vs_day} }
\end{figure}

The source that shows by far the strongest variation from day to day during the observation is Sco X-1, with an increase from about 1 to 67 mJy between 30 June and 1 July \color{black}(MJD 54281 - 54282)\color{black}, then back down to about 8 mJy on 4 July \color{black}(MJD 54285)\color{black} before increasing again to 19 mJy on 5 July \color{black}(MJD 54286)\color{black}, as shown on figure \ref{fig:sco_x-1_flux_vs_day}.
The radio variability of Sco X-1 has been studied by \citet{Fomalont_etal_2001}.
Cir X-1 also shows significant variations at 8.64 and 4.8 GHz, as shown on table \ref{tab:All_Radio}.  In this case the source is much weaker, it increases from about
1 mJy to over 6 mJy at 8.64 GHz from 29 June to 1 July \color{black}(MJD 54280 - 54282)\color{black}, then drops below the detection limit ($\sim$1 mJy) from 3 July on \color{black}(MJD $>$54283)\color{black}.



\FloatBarrier
\startlongtable
\begin{deluxetable}{|l|l|l|l|l|l|l|l|l|}
\tablecaption{Radio Frequency Detections and Upper Limits \label{tab:All_Radio}}
\tablehead{
\colhead{Source} & \colhead{Dwell} & \colhead{Freq} & \colhead{S$_{peak}$\tablenotemark{a}} & \colhead{RA}& \colhead{Dec} & \colhead{MJD} & \colhead{S$_{int}$\tablenotemark{b}} & \colhead{Diam\tablenotemark{c}} \\ 
\colhead{ }& \colhead{sec} & \colhead{GHz} & \colhead{mJy/beam} &\colhead{J2000} &\colhead{J2000} &\colhead{} & \colhead{mJy} & \colhead{\arcsec} }
\startdata
\hline
\multicolumn{9}{|c|}
{\bf Black Holes} \\
\hline
LMC X-3 & 590 & 8640 & $<0.3$ & & & & & \\
LMC X-3 & 590 & 4800 & $<0.3$ & & & & & \\
LMC X-1 & 1080 & 8640 & $<18$\tablenotemark{d} & & & & & \\
LMC X-1 & 1080 & 4800 & $<6.3$\tablenotemark{d}& & & & & \\
XTE J1550-564  & 527 & 8640 & $<0.3$& & & & & \\
XTE J1550-564  & 527 & 8640 & $<0.4$& & & & & \\
4U 1630-47  & 510 & 8640 & $<0.7$& & & & & \\
4U 1630-47  & 510 & 4800 & $<2.0$& & & & & \\
GRO J1655-40  & 823 & 8640 & $<$ 1.5 & & & & &\\ 
GRO J1655-40  & 823 & 4800 & $<$ 0.6 & & & & &\\ 
GX 339-4  &  1923 & 8640 & 1.03$\pm$0.09 & 17:02:49.38 & -48:47:23.2 & all & 1.02$\pm$0.14 & $<$4 \\
GX 339-4  &  1923 & 4800 & 1.57$\pm$0.12 & 17:02:49.38 & -48:47:22.8 & all & 1.59$\pm$0.17 & $<$5 \\
1E 1740.7-2942 & 1189 & 8640 & $<$0.4 & &&&& \\
1E 1740.7-2942 & 1189 & 4800 & $<$2.1 & &&& &\\
GRS 1758-258  & 630 & 8640 & $<$0.2 & & & & &\\
GRS 1758-258  & 630 & 4800 &$<$0.3 & & & & &\\
V4641 Sgr  & 630 & 8640 &$<$0.5 & & & & & \\
V4641 Sgr  & 630 & 4800 &$<$0.2 & & & & & \\
\hline
\multicolumn{9}{|c|}
{\bf Z Sources} \\
\hline
LMC X-2  & 450 & 8640 & $<$0.2 & & & & & \\
LMC X-2  & 450 & 4800 & $<$0.4 & & & & & \\
Cir X-1  & 1630 & 8640 & 2.1$\pm$0.5& 15:20:40.93& -57:09:59.9& all & 6.7$\pm$3.4&  $<$5 \\
Cir X-1  & 1630 & 4800 & 3.5$\pm$0.5& 15:20:40.83& -57:10:00.0& all & 6.7$\pm$1.9&  $<$5 \\
Cir X-1  & 660 & 8640 & 1.0 $\pm$ 0.32 & \dots & \dots& 54279.3 & \dots &  \dots \\
Cir X-1  & 450 & 8640 & 22.5 $\pm$ 2.8 & \dots & \dots& 54280.2 & \dots & \dots \\
Cir X-1  & 1500 & 8640 & 12.6 $\pm$ 0.4 & \dots & \dots& 54281.2 & \dots & \dots \\
Cir X-1  & 690 & 8640 & 1.2 $\pm$ 0.5 & \dots & \dots& 54283.5 & \dots & \dots \\
Cir X-1  & 180 & 8640 & 0.3 $\pm$ 0.6 & \dots & \dots& 54284.5 & \dots &  \dots \\
Cir X-1  & 330 & 8640 & 1.3 $\pm$ 0.4 & \dots & \dots& 54285.4 & \dots &  \dots \\
Cir X-1  & 660 & 4800 & 2.5 $\pm$ 0.5 & \dots & \dots& 54279.3 & \dots &  \dots \\
Cir X-1  & 450 & 4800 & 22.2 $\pm$ 1.0 & \dots & \dots& 54280.2 & \dots & \dots \\
Cir X-1  & 1500 & 4800 & 14.3 $\pm$ 0.4 & \dots & \dots& 54281.2 & \dots & \dots \\
Cir X-1  & 690 & 4800 & 2.4 $\pm$ 0.6 & \dots & \dots& 54283.5 & \dots & \dots \\
Cir X-1  & 180 & 4800 & 1.9 $\pm$ 0.6 & \dots & \dots& 54284.5 & \dots &  \dots \\
Cir X-1  & 330 & 4800 & 1.0 $\pm$ 0.5 & \dots & \dots& 54285.4 & \dots &  \dots \\
Sco X-1\tablenotemark{e}  &  1020 & 8640 & 11.96$\pm$2.31 & 16:19:55.06 & -15:38:25.9 & all & 13.86$\pm$12.01  & $<$4\\
Sco X-1\tablenotemark{e}  &  1020 & 4800 & 15.95$\pm$0.98 & 16:19:55.07 & -15:38:25.4 & all & 19.88$\pm$1.77  & $<$9  \\
Sco X-1 & 540 & 8640 & $<$1.5 & \dots & \dots & 54280.4 & \dots & \dots \\
Sco X-1 & 450 & 8640 & 67.5 $\pm$ 1.1 & \dots & \dots & 54281.5 & \dots & \dots \\
Sco X-1 & 600 & 8640 & 17.3 $\pm$ 1.2 & \dots & \dots & 54283.6 & \dots & \dots \\
Sco X-1 & 210 & 8640 & 7.7 $\pm$ 0.6 & \dots & \dots & 54284.6 & \dots & \dots \\
Sco X-1 & 360 & 8640 & 19.1 $\pm$ 0.6 & \dots & \dots & 54285.6 & \dots & \dots \\
Sco X-1 & 540 & 4800 & 5.0 $\pm$ 0.6 & \dots & \dots & 54280.4 & \dots & \dots \\
Sco X-1 & 450 & 4800 & 90.7 $\pm$ 1.8 & \dots & \dots & 54281.5 & \dots & \dots \\
Sco X-1 & 600 & 4800 & 13.2 $\pm$ 1.5 & \dots & \dots & 54283.6 & \dots & \dots \\
Sco X-1 & 210 & 4800 & 10.2 $\pm$ 0.7 & \dots & \dots & 54284.6 & \dots & \dots \\
Sco X-1 & 360 & 4800 & 24.2 $\pm$ 0.6 & \dots & \dots & 54285.6 & \dots & \dots \\
GX 340+0  & 527 & 8640 &  $<$1.4 &
\dots & \dots & \dots & \dots & \dots \\
GX 340+0  & 527  &4800  &$<$1.7 & 
\dots & \dots & \dots & \dots & \dots \\
XTE J1701-462  & 1073 &  8640 & $<$0.4 &
\dots & \dots & \dots & \dots & \dots \\
XTE J1701-462  &  1073 &4800 &  $<$0.5&
\dots & \dots & \dots & \dots & \dots \\
GX 349+2  & 920 &  8640 & $<$0.3&
\dots & \dots & \dots & \dots & \dots \\
GX 349+2  & 920 & 4800 & $<$0.4& 
\dots & \dots & \dots & \dots & \dots \\
GX 5-1  &  760 & 8640 & 0.94$\pm$0.08 & 18:01:08.22 & -25:04:42.4 & all & 0.95$\pm$0.90 & $<$11 \\
GX 5-1  &  760 & 4800 & 1.22$\pm$0.21  & 18:01:08.22 & -25:04:43.1 & all & 1.67$\pm$0.44 & $<$8 \\
GX 17+2  &  2123 & 8640 & 0.78$\pm$0.07 & 18:16:01.38 & -14:02:11.0 & all & 0.89$\pm$0.35 & $<$6 \\
GX 17+2  &  2123 & 4800 & 1.24$\pm$0.14 & 18:16:01.39 & -14:02:11.7 & all & 1.36$\pm$0.23 & 4 \\
\hline
\multicolumn{9}{|c|}
{\bf Atoll Sources} \\
\hline
EXO 0748-676  &  1160 & 8640 & $<$0.6 &
\dots & \dots & \dots & \dots & \dots \\
EXO 0748-676  &  1160 & 4800 & $<$1.0 &
\dots & \dots & \dots & \dots & \dots \\
2S 0921-630  & 510 & 8640 & $<$0.4 & 
\dots & \dots & \dots & \dots & \dots \\
2S 0921-630  & 510 & 4800 & $<$1.2 & 
\dots & \dots & \dots & \dots & \dots \\
4U 1254-690  & 1017 & 8640 & $<$0.6 &
\dots & \dots & \dots & \dots & \dots \\
4U 1254-690  & 1017 & 4800 & $<$0.6 &
\dots & \dots & \dots & \dots & \dots \\
Cen X-4  &  330 & 8640 & $<$0.3 &
\dots & \dots & \dots & \dots & \dots \\
Cen X-4  &  330 & 4800 & $<$0.3 &
\dots & \dots & \dots & \dots & \dots \\
4U 1543-624  &  660 & 8640 & $<$0.5 &
\dots & \dots & \dots & \dots & \dots \\
4U 1543-624  &  660 & 4800 & $<$0.5 & 
\dots & \dots & \dots & \dots & \dots \\
4U 1556-60  &  527 & 8640 & $<$0.4 &
\dots & \dots & \dots & \dots & \dots \\
4U 1556-60  &  527 & 4800 & $<$0.6 & 
\dots & \dots & \dots & \dots & \dots \\
4U 1608-52  &  920 & 8640 & $<$0.4 &
\dots & \dots & \dots & \dots & \dots \\
4U 1608-52  &  920 & 4800 & $<$0.4 & 
\dots & \dots & \dots & \dots & \dots \\
4U 1636-536  &  987 & 8640 & $<$0.4 &
\dots & \dots & \dots & \dots & \dots \\
4U 1636-536  &  987 & 4800 & $<$0.4 & 
\dots & \dots & \dots & \dots & \dots \\
MXB 1658-298  &  310 & 8640 & $<$0.3 &
\dots & \dots & \dots & \dots & \dots \\
MXB 1658-298  &  310 & 4800 & $<$0.3 & 
\dots & \dots & \dots & \dots & \dots \\
4U 1702-429  &  607 & 8640 & $<$0.4 &
\dots & \dots & \dots & \dots & \dots \\
4U 1702-429  &  607 & 4800 & $<$0.4 & 
\dots & \dots & \dots & \dots & \dots \\
4U 1705-44  &  570 & 8640 & $<$0.3 &
\dots & \dots & \dots & \dots & \dots \\
4U 1705-44  &  570 & 4800 & $<$0.4 &
\dots & \dots & \dots & \dots & \dots \\
GX 9+9  &  1577 & 8640 & $<$0.4 &
\dots & \dots & \dots & \dots & \dots \\
GX 9+9  &  1577 & 4800 & $<$0.4 &
\dots & \dots & \dots & \dots & \dots \\
GX 354+0  & 881 & 8640 & 0.61$\pm$0.09 & 17:31:57.73 & -33:50:00.4 & all & 0.69$\pm$0.15  & $<$6 \\ 
GX 354+0  & 881 & 4800 & $<$0.5 & 
\dots & \dots & \dots & \dots & \dots \\
4U 1735-444  & 1020 & 8640 & $<$0.4 &
\dots & \dots & \dots & \dots & \dots \\
4U 1735-444  & 1020 & 4800 & $<$0.5 &
\dots & \dots & \dots & \dots & \dots \\
GX 3+1  & 340 & 8640 & $<$0.9 &
\dots & \dots & \dots & \dots & \dots \\
GX 3+1  & 340 & 4800 & $<$0.5 &
\dots & \dots & \dots & \dots & \dots \\
GX 9+1  & 440 & 8640 & $<$0.8 &
\dots & \dots & \dots & \dots & \dots \\
GX 9+1  & 440 & 4800 & $<$0.6 &
\dots & \dots & \dots & \dots & \dots \\
Swift J1756.9-2508  & 1400 & 8640 & $<$0.4 & 
\dots & \dots & \dots & \dots & \dots \\
Swift J1756.9-2508  & 1400 & 4800 & $<$1.7 & 
\dots & \dots & \dots & \dots & \dots \\
GX 13+1  &  310 & 8640 & $<$0.7 &
\dots & \dots & \dots & \dots & \dots \\
GX 13+1  &  310 & 4800 & $<$3.6 &
\dots & \dots & \dots & \dots & \dots \\
4U 1822-371  & 310 & 8640 & $<$0.2 & 
\dots & \dots & \dots & \dots & \dots \\
4U 1822-371  & 310 & 4800 & $<$0.6 & 
\dots & \dots & \dots & \dots & \dots \\
\hline
\multicolumn{9}{|c|}
{\bf Pulsars} \\
\hline
SMC X-3  & 464 & 8640 & $<$0.9 & 
\dots & \dots & \dots & \dots & \dots \\
SMC X-3  & 464 & 4800 & $<$2.1 & 
\dots & \dots & \dots & \dots & \dots \\
SMC X-1  &  460 & 8640 & $<$0.4&
\dots & \dots & \dots & \dots & \dots \\
SMC X-1  &  460 & 4800 & $<$0.4&
\dots & \dots & \dots & \dots & \dots \\
Vela X-1  &  490 & 8640 & $<$0.3 &
\dots & \dots & \dots & \dots & \dots \\
Vela X-1  &  490 & 4800 & $<$0.3 &
\dots & \dots & \dots & \dots & \dots \\
Cen X-3  &  520 & 8640 & $<$0.3 &
\dots & \dots & \dots & \dots & \dots \\
Cen X-3  &  520 & 4800 & $<$0.3 &
\dots & \dots & \dots & \dots & \dots \\
GX 301-2  & 537 & 8640 & $<$0.7 & 
\dots & \dots & \dots & \dots & \dots \\
GX 301-2  & 537 & 4800 & $<$1.0 &
\dots & \dots & \dots & \dots & \dots \\
4U 1538-52  &  510 & 8640 & $<$0.3 &
\dots & \dots & \dots & \dots & \dots \\
4U 1538-52  &  510 & 4800 & $<$0.5 &
\dots & \dots & \dots & \dots & \dots \\
4U 1700-37  & 540 & 8640 & $<$0.5 &
\dots & \dots & \dots & \dots & \dots \\
4U 1700-37  & 540 & 4800 & $<$1.1 &
\dots & \dots & \dots & \dots & \dots \\
GX 1+4  &  197 & 8640 & $<$0.7 &
\dots & \dots & \dots & \dots & \dots \\
GX 1+4  &  197 & 4800 & $<$1.9 &
\dots & \dots & \dots & \dots & \dots \\
\hline
\enddata
\color{black}
\tablenotetext{a} {$S_{peak}$ is the peak brightness of the fitted Gaussian.}
\tablenotetext{b} {$S_{int}$ is the integrated flux density of the fitted Gaussian.}
\tablenotetext{c} {Diam is the full width to half maximum of the fitted Gaussian; these are mostly upper limits.}
\color{black}
\tablenotetext{d}{Confused by nearby sources in field.}
\tablenotetext{e}{See Fig. \ref{fig:sco_x-1_flux_vs_day}.}
\end{deluxetable}
\eject


\subsection{Infrared and Optical Observations}\label{sec:IR_Opt}

\begin{figure}[ht]
\hspace{.4in} \includegraphics[width=6in]{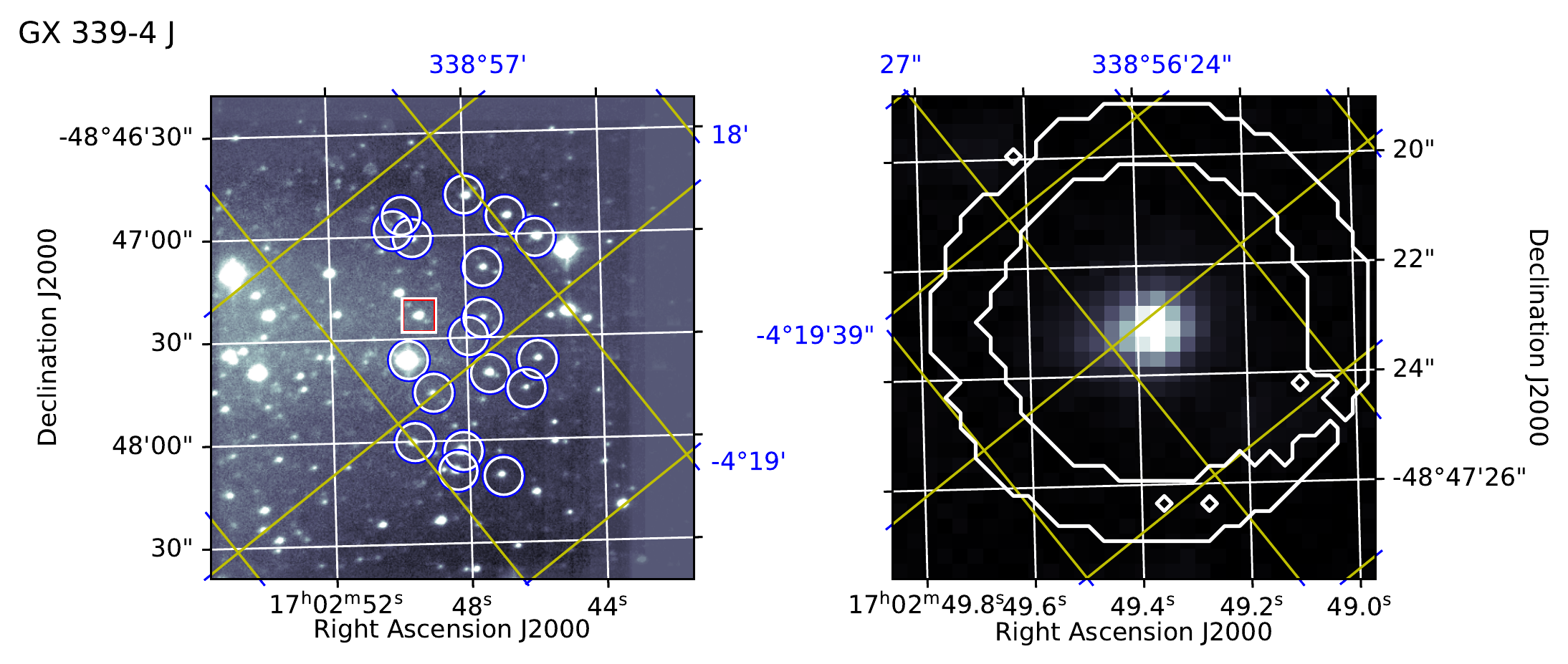}
\caption{SMARTS Andicam image of the field of GX 339-4 at J band.  On the left is the full field image, with photometric reference stars indicated by
blue circles and the X-ray source by a red square with side length 32 pixels=8.8\arcsec.  On the right is a blow-up of
the area inside the red square.  ICRF coordinates are indicated in black with white grid lines.  Galactic
coordinates are shown in blue, with yellow grid lines.  The source is fitted by a two dimensional Gaussian after
subtracting the background using a bi-linear function fitted to pixels inside the annulus indicated by the
white contour.  Confusing stars or other pixels more than 1.5 sigma above the mean in the annulus are dropped
before fitting, as indicated by the ragged boundaries of the white contour.  The complete figure set (37 images) is available in the online journal.
\label{fig:IR_images}}
\end{figure}

Observations in the near infrared and optical bands were made with the SMARTS (Small and Moderate Aperture Research Telescope System) 1.3m telescope at CTIO and the ANDICAM-IR instrument with a NICMOS 4 detector.  The near infra-red channel (NIR) has pixel scale 0.275\arcsec \xspace and field of view (FoV) 2.4\arcmin, while the optical channel has pixel scale 0.372\arcsec and FoV 6\arcmin\xspace \citep{Muzerolle_et_al_2019}.
The infrared observations were first flat-fielded using flat-field exposures taken one per night at J, H, and/or K band.  These are used to flag bad pixels and to determine a scale 
factor for each good pixel that is applied to every exposure before aligning all exposures to a common center.  All exposures for each night are then combined by taking the median
for each pixel.  A single frame for each night is saved, and an overall median image is computed for each source, an example is shown in figure \ref{fig:IR_images}.  Coordinates and
magnitudes are taken from comparison with star images in each band from the 2MASS survey (Skrutskie et al 2006).  Simple geometric scale and rotation parameters
were derived using the \textsc{KOORDS} utility in the KARMA package \citep{Gooch_1996}. 
Comparison of the derived star centroid positions with those in the 2MASS catalog gives an estimate for the precision of
the coordinates of better than 1\arcsec\xspace in most cases.  Magnitudes and fluxes for each source observed in the IR and optical
are given on Table \ref{tab:IRflux}.
In a few cases, source positions in the 2MASS catalog differ from 
those in the Gaia eDR3 \citep{Gaia_dr2} catalog (listed on Table \ref{tab:spreadsheet} columns 2 and 3) by more than 3\arcsec.  These are noted in the footnotes
to Table \ref{tab:IRflux}.  

\figsetstart
\figsetnum{\ref{fig:IR_images}}
\figsettitle{SMARTS K,H,J band images}

\figsetgrpstart
\figsetgrpnum{\ref{fig:IR_images}.1}
\figsetgrptitle{XTE J1550-564 J}
\figsetplot{figures/IRfields/IR_field_plots/apr7_XTE_J1550-564_J.pdf}
\figsetgrpnote{IR images of XRB sources.  The full field is on the left, on the right is a magnified image of the red square box centred on the X-ray position.}
\figsetgrpend

\figsetgrpstart
\figsetgrpnum{\ref{fig:IR_images}.2}
\figsetgrptitle{GRO J1655-40 K}
\figsetplot{figures/IRfields/IR_field_plots/apr7_J1655-40_K.pdf}
\figsetgrpnote{IR images of XRB sources.  The full field is on the left, on the right is a magnified image of the red square box centred on the X-ray position.}
\figsetgrpend

\figsetgrpstart
\figsetgrpnum{\ref{fig:IR_images}.3}
\figsetgrptitle{GRO J1655-40 J}
\figsetplot{figures/IRfields/IR_field_plots/apr7_J1655-40_J.pdf}
\figsetgrpnote{IR images of XRB sources.  The full field is on the left, on the right is a magnified image of the red square box centred on the X-ray position.}
\figsetgrpend

\figsetgrpstart
\figsetgrpnum{\ref{fig:IR_images}.4}
\figsetgrptitle{GX 339-4 H}
\figsetplot{figures/IRfields/IR_field_plots/apr7_GX_339-4_H.pdf}
\figsetgrpnote{IR images of XRB sources.  The full field is on the left, on the right is a magnified image of the red square box centred on the X-ray position.}
\figsetgrpend

\figsetgrpstart
\figsetgrpnum{\ref{fig:IR_images}.5}
\figsetgrptitle{GX 339-4 J}
\figsetplot{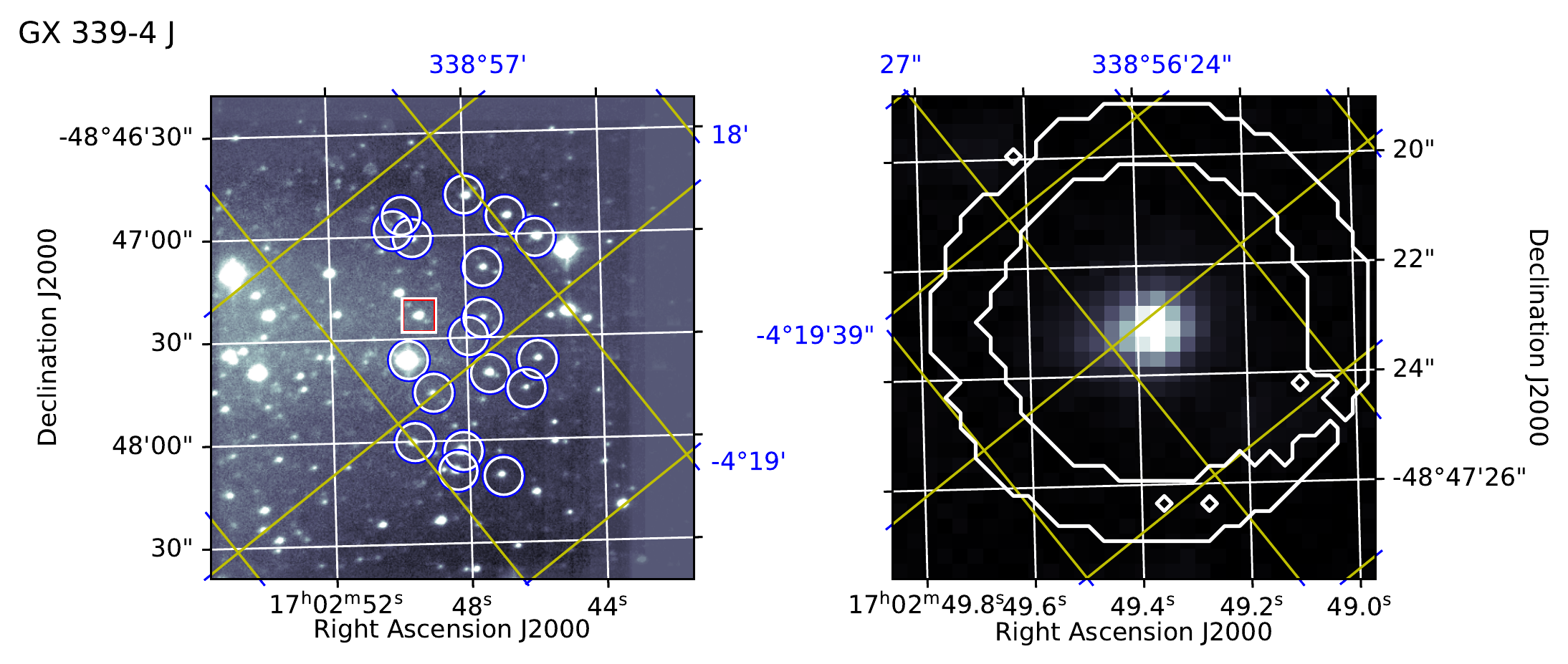}
\figsetgrpnote{IR images of XRB sources.  The full field is on the left, on the right is a magnified image of the red square box centred on the X-ray position.}
\figsetgrpend

\figsetgrpstart
\figsetgrpnum{\ref{fig:IR_images}.6}
\figsetgrptitle{V4641 Sgr K}
\figsetplot{figures/IRfields/IR_field_plots/apr7_V4641_K.pdf}
\figsetgrpnote{IR images of XRB sources.  The full field is on the left, on the right is a magnified image of the red square box centred on the X-ray position.}
\figsetgrpend

\figsetgrpstart
\figsetgrpnum{\ref{fig:IR_images}.7}
\figsetgrptitle{V4641 Sgr H}
\figsetplot{figures/IRfields/IR_field_plots/apr7_V4641_H.pdf}
\figsetgrpnote{IR images of XRB sources.  The full field is on the left, on the right is a magnified image of the red square box centred on the X-ray position.}
\figsetgrpend

\figsetgrpstart
\figsetgrpnum{\ref{fig:IR_images}.8}
\figsetgrptitle{V4641 Sgr J}
\figsetplot{figures/IRfields/IR_field_plots/apr7_V4641_J.pdf}
\figsetgrpnote{IR images of XRB sources.  The full field is on the left, on the right is a magnified image of the red square box centred on the X-ray position.}
\figsetgrpend

\figsetgrpstart
\figsetgrpnum{\ref{fig:IR_images}.9}
\figsetgrptitle{GRS1915+105 K}
\figsetplot{figures/IRfields/IR_field_plots/apr9_GRS1915+105_K.pdf}
\figsetgrpnote{IR images of XRB sources.  The full field is on the left, on the right is a magnified image of the red square box centred on the X-ray position.}
\figsetgrpend

\figsetgrpstart
\figsetgrpnum{\ref{fig:IR_images}.10}
\figsetgrptitle{Cir X-1 K}
\figsetplot{figures/IRfields/IR_field_plots/apr7_Cir_X-1_K.pdf}
\figsetgrpnote{IR images of XRB sources.  The full field is on the left, on the right is a magnified image of the red square box centred on the X-ray position.}
\figsetgrpend

\figsetgrpstart
\figsetgrpnum{\ref{fig:IR_images}.11}
\figsetgrptitle{Cir X-1 J}
\figsetplot{figures/IRfields/IR_field_plots/apr7_Cir_X-1_J.pdf}
\figsetgrpnote{IR images of XRB sources.  The full field is on the left, on the right is a magnified image of the red square box centred on the X-ray position.}
\figsetgrpend

\figsetgrpstart
\figsetgrpnum{\ref{fig:IR_images}.12}
\figsetgrptitle{Sco X-1 J}
\figsetplot{figures/IRfields/IR_field_plots/apr7_Sco_X-1_J.pdf}
\figsetgrpnote{IR images of XRB sources.  The full field is on the left, on the right is a magnified image of the red square box centred on the X-ray position.}
\figsetgrpend

\figsetgrpstart
\figsetgrpnum{\ref{fig:IR_images}.13}
\figsetgrptitle{XTE J1701-462 K}
\figsetplot{figures/IRfields/IR_field_plots/apr9_J1701-462_K.pdf}
\figsetgrpnote{IR images of XRB sources.  The full field is on the left, on the right is a magnified image of the red square box centred on the X-ray position.}
\figsetgrpend

\figsetgrpstart
\figsetgrpnum{\ref{fig:IR_images}.14}
\figsetgrptitle{XTE J1701-462 H}
\figsetplot{figures/IRfields/IR_field_plots/apr9_J1701-462_H.pdf}
\figsetgrpnote{IR images of XRB sources.  The full field is on the left, on the right is a magnified image of the red square box centred on the X-ray position.}
\figsetgrpend

\figsetgrpstart
\figsetgrpnum{\ref{fig:IR_images}.15}
\figsetgrptitle{XTE J1701-462 J}
\figsetplot{figures/IRfields/IR_field_plots/apr9_J1701-462_J.pdf}
\figsetgrpnote{IR images of XRB sources.  The full field is on the left, on the right is a magnified image of the red square box centred on the X-ray position.}
\figsetgrpend

\figsetgrpstart
\figsetgrpnum{\ref{fig:IR_images}.16}
\figsetgrptitle{GX 349+2 J}
\figsetplot{figures/IRfields/IR_field_plots/apr7_GX_349+2_J.pdf}
\figsetgrpnote{IR images of XRB sources.  The full field is on the left, on the right is a magnified image of the red square box centred on the X-ray position.}
\figsetgrpend

\figsetgrpstart
\figsetgrpnum{\ref{fig:IR_images}.17}
\figsetgrptitle{GX 5-1 J}
\figsetplot{figures/IRfields/IR_field_plots/apr7_GX_5-1_J.pdf}
\figsetgrpnote{IR images of XRB sources.  The full field is on the left, on the right is a magnified image of the red square box centred on the X-ray position.}
\figsetgrpend

\figsetgrpstart
\figsetgrpnum{\ref{fig:IR_images}.18}
\figsetgrptitle{GX 17+2 J}
\figsetplot{figures/IRfields/IR_field_plots/apr7_GX_17+2_J.pdf}
\figsetgrpnote{IR images of XRB sources.  The full field is on the left, on the right is a magnified image of the red square box centred on the X-ray position.}
\figsetgrpend

\figsetgrpstart
\figsetgrpnum{\ref{fig:IR_images}.19}
\figsetgrptitle{4U 1254-690 J}
\figsetplot{figures/IRfields/IR_field_plots/apr7_4u_1254-690_J.pdf}
\figsetgrpnote{IR images of XRB sources.  The full field is on the left, on the right is a magnified image of the red square box centred on the X-ray position.}
\figsetgrpend

\figsetgrpstart
\figsetgrpnum{\ref{fig:IR_images}.20}
\figsetgrptitle{Cen X-4 K}
\figsetplot{figures/IRfields/IR_field_plots/apr7_Cen_X-4_K.pdf}
\figsetgrpnote{IR images of XRB sources.  The full field is on the left, on the right is a magnified image of the red square box centred on the X-ray position.}
\figsetgrpend

\figsetgrpstart
\figsetgrpnum{\ref{fig:IR_images}.21}
\figsetgrptitle{Cen X-4 H}
\figsetplot{figures/IRfields/IR_field_plots/apr7_Cen_X-4_H.pdf}
\figsetgrpnote{IR images of XRB sources.  The full field is on the left, on the right is a magnified image of the red square box centred on the X-ray position.}
\figsetgrpend

\figsetgrpstart
\figsetgrpnum{\ref{fig:IR_images}.22}
\figsetgrptitle{4U 1543-624 J}
\figsetplot{figures/IRfields/IR_field_plots/apr9_4u_1543-624_J.pdf}
\figsetgrpnote{IR images of XRB sources.  The full field is on the left, on the right is a magnified image of the red square box centred on the X-ray position.}
\figsetgrpend

\figsetgrpstart
\figsetgrpnum{\ref{fig:IR_images}.23}
\figsetgrptitle{4U 1608-52 J}
\figsetplot{figures/IRfields/IR_field_plots/apr9_4u_1608-52_J.pdf}
\figsetgrpnote{IR images of XRB sources.  The full field is on the left, on the right is a magnified image of the red square box centred on the X-ray position.}
\figsetgrpend

\figsetgrpstart
\figsetgrpnum{\ref{fig:IR_images}.23}
\figsetgrptitle{4U 1636-536 J}
\figsetplot{figures/IRfields/IR_field_plots/apr7_4u_1636-536_J.pdf}
\figsetgrpnote{IR images of XRB sources.  The full field is on the left, on the right is a magnified image of the red square box centred on the X-ray position.}
\figsetgrpend

\figsetgrpstart
\figsetgrpnum{\ref{fig:IR_images}.24}
\figsetgrptitle{4U 1658-298 J}
\figsetplot{figures/IRfields/IR_field_plots/oct6_4u1658-298_J.pdf}
\figsetgrpnote{IR images of XRB sources.  The full field is on the left, on the right is a magnified image of the red square box centred on the X-ray position.}
\figsetgrpend

\figsetgrpstart
\figsetgrpnum{\ref{fig:IR_images}.25}
\figsetgrptitle{GX 9+9 J}
\figsetplot{figures/IRfields/IR_field_plots/apr7_GX_9+9_J.pdf}
\figsetgrpnote{IR images of XRB sources.  The full field is on the left, on the right is a magnified image of the red square box centred on the X-ray position.}
\figsetgrpend

\figsetgrpstart
\figsetgrpnum{\ref{fig:IR_images}.26}
\figsetgrptitle{GX 1+4 J}
\figsetplot{figures/IRfields/IR_field_plots/apr7_GX_1+4_J.pdf}
\figsetgrpnote{IR images of XRB sources.  The full field is on the left, on the right is a magnified image of the red square box centred on the X-ray position.}
\figsetgrpend

\figsetgrpstart
\figsetgrpnum{\ref{fig:IR_images}.27}
\figsetgrptitle{4U 1735-444 J}
\figsetplot{figures/IRfields/IR_field_plots/apr7_4u_1734-444_J.pdf}
\figsetgrpnote{IR images of XRB sources.  The full field is on the left, on the right is a magnified image of the red square box centred on the X-ray position.}
\figsetgrpend

\figsetgrpstart
\figsetgrpnum{\ref{fig:IR_images}.28}
\figsetgrptitle{Aql X-1 J}
\figsetplot{figures/IRfields/IR_field_plots/apr7_Aql_X-1_J.pdf}
\figsetgrpnote{IR images of XRB sources.  The full field is on the left, on the right is a magnified image of the red square box centred on the X-ray position.}
\figsetgrpend

\figsetgrpstart
\figsetgrpnum{\ref{fig:IR_images}.29}
\figsetgrptitle{SMC X-3 J}
\figsetplot{figures/IRfields/IR_field_plots/apr7_SMC_X-3_J.pdf}
\figsetgrpnote{IR images of XRB sources.  The full field is on the left, on the right is a magnified image of the red square box centred on the X-ray position.}
\figsetgrpend

\figsetgrpstart
\figsetgrpnum{\ref{fig:IR_images}.30}
\figsetgrptitle{SMC X-1 J}
\figsetplot{figures/IRfields/IR_field_plots/apr7_SMC_X-1_J.pdf}
\figsetgrpnote{IR images of XRB sources.  The full field is on the left, on the right is a magnified image of the red square box centred on the X-ray position.}
\figsetgrpend

\figsetgrpstart
\figsetgrpnum{\ref{fig:IR_images}.31}
\figsetgrptitle{Cen X-3 J}
\figsetplot{figures/IRfields/IR_field_plots/apr7_Cen_X-3_J.pdf}
\figsetgrpnote{IR images of XRB sources.  The full field is on the left, on the right is a magnified image of the red square box centred on the X-ray position.}
\figsetgrpend

\figsetgrpstart
\figsetgrpnum{\ref{fig:IR_images}.32}
\figsetgrptitle{GX 301-2 J}
\figsetplot{figures/IRfields/IR_field_plots/apr7_gx301-2_J.pdf}
\figsetgrpnote{IR images of XRB sources.  The full field is on the left, on the right is a magnified image of the red square box centred on the X-ray position.}
\figsetgrpend

\figsetgrpstart
\figsetgrpnum{\ref{fig:IR_images}.33}
\figsetgrptitle{4U 1538-522 J}
\figsetplot{figures/IRfields/IR_field_plots/apr9_4u_1538-52_J.pdf}
\figsetgrpnote{IR images of XRB sources.  The full field is on the left, on the right is a magnified image of the red square box centred on the X-ray position.}
\figsetgrpend

\figsetgrpstart
\figsetgrpnum{\ref{fig:IR_images}.34}
\figsetgrptitle{4U 1822-371 J}
\figsetplot{figures/IRfields/IR_field_plots/apr7_4u_1822-371_J.pdf}
\figsetgrpnote{IR images of XRB sources.  The full field is on the left, on the right is a magnified image of the red square box centred on the X-ray position.}
\figsetgrpend

\figsetgrpstart
\figsetgrpnum{\ref{fig:IR_images}.35}
\figsetgrptitle{4U 1907+097 J}
\figsetplot{figures/IRfields/IR_field_plots/oct6_4u_1907+097_J.pdf}
\figsetgrpnote{IR images of XRB sources.  The full field is on the left, on the right is a magnified image of the red square box centred on the X-ray position.}
\figsetgrpend

\figsetend

\begin{figure}[ht]
\hspace{1.7in} \includegraphics[width=3in]{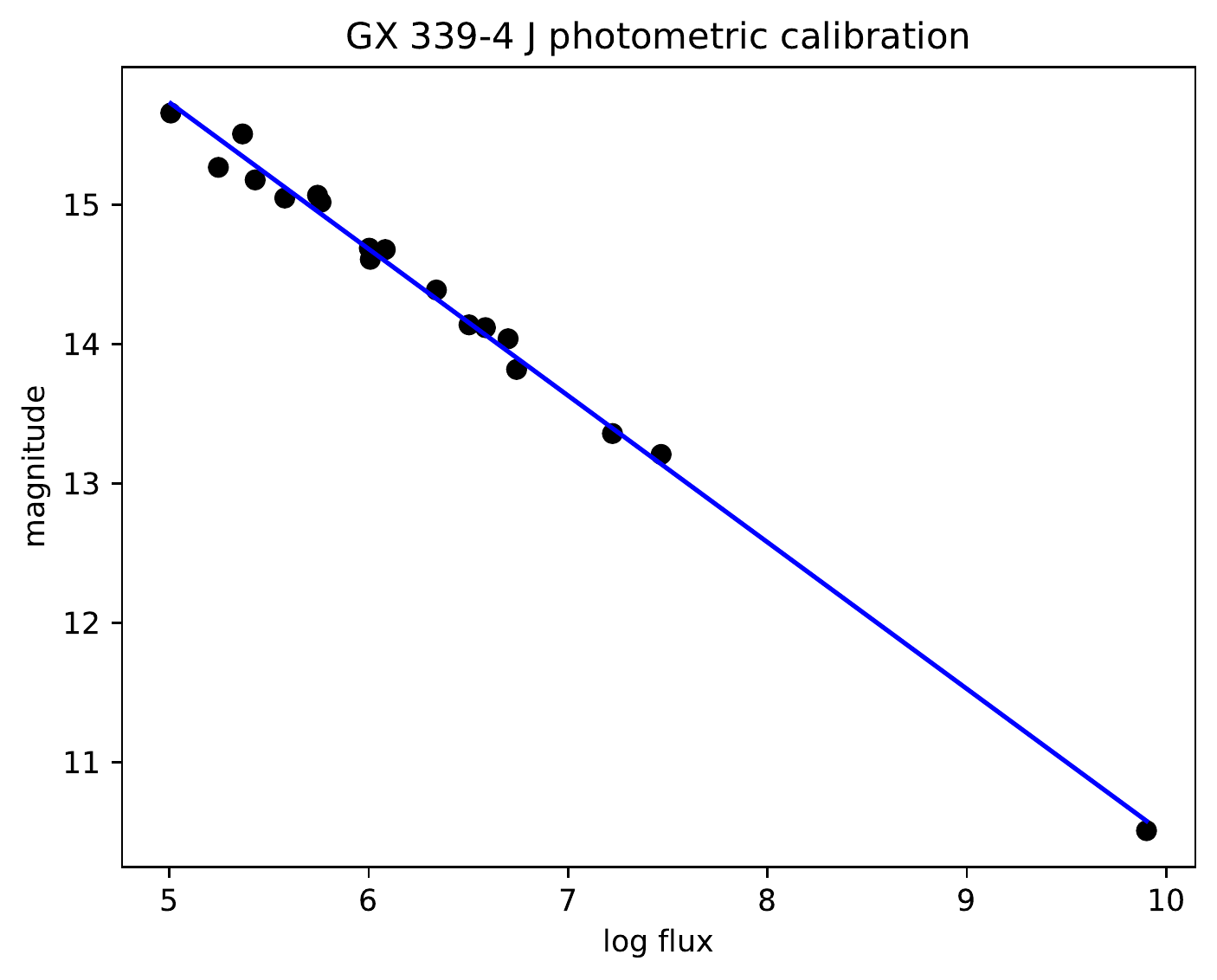}
\caption{SMARTS photometry of GX 339-4 at J band.  The reference stars are fitted with
two dimensional Gaussians, whose integrals are assumed to be proportional to the stars' flux.
The log of the flux is fitted to the catalog magnitudes of the stars from the 2MASS survey.
The scatter around this relation gives the 
\color{black}
magnitude error in column 4 of Table \ref{tab:IRflux}.
\color{black}
\label{fig:IR_photometry}}
\end{figure}

In each field, 50 comparison stars from the 2MASS point source catalog \citep{Cutri_etal_2003}\footnote{
\url{https://irsa.ipac.caltech.edu/cgi-bin/Gator/nph-dd}}
 are considered, and those that are unconfused are used to calibrate the
photometry of the program sources.  Two dimensional Gaussians are fitted to the star images, and their
integrals are assumed to be proportional to the flux in the band.  The catalog magnitudes are fitted by
a linear function of the logarithm of the flux, and the result is used to convert the program source
flux to magnitude.  An example of the flux vs. magnitude calibration is shown in Fig. \ref{fig:IR_photometry}. 
The rms of the residuals after subtracting the best fit power law (linear fit on the log scale of 
Fig. \ref{fig:IR_photometry}) gives the 
\color{black} 
the magnitude errors listed with the magnitudes on 
column 4, Table \ref{tab:IRflux}.  
\color{black}
Columns 5 and 6 give the corresponding flux density and flux error, using the zero point flux 
densities for the J, H, and K bands of 1594, 1024, and 667 Jy from \citet{Cohen_etal_2003}.
In the J and H bands both errors can be smaller than 0.1 magnitudes.

\begin{figure}[ht]

\hspace{1.in}\includegraphics[width=4in]{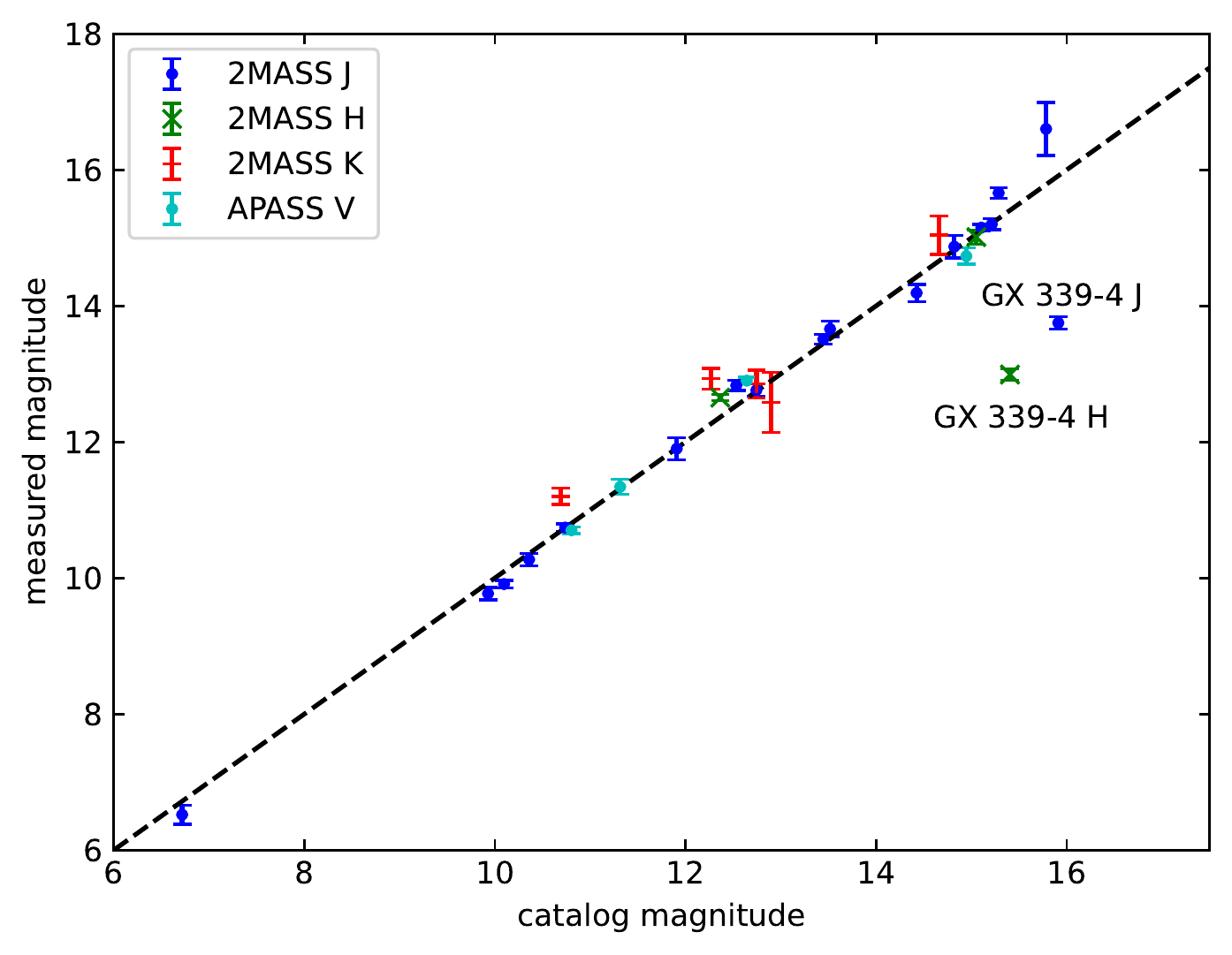}

\caption{ \color{black}
Comparison between the measured magnitudes of the X-ray sources vs. their catalog
values.  
GX 339-4 shows very significant difference from the 2MASS result, at the level of 30$\sigma$.  The other sources are all consistent within 
$3\sigma$. Only sources with positions within 2$\arcsec$ of the catalog values are included.\color{black}
\label{fig:photoIR_check}}
\end{figure}

\color{black}Although XRBs are variable stars, both due to rapid changes in the optical and infrared luminosities of the X-ray emission region, and to variability in the companion star, a comparison between the magnitudes measured in this project with the 2MASS catalog values shows that there is generally not much change between the two epochs.
We compare the observed J band magnitudes with the values in the 2MASS point source catalog and the few available V band magnitudes from the APASS catalog
for all sources on figure \ref{fig:photoIR_check}.
 There has been surprisingly little variation between the 2MASS epoch and
the date of these observations, roughly ten years later, except for
GX339-4, which is
much brighter than in the catalog (J=13.75 vs. 15.91 in the point source catalog, and H=12.99 vs. 15.40).
A similar result is found
by \citet{Shidatsu_etal_2011} who measure J=13.66 with a slow brightening to J=12.9 over a month of observations in March, 2009. \color{black}

\subsection{Optical Photometry}\label{sec:optical}

\begin{figure}[ht]
\hspace{.1in} \includegraphics[width=6in]{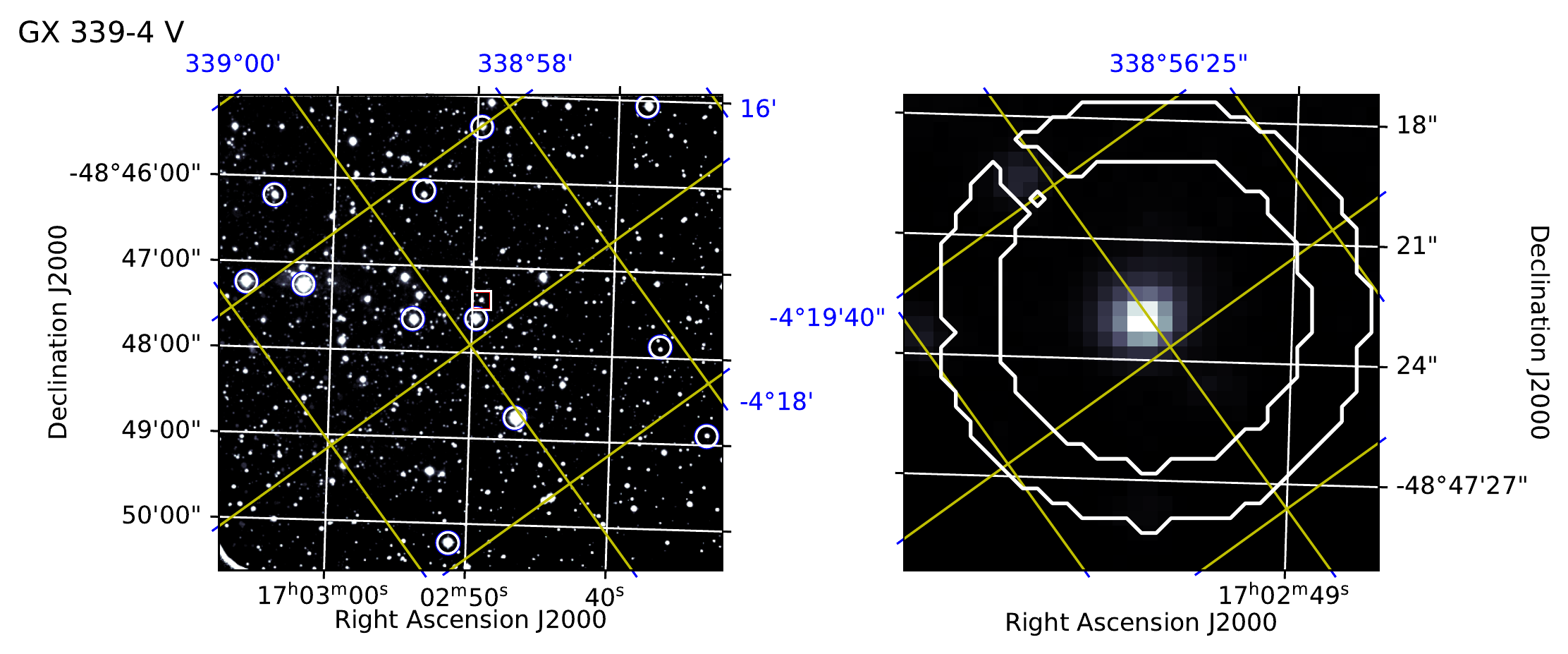}
\caption{SMARTS image of the field of GX 339-4 at V band.  Photometric reference stars are indicated by
blue circles, the X-ray source by a red square, with side lengths 32 pixels = 11.9\arcsec. 
The right panel shows an expanded image of the red square area, with the white contour 
indicating the area used to fit the background.
The complete figure set (28 images) is available in the online journal.
\label{fig:Opt_images}}
\end{figure}

\begin{figure}[ht]
\hspace{1in} \includegraphics[width=4in]{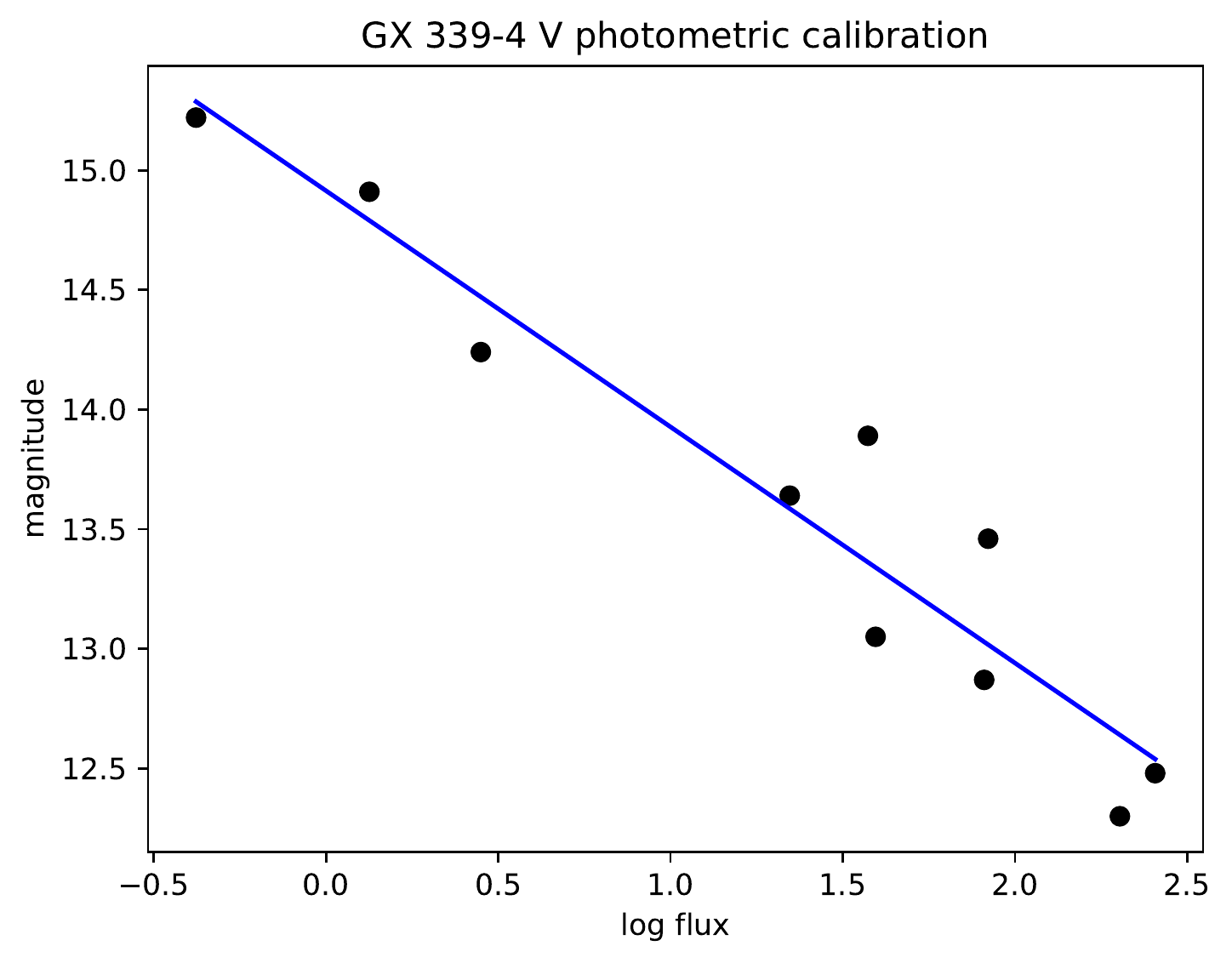}
\caption{SMARTS photometry of GX 339-4 at V band, similar to figure \ref{fig:IR_photometry},
but using reference stars from the APASS catalog.
\label{fig:Opt_photometry}}
\end{figure}

The optical V band data were also taken with the SMARTS 1.3m at CTIO, with the
ANDICAM-CCD based on the Fairchild 447 detector.  The images were flat fielded and
bad pixels flagged before aligning, and median filtered for each night.  Coordinates
were determined using five or six reference stars visible on the 2MASS J or H band images
in the KARMA task KOORDS.  Comparison stars near the X-ray source were used for
photometric calibration.  Magnitudes at V band for these stars were taken from 
\color{black} the APASS catalog \citep{APASS}. \color{black}
Stars were checked and rejected if there were nearby objects
that could confuse the Gaussian fitting.  
Reference stars in the optical are illustrated on Fig. \ref{fig:Opt_images}.  The photometric
calibration is done using these stars similarly to the 2MASS calibration for the IR observations
described above.  Fig. \ref{fig:Opt_photometry} illustrates the star flux vs. APASS catalog magnitudes for GX 339-4, similar to Fig. \ref{fig:IR_photometry}.
The errors following the magnitudes in column 4 are the 
\color{black} rms scatter of the reference stars about the best fit linear
relation between their magnitudes and the integral of their fitted Gaussians on Fig. \ref{fig:Opt_photometry}, \color{black} or the formal error in the constant term
of the fit to the reference star magnitudes, whichever is greater. \color{black}  
Fluxes and flux errors are derived from the magnitudes and their errors
assuming the Johnson V band zero magnitude flux density of 3780 Jy \citep{Cox_2001}.

\figsetstart
\figsetnum{\ref{fig:Opt_images}}
\figsettitle{SMARTS V band images}

\figsetgrpstart
\figsetgrpnum{\ref{fig:Opt_images}.1}
\figsetgrptitle{XTE J1550-564 V}
\figsetplot{figures/IRfields/opt_field_plots/XTE_J1550-564_plot.pdf}
\figsetgrpnote{Optical V band images of XRB sources.  The full field is on the left, on the right is a magnified image of the red square box centred on the X-ray position.}
\figsetgrpend

\figsetgrpstart
\figsetgrpnum{\ref{fig:Opt_images}.2}
\figsetgrptitle{GRO J1655-40 V}
\figsetplot{figures/IRfields/opt_field_plots/J1655-40_plot.pdf}
\figsetgrpnote{Optical V band images of XRB sources.  The full field is on the left, on the right is a magnified image of the red square box centred on the X-ray position.}
\figsetgrpend

\figsetgrpstart
\figsetgrpnum{\ref{fig:Opt_images}.3}
\figsetgrptitle{GX 339-4 V}
\figsetplot{figures/IRfields/opt_field_plots/GX_339-4_plot.pdf}
\figsetgrpnote{Optical V band images of XRB sources.  The full field is on the left, on the right is a magnified image of the red square box centred on the X-ray position.}
\figsetgrpend

\figsetgrpstart
\figsetgrpnum{\ref{fig:Opt_images}.4}
\figsetgrptitle{V4641 Sgr V}
\figsetplot{figures/IRfields/opt_field_plots/V4641_plot.pdf}
\figsetgrpnote{Optical V band images of XRB sources.  The full field is on the left, on the right is a magnified image of the red square box centred on the X-ray position.}
\figsetgrpend

\figsetgrpstart
\figsetgrpnum{\ref{fig:Opt_images}.5}
\figsetgrptitle{Cir X-1 V}
\figsetplot{figures/IRfields/opt_field_plots/Cir_X-1_plot.pdf}
\figsetgrpnote{Optical V band images of XRB sources.  The full field is on the left, on the right is a magnified image of the red square box centred on the X-ray position.}
\figsetgrpend

\figsetgrpstart
\figsetgrpnum{\ref{fig:Opt_images}.6}
\figsetgrptitle{Sco X-1 V}
\figsetplot{figures/IRfields/opt_field_plots/Sco_X-1_plot.pdf}
\figsetgrpnote{Optical V band images of XRB sources.  The full field is on the left, on the right is a magnified image of the red square box centred on the X-ray position.}
\figsetgrpend

\figsetgrpstart
\figsetgrpnum{\ref{fig:Opt_images}.7}
\figsetgrptitle{XTE J1701-462 V}
\figsetplot{figures/IRfields/opt_field_plots/J1701-462_plot.pdf}
\figsetgrpnote{Optical V band images of XRB sources.  The full field is on the left, on the right is a magnified image of the red square box centred on the X-ray position.}
\figsetgrpend

\figsetgrpstart
\figsetgrpnum{\ref{fig:Opt_images}.8}
\figsetgrptitle{GX 349+2 V}
\figsetplot{figures/IRfields/opt_field_plots/GX_349+2_plot.pdf}
\figsetgrpnote{Optical V band images of XRB sources.  The full field is on the left, on the right is a magnified image of the red square box centred on the X-ray position.}
\figsetgrpend

\figsetgrpstart
\figsetgrpnum{\ref{fig:Opt_images}.9}
\figsetgrptitle{GX 5-1 V}
\figsetplot{figures/IRfields/opt_field_plots/GX_5-1_plot.pdf}
\figsetgrpnote{Optical V band images of XRB sources.  The full field is on the left, on the right is a magnified image of the red square box centred on the X-ray position.}
\figsetgrpend

\figsetgrpstart
\figsetgrpnum{\ref{fig:Opt_images}.10}
\figsetgrptitle{GX 17+2 V}
\figsetplot{figures/IRfields/opt_field_plots/GX_17+2_plot.pdf}
\figsetgrpnote{Optical V band images of XRB sources.  The full field is on the left, on the right is a magnified image of the red square box centred on the X-ray position.}
\figsetgrpend

\figsetgrpstart
\figsetgrpnum{\ref{fig:Opt_images}.11}
\figsetgrptitle{4U 1254-690 V}
\figsetplot{figures/IRfields/opt_field_plots/4u_1254-690_plot.pdf}
\figsetgrpnote{Optical V band images of XRB sources.  The full field is on the left, on the right is a magnified image of the red square box centred on the X-ray position.}
\figsetgrpend

\figsetgrpstart
\figsetgrpnum{\ref{fig:Opt_images}.12}
\figsetgrptitle{Cen X-4 V}
\figsetplot{figures/IRfields/opt_field_plots/Cen_X-4_plot.pdf}
\figsetgrpnote{Optical V band images of XRB sources.  The full field is on the left, on the right is a magnified image of the red square box centred on the X-ray position.}
\figsetgrpend

\figsetgrpstart
\figsetgrpnum{\ref{fig:Opt_images}.13}
\figsetgrptitle{4U 1543-624 V}
\figsetplot{figures/IRfields/opt_field_plots/4u_1543-624_plot.pdf}
\figsetgrpnote{Optical V band images of XRB sources.  The full field is on the left, on the right is a magnified image of the red square box centred on the X-ray position.}
\figsetgrpend

\figsetgrpstart
\figsetgrpnum{\ref{fig:Opt_images}.14}
\figsetgrptitle{4U 1608-52 V}
\figsetplot{figures/IRfields/opt_field_plots/4u_1608-52_plot.pdf}
\figsetgrpnote{Optical V band images of XRB sources.  The full field is on the left, on the right is a magnified image of the red square box centred on the X-ray position.}
\figsetgrpend

\figsetgrpstart
\figsetgrpnum{\ref{fig:Opt_images}.15}
\figsetgrptitle{4U 1636-536 V}
\figsetplot{figures/IRfields/opt_field_plots/4u_1636-536_plot.pdf}
\figsetgrpnote{Optical V band images of XRB sources.  The full field is on the left, on the right is a magnified image of the red square box centred on the X-ray position.}
\figsetgrpend

\figsetgrpstart
\figsetgrpnum{\ref{fig:Opt_images}.16}
\figsetgrptitle{MXB 1658-298 V}
\figsetplot{figures/IRfields/opt_field_plots/MXB_1658-298_plot.pdf}
\figsetgrpnote{Optical V band images of XRB sources.  The full field is on the left, on the right is a magnified image of the red square box centred on the X-ray position.}
\figsetgrpend

\figsetgrpstart
\figsetgrpnum{\ref{fig:Opt_images}.17}
\figsetgrptitle{GX 9+9 V}
\figsetplot{figures/IRfields/opt_field_plots/GX_9+9_plot.pdf}
\figsetgrpnote{Optical V band images of XRB sources.  The full field is on the left, on the right is a magnified image of the red square box centred on the X-ray position.}
\figsetgrpend

\figsetgrpstart
\figsetgrpnum{\ref{fig:Opt_images}.18}
\figsetgrptitle{GX 1+4 V}
\figsetplot{figures/IRfields/opt_field_plots/GX_1+4_plot.pdf}
\figsetgrpnote{Optical V band images of XRB sources.  The full field is on the left, on the right is a magnified image of the red square box centred on the X-ray position.}
\figsetgrpend

\figsetgrpstart
\figsetgrpnum{\ref{fig:Opt_images}.19}
\figsetgrptitle{4U 1735-444 V}
\figsetplot{figures/IRfields/opt_field_plots/4u_1735-444_plot.pdf}
\figsetgrpnote{Optical V band images of XRB sources.  The full field is on the left, on the right is a magnified image of the red square box centred on the X-ray position.}
\figsetgrpend

\figsetgrpstart
\figsetgrpnum{\ref{fig:Opt_images}.20}
\figsetgrptitle{Aql X-1 V}
\figsetplot{figures/IRfields/opt_field_plots/Aql_X-1_plot.pdf}
\figsetgrpnote{Optical V band images of XRB sources.  The full field is on the left, on the right is a magnified image of the red square box centred on the X-ray position.}
\figsetgrpend

\figsetgrpstart
\figsetgrpnum{\ref{fig:Opt_images}.21}
\figsetgrptitle{SMC X-3 V}
\figsetplot{figures/IRfields/opt_field_plots/SMC_X-3_plot.pdf}
\figsetgrpnote{Optical V band images of XRB sources.  The full field is on the left, on the right is a magnified image of the red square box centred on the X-ray position.}
\figsetgrpend

\figsetgrpstart
\figsetgrpnum{\ref{fig:Opt_images}.22}
\figsetgrptitle{SMC X-1 V}
\figsetplot{figures/IRfields/opt_field_plots/SMC_X-1_plot.pdf}
\figsetgrpnote{Optical V band images of XRB sources.  The full field is on the left, on the right is a magnified image of the red square box centred on the X-ray position.}
\figsetgrpend

\figsetgrpstart
\figsetgrpnum{\ref{fig:Opt_images}.23}
\figsetgrptitle{Cen X-3 V}
\figsetplot{figures/IRfields/opt_field_plots/Cen_X-3_plot.pdf}
\figsetgrpnote{Optical V band images of XRB sources.  The full field is on the left, on the right is a magnified image of the red square box centred on the X-ray position.}
\figsetgrpend

\figsetgrpstart
\figsetgrpnum{\ref{fig:Opt_images}.24}
\figsetgrptitle{GX 301-2 V}
\figsetplot{figures/IRfields/opt_field_plots/gx301-2_plot.pdf}
\figsetgrpnote{Optical V band images of XRB sources.  The full field is on the left, on the right is a magnified image of the red square box centred on the X-ray position.}
\figsetgrpend

\figsetgrpstart
\figsetgrpnum{\ref{fig:Opt_images}.25}
\figsetgrptitle{4U 1538-522 V}
\figsetplot{figures/IRfields/opt_field_plots/4u_1538-52_plot.pdf}
\figsetgrpnote{Optical V band images of XRB sources.  The full field is on the left, on the right is a magnified image of the red square box centred on the X-ray position.}
\figsetgrpend

\figsetgrpstart
\figsetgrpnum{\ref{fig:Opt_images}.26}
\figsetgrptitle{4U 1822-371 V}
\figsetplot{figures/IRfields/opt_field_plots/4u_1822-371_plot.pdf}
\figsetgrpnote{Optical V band images of XRB sources.  The full field is on the left, on the right is a magnified image of the red square box centred on the X-ray position.}
\figsetgrpend

\figsetgrpstart
\figsetgrpnum{\ref{fig:Opt_images}.27}
\figsetgrptitle{4U 1907+097 V}
\figsetplot{figures/IRfields/opt_field_plots/4u_1907+097_plot.pdf}
\figsetgrpnote{Optical V band images of XRB sources.  The full field is on the left, on the right is a magnified image of the red square box centred on the X-ray position.}
\figsetgrpend

\figsetend

%
%
%

\clearpage


\clearpage


\startlongtable
\begin{deluxetable*}{|c|c|c|c|c|c|}
\tablecaption{Infra-Red and Optical Magnitudes \label{tab:IRflux}}
\tablewidth{6.5in}
\tablehead{
\colhead{Source} &
\colhead{Band} &
\colhead{MJD} &
\colhead{Magnitude} &
\colhead{Flux Density} &
\colhead{Flux Error}  \\
\colhead{}  &\colhead{} &\colhead{} &\colhead{} &  \colhead{mJy}  & \colhead{mJy}      }
\startdata
\multicolumn{6}{|c|}
 {\bf Black Holes}   \\
\hline
\hline
XTE J1550-564  &  J\tablenotemark{a}  &  54281  & 18.5$\pm$1.2 & 0.06 &  0.04\\ 
 \hline
 \hline
GRO J1655-40  &  K  &  54281--54286  & 12.85$\pm$0.2 &   4.8 &  0.8\\
GRO J1655-40  &  J  &  54281--54286  & 13.66$\pm$0.11 &   5.48 &  0.53\\
GRO J1655-40   &    V  &  all days  & 16.48$\pm$0.17 &   1.0 &  0.1\\
 \hline
GX 339-4  &  H  &  54286--54286  & 12.99$\pm$0.08 &   6.52 &  0.46\\
GX 339-4  &  J  &  54281--54286  & 13.75$\pm$0.09 &   5.04 &  0.40\\
GX 339-4   &    V  &  all days  & 16.24$\pm$0.26 &   1.3  &  0.3\\
\hline
V4641 Sgr &  K  &  54283-54286  & 12.93$\pm$0.14 &   25 &  3\\
V4641 Sgr &  H  &  54283-54286  & 12.65$\pm$0.04 &   8.92 &  0.32\\
V4641 Sgr &  J  &  54283-54286  & 12.83$\pm$0.07 &   11.76 &  0.84\\
V4641 Sgr  &  V  &  all days  & 14.88$\pm$0.14 &   4.2 &  0.4\\
 \hline
GRS1915+105  &  K  &  54281-54286  &   12.6$\pm$0.4 &   6.2  &  2.1 \\
 \hline
\multicolumn{6}{|c|}
 {\bf Z Sources}   \\
 \hline
Cir X-1  &  K  &  54282+54286  & 11.20$\pm$0.12 &   22 &  2\\
Cir X-1  &  J  &  54282+54286  & 12.76$\pm$0.09 &   12.5 &  1.0\\
Cir X-1   &   V  &  54285   & 19.5$\pm$0.2 &   0.06 & 0.02\\
 \hline
Sco X-1  &  J  &  54282  & 11.9$\pm$0.2 &   28 &  4.7\\
Sco X-1   &   V  &  54281   & 12.90$\pm$0.05 &   26 &  1\\
 \hline
XTE J1701-462  &  K  &  54282--54284  & 15.04$\pm$0.30 &   0.64 &  0.16\\
XTE J1701-462  &  H &  54282--54284  & 15.12$\pm$0.15 &   0.92 &  0.13\\
XTE J1701-462  &  J &  54282--54284  & 16.8$\pm$0.2 &   0.30 &  0.07\\
XTE J1701-462  &  V  &  54282+54283  & 19$\pm$0.4 &   0.09 & 0.04\\
 \hline
GX 349+2  &  J  &  54282+54286  & 15.20$\pm$0.07 &   1.33 &  0.09\\ 
GX 349+2  &   V  &  54285  & 17.37$\pm$0.16 &   0.43 &  0.03\\
 \hline
GX 5-1\tablenotemark{b}  &  J  &  54282  &  9.77$\pm$0.09 &   197 &  16\\
GX 5-1\tablenotemark{b}  &  V  &  54280  & 11.34$\pm$0.20 &   110 &  11\\
 \hline
GX 17+2  &  J  &  54282+54285  & 15.15$\pm$0.04 &   1.39 &  0.05\\
GX 17+2   &   V   &  54284  & 17.27$\pm$0.09 &   0.47 &  0.04\\
\hline
\multicolumn{6}{|c|}
 {\bf Atoll Sources}          \\
 \hline
4U 1254-690  &  J\tablenotemark{c}  &  54282  & 16.9$\pm$0.4 &   0.28 &  0.10\\
4U 1254-690  &  V  &  54281  & 18.9$\pm$0.1 &   0.10 &  0.02\\
 \hline
Cen X-4  &  K  &  54282--54286  & 15.0$\pm$0.2 &   0.67 &  0.11\\
Cen X-4  &  H  &  54282--54286  & 15.01$\pm$0.10 &   1.01 &  0.09\\
Cen X-4  &  V  &  54286  & 17.63$\pm$0.16 &   0.34 &   0.05\\
 \hline
4U 1543-624\tablenotemark{b}  &  J  &  54283+54284  & $>$17 &   $<$0.25 &  \dots\\
4U 1543-624\tablenotemark{b}   &   V   &  54282+54283 & 15.23$\pm$0.09 &   3.1 & 0.2 \\
 \hline
4U 1608-52\tablenotemark{b}  &  J  &  54286  & $>17$ &    $<0.25$ &  \dots\\
4U 1608-52\tablenotemark{b}  &  V &  54285  & 17.5$\pm$0.2 &   0.38 &  0.06\\
 \hline
4U 1636-536\tablenotemark{b}  &  J\tablenotemark{d} &  54284  & 15.9$\pm$0.2 &   0.70 &  0.12\\
4U 1636-536   &   V   &  54283   & 18.2$\pm$0.09 &   0.2 &  0.03\\
 \hline
MXB 1658-298  &  J  &  54284  & $>$17 &  $<$0.25 &  \dots\\
MXB 1658-298   &   V  &  54283   & $>$19 &   $<$0.09 &  \dots\\
 \hline
GX 9+9\tablenotemark{b}  &  J  &  54284+54285  & 14.19$\pm$0.12 &   3.36 &  0.35\\
GX 9+9\tablenotemark{b} &   V  &  54283+54284  & 15.31$\pm$0.18 &   2.8 &  0.4\\
 \hline
4U 1735-444  &  J\tablenotemark{c}  &  54285  & 16.52$\pm$0.14 &   0.39 &  0.06\\
4U 1735-444   &   V  &  54284   & 17.2$\pm$0.5 &   0.5 &  0.1\\
 \hline
4U 1822-371  &  J  &  54285  & 15.66$\pm$0.06 &   0.87 &  0.05\\
4U 1822-371   &   V  &  54284  & 16.2$\pm$0.9 &   1.3 &  0.5\\
 \hline
Aql X-1  &  J  &  54286  & 16.6$\pm$0.4 &   0.37 &  0.11\\
Aql X-1   &    V  & 54283+54285   & 19.4$\pm$0.4 &   0.07 &  0.02\\
\hline
\multicolumn{6}{|c|}
 {\bf Pulsars}   \\
 \hline
SMC X-3  &  J  &  54282  & 14.87$\pm$0.16 &   1.80 &  0.25\\
SMC X-3   &   V  &  54280+54281   & 14.73$\pm$0.19 &   4.8 &  0.5\\
 \hline
SMC X-1  &  J  &  54281+54282  & 13.51$\pm$0.11 &   6.3 &  0.4\\
SMC X-1   &   V   &   54280+54281  & 13.16$\pm$0.08 &   20.6 &  2.0\\
 \hline
Cen X-3  &  J  &  54283  & 10.74$\pm$0.05 &   80.6 &  3.6\\
Cen X-3   &    V  &  54282   & 13.32$\pm$0.08 &   17.8 & 0.8\\
 \hline
GX 301-2  &  J  &  54283  &  6.52$\pm$0.10 &   3930 &  480\\
GX 301-2   &    V   &  54282  & 10.70$\pm$0.07 &   198&  9\\
 \hline
4U 1538-52  &  J  &  54283  & 10.27$\pm$0.09 &   124 &  10\\
4U 1538-52   &   V  &  54282  & 13.6$\pm$0.3 &   14 &  2\\
 \hline
GX 1+4  &  J  &  54284  &  9.91$\pm$0.05 &   173 &  8\\
GX 1+4  &   V  & 54283   & 17.07$\pm$0.14 &   0.56 &  0.03\\
 \hline
4U 1907+097  &  J &  54281  & 12.66$\pm$0.10 &   13.8 &  1.2\\
4U 1907+097  &  V  &  54281  & 18.0$\pm$0.08 &    0.24 &  0.02 \\
 \hline
\enddata
\tablenotetext{a}{position discrepant by 9\arcsec}
\tablenotetext{b}{confused by nearby bright star}
\tablenotetext{c}{position discrepant by 8\arcsec}
\tablenotetext{d}{position discrepant by 6\arcsec}
\end{deluxetable*}

\eject


\subsection{UV Observations --  Swift UVOT} \label{sec:Obs_uv}

{\it Neil Gehrels Swift Observatory} is a multi-wavelength observatory dedicated to the study of gamma-ray burst (GRB) science and transient sources. It has three instruments that observe simultaneously in the gamma-ray, X-ray, ultraviolet, and optical wavebands,
which work in tandem to provide rapid identification and multi-wavelength follow-up of GRBs and transient sources.  The three instruments are the (1) Burst Alert Telescope (BAT) observing at 15 - 150 ~keV, (2) X-ray Telescope (XRT) observing at 0.3 - 10 ~keV, and (3) UV/Optical Telescope (UVOT) observing at 170 - 600 nm.  More details about the mission can be found in \cite{Gehrels04}.

We requested and acquired multiple pointed Swift observations for 15 sources during the week of our observing campaign. In addition, for all of the sources observed in our campaign, we made BAT lightcurves from the Swift/BAT hard X-ray transient monitor data.

The UltraViolet and Optical Telescope (UVOT) is a diffraction-limited 30 cm (12 inch aperture) Ritchey-Chr\'etien reflector, sensitive to magnitude 22.3 in a 17 minute exposure. The Ritchey-Chr\'etien telescope has a f/2.0 primary that is re-imaged to f/13 by the secondary. This results in pixels that are 0.502\arcsec ~over its 17\arcmin ~square FoV. The detector is an intensified CCD with seven broadband UVOT filters (white, U, V, B, UVW1, UVM2, and UVW2) covering the 170 - 600 nm wavelength range. For more details on UVOT see \cite{Roming05}.

For most of the Swift observations a UVOT observation in image mode was made in one or more of the filters.  For these observations we analyzed the data using the standard UVOT analysis threads \footnote{https://www.swift.ac.uk/analysis/uvot/} using the tools for Swift data analysis found in the Heasoft software package \footnote{https://heasarc.gsfc.nasa.gov/docs/software/heasoft/}.
For most datasets we used the pipeline-reduced summed image in the data products.  For a few observations for which pipeline-reduced products were not complete it was necessary to sum the images using the tool {\it uvotimsum}. For each source we created a circular 5\arcsec ~radius extraction region centered on the source and a nearby background extraction region (with no obvious sources) with at least four times the area of the source. For each band observed we ran {\it uvotsource} using a background threshold of 3 sigma to obtain the fluxes and their errors. \color{black}  
For the Swift/UVOT the PSF is 2.5 arcsec and the pixel scale is 0.502.  This limits the resolution and sources that we include or exclude from the source detection region. Most of the Swift/UVOT observations are in one of the ultraviolet filters.  In the ultraviolet most stars drop in flux and most of the XRBs have significant emission because of the accretion disk. This limits the number of sources that are detected. For the significant Swift/UVOT detections we find well defined sources at the positions of the XRBs.  For fainter sources we exclude any confusing sources that overlap the source detection region. \color{black}
For detected sources we determine flux densities and errors given in table \ref{tab:Swift_uvot}.  For sources not detected we give upper limits.  The conversion between UVOT band and frequency are given in table \ref{tab:uvot_freqs}.

\startlongtable
\begin{deluxetable*}{|c|c|c|c|c|c|c|c|c|c|}
\tablewidth{6.5in}
\tablecaption{SWIFT UVOT Results \label{tab:Swift_uvot}}
\tablehead{
\colhead{Source }& \colhead{Obsid\tablenotemark{a}}& \colhead{Band }& \colhead{Exposure }& \colhead{Mag. }& \colhead{Error }& \colhead{Error }& \colhead{Flux }& \colhead{Flux Error }& \colhead{Flux Error }\\   [0.5ex]
\colhead{ }&\colhead{  }&\colhead{ }&\colhead{ (seconds) }& \colhead{ }&\colhead{ (stat.) }& \colhead{(sys.) }& \colhead{(Jy)  }& \colhead{Jy (stat.) }& \colhead{Jy (sys.)} 
}
\startdata
\multicolumn{10}{|c|}
 {\bf Black Holes} \\
 \hline
 LMC X-3 & 00037080001 & V & 188.62 & 16.71 & 0.08 & 0.01   &  $\rm 7.51 \times  10^{-4}$ &  $\rm 5.5 \times 10^{-5}$  & $\rm 2.0 \times 10^{-6}$  \\
 LMC X-3 & 00037080001 & B & 188.63 & 16.79  & 0.04 & 0.02   &  $\rm 7.77 \times  10^{-4}$ &  $\rm 3.2 \times 10^{-5}$  & $\rm 3.0 \times 10^{-6}$  \\
 LMC X-3 & 00037080001 & U & 188.62 & 15.91 & 0.04 & 0.02   &  $\rm 6.22 \times  10^{-4}$ &  $\rm 2.3 \times 10^{-5}$  & $\rm 1.0 \times 10^{-5}$  \\
 LMC X-3 & 00037080001 & UVW1 & 398.90 & 15.65 & 0.03 & 0.03   &  $\rm 4.90 \times  10^{-4}$ &  $\rm 1.5 \times 10^{-5}$  & $\rm 1.5 \times 10^{-5}$  \\
 LMC X-3 & 00037080001 & UVM2 & 492.79 & 15.58 & 0.04 & 0.03  &  $\rm 4.50 \times  10^{-4}$ &  $\rm 1.5 \times 10^{-5}$  & $\rm 3.0 \times 10^{-6}$  \\
 LMC X-3 & 00037080001 & UVW2 & 797.64 & 15.64 & 0.03 & 0.03  &   $\rm 4.08 \times  10^{-4}$ &  $\rm 1.0 \times 10^{-5}$  & $\rm 9.0 \times 10^{-6}$  \\
 LMC X-3 & 00037080002 & V & 176.45 & 16.82 & 0.08 & 0.01   &  $\rm 6.80 \times  10^{-4}$ &  $\rm 5.0 \times 10^{-5}$  & $\rm 2.0 \times 10^{-6}$  \\
 LMC X-3 & 00037080002 & B & 176.62 & 16.93  & 0.05 & 0.02   &  $\rm 6.88 \times  10^{-4}$ &  $\rm 3.1 \times 10^{-5}$  & $\rm 3.0 \times 10^{-6}$  \\
 LMC X-3 & 00037080002 & U & 176.50 & 16.01 & 0.04 & 0.02   &  $\rm 5.69 \times  10^{-4}$ &  $\rm 2.2 \times 10^{-5}$  & $\rm 9.0 \times 10^{-6}$  \\
 LMC X-3 & 00037080002 & UVW1 & 350.68 & 15.72 & 0.03 & 0.03   &  $\rm 4.62 \times  10^{-4}$ &  $\rm 1.5 \times 10^{-5}$  & $\rm 1.4 \times 10^{-5}$  \\
 LMC X-3 & 00037080002 & UVM2 & 470.71 & 15.64 & 0.04 & 0.03  &  $\rm 4.27 \times  10^{-4}$ &  $\rm 1.5 \times 10^{-5}$  & $\rm 3.0 \times 10^{-6}$  \\
 LMC X-3 & 00037080002 & UVW2 & 705.04 & 15.64 & 0.03 & 0.03  &   $\rm 4.09 \times  10^{-4}$ &  $\rm 1.1 \times 10^{-5}$  & $\rm 9.0 \times 10^{-6}$  \\
 LMC X-3 & 00037080003 & V & 191.20 & 16.70 & 0.07 & 0.01   &  $\rm 7.63 \times  10^{-4}$ &  $\rm 4.9 \times 10^{-5}$  & $\rm 3.0 \times 10^{-6}$  \\
 LMC X-3 & 00037080003 & B & 191.25 & 16.80  & 0.04 & 0.02   &  $\rm 7.70 \times  10^{-4}$ &  $\rm 3.0 \times 10^{-5}$  & $\rm 3.0 \times 10^{-6}$  \\
 LMC X-3 & 00037080003 & U & 191.28 & 15.93 & 0.04 & 0.02   &  $\rm 6.12 \times  10^{-4}$ &  $\rm 2.2 \times 10^{-5}$  & $\rm 9.0 \times 10^{-6}$  \\
 LMC X-3 & 00037080003 & UVW1 & 383.15 & 15.67 & 0.03 & 0.03   &  $\rm 4.81 \times  10^{-4}$ &  $\rm 1.5 \times 10^{-5}$  & $\rm 1.5 \times 10^{-5}$  \\
 LMC X-3 & 00037080003 & UVM2 & 518.15 & 15.61 & 0.04 & 0.03  &  $\rm 4.37 \times  10^{-4}$ &  $\rm 1.5 \times 10^{-5}$  & $\rm 3.0 \times 10^{-6}$  \\
 LMC X-3 & 00037080003 & UVW2 & 769.09 & 15.70 & 0.03 & 0.03  &   $\rm 3.85 \times  10^{-4}$ &  $\rm 1.0 \times 10^{-5}$  & $\rm 8.0 \times 10^{-6}$  \\
 \hline
 GX339-4 & 00030953004 & B & 7.12 & 17.42 & 0.25 &  0.02  &  $\rm 4.35 \times  10^{-4}$ &  $\rm 9.9 \times 10^{-5}$  & $\rm 2.0 \times 10^{-6}$  \\
 GX339-4 & 00030953004 & U & 83.39 & 17.66 & 0.13 & 0.02   &  $\rm 1.25 \times  10^{-4}$ &  $\rm 1.5 \times 10^{-5}$  & $\rm 2.0 \times 10^{-6}$  \\
 GX339-4 & 00030953004 & UVW1 & 382.41 & 18.56 & 0.14 & 0.03   &  $\rm 3.36 \times  10^{-5}$ &  $\rm 4.3 \times 10^{-6}$  & $\rm 1.0 \times 10^{-6}$  \\
 GX339-4 & 00030953005 & V & 136.90 & 16.45 & 0.07 & 0.01   &  $\rm 9.62 \times  10^{-4}$ &  $\rm 6.2 \times 10^{-5}$  & $\rm 3.0 \times 10^{-6}$  \\
 GX339-4 & 00030953005 & B & 136.91 & 17.61 & 0.07 & 0.02   &  $\rm 3.68 \times  10^{-4}$ &  $\rm 2.4 \times 10^{-5}$  & $\rm 1.0 \times 10^{-6}$  \\
 GX339-4 & 00030953005 & U & 136.91 & 17.37 & 0.08 & 0.02   &  $\rm 1.63 \times  10^{-4}$ &  $\rm 1.3 \times 10^{-5}$  & $\rm 3.0 \times 10^{-6}$  \\
 GX339-4 & 00030953005 & UVW1 & 274.701 & 18.41 & 0.14 & 0.03   &  $\rm 3.86 \times  10^{-5}$ &  $\rm 5.1 \times 10^{-6}$  & $\rm 1.2 \times 10^{-6}$  \\
 GX339-4 & 00030953005 & UVM2 & 297.36 & 19.71 & 0.35 & 0.03  &  $\rm 1.01 \times  10^{-5}$ &  $\rm 3.3 \times 10^{-6}$  & $\rm 1.0 \times 10^{-7}$  \\
 GX339-4 & 00030953005 & UVW2 & 550.28 & 19.99 & 0.29 & 0.03  &   $\rm 7.44 \times  10^{-6}$ &  $\rm 2.00 \times 10^{-6}$  & $\rm 1.6 \times 10^{-7}$  \\
 GX339-4 & 00030953006 & V & 52.94 & 16.54 & 0.11 & 0.01   &  $\rm 8.83 \times  10^{-4}$ &  $\rm 9.3 \times 10^{-5}$  & $\rm 3.0 \times 10^{-6}$  \\
 GX339-4 & 00030953006 & B & 52.93 & 17.50 & 0.11 & 0.02   &  $\rm 4.04 \times  10^{-4}$ &  $\rm 4.0 \times 10^{-5}$  & $\rm 2.0 \times 10^{-6}$  \\
 GX339-4 & 00030953006 & U & 52.93 & 17.31 & 0.13 & 0.02   &  $\rm 1.72 \times  10^{-4}$ &  $\rm 2.0 \times 10^{-5}$  & $\rm 3.0 \times 10^{-6}$  \\
 GX339-4 & 00030953006 & UVW1 & 105.08 & 18.50 & 0.25 & 0.03   &  $\rm 3.55 \times  10^{-5}$ &  $\rm 8.3 \times 10^{-6}$  & $\rm 1.1 \times 10^{-6}$  \\
 GX339-4 & 00030953006 & UVM2 & 135.91 & 19.36 & 0.36 & 0.03  &  $\rm 1.39 \times  10^{-5}$ &  $\rm 4.6 \times 10^{-6}$  & $\rm 1.0 \times 10^{-7}$  \\
 GX339-4 & 00030953006 & UVW2 & 211.39 & 19.16 & 0.24 & 0.03  &   $\rm 1.60 \times  10^{-5}$ &  $\rm 3.5 \times 10^{-6}$  & $\rm 3.0 \times 10^{-7}$  \\
 \hline
 1E1740-29 & 00030960001 & U & 3785.47 & 21.19 & $\ast$ &$\ast$   &  $\rm 4.81 \times  10^{-6}$ & $\ast$   & $\ast$  \\ 
 \hline
 GRS1758-258 & 00030961001 & U & 3257.18 & 21.03 & $\ast$ & $\ast$   &  $\rm 5.56 \times  10^{-6}$ & $\ast$   & $\ast$  \\ 
 \hline
 4U 1957+11 & 00030960001 & UVM2 & 3843.32 & 18.01 &  0.04 & 0.03   &  $\rm 4.81 \times  10^{-5}$ &  $\rm 2.0 \times  10^{-6}$  & $\rm 3.0 \times  10^{-7}$  \\ 
 \hline
\multicolumn{10}{|c|}
 {\bf Z Sources} \\
 \hline
 Cir X-1 & 00030268024 & UVM2 & 1024.19 & 16.02 & $\ast$ & $\ast$ &  $\rm 2.99 \times  10^{-4}$ & $\ast$   &  $\ast$  \\ 
 Cir X-1 & 00030268025 & UVW2 & 1208.71 & 20.68 & $\ast$ & $\ast$ &   $\rm 3.93 \times  10^{-6}$ &  $\ast$   &  $\ast$  \\ 
 \hline
 GX17+2 & 00035340006 & UVW2 & 3353.89 & 21.28 & $\ast$ & $\ast$ &   $\rm 2.27 \times  10^{-6}$ &  $\ast$   &  $\ast$  \\ 
 \hline
 GX349+2 & 00036690002 & UVM2 & 4533.23 & 21.27 & $\ast$ & $\ast$ &  $\rm 2.38 \times  10^{-6}$ &  $\ast$   &  $\ast$  \\ 
 \hline
\multicolumn{10}{|c|}
 {\bf Atoll Sources} \\
 \hline
 4U 1608-52 & 00030791018 & UVW1 & 1240.11 & 20.83 & $\ast$ & $\ast$  &  $\rm 4.15 \times  10^{-6}$ & $\ast$   & $\ast$  \\ 
 4U 1608-52 & 00030791019 & U  & 1005.07 & 20.57 & 0.29 & 0.02  & $\rm 8.52 \times 10^{-6}$ & $\rm 2.25 \times 10^{-6}$ & $\rm 1.3 \times 10^{-7}$ \\
 4U 1608-52 & 00030791020 & UVM2 & 902.65 & 20.55 & $\ast$ & $\ast$ &  $\rm 4.63 \times  10^{-6}$ & $\ast$   & $\ast$  \\ 
 4U 1608-52 & 00030791021 & U & 1119.99 & 20.32 & 0.33 & 0.02  &  $\rm 1.08 \times  10^{-5}$ & $\rm 2.20 \times 10^{-6}$   & $\rm 2.0 \times 10^{-7}$  \\ 
 4U 1608-52 & 00030791022 & UVM2 & 1475.56 & 20.92 & $\ast$ & $\ast$ &  $\rm 3.29 \times  10^{-6}$ & $\ast$   & $\ast$  \\ 
 4U 1608-52 & 00030791023 & U & 945.80 & 20.41 & 0.26 & 0.02  &  $\rm 9.93 \times  10^{-6}$ & $\rm 2.39 \times 10^{-6}$   & $\rm 1.5 \times 10^{-7}$  \\ 
 \hline
 GX9+9 & 00030965001 & UVW2 & 4086.32 & 16.32 & 0.02 & 0.03 &   $\rm 2.18 \times  10^{-4}$ &  $\rm 5.0 \times 10^{-6}$  & $\rm 5.0 \times 10^{-6}$   \\ 
 GX9+9 & 00030965003 & UVM2 & 195.28 & 16.21 & 0.07 & 0.03 &  $\rm 2.52 \times  10^{-4}$ &  $\rm 1.5 \times 10^{-5}$  & $\rm 2.0 \times 10^{-6}$   \\ 
 \hline
 GX354+0 & 00030964001 & UVM2 & 3770.68 & 21.11 & $\ast$ & $\ast$ &  $\rm 2.77 \times  10^{-6}$ & $\ast$   & $\ast$  \\ 
 \hline
 4U 1735-44 & 00035338003 & UVM2 & 3728.44 & 17.22 & 0.03 & 0.03 &  $\rm 9.92 \times  10^{-5}$ & $\rm 3.1 \times 10^{-6}$   & $\rm 6.0 \times 10^{-7}$  \\ 
 \hline
 Swift 1756-2508 & 00030952011 & U & 517.43 & 19.88 & $\ast$ & $\ast$  & $\rm 1.61 \times  10^{-5}$& $\ast$ & $\ast$ \\
 Swift 1756-2508 & 00030952011 & UVW1 & 6231.58 & 21.14 & $\ast$ & $\ast$  & $\rm 3.13 \times 10^{-6}$ & $\ast$ & $\ast$ \\
 Swift 1756-2508 & 00030952012 & U & 6390.68 & 21.25 & $\ast$ & $\ast$  & $\rm 4.57 \times  10^{-6}$ & $\ast$ & $\ast$ \\
 Swift 1756-2508 & 00030952013 & UVM2 & 438.47 & 19.59 &  $\ast$ &  $\ast$ & $\rm 1.12 \times  10^{-5}$ &  $\ast$ &  $\ast$ \\
 Swift 1756-2508 & 00030952013 & UVW2 & 6048.35 & 21.21 &  $\ast$ &  $\ast$ &   $\rm 2.42 \times  10^{-6}$  &  $\ast$ &  $\ast$ \\ 
 \hline
 4U 1822-37 & 00036691001 & U & 1660.69 & 14.40 & 0.02 & 0.02  &  $\rm 2.50 \times  10^{-3}$ & $\rm 5.0 \times 10^{-5}$   & $\rm 4.0 \times 10^{-5}$  \\ 
 \hline
\multicolumn{10}{|c|}
 {\bf Pulsars} \\
 \hline
 SMC X-1 & 00035216002 & UVW1 & 1569.52 & 11.11 & 0.03 & 0.03  &  $\rm 3.20 \times  10^{-2}$ &  $\rm 9.0 \times  10^{-4}$   & $\rm 1.0 \times  10^{-3}$   \\ 
 \hline
\enddata

\tablenotetext{a}{For corresponding MJD start times see Table \ref{tab:Swift_xrt}.}
\tablenotetext{*}{the magnitude of the flux is an upper limit.}
\end{deluxetable*}



\begin{table}[h!]
\centering
\begin{tabular}{|c|c|} 
 \hline
 Swift/UVOT Filter &  Freq. (Hz) \\   [0.5ex]
 \hline\hline
 V    &  $\rm 5.550 \times 10^{14}$ \\  
 B    &  $\rm 6.925 \times 10^{14}$ \\    
 U    &  $\rm 8.563 \times 10^{14}$ \\
 UVW1 &  $\rm 1.138 \times 10^{15}$ \\ 
 UVM2 &  $\rm 1.344 \times 10^{15}$ \\ 
 UVW2 &  $\rm 1.477 \times 10^{15}$ \\ 
 \hline
\end{tabular}
\caption{Swift/UVOT to Frequency conversion}\label{tab:uvot_freqs}
\end{table}


\subsection{X-ray Observations}\label{sec:Obs_xray}

\subsubsection{Swift XRT}

The XRT is a focusing X-ray telescope with a $\rm 110 ~cm^{2}$ effective area, $\rm 23.6 \arcmin \times 23.6 \arcmin$  ~FOV, 18\arcsec ~resolution (half-power diameter), and $\rm 0.2-10 ~keV$ energy range. The XRT uses a grazing incidence Wolter 1 telescope to focus X-rays onto a CCD detector.  Further information on the XRT is given by \citet{Burrows05}. 

The Swift observations were done in window timing mode (WT) for bright sources or photon counting mode (PC) for fainter sources.  For bright sources the WT data consists of a 1-dimensional line.  The spectral data extraction region used is a 40 pixel width box with a height large enough to include all of the photons centered on the source.  One or more boxes are created at either end of the line to collect the background spectra.  For fainter sources done in PC mode a circle with a 20 pixel radius was used along with a circular background having at least four times the source extraction area, at a position away from any sources on the detector.  The Heasoft tool {\it xselect} is then used to extract the spectrum and filter the data. 

The results of modelling the Swift/XRT spectra are shown on Table \ref{tab:Swift_xrt}.  
\color{black}All of the data sets were fitted with both disk black-body (DBB) and power-law (PL) models, as indicated in column 5. 
Comparing the $\chi^2$ per degree of freedom (column 10) gives a good measure of the relative quality of the fits
using the two
models.  In some cases they are nearly the same, e.g. Aql X-1 after MJD 54286.  Column 8 gives the derived value
of the hydrogen column density along the path to the source.  For comparison, column 9 gives the total column 
density of Milky Way HI from the HEASARC Tools\footnote{\url{https://heasarc.gsfc.nasa.gov/cgi-bin/Tools/w3nh/w3nh.pl}} using data from the HI4PI survey \citep{HI4PI}.
The quality of the fits (statistics and residuals) were such that fitting with more complicated models was not warranted. The corresponding best fit temperature is in column 7 and the power-law index in column 6.  The quality of the fit is indicated by chi-square and the corresponding number of
degrees of freedom is the denominator in column 10.  The derived flux in the 0.5 to 10 keV band is given in column 10.  The Z source Cir X-1 and the pulsar 4U1822-37 have such shallow slopes (Gamma = 0.56 for the PL), that it is difficult to fit the temperature in a DBB model.
\color{black}

The issue of pile-up is examined for each spectrum.  If the PC mode is less than $\rm \sim 0.5 ~cts~s^{-1}$ or the WT mode is less than $\rm 100 ~cts~s^{-1}$ then pile-up is not a significant issue.  If pile-up is a possible problem we use standard procedures\footnote{https://www.swift.ac.uk/analysis/xrt/pileup.php} to determine the impact and how to mitigate it.  The main approach to correct pile-up is to eliminate inner pixels of the extraction region and re-extract the spectrum and use only grade 0 events.

\color{black} XRT data were taken during our campaign week (54280 $\leq$ MJD $\leq$ 54287) for 13 sources on Table \ref{tab:Swift_xrt}.  Data for three other sources were taken a few days before or after our campaign, and they are included on the Table and on the spectral energy distributions plotted in Sec. \ref{sec:SEDs} below.  These are LMC X-3, observed some 12 to 15 days after the end of the campaign,  4U 1822-37, observed four days before the start, and SMC X-1, observed nine days before the start.  For LMC X-3, this time offset spans a period of rapid variation in the source properties, as discussed in Sec. \ref{sec:SEDs}.\color{black}

\begin{longrotatetable}
\begin{deluxetable}{|c|c|c|c|c|c|c|c|c|c|c|}
\tablecaption{Swift XRT Results  \label{tab:Swift_xrt}}
\tablehead{ \colhead{Source} & \colhead{Obsid} & \colhead{Start} & \colhead{Exposure} & \colhead{Model} & \colhead{Gamma} &
 \colhead{T} & \colhead{$\rm N_H$} & \colhead{\textcolor{black}{$\rm N_H$ (Gal.)}} &
 \colhead{$\chi^2$} & \colhead{$\rm F_x$ ($\rm 0.5 - 10~keV$)} \\
&&\colhead{MJD} & \colhead{sec} & &&\colhead{keV} & \colhead{$\rm 10^{22}~ cm^{-2}$} &
 \colhead{\textcolor{black}{$\rm 10^{22}~ cm^{-2}$}} & \colhead{per d.o.f.} &
 \colhead{$\rm ergs~cm^{-2}~s^{-1}$}}
\startdata
 {\bf Black Holes} & & & & & & & & & &  \\
 \hline
 1E1740-29 & 00030960001 & 54284.414 &  3803 & PL & $\rm 1.86^{+0.31}_{-0.28}$ &  $\ast$ & $\rm 9.62^{+1.57}_{-1.32}$ & \textcolor{black}{1.50} & 68.83/75 &   $\rm 1.14^{+0.02}_{-0.12} \times  10^{-10}$ \\
 \textcolor{black}{1E1740-29} & \textcolor{black}{00030960001} & \textcolor{black}{54284.414} &  \textcolor{black}{3803} & \textcolor{black}{DBB} & \textcolor{black}{$\ast$} &  
 \textcolor{black}{$\rm 2.77^{+0.57}_{-0.41}$} & \textcolor{black}{$\rm 7.83^{+1.07}_{-0.90}$} & \textcolor{black}{1.50} & \textcolor{black}{64.93/75} &    \textcolor{black}{$\rm 1.11^{+0.01}_{-0.29} \times 10^{-10}$}  \\
  \hline  
 GRS1758-258 & 00030961001 & 54284.547 & 1951 & PL & $\rm 2.65^{+0.05}_{-0.05}$ &  $\ast$ & $\rm 1.59^{+0.05}_{-0.05}$ &  \textcolor{black}{0.60}  &   538.88/504 &   $\rm 1.37^{+0.01}_{-0.02} \times10^{-9}$ \\
\textcolor{black}{ GRS1758-258} & \textcolor{black}{00030961001} & \textcolor{black}{54284.547} & 
\textcolor{black}{1951} & \textcolor{black}{DBB} &
\textcolor{black}{$\ast$} &  \textcolor{black}{$\rm 1.17^{+0.03}_{-0.03}$} &
\textcolor{black}{$\rm 0.95^{+0.03}_{-0.03}$} &  \textcolor{black}{0.60}  &   \textcolor{black}{900.42/504} &   
\textcolor{black}{$\rm 1.22^{+0.01}_{-0.01} \times10^{-9}$} \\
 \hline
 4U1957+115 & 00030959001 & 54282.547 & 1744 & DBB & $\ast$ & $\rm 1.43^{+0.01}_{-0.01}$ & $\rm 0.08^{+0.01}_{-0.01}$ &  \textcolor{black}{0.12} &   731.42/626 &   $\rm 7.84^{+0.04}_{-0.04} \times  10^{-10}$ \\
\textcolor{black}{ 4U1957+115} & \textcolor{black}{00030959001} & \textcolor{black}{54282.547} & \textcolor{black}{1744} &
\textcolor{black}{PL} & \textcolor{black}{$\rm 1.96^{+0.09}_{-0.09}$} & \textcolor{black}{$\ast$} & 
\textcolor{black}{$\rm 0.27^{+0.004}_{-0.004}$} &  \textcolor{black}{0.12} &
\textcolor{black}{2495.85/626} &   \textcolor{black}{$\rm 8.41^{+0.03}_{-0.04} \times  10^{-10}$} \\
\hline
GX339-4 & 00030953004 & 54281.785 & 493.4 & PL & $\rm 1.62^{+0.13}_{-0.12}$ &  $\ast$ & $\rm 0.42^{+0.07}_{-0.06}$ & \textcolor{black}{0.39} &   84.54/91 &   $\rm 1.78^{+0.07}_{-0.07} \times  10^{-10}$ \\
 \textcolor{black}{ GX339-4} & \textcolor{black}{00030953004} & \textcolor{black}{54281.785} & \textcolor{black}{493.4} & 
\textcolor{black}{DBB} &  \textcolor{black}{$\ast$} & \textcolor{black}{$\rm 1.86^{+0.21}_{-0.18}$}  & 
\textcolor{black}{$\rm 0.23^{+0.04}_{-0.04}$} & \textcolor{black}{0.39} &
\textcolor{black}{96.29/91} &   \textcolor{black}{$\rm 1.56^{+0.01}_{-0.17} \times  10^{-10}$} \\ 
 GX339-4 & 00030953005 & 54285.195 & 1659 & PL & $\rm 1.56^{+0.06}_{-0.05}$ &  $\ast$ & $\rm 0.38^{+0.03}_{-0.03}$ & \textcolor{black}{0.39} &   307.36/315 &   $\rm 2.47^{+0.04}_{-0.05} \times  10^{-10}$ \\  
\textcolor{black}{ GX339-4} & \textcolor{black}{00030953005} & \textcolor{black}{54285.195} & \textcolor{black}{1659} & 
\textcolor{black}{DBB} &  \textcolor{black}{$\ast$} & \textcolor{black}{$\rm 2.04^{+0.11}_{-0.10}$}  & 
\textcolor{black}{$\rm 0.21^{+0.02}_{-0.02}$} & \textcolor{black}{0.39} &   
\textcolor{black}{334.20/315} &   \textcolor{black}{$\rm 2.23^{+0.13}_{-0.04} \times  10^{-10}$} \\
 GX339-4 & 00030953006 & 54288.14 & 1149 & PL & $\rm 1.63^{+0.06}_{-0.06}$ &  $\ast$ & $\rm 0.39^{+0.03}_{-0.03}$ & \textcolor{black}{0.39} &   225.10/267 &   $\rm 2.60^{+0.06}_{-0.05} \times  10^{-10}$ \\
\textcolor{black}{ GX339-4} & \textcolor{black}{00030953006} & \textcolor{black}{54288.14} & \textcolor{black}{1149} &  
\textcolor{black}{DBB} & \textcolor{black}{$\ast$} & \textcolor{black}{$\rm 1.90^{+0.11}_{-0.10}$} &  
\textcolor{black}{$\rm 0.21^{+0.02}_{-0.02}$} & \textcolor{black}{0.39} &
\textcolor{black}{264.41/267} &   \textcolor{black}{$\rm 2.33^{+0.04}_{-0.07} \times  10^{-10}$} \\
 \hline
 LMC X-3 & 00037080001 & 54299.574 & 2405 & DBB & $\ast$ & $\rm 1.18^{+0.03}_{-0.03}$ & $\rm 0.03^{+0.01}_{-0.01}$ & \textcolor{black}{0.04} &   367.23/321 &   $\rm 5.36^{+0.05}_{-0.07} \times  10^{-10}$ \\  
\textcolor{black}{LMC X-3} & \textcolor{black}{00037080001} & \textcolor{black}{54299.574} & \textcolor{black}{2405} &
\textcolor{black}{PL}  & \textcolor{black}{$\rm 2.16^{+0.04}_{-0.04}$} & \textcolor{black}{$\ast$} & 
\textcolor{black}{$\rm 0.21^{+0.01}_{-0.01}$} & \textcolor{black}{0.04} &
\textcolor{black}{542.66/321} &   \textcolor{black}{$\rm 5.81^{+0.07}_{-0.09} \times  10^{-10}$} \\  
LMC X-3 & 00037080002 & 54300.773 &  2133  & DBB & $\ast$ & $\rm 1.17^{+0.03}_{-0.03}$ & $\rm 0.03^{+0.01}_{-0.01}$ & \textcolor{black}{0.04} &   351.46/303 &   $\rm 5.58^{+0.04}_{-0.07} \times  10^{-10}$ \\  
\textcolor{black}{LMC X-3} & \textcolor{black}{00037080002} & \textcolor{black}{54300.773} &  \textcolor{black}{2133}  &
\textcolor{black}{PL}  & \textcolor{black}{$\rm 2.10^{+0.05}_{-0.05}$} & \textcolor{black}{$\ast$} &
\textcolor{black}{$\rm 0.19^{+0.02}_{-0.01}$} & \textcolor{black}{0.04} &
\textcolor{black}{510.27/303} &   \textcolor{black}{$\rm 6.18^{+0.10}_{-0.07} \times  10^{-10}$} \\  
LMC X-3 & 00037080003 & 54302.254 & 2303 & DBB & $\ast$ & $\rm 1.16^{+0.03}_{-0.03}$ & $\rm 0.03^{+0.01}_{-0.01}$ & \textcolor{black}{0.04} &   307.34/336 &   $\rm 5.61^{+0.10}_{-0.05} \times  10^{-10}$ \\
 \textcolor{black}{LMC X-3} & \textcolor{black}{00037080003} & \textcolor{black}{54302.254} & \textcolor{black}{2303} & 
 \textcolor{black}{PL}  & \textcolor{black}{$\rm 2.09^{+0.04}_{-0.04}$} & \textcolor{black}{$\ast$} & 
 \textcolor{black}{$\rm 0.18^{+0.01}_{-0.01}$} & \textcolor{black}{0.04} &
 \textcolor{black}{573.16/336} &   \textcolor{black}{$\rm 6.22^{+0.11}_{-0.08} \times  10^{-10}$} \\
 \hline\hline
 {\bf Atoll Sources} & & & & & & & & & &  \\
 \hline
 Aql X-1 & 00030796021 & 54279.777 & 1935 & PL & 
 $\rm 1.83^{+0.09}_{-0.09}$ &  $\ast$ & $\rm 0.45^{+0.05}_{-0.05}$ & \textcolor{black}{0.31} &  167.75/170 &   $\rm 2.47^{+0.06}_{-0.07} \times  10^{-10}$ \\ 
\textcolor{black}{ Aql X-1} & \textcolor{black}{00030796021} & \textcolor{black}{54279.777} & \textcolor{black}{1935} & 
\textcolor{black}{DBB} &  \textcolor{black}{$\ast$} & \textcolor{black}{$\rm 1.55^{+0.11}_{-0.10}$}  & 
\textcolor{black}{$\rm 0.21^{+0.03}_{-0.03}$} & \textcolor{black}{0.31} &  
\textcolor{black}{201.30/170} &  \textcolor{black}{$\rm 2.12^{+0.05}_{-0.08} \times  10^{-10}$} \\ 
 Aql X-1 & 00030796022 & 54283.473 & 2228 & PL & 
  $\rm 2.55^{+0.22}_{-0.20}$ &  $\ast$ & $\rm 0.63^{+0.10}_{-0.09}$ &  \textcolor{black}{0.31} &   57.53/57 &   $\rm 1.78^{+0.08}_{-0.10} \times  10^{-11}$ \\ 
\textcolor{black}{ Aql X-1} & \textcolor{black}{00030796022} & \textcolor{black}{54283.473} & \textcolor{black}{2228} & 
\textcolor{black}{DBB} &  \textcolor{black}{$\ast$} & \textcolor{black}{$\rm 1.00^{+0.11}_{-0.10}$}  & 
\textcolor{black}{$\rm 0.28^{+0.07}_{-0.06}$} & \textcolor{black}{0.31} &  
\textcolor{black}{89.00/57} &  \textcolor{black}{$\rm 1.53^{+0.03}_{-0.13} \times  10^{-10}$} \\ 
 Aql X-1 & 00030796023 & 54286.133 & 2462 & PL & 
  $\rm 4.67^{+1.32}_{-1.08}$ &  $\ast$ & $\rm 0.96^{+0.42}_{-0.33}$ &  \textcolor{black}{0.31} &   $\rm 191.11/971^{\dagger}$ &   $\rm 6.72^{+0.56}_{-1.84} \times  10^{-13}$ \\ 
\textcolor{black}{ Aql X-1} & \textcolor{black}{00030796023} & \textcolor{black}{54286.133} & \textcolor{black}{2462} & 
\textcolor{black}{DBB} &  \textcolor{black}{$\ast$} & \textcolor{black}{$\rm 0.48^{+0.16}_{-0.11}$}  & 
\textcolor{black}{$\rm 0.38^{+0.26}_{-0.19}$} & \textcolor{black}{0.31} &  
\textcolor{black}{$\rm {193.05/971}^{\dagger}$} &  \textcolor{black}{$\rm 6.73^{+0.01}_{-0.40} \times  10^{-13}$} \\ 
Aql X-1 & 00030796024 & 54287.492 & 0 & PL & $\ast$ &  $\ast$ & $\ast$ & \textcolor{black}{0.31} & $\ast$ &  $\ast$ \\   
\textcolor{black}{Aql X-1} & \textcolor{black}{00030796024} & \textcolor{black}{54287.492} & \textcolor{black}{0} & 
\textcolor{black}{DBB} & \textcolor{black}{$\ast$} &  \textcolor{black}{$\ast$} & \textcolor{black}{$\ast$} & \textcolor{black}{0.31} & 
\textcolor{black}{$\ast$} &  \textcolor{black}{$\ast$} \\   
 Aql X-1 & 00030796025 & 54288.55 & 2168 & PL & 
  $\rm 3.74^{+0.96}_{-0.82}$ &  $\ast$ & $\rm 0.72^{+0.30}_{-0.25}$ &  \textcolor{black}{0.31} &   $\rm 213.13/971^{\dagger}$ &   $\rm 9.63^{+0.47}_{-2.46} \times  10^{-13}$ \\  
\textcolor{black}{Aql X-1} & \textcolor{black}{00030796025} & \textcolor{black}{54288.55} & \textcolor{black}{2168} & 
\textcolor{black}{DBB} & \textcolor{black}{$\ast$}  &  \textcolor{black}{$\rm 0.65^{+0.20}_{-0.14}$} & 
\textcolor{black}{$\rm 0.26^{+0.18}_{-0.14}$} &  \textcolor{black}{0.31} &   
\textcolor{black}{$\rm 218.49/971^{\dagger}$} &   \textcolor{black}{$\rm 9.63^{+0.55}_{-4.33} \times  10^{-13}$} \\  
 Aql X-1 & 00030796026 & 54289.773 & 3376 & PL & 
  $\rm 3.07^{+0.66}_{-0.58}$ &  $\ast$ & $\rm 0.43^{+0.19}_{-0.16}$ &  \textcolor{black}{0.31} &   
  $\rm 266.89/971^{\dagger}$ &   $\rm 9.77^{+0.73}_{-1.59} \times  10^{-13}$ \\  
 \textcolor{black}{Aql X-1} & \textcolor{black}{00030796026} & \textcolor{black}{54289.773} & \textcolor{black}{3376} & 
 \textcolor{black}{DBB} & \textcolor{black}{$\ast$} & \textcolor{black}{$\rm 0.85^{+0.24}_{-0.16}$} & 
 \textcolor{black}{$\rm 0.09^{+0.10}_{-0.08}$} &  \textcolor{black}{0.31} &   
 \textcolor{black}{$\rm 279.59/971^{\dagger}$} &   \textcolor{black}{$\rm 9.71^{+0.57}_{-2.31} \times  10^{-13}$} \\  
 \hline
 4U1608-52 & 00030791018 & 54279.91 & 1262 & DBB & $\ast$ & $\rm 1.83^{+0.03}_{-0.03}$ & $\rm 0.86^{+0.02}_{-0.02}$ & \textcolor{black}{1.66} &   878.54/706 &   $\rm 3.70^{+0.02}_{-0.02} \times  10^{-9}$ \\ 
\textcolor{black}{4U1608-52} & \textcolor{black}{00030791018} & \textcolor{black}{54279.91} & \textcolor{black}{1262} & 
\textcolor{black}{PL} & \textcolor{black}{$\rm 1.93^{+0.03}_{-0.03}$} & \textcolor{black}{$\ast$}  & 
\textcolor{black}{$\rm 1.36^{+0.03}_{-0.03}$} & \textcolor{black}{1.66} &   
\textcolor{black}{1233.45/706} &   \textcolor{black}{$\rm 4.03^{+0.03}_{-0.01} \times  10^{-9}$} \\ 
 4U1608-52 & 00030791019 & 54280.926 & 1006 & DBB & $\ast$ & $\rm 1.89^{+0.05}_{-0.05}$ & $\rm 0.83^{+0.03}_{-0.03}$ & 
 \textcolor{black}{1.66} &   683.90/597 &   $\rm 1.91^{+0.02}_{-0.02} \times  10^{-9}$ \\ 
 \textcolor{black}{4U1608-52} & \textcolor{black}{00030791019} & \textcolor{black}{54280.926} & \textcolor{black}{1006} & 
 \textcolor{black}{PL} & \textcolor{black}{$\rm 1.87^{+0.04}_{-0.04}$} &  \textcolor{black}{$\ast$} & 
 \textcolor{black}{$\rm 1.31^{+0.05}_{-0.04}$} & \textcolor{black}{1.66} &   
 \textcolor{black}{719.83/597} &   \textcolor{black}{$\rm 2.10^{+0.01}_{-0.02} \times  10^{-9}$} \\ 
 4U1608-52 & 00030791020 & 54282.05 & 905.7 & DBB & $\ast$ & $\rm 1.62^{+0.03}_{-0.03}$ & $\rm 0.82^{+0.02}_{-0.02}$ & \textcolor{black}{1.66} &   707.25/607 &   $\rm 2.46^{+0.02}_{-0.02} \times  10^{-9}$ \\ 
\textcolor{black}{4U1608-52} & \textcolor{black}{00030791020} & \textcolor{black}{54282.05} & \textcolor{black}{905.7} & 
\textcolor{black}{PL} & \textcolor{black}{$\rm 2.08^{+0.04}_{-0.03}$} & \textcolor{black}{$\ast$}  & 
\textcolor{black}{$\rm 1.33^{+0.04}_{-0.04}$} & \textcolor{black}{1.66} &   
\textcolor{black}{758.20/607} & \textcolor{black}{$\rm 2.71^{+0.02}_{-0.02} \times  10^{-9}$} \\ 
 4U1608-52 & 00030791021 & 54284.863 & 1177 & PL & $\rm 2.27^{+0.06}_{-0.05}$ &  $\ast$ & $\rm 1.24^{+0.05}_{-0.05}$ & \textcolor{black}{1.66} &   437.94/462 &   $\rm 8.26^{+0.09}_{-0.11} \times  10^{-10}$ \\
\textcolor{black}{4U1608-52} & \textcolor{black}{00030791021} & \textcolor{black}{54284.863} & \textcolor{black}{1177} & 
\textcolor{black}{DBB} &  \textcolor{black}{$\ast$} &  \textcolor{black}{$\rm 1.39^{+0.04}_{-0.04}$} & 
\textcolor{black}{$\rm 0.73^{+0.03}_{-0.03}$} & \textcolor{black}{1.66} &   
\textcolor{black}{571.69/462} &   \textcolor{black}{$\rm 7.32^{+0.06}_{-0.11} \times  10^{-10}$} \\
4U1608-52 & 00030791022 & 54286.336 & 1483 & PL & $\rm 1.90^{+0.03}_{-0.03}$ &  $\ast$ & $\rm 0.89^{+0.02}_{-0.02}$ & \textcolor{black}{1.66} &   690.01/617 &   $\rm 1.48^{+0.01}_{-0.01} \times  10^{-9}$ \\
\textcolor{black}{4U1608-52} & \textcolor{black}{00030791022} & \textcolor{black}{54286.336} & \textcolor{black}{1483} & 
\textcolor{black}{DBB} & \textcolor{black}{$\ast$} &  \textcolor{black}{$\rm 1.75^{+0.02}_{-0.04}$} & 
\textcolor{black}{$\rm 0.52^{+0.02}_{-0.01}$} & \textcolor{black}{1.66} &   
\textcolor{black}{1134.30/617} &   \textcolor{black}{$\rm 1.34^{+0.01}_{-0.01} \times  10^{-9}$} \\
 4U1608-52 & 00030791023 & 54288.004 & 950 & DBB & $\ast$ & $\rm 2.15^{+0.10}_{-0.09}$ & $\rm 0.69^{+0.04}_{-0.04}$ & \textcolor{black}{1.66} &   386.75/433 &   $\rm 8.09^{+0.06}_{-0.19} \times  10^{-10}$ \\ 
\textcolor{black}{4U1608-52} & \textcolor{black}{00030791023} & \textcolor{black}{54288.004} & \textcolor{black}{950} & 
\textcolor{black}{PL} & \textcolor{black}{$\rm 1.66^{+0.06}_{-0.06}$} & \textcolor{black}{$\ast$} & 
\textcolor{black}{$\rm 1.06^{+0.06}_{-0.06}$} & \textcolor{black}{1.66} &   
\textcolor{black}{410.99/433} &   \textcolor{black}{$\rm 8.82^{+0.13}_{-0.14} \times  10^{-10}$} \\ 
\hline
 4U1735-44 & 00035338003 & 54282.32 & 1818 & PL & $\rm 1.62^{+0.02}_{-0.02}$ &  $\ast$ & $\rm 0.30^{+0.01}_{-0.01}$ & \textcolor{black}{0.28} &   779.33/652 &   $\rm 3.66^{+0.02}_{-0.01} \times  10^{-9}$ \\
\textcolor{black}{4U1735-44} & \textcolor{black}{00035338003} & \textcolor{black}{54282.32} & \textcolor{black}{1818} & 
\textcolor{black}{DBB} & \textcolor{black}{$\ast$} & \textcolor{black}{$\rm 1.87^{+0.03}_{-0.03}$} &   
\textcolor{black}{$\rm 0.14^{+0.01}_{-0.01}$} & \textcolor{black}{0.28} &   
\textcolor{black}{936.38/652} &   \textcolor{black}{$\rm 3.27^{+0.02}_{-0.03} \times  10^{-9}$} \\
\hline
 GX354+0 & 00030964001 & 54286.4 & 3505 & PL & $\rm 1.82^{+0.04}_{-0.04}$ &  $\ast$ & $\rm 2.88^{+0.07}_{-0.07}$ & \textcolor{black}{1.30} &   816.74/727 &   $\rm 1.25^{+0.01}_{-0.01} \times  10^{-9}$ \\
\textcolor{black}{GX354+0} & \textcolor{black}{00030964001} & \textcolor{black}{54286.4} & \textcolor{black}{3505} & 
\textcolor{black}{DBB} &   \textcolor{black}{$\ast$} & \textcolor{black}{$\rm 2.36^{+0.06}_{-0.05}$} &
\textcolor{black}{$\rm 2.04^{+0.04}_{-0.04}$} & \textcolor{black}{1.30} &   
\textcolor{black}{793.41/727} &   \textcolor{black}{$\rm 1.19^{+0.01}_{-0.01} \times  10^{-9}$} \\
 \hline
 GX9+9 & 00030965001 & 54285.74 & 4154 & DBB & $\ast$ & $\rm 1.84^{+0.02}_{-0.02}$ & $\rm 0.17^{+0.01}_{-0.01}$ & \textcolor{black}{0.19} &   936.83/781 &   $\rm 3.90^{+0.01}_{-0.01} \times  10^{-9}$ \\
\textcolor{black}{GX9+9} & \textcolor{black}{00030965001} & \textcolor{black}{54285.74} & \textcolor{black}{4154} & 
\textcolor{black}{PL} & \textcolor{black}{$\rm 1.68^{+0.01}_{-0.01}$} & \textcolor{black}{$\ast$} & 
\textcolor{black}{$\rm 0.36^{+0.01}_{-0.01}$} & \textcolor{black}{0.19} &   
\textcolor{black}{2645.35/781} &   \textcolor{black}{$\rm 4.26^{+0.01}_{-0.01} \times  10^{-9}$} \\
GX9+9 & 00030965002 & 54285.137 & 1205 & DBB & $\ast$ & $\rm 1.97^{+0.03}_{-0.03}$ & $\rm 0.17^{+0.01}_{-0.01}$ & \textcolor{black}{0.19} &   748.38/705 &   $\rm 4.68^{+0.03}_{-0.03} \times  10^{-9}$ \\  
\textcolor{black}{GX9+9} & \textcolor{black}{00030965002} & \textcolor{black}{54285.137} & \textcolor{black}{1205} & 
\textcolor{black}{PL} & \textcolor{black}{$\rm 1.58^{+0.02}_{-0.02}$} & \textcolor{black}{$\ast$} & 
\textcolor{black}{$\rm 0.34^{+0.01}_{-0.01}$} & \textcolor{black}{0.19} &   
\textcolor{black}{1333.34/705} &  \textcolor{black}{$\rm 5.16^{+0.03}_{-0.02} \times  10^{-9}$} \\  
GX9+9 & 00030965003 & 54286.613 & 186.9 & DBB & $\ast$ & $\rm 1.96^{+0.08}_{-0.07}$ & $\rm 0.18^{+0.01}_{-0.01}$ & \textcolor{black}{0.19} &   427.19/403 &   $\rm 4.26^{+0.05}_{-0.07} \times  10^{-9}$ \\ 
\textcolor{black}{GX9+9} & \textcolor{black}{00030965003} & \textcolor{black}{54286.613} & \textcolor{black}{186.9} & 
\textcolor{black}{PL} & \textcolor{black}{$\rm 1.58^{+0.04}_{-0.04}$} & \textcolor{black}{$\ast$} & 
\textcolor{black}{$\rm 0.35^{+0.02}_{-0.02}$} & \textcolor{black}{0.19} &   
\textcolor{black}{530.08/403} &   \textcolor{black}{$\rm 4.69^{+0.06}_{-0.06} \times  10^{-9}$} \\ 
 \hline   
 {\bf Z Sources} & & & & & & & & &  \\
 \hline
 Cir X-1 & 00030268024 & 54278.906 & 1016 & PL & $\rm 0.85^{+0.38}_{-0.35}$ &  $\ast$ & $\rm 2.09^{+0.73}_{-0.59}$ & \textcolor{black}{1.63} &   25.51/27 &   $\rm 4.98^{+0.15}_{-0.88} \times  10^{-11}$ \\ 
 \textcolor{black}{Cir X-1} & \textcolor{black}{00030268024} & \textcolor{black}{54278.906} & \textcolor{black}{1016} & 
 \textcolor{black}{DBB} &   \textcolor{black}{$\ast$} & \textcolor{black}{$\rm 14.3^{\dagger \dagger}$} &
 \textcolor{black}{$\rm 1.88^{+0.39}_{-0.31}$} & \textcolor{black}{1.63} &   
 \textcolor{black}{26.07/27} &   \textcolor{black}{$\rm 5.08^{+0.28}_{-0.30} \times  10^{-11}$} \\
 Cir X-1 & 00030268025 & 54285.062 & 1194 & PL & $\rm 0.79^{+0.69}_{-0.55}$ &  $\ast$ & $\rm 0.62^{+0.76}_{-0.55}$ & \textcolor{black}{1.63} &   3.69/9 &   $\rm 1.22^{+0.11}_{-0.35} \times  10^{-11}$ \\
\textcolor{black}{Cir X-1} & \textcolor{black}{00030268025} & \textcolor{black}{54285.062} & \textcolor{black}{1194} & 
 \textcolor{black}{DBB} &  \textcolor{black}{$\ast$} & \textcolor{black}{$\rm 12.79^{\dagger \dagger}$} &
 \textcolor{black}{$\rm 0.55^{+0.50}_{-0.23}$} & \textcolor{black}{1.63} &   
 \textcolor{black}{3.69/9} &   \textcolor{black}{$\rm 1.23^{+0.12}_{-0.09} \times  10^{-11}$} \\
\hline
GX17+2 & 00035340006 & 54285.605 & 3397 & DBB & $\ast$ & $\rm 2.29^{+0.03}_{-0.03}$ & $\rm 2.16^{+0.03}_{-0.03}$ & \textcolor{black}{0.85} &   943.01/845 &   $\rm 5.08^{+0.02}_{-0.02} \times  10^{-9}$ \\ 
\textcolor{black}{GX17+2} & \textcolor{black}{00035340006} & \textcolor{black}{54285.605} & \textcolor{black}{3397} & 
\textcolor{black}{PL} & \textcolor{black}{$\rm 1.85^{+0.01}_{-0.01}$} & \textcolor{black}{$\ast$} & 
\textcolor{black}{$\rm 3.02^{+0.03}_{-0.03}$} & \textcolor{black}{0.85} &   
\textcolor{black}{1820.66/845} &   \textcolor{black}{$\rm 5.35^{+0.02}_{-0.02} \times  10^{-9}$} \\ 
GX17+2 & 00035340007 & 54285.89 & 284.5 & DBB & $\ast$ & $\rm 2.00^{+0.09}_{-0.08}$ & $\rm 2.20^{+0.10}_{-0.09}$ & \textcolor{black}{0.85} &   401.96/465 &   $\rm 3.70^{+0.02}_{-0.08} \times  10^{-9}$ \\
\textcolor{black}{GX17+2} & \textcolor{black}{00035340007} & \textcolor{black}{54285.89} & \textcolor{black}{284.5} & 
\textcolor{black}{PL} & \textcolor{black}{$\rm 2.04^{+0.08}_{-0.08}$} & \textcolor{black}{$\ast$} &  
\textcolor{black}{$\rm 3.18^{+0.16}_{-0.15}$} & \textcolor{black}{0.85} &   
\textcolor{black}{464.77/465} &   \textcolor{black}{$\rm 3.95^{+0.04}_{-0.07} \times  10^{-9}$} \\
 \hline
GX349+2 & 00036690002 & 54286.406 & 4604 & DBB & $\ast$ & $\rm 2.39^{+0.02}_{-0.02}$ & $\rm 0.70^{+0.01}_{-0.01}$ & \textcolor{black}{0.47} &  934.99/868 &   $\rm 8.42^{+0.02}_{-0.03} \times  10^{-9}$ \\
\textcolor{black}{GX349+2} & \textcolor{black}{00036690002} & \textcolor{black}{54286.406} & \textcolor{black}{4604} & 
\textcolor{black}{PL} & \textcolor{black}{$\rm 1.57^{+0.01}_{-0.01}$} & \textcolor{black}{$\ast$} & 
\textcolor{black}{$\rm 1.02^{+0.01}_{-0.01}$} & \textcolor{black}{0.47} &  
\textcolor{black}{3201.71/868} &   \textcolor{black}{$\rm 8.96^{+0.02}_{-0.03} \times  10^{-9}$} \\
\hline
XTEJ1701-462 & 00030383022 & 54283.254 & 3991 & PL & $\rm 1.78^{+0.05}_{-0.05}$ &  $\ast$ & $\rm 2.10^{+0.07}_{-0.07}$ & \textcolor{black}{0.69} &   568.95/580 &   $\rm 9.59^{+0.07}_{-0.11} \times  10^{-9}$ \\
\textcolor{black}{XTEJ1701-462} & \textcolor{black}{00030383022} & \textcolor{black}{54283.254} & \textcolor{black}{3991} & 
\textcolor{black}{DBB} & \textcolor{black}{$\ast$} & \textcolor{black}{$\rm 2.29^{+0.08}_{-0.08}$} &
\textcolor{black}{$\rm 2.10^{+0.07}_{-0.07}$} & \textcolor{black}{0.69} &   
\textcolor{black}{670.24/580} &   \textcolor{black}{$\rm 8.99^{+0.06}_{-0.17} \times  10^{-9}$} \\
\hline
 {\bf Pulsars} & & & & & & & & &  \\
 \hline
 4U1822-37 & 00036691001 & 54276.16 & 1689 & PL & $\rm 0.56^{+0.03}_{-0.03}$ &  $\ast$ & $\rm 0.01^{+0.01}_{-0.01}$ & \textcolor{black}{0.10} &   584.55/445 &   $\rm 4.87^{+0.08}_{-0.06} \times  10^{-10}$ \\
 \textcolor{black}{4U1822-37} & \textcolor{black}{00036691001} & \textcolor{black}{54276.16} & \textcolor{black}{1689} & 
 \textcolor{black}{DBB} & \textcolor{black}{$\ast$} & \textcolor{black}{$\rm 50.98^{\dagger \dagger}$} &
 \textcolor{black}{$\rm 0.04^{+0.01}_{-0.01}$} & \textcolor{black}{0.10} &   
 \textcolor{black}{612.59/445} &   \textcolor{black}{$\rm 4.58^{+0.06}_{-0.04} \times  10^{-10}$} \\
 \hline
SMC X-1 & 00035216002 & 54271.78 & 1583 & PL & $\rm 0.94^{+0.07}_{-0.07}$ &  $\ast$ & $\rm 0.10^{+0.03}_{-0.03}$ &  \textcolor{black}{0.46} &  258.03/188 &   $\rm 7.20^{+0.12}_{-0.20} \times  10^{-10}$ \\ 
\textcolor{black}{SMC X-1} & \textcolor{black}{00035216002} & \textcolor{black}{54271.78} & \textcolor{black}{1583} & 
\textcolor{black}{DBB} &  \textcolor{black}{$\ast$} & \textcolor{black}{$\rm 5.13^{+1.34}_{-0.85}$} & 
\textcolor{black}{$\rm 0.05^{+0.02}_{-0.02}$} &  \textcolor{black}{0.46} &  
\textcolor{black}{268.09/188} &   \textcolor{black}{$\rm 6.97^{+0.01}_{-1.03} \times  10^{-10}$} \\ 
 \hline
 \hline
\enddata
\vspace*{.2in}
\tablenotetext{a}{
\color{black}$\ast$ denotes no value for this parameter. PL = power-law and DBB = disk black body.  $\dagger$ indicates that, because of a low count rate, cstat was used for fitting instead of chi-squared, and the cstat result is  given in place of $\chi^2$. \color{black}$\dagger\dagger$ This parameter is unconstrained by the fit for this model. \color{black}}
\end{deluxetable}
\end{longrotatetable}

\startlongtable
\begin{deluxetable*}{|c|c|c|c|c|}
\tablewidth{5.5in}
\tablecaption{SWIFT XRT Unabsorbed Fluxes \label{tab:xrt_unabs}}
\tablehead{
 \colhead{Source }& \colhead{Obsid }& \colhead{MJD Start }& \colhead{$ \rm F_x$ ($\rm 0.5 - 10~keV$)}& \colhead{$\rm F_x$ ($\rm 2 - 10~keV$)} \\ \colhead{} & \colhead{}  &\colhead{}  & \colhead{$\times 10^{-10}$ ergs cm$^{-2}$ s$^{-1}$} & \colhead{$\times 10^{-10}$ ergs cm$^{-2}$ s$^{-1}$} }
\startdata
\multicolumn{5}{|c|}
{\bf Black Holes} \\
 \hline
\hline 
LMC X-3 & 00037080001 & 54299.574 & 5.51$^{+0.12}_{-0.15}$   & 3.21$^{+0.11}_{-0.11}$  \\  
LMC X-3 & 00037080002 & 54300.773 & 5.75$^{+0.14}_{-0.14}$   & 3.34$^{+0.13}_{-0.13}$  \\  
LMC X-3 & 00037080003 & 54302.254 & 5.77$^{+0.12}_{-0.12}$   & 3.33$^{+0.11}_{-0.11}$  \\  
\hline 
GX 339-4 & 00030953004 & 54281.785 & 2.27$^{+0.13}_{-0.15}$   & 1.53$^{+0.12}_{-0.12}$  \\  
GX 339-4 & 00030953005 & 54285.195 & 3.06$^{+0.08}_{-0.08}$   & 2.12$^{+0.07}_{-0.07}$  \\  
GX 339-4 & 00030953006 & 54288.14 & 3.29$^{+0.10}_{-0.09}$   & 2.20$^{+0.09}_{-0.09}$  \\  
\hline 
1E 1740-29 & 00030960001 & 54284.414 & 3.56$^{+1.33}_{-0.75}$   & 2.10$^{+0.27}_{-0.20}$  \\  
\hline 
GRS 1758-258 & 00030961001 & 54284.547 & 45.3$^{+2.4}_{-2.2}$   & 13.9$^{+0.2}_{-0.2}$  \\  
\hline 
4U 1957+11 & 00030959001 & 54282.547 & 8.39$^{+0.07}_{-0.07}$   & 5.51$^{+0.06}_{-0.06}$  \\  
\hline 
\multicolumn{5}{|c|}
{\bf  Z Sources}\\
\hline 
Cir X-1 & 00030268024 & 54278.906 & 0.63$^{+0.06}_{-0.06}$   & 0.55$^{+0.05}_{-0.06}$  \\  
Cir X-1 & 00030268025 & 54285.062 & 0.136$^{+0.005}_{-0.03}$   & 0.12$^{+0.04}_{-0.03}$  \\  
\hline 
XTE J1701-462 & 00030383022 & 54283.254 & 175.$^{+5.}_{-5.}$   & 108$^{+2.}_{-1.}$  \\  
\hline 
GX 349+2 & 00036690002 & 54286.406 & 103$^{+5.}_{-5.}$   & 80.9$^{+0.4}_{-0.4}$  \\  
\hline 
GX 17+2 & 00035340006 & 54285.605 & 74.3$^{+0.5}_{-0.5}$   & 58.1$^{+0.4}_{-0.4}$  \\  
GX 17+2 & 00035340007 & 54285.89 & 56.8$^{+1.3}_{-0.9}$   & 42.8$^{+1.0}_{-1.0}$  \\  
\hline 
\multicolumn{5}{|c|}
{\bf  Atoll Sources}\\
\hline 
4U 1608-52 & 00030791018 & 54279.91 & 48.9$^{+0.4}_{-0.5}$   & 35.7$^{+0.4}_{-0.4}$  \\  
4U 1608-52 & 00030791019 & 54280.926 & 24.9$^{+0.3}_{-0.3}$   & 18.4$^{+0.3}_{-0.3}$  \\  
4U 1608-52 & 00030791020 & 54282.05 & 33.3$^{+0.4}_{-0.4}$   & 23.2$^{+0.3}_{-0.3}$  \\  
4U 1608-52 & 00030791021 & 54284.863 & 18.0$^{+0.8}_{-0.7}$   & 7.91$^{+0.17}_{-0.17}$  \\  
4U 1608-52 & 00030791022 & 54286.336 & 23.8$^{+0.4}_{-0.4}$   & 13.7$^{+0.2}_{-0.2}$  \\  
4U 1608-52 & 00030791023 & 54288.004 & 9.98$^{+0.22}_{-0.21}$   & 7.67$^{+0.21}_{-0.21}$  \\  
\hline 
GX 354+0 & 00030964001 & 54286.4 & 25.5$^{+0.7}_{-0.6}$   & 15.4$^{+0.2}_{-0.2}$  \\  
\hline 
GX 9+9 & 00030965001 & 54285.74 & 42.7$^{+0.2}_{-0.2}$   & 31.3$^{+0.2}_{-0.2}$  \\  
GX 9+9 & 00030965002 & 54285.137 & 51.0$^{+0.5}_{-0.5}$   & 38.2$^{+0.5}_{-0.5}$  \\  
GX 9+9 & 00030965003 & 54286.613 & 46.5$^{+1.1}_{-1.1}$   & 34.8$^{+1.1}_{-1.1}$  \\  
\hline 
4U 1735-44 & 00035338003 & 54282.32 & 44.6$^{+0.4}_{-0.4}$   & 30.0$^{+0.4}_{-0.4}$  \\  
\hline 
Aql X-1 & 00030796021 & 54279.777 & 3.38$^{+0.15}_{-0.14}$   & 2.03$^{+0.11}_{-0.11}$  \\  
Aql X-1 & 00030796022 & 54283.473 & 0.382$^{+0.062}_{-0.046}$   & 0.129$^{+0.015}_{-0.014}$  \\  
Aql X-1 & 00030796023 & 54286.133 & 0.102$^{+0.022}_{-0.019}$   & 0.0025$^{+0.002}_{-0.001}$  \\  
Aql X-1 & 00030796025 & 54288.55  & 0.052$^{+0.089}_{-0.027}$   & 0.0044$^{+0.0025}_{-0.0016}$  \\  
Aql X-1 & 00030796026 & 54289.773 & 0.024$^{+0.018}_{-0.008}$   & 0.0047$^{+0.0022}_{-0.0015}$  \\  
\hline 
\multicolumn{5}{|c|}
{\bf  Pulsars}\\
\hline 
SMC X-1 & 00035216002 & 54271.78 & 7.45$^{+0.31}_{-0.30}$   & 6.36$^{+0.33}_{-0.33}$  \\  
\hline 
4U 1822-371 & 00036691001 & 54276.16 & 4.87$^{+0.12}_{-0.12}$   & 4.46$^{+0.12}_{-0.12}$  \\  
\hline 
\enddata 
\end{deluxetable*}


The energy range of the XRT is sufficiently broad to allow estimation of the intervening hydrogen column density,
$N_H$, given in column 8 of Table \ref{tab:Swift_xrt}.  Based on $N_H$, the measured flux density can be corrected for absorption giving an estimate for the unabsorbed flux over two bands, (0.5 - 10 keV) and (2 - 10 keV).  These are given for selected sources on Table \ref{tab:xrt_unabs}.

\subsubsection{RXTE ASM}

\FloatBarrier
\startlongtable
\begin{deluxetable}{|l|c|c|c|}
\tablecaption{RXTE-ASM Results \label{tab:ASM}}
\tablehead{
\colhead{Source} & \colhead{ASM Counts} & \colhead{Flux density} & \colhead{Flux 2-10 keV} \\
\colhead{} & \colhead{sec$^{-1}$} & \colhead{mJy at 5 keV} & \colhead {ergs cm$^{-2}$ s$^{-1}$}}
\startdata
\hline
\multicolumn{4}{|c|}
{\bf Black Holes} \\
\hline
LMC X-3 & $<$0.01 &&\\
LMC X-1 & 1.78 & 0.026 & 7  $\cdot 10^{-12}$ \\
XTE J1550-564 & $<$0.01 &&\\
4U1630-47 & $<$0.01 &&\\
GROJ1655-40 & 1.9& $<$0.03 & $<$5.52  $\cdot 10^{-10}$ \\
GX339-4 & $<$0.01 &&\\
1E 1740-29& $<$0.01 &&\\
GRS 1758-258 & 2.73 & 0.04 & 7.76  $\cdot 10^{-10}$\\
V4641 SGR& $<$0.01 &&\\
GRS1915+105 & 89.3 & 1.30 & 2.54  $\cdot 10^{-8}$\\
4U1957+115 &2.09 & 0.03 & 5.94  $\cdot 10^{-10}$\\
\hline
\multicolumn{4}{|c|}
{\bf Z Sources} \\
\hline
LMC X-2 & 1.90	 & 0.28 & 5.41  $\cdot 10^{-10}$\\
Cir X-1& $<$0.01 &&\\
Sco X-1 & 906.51 & 13.15 & 2.57  $\cdot 10^{-7}$\\
GX340+0 & 28.94 & 0.42 & 8.23  $\cdot 10^{-9}$\\
XTE J1701-462 & 10.42 & 0.15& 2.96  $\cdot 10^{-9}$\\
GX349+2 & 46.23 & 0.67	& 0.31  $\cdot 10^{-8}$\\
GX 5-1 & 66.61	 & 0.97 & 1.89  $\cdot 10^{-8}$\\
GX 17+2 & 41.67 & 0.60	 & 0.18  $\cdot 10^{-8}$\\
\hline
\multicolumn{4}{|c|}
{\bf Atoll Sources} \\
\hline
EXO 0748-67 & 0.72 & 	0.01 & 2.04  $\cdot 10^{-10}$\\
2S 0921-630& $<$0.01 &&\\
4U1254-69 & 2.30 & 0.03 & 6.53  $\cdot 10^{-10}$\\
Cen X-4& $<$0.01 &&\\
4U1543-624 & 2.04 & 0.03 & 5.80  $\cdot 10^{-10}$\\
4U1556-60 & 1.17 & 0.17 & 3.32  $\cdot 10^{-10}$\\
4U1608-522 & 12.90 & 0.19 & 3.66  $\cdot 10^{-9}$\\
4U1636-53 & 4.92 &0.71 & 1.40  $\cdot 10^{-9}$\\
4U1658-298 & 2.02 &0.03 & 5.74  $\cdot 10^{-10}$\\
4U1702-429 & 2.54 & 0.04 & 7.23  $\cdot 10^{-10}$\\
4U1705-44 & 8.04 &0.12 & 2.28  $\cdot 10^{-9}$\\
GX9+9 & 25.93 & 0.38 & 7.37  $\cdot 10^{-9}$ \\
GX354+0 & 6.320	 & 0.09 & 1.80  $\cdot 10^{-9}$\\
4U1735-444 & 11.90 & 0.17 & 3.38  $\cdot 10^{-9}$\\
GX 3+1 & 26.88	 & 0.39 & 7.64  $\cdot 10^{-9}$\\
GX 9+1 & 36.18	 & 0.53 & 1.03  $\cdot 10^{-8}$\\
GX 13+1 & 21.88 & 0.32 & 6.20  $\cdot 10^{-9}$\\
Aql X-1 & $<$0.01 &&\\
\hline
\multicolumn{4}{|c|}
{\bf Pulsars} \\
\hline
SMC X-3& $<$0.01 &&\\
SMC X-1 &2.368 & 0.03 & 	6.73  $\cdot 10^{-10}$\\
Vela X-1 &3.79 & 0.06 & 1.08  $\cdot 10^{-9}$\\
Cen X-3 &1.81 & 0.03 & 5.14  $\cdot 10^{-10}$\\
GX 301-2 &1.10 & 0.16 & 3.12  $\cdot 10^{-10}$\\
4U1538-522 &1.60 & 	0.02 & 4.55  $\cdot 10^{-10}$\\
4U1700-37 &5.26 &0.08 & 1.49  $\cdot 10^{-9}$\\
GX 1+4 & $<$0.01 &&\\
Swift J1756.9-2508 & $<$0.01 &&\\
4U1822-37 & 1.56 &0.22 & 4.43  $\cdot 10^{-10}$\\
4U1907+097&1.67 &0.24 & 4.75  $\cdot 10^{-10}$\\
\enddata
\end{deluxetable}
\vspace*{.2in}

The All Sky Monitor \citep[ASM:][]{Levine_etal_1996}
on board NASA's Rossi X-ray Timing Explorer (RXTE) operated continuously for nearly 15 years (1996-2012).  It
consisted of three wide-angle shadow cameras equipped with proportional counters with a total collecting area of 90 cm$^2$.  It
had an energy range of 2-10 keV and covered 80\% of the sky every 90 minutes with a sensitivity of 30 mCrab. The MIT
ASM \footnote{\url{http://xte.mit.edu/ASM_lc.html}} team provides data in three energy bins (1.3-3.0, 3.0-5.0, and 5.0-12.0 keV) sampled 4-8 times
a day, as well as one-day averages.
\color{black}
The 14-year lightcurves with the week of our observations highlighted are plotted in the Appendix (Sec. \ref{sec:appendix}) to show the context of this campaign in the history of each source. 
\color{black}
Averages of the X-ray flux of individual sources were extracted for each day of our campaign.
Only detections of 3 sigma or greater were accepted.  The 1.3-12 keV energy
band count rates were averaged over the week of the campaign to represent the intensity of the source, \color{black} shown on Table \ref{tab:ASM}.  \color{black} The
count rates were converted to flux density in Crabs by conversion factor 75 ASM counts sec$^{-1}$ equals 1 Crab.  For the SEDs, the flux density (in Jy)
was determined from the formula 1 Crab at 5 keV is 1.088 mJy\footnote{\url{https://space.mit.edu/~jonathan/xray_detect.html}}. 
For the SEDs (section \ref{sec:SEDs} below), values are plotted at frequency corresponding to 5 keV
= 1.21$\cdot 10^{18}$ Hz.


\subsubsection{RXTE PCA}

All PCA observations collected data in two standard modes, Standard1, with a time resolution of 0.125s and no energy information, and Std2 with a time resolution of 16s and 129 energy channels (numbered 0-128) from the full energy range. Our analysis used the Standard2 data files, backgrounds, and responses, downloaded from NASA's High Energy Archive Research Center (HEASARC).  Inspection of background light curves showed that the Standard mode extractions did not need to be revised, although for the dim source LMC X-3 we adjusted the scale of the background.

We fit the spectra using XSPEC (Arnaud 1996). We fit neutron star systems with a model of a Comptonized hard spectrum (Titarchuk 1994, COMPTT), plus a multicolor disk blackbody (DISKBB, Makishima et al. 1986), absorbed by a column density usually set to that either found by contemporaneous Swift observations or expected from interstellar gas along the sightline (calculated from HI data). For the absorbing component we used the TBabs model, based on Wilms et al. (2000). In some cases to obtain an acceptable fit (with $\chi^2_\nu < 2$) we allowed the interstellar absorbing column to vary. In some cases we included a Gaussian emission line near 6.4 keV. For 4U 1907+09, we included a cyclotron absorption line. For Sco X-1, we allowed a systematic error of 0.25

Errors on fluxes were determined using the XSPEC cflux multiplicative model component. In cases where this returned only the errors on the unabsorbed flux of model components, we scaled those errors to apply to the absorbed fluxes. Where power law, Comptonized, or disk blackbody components were absorbed not only by interstellar gas but
by a cyclotron feature, we report the flux after the model component has been subject to multiplication by the cyclotron feature.
Resulting model parameters are given on Table 
\ref{tab:PCA-2} for black holes, and 
\ref{tab:PCA-1} for neutron stars, and Table \ref{tab:PCA-3} gives the resulting fluxes (absorbed and unabsorbed). Tables \ref{tab:PCA-2} and \ref{tab:PCA-1} include break-downs for the disk black-body contribution and the contributions of power law or Comptonization components.

\begin{deluxetable*}{lrrrrrrrr}
\tablewidth{6.5in}
\tablecaption{RXTE PCA: Model components for each source, black holes fit with powerlaw and diskBB \label{tab:PCA-2}}
\tablehead{
\colhead{} & \colhead{} & \colhead{} & \colhead{} & \multicolumn{2}{c}{Power Law} & \multicolumn{2}{c}{diskBB}\\
\colhead{Name} & \colhead{Start} & \colhead{Exp.} & \colhead{N$_H$} & \colhead{$\gamma$} & \colhead{norm}
& \colhead{T$_{\rm in}$} & \colhead{norm} & \colhead{$\chi^2$}\\
\colhead{} & \colhead{MJD} & \colhead{(s)} & \colhead{10$^{22}$ cm$^{-2}$} & \colhead{} & \colhead{} & \colhead{keV} & \colhead{} & \colhead{per d.o.f.}}
\startdata
\hline
LMC X-3 
 & 54280.211- &  22926 & 0.03 &\color{black} 1.9$^{+0.1}_{-0.5}$ & \color{black}2.6$\pm0.1 \times10^{-3}$ & \dots\tablenotemark{a} & \dots &\color{black} 37.6/42 \color{black}\\
  & 54287.237\\
 \hline
GX 339-4 & 54280.101-  & 10160 & 0.51 & 1.3$^{+0.1}_{-0.2}$ & 0.10$^{+0.01}_{-0.05}$  & 1.7$^{+0.9}_{-0.3}$ & 2$\pm2$ & 51/37\\
 & 54287.237\\
\hline
GRS 1915+105 & 54286.124 & 1536 & 3$\pm2$ & 2.29$\pm0.04$ & 70$\pm10$ & 3.8$\pm0.1$ & 180$^{+50}_{-40}$ & 52.9/39\\
\hline
\color{black}
4U 1957+11 & \color{black} 54282.593 & \color{black} 6304 & \color{black} 0.08\tablenotemark{b} &\color{black}  -1.6$\pm$1.8 &\color{black}  1$\times 10^{-7}$ ($< 2\times10^{-5}$) &\color{black}  1.39$\pm$0.01 &\color{black}  13.4$\pm$0.7 & \color{black} 38.3/40 \\
\enddata
\color{black}
\vspace*{0.2in}
\color{black}\tablenotetext{a}{There is no evidence for a diskBB component.}  
\color{black}\tablenotetext{b}{This variable fixed not fitted, to match the XRT value.}
\end{deluxetable*}

\startlongtable
\begin{deluxetable*}{|l|r|r|c|c|c|c|c|c|r|}
\tablecaption{RXTE PCA: Model components for each source, neutron stars fit with compTT and diskBB \label{tab:PCA-1}}
\tablewidth{6.5in}
\tablehead{
\colhead{} & \colhead{} & \colhead{} & \multicolumn{3}{c}{CompTT} & \multicolumn{3}{c}{diskBB} & \colhead{} \\
\colhead{Start} & \colhead{Exp} & \colhead{N$_H$ (10$^{22}$)} & \colhead{T$_0$} & \colhead{kT} & \colhead{$\tau_{\rm p}$} & \colhead{norm} & \colhead{T$_{\rm in}$} & \colhead{norm} & \colhead{$\chi^2$}\\
\colhead{MJD} & 
\colhead{s} &
\colhead{cm$^{-2}$} &
\colhead{keV} &
\colhead{keV} &
\colhead{ } &
\colhead{ } &
\colhead{ } &
\colhead{ } &
\colhead{\color{black}per d.o.f. } 
} 
\startdata
\multicolumn{10}{|c|}{\bf Z Sources} \\
\hline
\multicolumn{10}{|l|}{Sco X-1}\\
\hline
54280.172 & 1344 & \nodata & \nodata & \nodata &
 \nodata & \nodata & \nodata & \nodata & \nodata \\
 54284.441 & 1104 & \nodata & \nodata & \nodata &
 \nodata & \nodata & \nodata & \nodata & \nodata \\
            Average & \nodata & 0.30 & 1.0$^{+0.1}_{-0.9}$ & 3.0$\pm0.03$ & 5.7$\pm0.3$ & 8$\pm7$ & 1.2$\pm0.2$ & $<10^4$  &
                     62.2/37\\
\hline
\hline
\multicolumn{10}{|l|}{XTE J1701-461}\\
\hline
\parbox{1in}{Average\\54280.000-\\54285.633} & 4944 & 2.1 & 1.18$^{+0.02}_{-0.1}$ & 6  $< 50$ & $< 7$$^{\dagger}$ & \dots & 3.21$^{+0.08}_{-0.04}$ & 0.17$^{+0.1}_{-0.7}$ &  39.1/37\\
\hline
\hline
\multicolumn{10}{|l|}{GX 349$+$2} \\
\hline
54286.453 & 2704 & 0.7 & 1.0$^{+1}_{-0.1}$ & $<70$ & $<20$ & $>0.02$ & 2.9$\pm0.1$ & 7$^{+2}_{-1}$ & 56.3/36 \\
\hline
\hline
\multicolumn{10}{|l|}{GX 17+2}\\
\hline
54285.668 & 3104 & 2.16 & 0.001 & 2.6$\pm0.2$ & 200$^{\dagger}$ & 0.2$\pm4$ & 1.64$^{+0.04}_{-0.02}$ & 174$\pm1$ & 70.9/44\\
\hline
\hline
\multicolumn{10}{|c|}{\bf Atoll Sources} \\
\hline
\multicolumn{10}{|l|}{EXO 0748-676} \\
\hline
\parbox{1in}{Average\\54286.531-\\54286.844}  & 9088  & 0.101 & 1.2$^{+5}_{-0.3}$ & $<70$ & 0.01$^{\dagger}$\tablenotemark{a}  & $<0.8$ & 7.8$^{+1}_{-0.3}$ & 7$^{+1}_{-2}\times10^{-3}$ & 36.0/44\\
\hline
\hline
\multicolumn{10}{|l|}{4U 1254-690}\\
\hline
54280.348 & 6000 & 2$^{+3}_{-1}$ & 0.001 & 2.4$^{0.2}_{-0.1}$ & $(>100)$ &
$>0.0086$ & 1.6$\pm0.1$ & 8$\pm2$  & 36.5/37\\
\hline
\hline
\multicolumn{10}{|l|}{4U 1608-52}\\
\hline 
54280.449 & 1072 & 0.84 & \color{black}0.96$\pm0.01$ & \color{black}2.77$^{+0.04}_{-0.04}$ &
\color{black}4.7$^{+0.2}_{-0.2}$ &\color{black} 0.67$^{+0.03}_{-0.02}$\color{black} & \dots & \dots
& 35.8/38\\ 
 54280.766 & 1104 & 0.84 & 8$\pm8$ & 2.20$^{+1.4}_{-0.04}$ &\color{black} $>$50 & 0.4$^{+2}_{-0.2}$ & 1.96$\pm0.03$ & 34$\pm2$ & 37.2/38\\
 54285.730 & 1680 & 0.84 & \color{black} 7 ($<$13) &\color{black}  2.1$^{+1.0}_{-0.05}$  & 200$^{\dagger}$ &\color{black}  0.4$^{+1.2}_{-0.2}$ &
\color{black} 2.12$\pm 0.05$ & \color{black} 25$\pm3$ &\color{black} 36.8/38\\ 
\hline
\hline
\multicolumn{10}{|l|}{4U 1636-53}\\
\hline
54285.731 &  3200 & 0.26 & 0.95$\pm0.02$ & 2.44$^{+0.05}_{-0.04}$ & 5.0$\pm0.2$ & 0.75$^{+0.03}_{-0.04}$ & $<0.9$
 & \dots & 34.0/38\\
\hline
\hline
\multicolumn{10}{|l|}{4U 1705-44}\\
\hline 
 54284.488 & 2688 & 0.808 & 0.001 & $<3$ & $>180$ &\color{black} $<30$ & 1.60$\pm0.01$ & $>30$ &
                47.2/36\\
54280.645 & 2912 & 0.808 & $<20$ & 2.54$^{+2}_{-0.1}$ & 200$^{\dagger}$ & $<1$ & 1.88$\pm0.02$ & 12.1$\pm0.1$ &
      30.4/35\\
average &  \nodata &  0.808 & 1.0$^{+0.2}_{-0.3}$ & 2.6$\pm0.1$ & 6$^{+4}_{-1}$ & 0.2$\pm0.1$ &\color{black} 1.3$^{+0.2}_{-1.3}$ & $<2000$ & 58.7/ 36\\
\hline
\hline
\multicolumn{10}{|l|}{GX 9+9}\\
\hline
54285.730 & 3200 & 0.21 & 7.11$\pm0.06$ & 2.09$^{+0.01}_{-0.08}$ & 200$^{\dagger}$ & $<0.4$ &
1.89$\pm0.01$ & 39$\pm1$ & 43.4/38\\
 \hline
\hline
\multicolumn{10}{|l|}{4U 1735-444}\\
\hline
\parbox{1in}{Average\\54280.699-\\542825.273}  & 
25984 & 0.3 &\color{black} 1.02$\pm$0.02 & 2.8$\pm0.1$ & 7$^{+5}_{-1}$ & 0.2$\pm0.2$ & $>0.2$ &\color{black} 16$\pm$3 & 25.2/38\\ 
\hline
\hline
\multicolumn{10}{|l|}{Aql X-1 }\\
\hline
 54281.223 & 1184 & 0.45 & 0.7$\pm0.2$ & 470$^{+30}_{-140}$ & 0.03$^{+0.02}_{-0.01}$ &
6$\pm2$(-5) & \nodata & \nodata & 44.4/ 40\\
 54282.133 & 1264 & 0.63 & $<1$ & $<200$ & 2.5$^{+14}_{-2.1}$ & $<0.01$ & \nodata
& \nodata &  25.4/40\\
 54283.516 & 2128 & 0.63 & $<1.3$ & 20$^{+100}_{-10}$ & $<5$ & $<0.002$ & \nodata & \nodata &  39.1/40\\ 
54284.230 & 1104 & 0.63 & $<2.3$ &\color{black} 500$\pm50$  & $<100$ & $<0.02$ & \nodata & \nodata & 13.1/40 \tablenotemark{c}\\ 
 54285.145 & 1072 & 0.96 & $<5$ & $>200$ & $<200$ & $<0.1$ & \nodata & \nodata & 23.4/40\\
\hline
\hline
\multicolumn{10}{|c|}{\bf Pulsars} \\
\hline
\multicolumn{10}{|l|}{4U 1907+09\tablenotemark{b}}\\
\hline
54282.398 & 1904 & 1.70 & 90 & 5.9$^{+1.5}_{-0.5}$ & 200$^{\dagger}$ & 35$^{+50}_{-30}$ & 3.7$^{+0.8}_{-0.2}$ &
           0.11$^{+0.2}_{-0.05}$ & 34.0/34\\
54284.363 & 1920 & 1.70 & 0.001 & 30$\pm10$ & 200$^{\dagger}$  & $<0.1$ & 2$\pm1$ & $<0.5$  & 23.2/35\\
 54281.223 & 1760 & 1.70 & $<1.4$ & 20$^{+100}_{-10}$  & 3$\pm3$ & $<0.08$ & \nodata & \nodata & 32.6/39\\
 Average & \nodata & 1.70 & 85 & 7$^{+1}_{-2}$ & 200$^{\dagger}$ & 7.1$^{+12}_{-7}$ & 3.7$^{+0.5}_{-0.1}$ & 0.05$^{+0.06}_{-0.02}$ & 30.2/34\\
\hline
\hline
\enddata
\vspace*{.2in}
\color{black}\tablenotetext{a}{ ${\dagger}$ indicates that the best fit is with this parameter at the edge of its allowed range.}  
\tablenotetext{b}{This source contains a cyclotron line.  Leaving the parameters
of the cyclotron line free along with the compTT high energy component of
the spectrum results in T0 for compTT being very poorly constrained.}
\tablenotetext{c}{The low value of $\chi^2$ results from large error bars on the spectrum.}
\color{black}
\end{deluxetable*}

\begin{sidewaystable}
\caption{RXTE PCA: Fluxes, absorbed and unabsorbed \label{tab:PCA-3}}
\resizebox{9in}{!}{%
\begin{tabular}{|lr|rr|rr|rrr|rrr|}
\hline
&& \multicolumn{2}{|c}{Total} & \multicolumn{2}{|c}{Total, N$_H$=0} & \multicolumn{2}{|c}{diskBB} & N$_H$=0
& \multicolumn{2}{|c}{PL\footnote{PL is power law or Comptonization component}} & N$_H$=0 \\
Source & Date & 2--10 keV & 10--20 keV & 2--10 keV & 10--20 keV & 2--10 keV & 10--20 keV & 2--10 keV\footnote{\color{black}The N$_H$=0 corrections for 10--20 keV band in the separate components are small, and similar to the correction to the total flux.\color{black}} & 2--10 keV & 10--20 keV & 2--10 keV \\

\hline
\multicolumn{12}{|l|}{\bf Black Holes}\\
\hline

LMC X-3 & all & 1.8$<8\times 10^{-11}$ & 3.8$\pm0.4\times 10^{-12}$ & 1.8$<8\times10^{-11}$ & 3.8$\pm0.4\times10^{-12}$ & $<4\times10^{-10}$ & $<6\times10^{-10}$ & $<4\times10^{-10}$  
& 7$\pm1\times 10^{-12}$ & 3.8$\pm0.04\times 10^{-12}$ & 7$\pm1\times 10^{-12}$ 
\\
\hline
 GX 339-4 & all & 7.6$^{+0.4}_{-0.3}\times 10^{-10}$ & 7.49$\pm0.06\times 10^{-10}$ & 1.3$\pm0.1\times 10^{-9}$ & 7.8$\pm0.1\times 10^{-10}$ & 2.9$\pm0.6\times 10^{-10}$ & 4$\pm2\times 10^{-11}$ & 5$\pm1\times 10^{-10}$ 
 & 4.7$\pm0.5\times 10^{-10}$ & 7.1$\pm0.2\times 10^{-10}$ &  7.5$\pm0.8\times 10^{-10}$ 
\\
\hline
GRS 1915+105 & 54285 & 4.7$^{+0.1}_{-0.3}\times 10^{-8}$ & 5.29$\pm0.03\times 10^{-9}$ & 6.8$\pm0.4\times 10^{-8}$ & 5.39$\pm0.03\times 10^{-9}$ & 2.1$\pm0.1\times 10^{-8}$ & 3.7$\times 10^{-9}$) & 2.7$\pm0.1\times 10^{-8}$ 
& 2.5$\pm0.1\times 10^{-8}$ & 1.59$\times 10^{-9}$ & 4.1$\pm0.2\times10^{-8}$ 
\\
\hline
4U 1957+11 & \color{black} all &\color{black}  6.38$\pm0.09\times10^{-10}$ &\color{black}  1.6$^{+0.03}_{-0.1}\times10^{-11}$ &\color{black}  6.44$\pm0.09\times10^{-10}$ &\color{black}   1.6$^{+0.03}_{-0.1}\times10^{-11}$ &\color{black}  6.38$\pm0.09\times10^{-10}$  &\color{black}  1.40$\pm0.05\times10^{-11}$ &\color{black}  6.44$\pm0.09\times10^{-10}$ 
&\color{black}  2$^{+12}_{-1}\times10^{-13}$ &\color{black}  2$\pm1\times10^{-12}$ &\color{black}  2$^{+12}_{-1}\times10^{-13}$ 
\\   

\hline
\multicolumn{12}{|l|}{\bf Z Sources}\\
\hline

Sco X-1 & all & 1.172$^{+0.004}_{-0.002}\times 10^{-7}$ & 3.115$\pm0.004\times 10^{-8}$ & 1.20$^{+0.03}_{-0.02}\times 10^{-7}$ & 3.120$\pm0.003\times 10^{-8}$ & 1.91$^{+0.04}_{-0.05}\times 10^{-8}$ & 1.92$^{+0.04}_{-0.03}\times 10^{-10}$
& 2.00$\pm0.05\times 10^{-8}$ 
& 1.0$\pm0.1\times 10^{-7}$ & 3.10$^{+0.01}_{-0.04}\times 10^{-8}$ & 1.0$\pm0.1\times 10^{-7}$  
\\
\hline
XTE J1701-461 & all &\color{black} 3.31$^{+0.1}_{-0.03}\times10^{-9}$ & \color{black} 8.69$\pm0.04\times10^{-10}$  & \color{black} 3.85$^{+0.1}_{-0.03}\times10^{-9}$ & \color{black} 8.69$\pm0.04\times10^{-10}$ & \color{black} 2.09$^{+0.06}_{-0.05}\times10^{-9}$ & \color{black} 8.17$^{+0.04}_{-0.03}\times10^{-10}$ &  \color{black} 2.41$^{+0.06}_{-0.05}\times10^{-9}$ 
&\color{black}  1.2$^{+2}_{-0.1}\times 10^{-9}$ &\color{black}  5$^{+20}_{-1} \times 10^{-11}$ &\color{black}1.4$^{+2}_{-0.1}\times 10^{-9}$ 
\\
\hline
GX 349+2 & 54286 & 1.32$^{+0.03}_{-0.02}\times10^{-8}$ & 2.51$\pm0.01\times10^{-9}$ & 1.39$^{0.03}_{-0.02}\times 10^{-8}$ & 2.52$\pm0.01\times10^{-9}$
& 6.7$^{+0.1}_{-0.4}\times 10^{-9}$ & 1.93$^{+0.01}_{-0.04}\times 10^{-9}$ & 7.1$^{+0.1}_{-0.5}\times 10^{-9}$ 
& 6.1$\pm0.3 \times 10^{-9}$ & 5.9$^{+1.3}_{-0.5}\times 10^{-10}$ & 6.5$\pm0.3\times 10^{-9}$ 
\\
\hline
GX 17+2 & 54285 & 1.64$\pm0.03\times 10^{-8}$ & 2.82$\pm0.01 \times 10^{-9}$ & 2.00$^{+0.04}_{-0.03}\times 10^{-8}$ & 2.86$\pm0.01 \times 10^{-9}$ & 1.4$\times 10^{-8}$ & 7.8$\times 10^{-10}$ & 1.7$\times 10^{-8}$ 
& 2.6$\times 10^{-9}$ & 2.0$\times 10^{-9}$ & 2.8$\times 10^{-9}$ 
\\

\hline
\multicolumn{12}{|l|}{\bf Atoll Sources}\\
\hline

EXO 0748-676 & all & 2.45$^{+0.07}_{-0.04}\times 10^{-10}$ & 1.80$\pm0.01\times10^{-10}$ & 2.45$^{+0.07}_{-0.04}\times10^{-10}$ & 1.80$\pm0.01 \times10^{-10}$ & 1.67$^{+0.05}_{-0.40}\times 10^{-10}$ & 1.76$^{+0.02}_{-0.30}\times 10^{-10}$ &
1.67$^{+0.05}_{-0.40}\times 10^{-10}$ 
& 8.0$^{+0.4}_{-0.3}\times10^{-11}$ & 4$^{+20}_{-1}\times 10^{-12}$ & 8.0$^{+0.4}_{-0.3}\times 10^{-11}$ 
\\
\hline
4U 1254-690 & 54280 & 7$\pm1\times10^{-10}$ & 1.2$\pm0.2\times 10^{-10}$ & 8.5$^{+0.7}_{-0.1}\times 10^{-10}$ & 1.21$\pm0.04\times 10^{-10}$ & 5.6$\times 10^{-10}$ & 2.9$\times 10^{-11}$ & 7.1$\times 10^{-10}$ 
& 1.3$>0.2\times 10^{-10}$ & 9.1$\times 10^{-11}$ & 1.4$\pm0.2\times 10^{-10}$ 
\\
\hline
4U 1608-52 & 54280 & 6.6$\pm0.2\times10^{-9}$ & 1.25$\pm0.01\times 10^{-9}$ & 7.0$^{+0.4}_{-0.1}\times10^{-9}$ & 1.26$\pm0.01\times 10^{-9}$ & $<3\times 10^{-10}$ & $8\times10^{-13}$ & \dots 
& 6$\pm1\times10^{-9}$ & 1.25$\pm 0.01\times10^{-9}$ &  6.5$\pm1.0\times10^{-9}$ 
\\

4U 1608-52 & 54280 & 7.0$\pm0.1\times10^{-9}$ & 1.22$\pm0.01\times10^{-9}$ & 7.5$\pm0.1\times10^{-9}$ & 1.23$\pm0.01\times10^{-9}$ & 6.7$\pm0.1\times10^{-9}$ & 6.6$\pm0.3\times10^{-10}$ & 7.3$\pm0.1\times10^{-9}$ 
& 2.6$^{+0.6}_{-0.2}\times10^{-10}$ & 5.6$^{+0.4}_{-0.2}\times10^{-10}$ & 2.7$^{+0.6}_{-0.2}\times10^{-10}$ 
\\

4U 1608-52 & 54281 & 7.2$\pm0.1\times10^{-9}$ & 1.58$\pm0.01\times10^{-9}$ & 7.8$\pm0.1\times10^{-9}$ & 1.6$\times10^{-9}$ & 6.77$\pm0.04\times10^{-9}$ & 8.7$\pm0.7\times 10^{-10}$ & 7.69$\pm0.04\times 10^{-9}$ 
& 3.9$^{+0.9}_{-0.6}\times 10^{-10}$ & 7.2$^{+0.8}_{-0.6}\times 10^{-10}$ & 4.0$\times 10^{-10}$ 
\\
\hline
4U 1636-53 & all & 6.5$\pm0.1\times10^{-9}$ & 9.52$^{+0.05}_{-0.04}\times10^{-10}$ & 6.6$\pm0.1\times10^{-9}$ & 9.52$^{+0.05}_{-0.04}\times 10^{-10}$ & 6.5$\pm0.1\times10^{-9}$ & 9.52$^{+0.05}_{-0.04}\times10^{-10}$ & \dots 
& \dots & \dots & \dots 
\\
\hline
4U 1705-44 & 54280 & 3.55$\pm0.01\times 10^{-9}$ & 6.36$\pm0.02\times 10^{-10}$ & 3.83$\pm0.01\times 10^{-9}$ & 6.40$\pm0.01\times 10^{-10}$ & 2.7$\times 10^{-9}$ & 1.2$\times 10^{-10}$ & 3.0$\times 10^{-9}$ 
& 1.1$\times 10^{-9}$ & 5.2$\times 10^{-10}$ & 8.5$\times 10^{-10}$ 
\\

4U 1705-44 & 54284 & 2.2$\pm0.1\times 10^{-9}$ & 4.0$\pm0.1\times 10^{-10}$ & 2.4$\pm0.2\times 10^{-9}$ & 4.1$\pm0.1\times 10^{-10}$ & 2.00$\pm0.01\times 10^{-9}$ & 1.70$^{+0.01}_{-0.02}\times 10^{-10}$ & 2.16$\pm0.01\times 10^{-9}$ 
&  1.8$^{+0.3}_{-0.6}\times 10^{-10}$  & 2.35$^{+0.05}_{-0.02}\times 10^{-10}$ & 1.9$^{+0.3}_{-0.6}\times 10^{-10}$ 
\\

4U 1705-44 & all & 2.7$\pm0.1\times 10^{-9}$ & 5.2$\pm0.3\times 10^{-10}$ & 2.9$\pm0.1\times 10^{-9}$ & 5.2$\pm0.3\times 10^{-10}$ & 6.2$\pm0.2\times 10^{-10}$ & 9.0$\pm0.4\times 10^{-12}$ & 6.9$\pm0.3\times 10^{-10}$ 
& 2.0$\pm0.1\times 10^{-9}$ & 5.1$^{+0.1}_{-0.8}\times 10^{-10}$ & 2.2$\pm0.1\times 10^{-9}$ 
\\
\hline
GX 9+9 & 54285 & 7.2$\pm0.1\times 10^{-9}$ & 9.5$^{+0.7}_{-0.3}\times 10^{-10}$ & 7.4$\pm0.14\times 10^{-9}$ & 9.5$\pm0.1\times 10^{-10}$ & 7.01$\pm0.04\times 10^{-9}$ & 5.7$^{+0.3}_{-0.5}\times 10^{-10}$& 7.16$\pm0.04\times 10^{-9}$ 
& 2.3$^{+0.7}_{-0.3}\times 10^{-10}$ & 3.8$^{+0.5}_{-0.3}\times 10^{-10}$ & 2.3$^{+0.7}_{-0.3}\times 10^{-10}$ 
\\
\hline
4U 1735-444 & all & 3.2$^{+0.5}_{-0.1}\times 10^{-9}$ & 1.02$\pm0.01\times 10^{-9}$ & 3.3$\pm0.2\times 10^{-9}$ & 1.02$\pm0.01\times 10^{-9}$ & 7.3$\times 10^{-10}$ & 1.6$\times10^{-11}$ & 7.6$\times10^{-10}$ 
& 2.5$(>1.5)\times 10^{-9}$ & 1.0$\pm0.1\times 10^{-9}$ & 2.5$(>1.5)\times 10^{-9}$ 
\\
\hline
Aql X-1 &  54281 & 9.5$\pm1.1\times 10^{-11}$ & 5.7$^{+0.4}_{-0.2}\times 10^{-11}$ & 9.9$\pm1.2\times 10^{-11}$ & 5.7$\times 10^{-11}$ &\dots &\dots &\dots 
& 9.5$\pm1.1\times 10^{-11}$ & 5.7$^{+0.4}_{-0.2}\times 10^{-11}$ &9.5$\pm1.1 \times 10^{-11}$  
\\

Aql X-1 & 54282 &  5.3$\pm0.5\times 10^{-11}$ & 3.7$\pm0.2\times 10^{-11}$ & 5.6$^{+0.7}_{-0.6}\times 10^{-11}$ & 3.8$^{+0.2}_{-0.3}\times 10^{-11}$ & \dots & \dots & \dots 
& 5.3$\pm0.5\times 10^{-11}$ & 3.7$\pm0.2\times 10^{-11}$ & 5.6$^{+0.7}_{-0.6}\times 10^{-11}$ 
\\

Aql X-1 & 54283 &  1.5$^{+0.4}_{-0.2}\times 10^{-11}$ & 1.7$\pm0.2\times 10^{-11}$ & 1.6$^{+0.4}_{-0.3}\times 10^{-11}$ & 1.7$\pm2\times 10^{-11}$ & \dots & \dots & \dots 
& 1.5$^{+0.4}_{-0.2}\times 10^{-11}$ & 1.7$\pm0.2\times 10^{-11}$ &
 1.6$^{+0.4}_{-0.3}\times 10^{-11}$ 
\\

Aql X-1 & 54284 &   5.7$\pm0.2\times 10^{-12}$ & 8.8$\pm0.1\times 10^{-12}$ & 5.9$^{+0.3}_{-0.1}\times 10^{-12}$ & 8.8$\pm0.2\times 10^{-12}$ & \dots & \dots & \dots 
& 5.7$\pm0.2\times 10^{-12}$ & 8.8$\pm0.1\times 10^{-12}$ & 5.9$^{+0.3}_{-0.1}\times 10^{-12}$ 
\\

Aql X-1 & 54285 & 3.7$^{+2}_{-1}\times 10^{-12}$ & 6$\pm 1 \times 10^{-12}$ & 3.9$^{+4}_{-3}\times 10^{-12}$ & 6$^{+1}_{-2}\times 10^{-12}$  &  \dots &  \dots & \dots 
& 3.7$^{+2}_{-1}\times 10^{-12}$ & 6$\pm1\times 10^{-12}$ & 3.9$^{+4}_{-3}\times 10^{-12}$ 
\\

\hline
\multicolumn{12}{|l|}{\bf Pulsars}\\
\hline

4U 1907+09 & 54280 & 2.32$\pm5\times 10^{-10}$  & 1.5$^{+0.3}_{-0.1}\times 10^{-10}$ & 2.6$\pm0.1\times 10^{-10}$ & 1.6$^{+0.3}_{-0.2}\times 10^{-10}$
& 1.8$\pm0.1\times 10^{-10}$ & 2.6($^{+0.9}_{-0.4}\times 10^{-11}$ & 2.1$\pm0.1\times 10^{-10}$ 
& 3$\pm1\times 10^{-11}$ & 1.3$^{+0.3}_{-0.5}\times 10^{-10}$& 3$\pm1\times 10^{-11}$ 
\\

4U 1907+09 & 54282 & 3.8$^{+0.2}_{-0.4}\times 10^{-11}$ & 1.8$^{+0.4}_{-0.1}\times 10^{-11}$ & 4.2$\pm0.3\times 10^{-11}$ & 1.8$^{+0.7}_{-0.1}\times 10^{-11}$ &
 2.1$^{+0.3}_{-0.4}\times 10^{-11}$ & 1.9$^{+0.2}_{-0.6}\times 10^{-12}$ & 3.1$^{+0.4}_{-0.5}\times 10^{-11}$ 
& 8.4$\pm0.5\times 10^{-12}$ & 1.8$\pm0.8\times 10^{-11}$ & 1.1$>0.6\times 10^{-11}$ 
\\

4U 1907+09 & 54285 & 1.7$^{+0.5}_{-0.4}\times 10^{-11}$ & 1.8$\pm0.3\times 10^{-12}$ & 2.0$\pm0.6\times 10^{-11}$ & 1.8$\pm0.3\times 10^{-12}$ & 0 & 0 & 0 
& 1.7$^{+0.5}_{-0.4}\times 10^{-11}$ & 1.8$\pm0.3\times 10^{-12}$ & 2.0$\pm0.6\times 10^{-11}$ 
\\

4U 1907+09 & all & 1.11$^{+0.14}_{-0.03}\times 10^{-10}$ & 6.8$^{+2}_{-0.3}\times 10^{-11}$ & 1.24$^{+0.16}_{-0.04}\times 10^{-10}$& 6.9$^{+2}_{-0.1}\times 10^{-11}$ & 1.1$^{+0.08}_{-0.02}\times 10^{-10}$ &
5.5$\times 10^{-11}$ & 1.24$^{+0.08}_{-0.02}\times 10^{-10}$ 
& 2.2$^{+3}_{-0.5}\times 10^{-11}$ & 7.6$^{+1.4}_{-6.6}\times 10^{-10}$ & 2.3$^{+3}_{-0.5}\times 10^{-11}$ 
\\
\hline
\end{tabular}
}
\end{sidewaystable}

\subsubsection{Swift BAT observations}\label{sec:BAT}

The Burst Alert Telescope (BAT) is a highly sensitive, wide field, coded aperture imaging instrument with a 1.4 steradian field-of-view (half coded) with CdZnTe detectors. The energy range is $\rm 15-150 ~keV$ for imaging with a non-coded response up to $\rm 500 ~keV$, \color{black} but here we use only 15-50 \color{black}keV \color{black} as the higher energy values are less reliable. We divide the flux in this energy range by the bandwidth to get the flux density at the nominal band center of 32.5 \color{black}keV. \color{black}  Further information on the BAT is given by \cite{Barthelmy05}.  
For the model a nominal column density of $\rm 1 \times 10^{21} ~cm^{-2}$ is used for $\rm N_{H}$ \footnote{\color{black} For these energy bands the estimated flux is insensitive to absorption.  At $\rm 1 \times 10^{21} ~cm^{-2}$ there is no difference between the absorbed and unabsorbed fluxes.  For $\rm 1 \times 10^{24} ~cm^{-2}$ the difference between the absorbed and unabsorbed fluxes is only 2\%.\color{black}}, and a power-law with index=1.5 for the emission spectrum.  
This provides a flux in $\rm ergs~cm^{-2}~s^{-1}$, 
that is converted to a flux density by dividing by the bandwidth in Hz.
For the SEDs (section \ref{sec:SEDs} below) the values are plotted at the
central energy of  $\rm 32.5 ~keV$ which corresponds to 7.86$\cdot 10^{18}$ Hz.  Flux densities and upper limits from the BAT are given on Table \ref{tab:BAT}.


\FloatBarrier
\color{black}
\startlongtable
\begin{deluxetable}{|l|c|c|}
\tablecaption{Swift BAT Results \label{tab:BAT}}
\tablehead{
\colhead{Source} & \colhead{BAT Counts}  & \colhead{Flux at 31.5 keV} \\
\colhead{} & \colhead{sec$^{-1}$} &  \colhead {ergs cm$^{-2}$ s$^{-1}$}}
\startdata
\hline
\multicolumn{3}{|c|}
{\bf Black Holes} \\
\hline
LMC X-3 & $<$0.01 &$<$1.0$\times 10^{-10}$\\
LMC X-1 & $<$0.01 &$<$1.0$\times 10^{-10}$\\
XTE J1550-564 & $<$0.01 &$<$1.0$\times 10^{-10}$\\
4U1630-47 & $<$0.01 &$<$1.0$\times 10^{-10}$\\
GROJ1655-40 & $<$0.01 &$<$1.0$\times 10^{-10}$\\
GX339-4 & 0.0063 & 6.49$\times 10^{-11}$\\
1E 1740-29& 0.0032 &3.30$\times 10^{-11}$\\
GRS 1758-258 & 0.0054 & 5.56$\times 10^{-11}$\\
V4641 SGR& $<$0.01 &$<$1.0$\times 10^{-10}$\\
GRS1915+105 & 0.035 & 3.63$\times 10^{-10}$\\
4U1957+115 &$<$0.0023 & $<$2.4$\times 10^{-11}$\\
\hline
\multicolumn{3}{|c|}
{\bf Z Sources} \\
\hline
LMC X-2 & $<$0.01 &$<$1.0$\times 10^{-10}$\\
Cir X-1& $<$0.01 &$<$1.0$\times 10^{-10}$\\
Sco X-1 & 0.3575 &  3.68$\times 10^{-9}$\\
GX340+0 & $<$0.01 &$<$1.0$\times 10^{-10}$\\
XTE J1701-462 & 0.0058 & 5.98$\times 10^{-11}$\\
GX349+2 & 0.018 & 1.85$\times 10^{-10}$\\
GX 5-1 & 0.028 & 2.84$\times 10^{-10}$\\
GX 17+2 & 0.0142 & 1.46$\times 10^{-10}$\\
\hline
\multicolumn{3}{|c|}
{\bf Atoll Sources} \\
\hline
EXO 0748-67 & 0.0088 & 9.07$\times 10^{-11}$\\
4U1254-69 & $<$0.0033 & $<$ 9.1$\times 10^{-11}$\\
Cen X-4&$<$0.01 &$<$1.0$\times 10^{-10}$\\
4U1543-624 &$<$0.01 &$<$1.0$\times 10^{-10}$\\
4U1608-522 & 0.0074 & 7.63$\times 10^{-11}$\\
4U1636-53 & 0.0082 & 8.45$\times 10^{-11}$\\
4U1658-298 & $<$0.01 &$<$1.0$\times 10^{-10}$\\
4U1702-429 & 0.0026 & 2.68$\times 10^{-11}$\\
4U1705-44 & 0.0023 & 2.37$\times 10^{-11}$\\
GX9+9 & 0.0066 & 6.80$\times 10^{-11}$\\
GX354+0 & 0.0176 & 1.81$\times 10^{-10}$\\
4U1735-444 & 0.0082 & 8.45$\times 10^{-11}$\\ 
GX 3+1 &  0.0062 & 6.39$\times 10^{-11}$\\
GX 9+1 & 0.0071 & 7.32$\times 10^{-11}$\\
GX 13+1 & 0.0035 & 3.61$\times 10^{-11}$\\
Aql X-1 & $<$0.01 &$<$1.0$\times 10^{-10}$\\
\hline
\multicolumn{3}{|c|}
{\bf Pulsars} \\
\hline
SMC X-3 & $<$0.01 &$<$1.0$\times 10^{-10}$\\
SMC X-1 & $<$0.002 & $<$2.27$\times 10^{-11}$\\
Vela X-1 & 0.0706 & 7.28$\times 10^{-10}$\\
Cen X-3 & 0.0144 & 1.48$\times 10^{-10}$\\
GX 301-2 & 0.0451 & 4.65$\times 10^{-10}$\\
4U1538-522 & 0.0054 & 5.56$\times 10^{-11}$\\
4U1700-37 & 0.0269 & 2.77$\times 10^{-10}$\\
GX 1+4 & 0.0173 & 1.78$\times 10^{-10}$\\
Swift J1756.9-2508 &$<$0.01 &$<$1.0$\times 10^{-10}$\\
4U1822-37 & 0.0084 & 8.66$\times 10^{-11}$\\
4U1907+097&0.0049 & 5.05$\times 10^{-11}$\\
\hline
\enddata
\end{deluxetable}

\section{Spectral Energy Distributions \label{sec:SEDs}}

Here we summarize the results and
interpret the SEDs 
of each of the X-ray binaries observed.  While most sources have detections at multiple wavelengths during the week of our campaign, we only plot the SEDs of those that had at least one X-ray spectrum, rather than just single frequency observations.  Overall, 23 individual sources have X-ray spectra: 16 from 2-20 keV obtained with the RXTE PCA, and 16 from 0.2 to 10 keV obtained with the Swift XRT, with 9 detected by both instruments.  

To determine the states of individual sources during our week of observations we obtained the XTE Mission-Long Source Catalog file \footnote{\url{https://heasarc.gsfc.nasa.gov/docs/xte/recipes/mllc_start.html}} for each source from the HEASARC archive \footnote{\url{https://heasarc.gsfc.nasa.gov/cgi-bin/W3Browse/w3browse.pl}}\color{black}, when available.\color{black} Each row in the file is a separate RXTE observation (obsid), drawn from all proposals that included the source as a target.  For each observation this gives the time of the observation, PCU Std 2 data rates (in three bands: 2-4 keV, 4-9 keV, 9-20 keV), and errors.  The PCA data are normalized to 1 PCU (ie., rates are in units of c/s/PCU).  There is a single value for each pointed RXTE observation. \color{black}For sources
for which these files exist we created a color-intensity (CI) \color{black} plot for each source by summing the three bands and determining an X-ray color or hardness ratio (HR, normalized to $\pm$ 1) from the 2-4 keV and 9-20 keV bands. \color{black}
The X-ray color is expressed as
a ratio of the difference divided by the sum of bands at 9-20 kev (p920) and 2-4 keV (p24).  The color is shown on the x-axis, while the intensity over the full range 2 to 20 keV is plotted on the y-axis.
\color{black}
 When Mission-Long Source Catalog files are not available for a source the
CI diagrams are made using RXTE ASM data which are provided in three energy
bands: a (1.5-3 keV), b (3-5 keV), and c (5-12 keV). Color is determined using bands
a and c, and intensity is determined by the sum of all three bands.
\color{black}RXTE observations \color{black}(either PCA or ASM) made \color{black}
during our week of observations are denoted in red. \color{black} Blue points indicate
the range of values measured for the source at other epochs using the same instrument.

\color{black} Sources are described as persistent
if they are continuously active and transient if they have periods of
quiescence when little or no emission is detected.  Persistent sources can be variable,
i.e. show changes in intensity, or steady, i.e. show a nearly constant count rate.
\color{black}
ASM light curves for all sources for most of the 15 years that the RXTE was active
are shown in Sec. \ref{sec:appendix}.\color{black}

\subsection{Black holes (11 sources; 6 with SEDs) \label{subsec:BH_seds}}

\begin{figure}[ht]
\hspace{.2in} \includegraphics[width=3.4in]{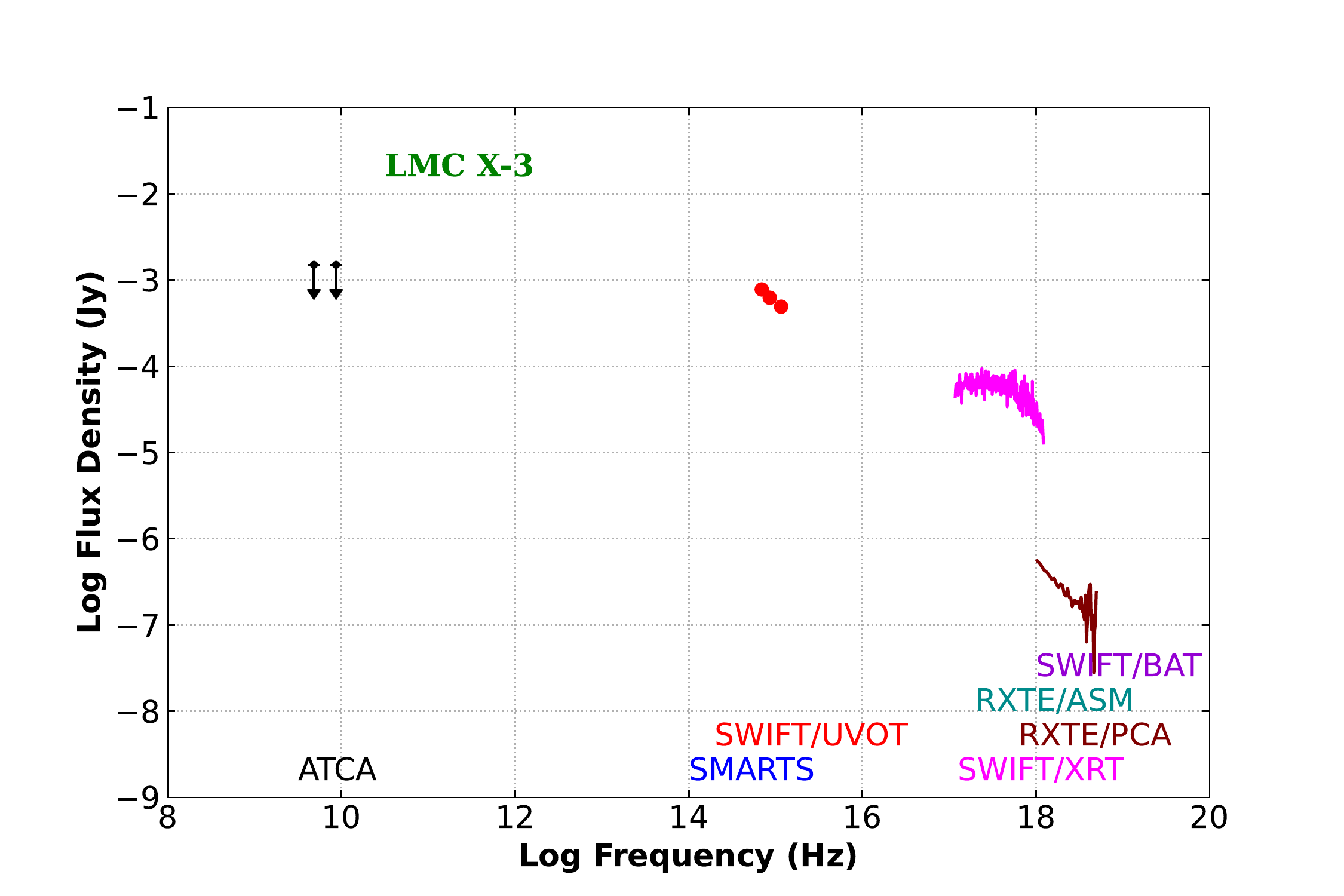}
\hspace{-.5in}
\hspace{.5in} \includegraphics[width=3in]{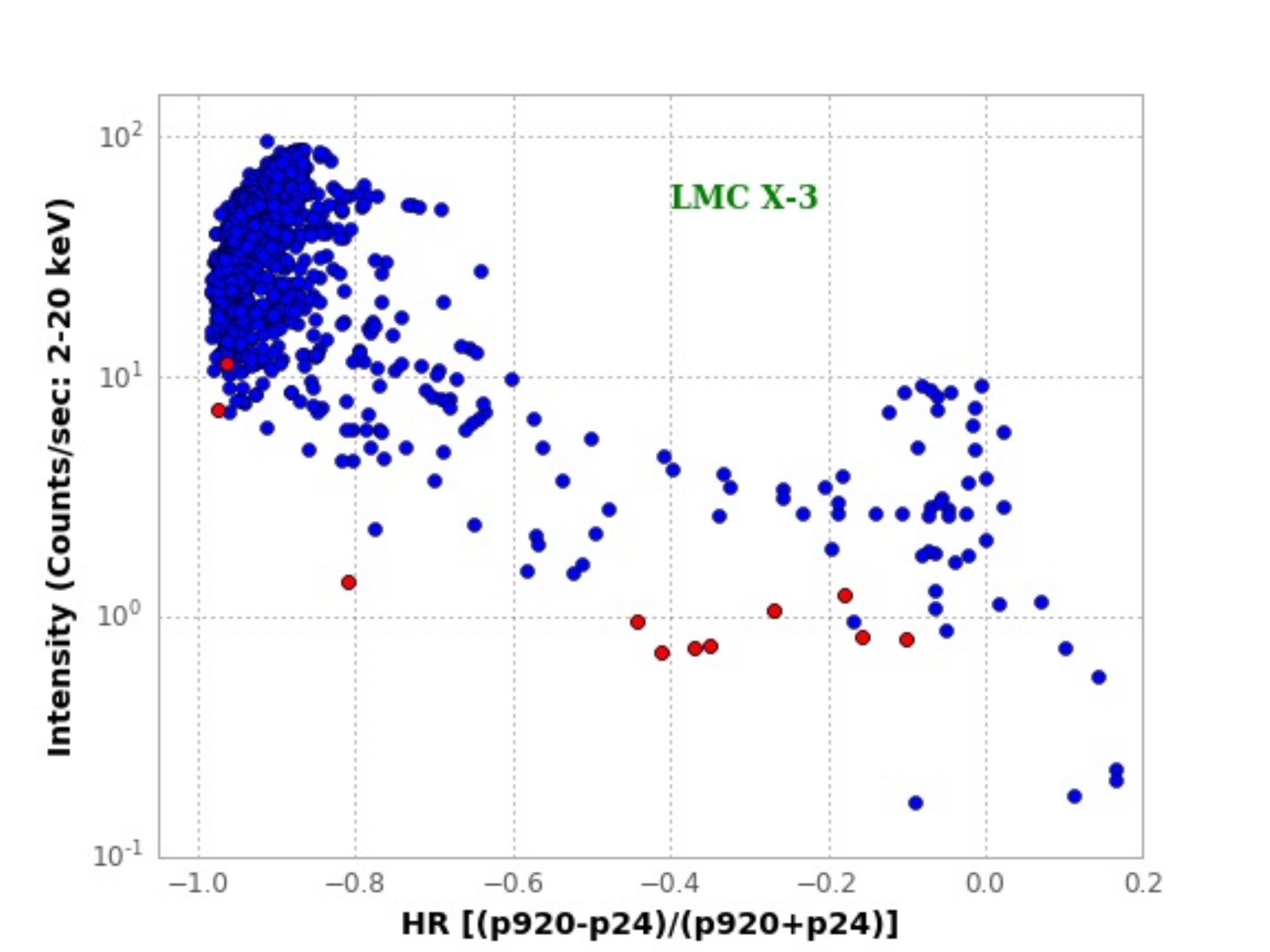}
\caption{Left panel: spectral energy distribution of LMC X-3.
The y-axis is the log$_{10}$ of the flux density in Jy and the x-axis is log$_{10}$ of the frequency in Hz.  The various instruments contributing are indicated across the bottom with colors matching the corresponding data.
Right panel: the color-intensity (CI) plot for LMC X-3  \color{black}
using RXTE/PCA data.
\color{black} The X-ray color expressed as
a ratio of the difference divided by the sum of bands at 9-20 kev (p920) and 2-4 keV (p24) is shown
on the x-axis, while the intensity over the full range 2 to 20 keV is plotted on the
y-axis.\color{black}  The red points indicate the values obtained for the week of the
\color{black}campaign, \color{black} and
the blue points indicate the range of values measured for the source at other epochs.
\label{fig:SED_LMCX3}}
\end{figure}

LMC X-3 (4U 0538-64; Fig. \ref{fig:SED_LMCX3}) is a persistent, variable source with a high mass companion star.  \color{black} It is found mainly in a soft X-ray state with rare excursions into hard
state \citep{Bhuvana_etal_2021}.  During the week of our campaign it was at very low X-ray flux (blue arrows in Fig. \ref{fig:appendix}.28.\color{black}).  It gives upper limits in the ATCA data.  It was not observed by SMARTS.   No significant detections were made with the Swift BAT during the week of the campaign.  \color{black}While there is no Swift XRT data during the week of the campaign we have included a spectrum taken 12 days later (green arrows in Fig. \ref{fig:appendix}.28.\color{black}) when the source was rising in flux (first entry on Table \ref{tab:Swift_xrt}).  The significant discrepancy between the XRT and PCA spectra taken a few days apart demonstrates the need for simultaneous, multiwavelength observations when observing XRBs.  The CI diagram (Fig. \ref{fig:SED_LMCX3} right panel) constructed with PCA data shows that the source was in the faint intermediate state (segment labeled D in Fig. \ref{fig:BHschematic}). 
No radio emission is expected in the D state, consistent with the upper limit found by the ATCA.  

\begin{figure}[ht]
\hspace{.2in} \includegraphics[width=3.4in]{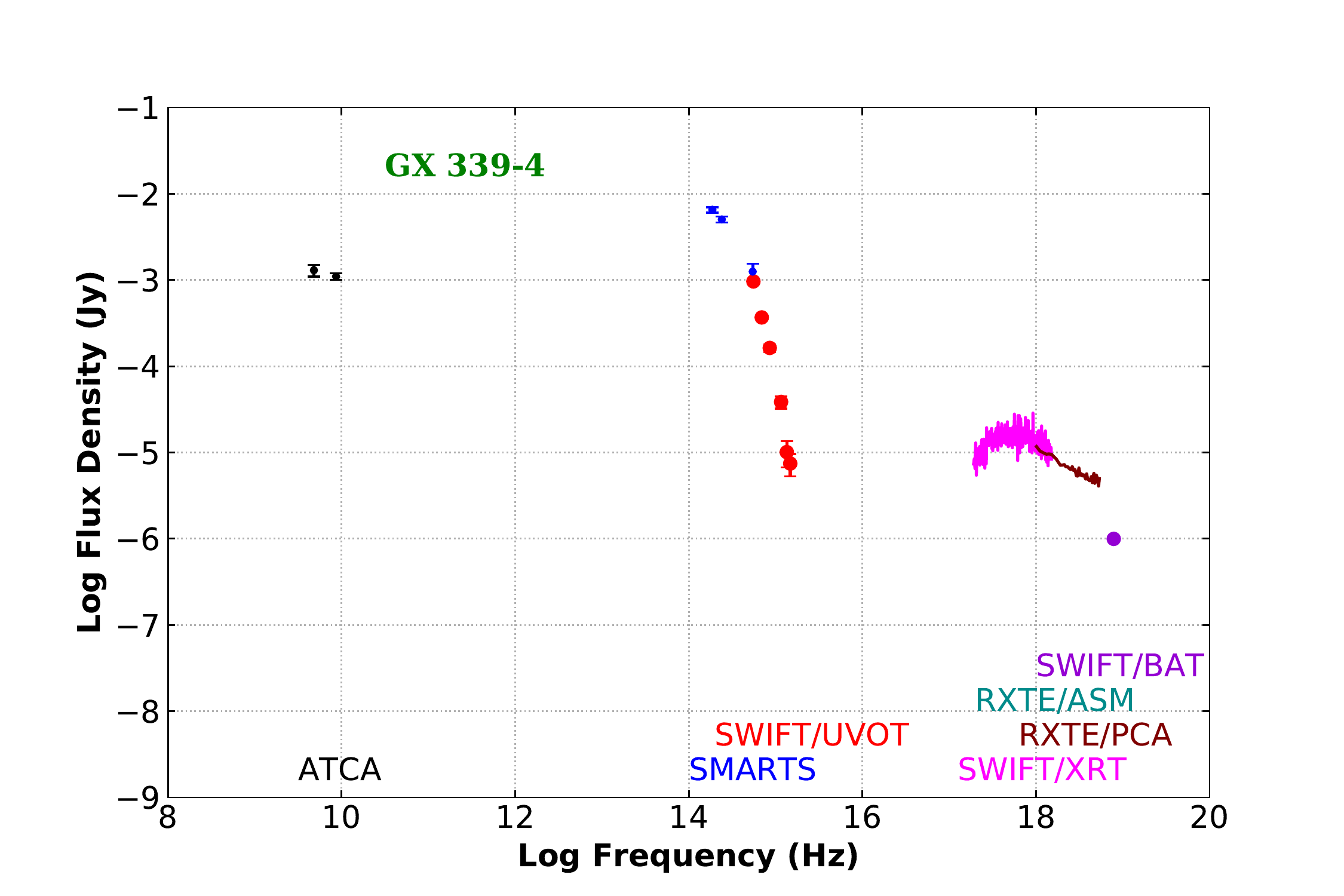}
\hspace{-.5in}
\hspace{.5in} \includegraphics[width=3in]{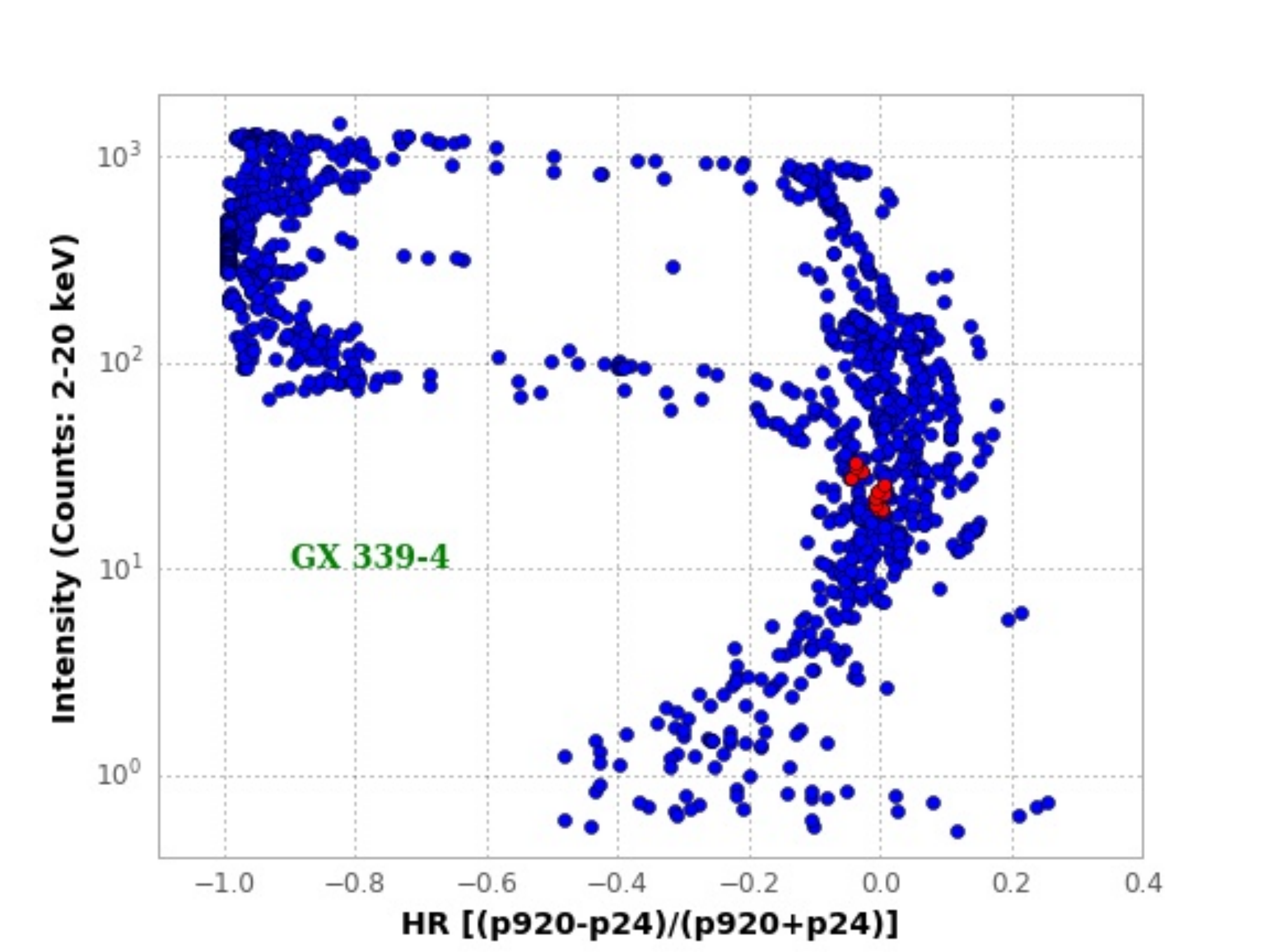}
\caption{Spectral Energy Distribution and color-intensity plots of GX 339-4. \color{black}
The X-ray spectra are from obsid 00030953005 on Table \ref{tab:Swift_xrt} (SWIFT/XRT) and the time average shown on Table \ref{tab:PCA-2} (RXTE/PCA).
The panels and colors are
\color{black} as in figure \ref{fig:SED_LMCX3}.
\label{fig:SED_GX339}}
\end{figure}

GX 339-4 (4U 1659-487; V821 Ara; Fig. \ref{fig:SED_GX339}; Fig. \ref{fig:appendix}.19.\color{black}) is an X-ray transient caught 
\color{black} in the lower part of the low/hard state. \color{black}
It is a dipping source implying a relatively high inclination angle \citep{Sasano_etal_2014}. \color{black}
There are ATCA detections at the mJy level. There are detections by SMARTS, Swift UVOT, Swift XRT, RXTE PCA, and Swift BAT. There is no significant detection by the RXTE ASM during the week of the campaign. The PCA CI diagram shows that the source is\color{black} at the cusp between C, D, and E states on Fig. \ref{fig:BHschematic}. The radio detection implies that the flat-spectrum, compact radio source is present, as expected in the C and E states. \color{black} The Swift BAT points suggest a sharp cutoff around 20 to 30 keV. 

\begin{figure}[ht]
\hspace{.2in} \includegraphics[width=3.4in]{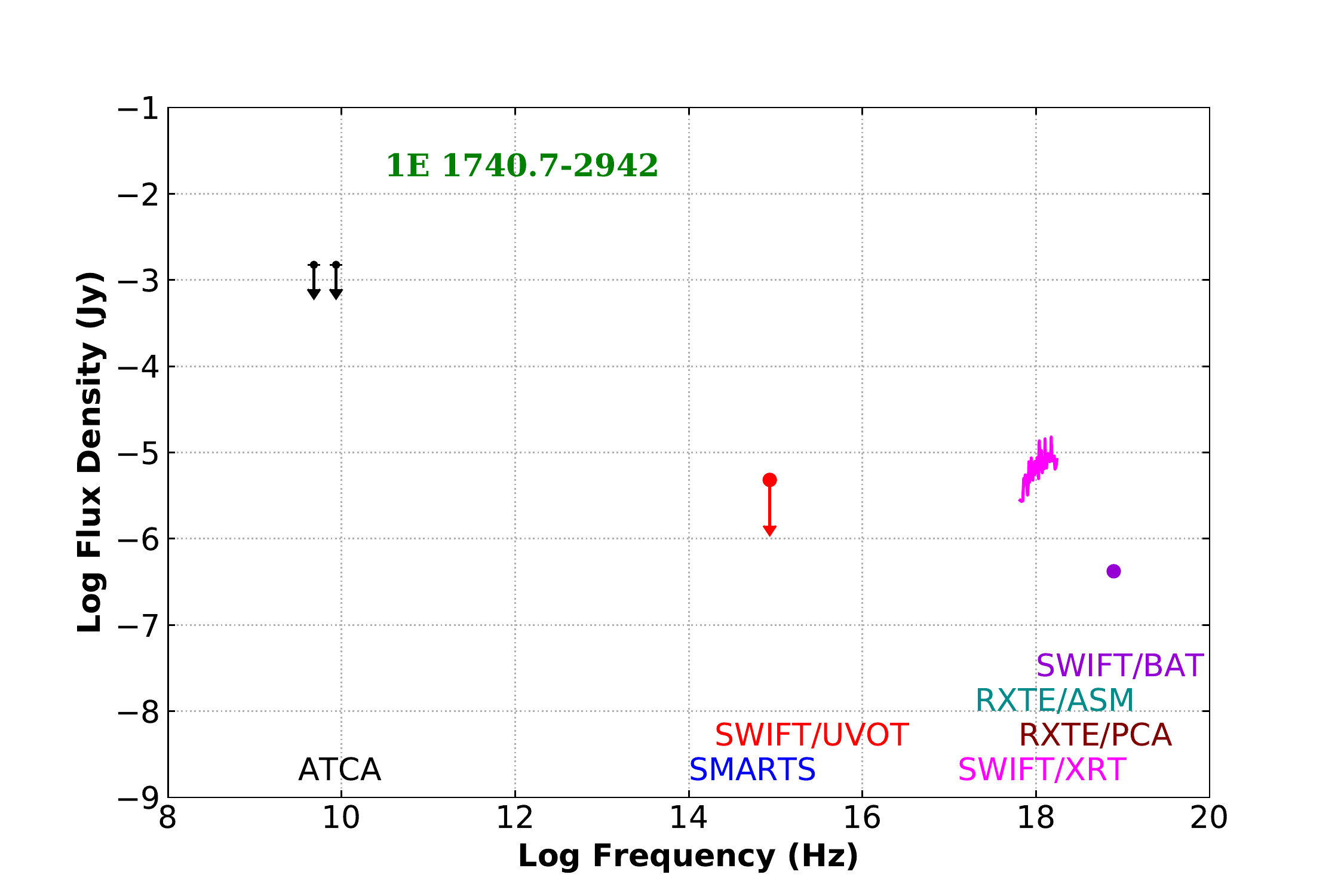}
\hspace{-.5in}
\hspace{.5in} \includegraphics[width=3in]{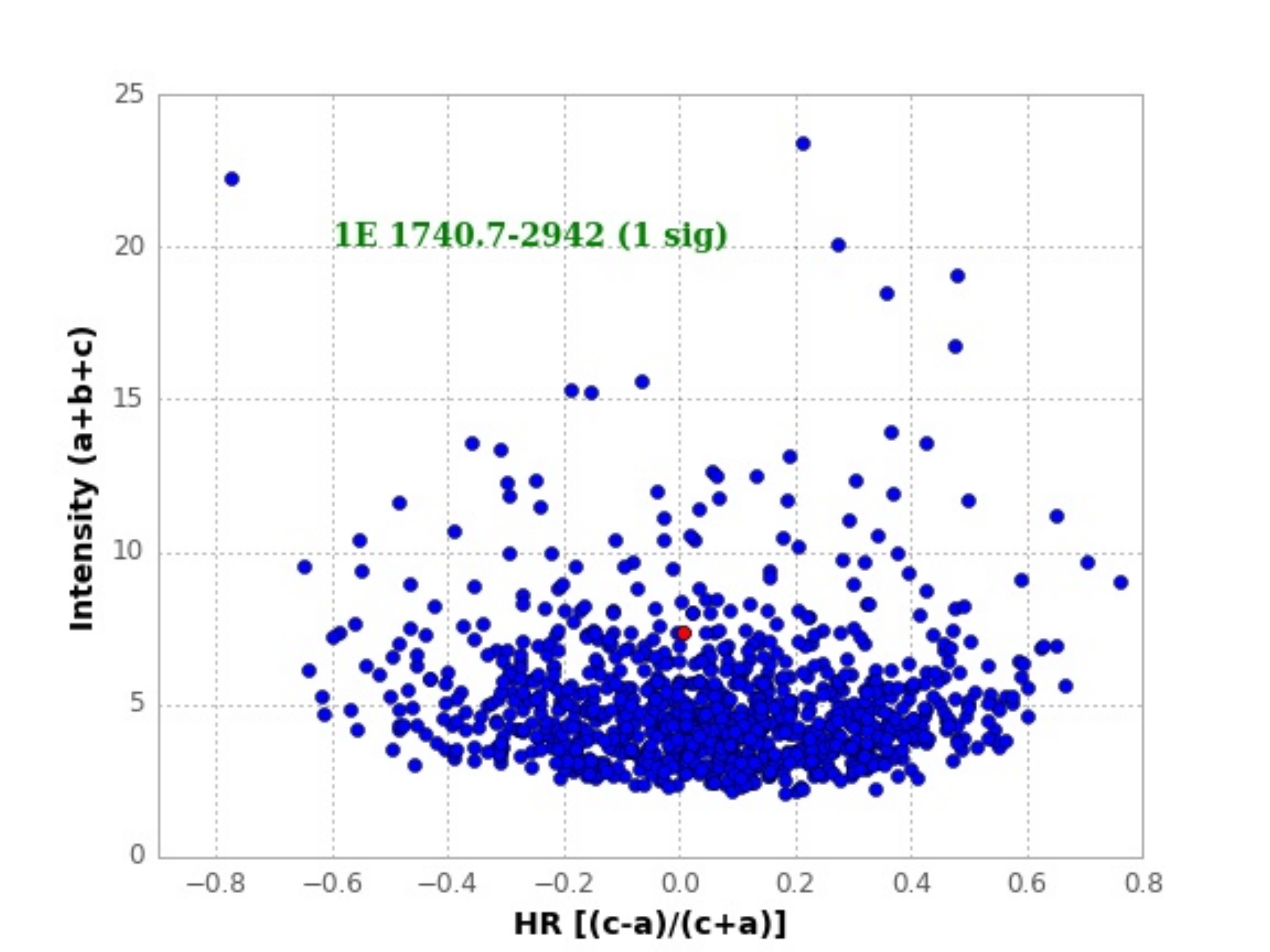}
\caption{Spectral Energy Distribution and CI plots of 1E 1740.7-2942
\color{black} using RXTE/ASM data.
\color{black}The left panel axes are as in figure \ref{fig:SED_LMCX3}.
The right panel has color (hardness ratio)
defined using the ASM bands c \color{black}(5-12 keV) \color{black} and a \color{black}(1.5-3 keV), \color{black} and intensity using
the sum of ASM bands a, b, and c \color{black} (1.5-12 keV)\color{black}. The red point
indicates the values \color{black} during the campaign, and the blue points
indicate the range of values measured for the source at other epochs.
 \label{fig:SED_1E1740}}
\end{figure}

1E 1740.7-2942 (Great annihilator; Fig. \ref{fig:SED_1E1740}; Fig. \ref{fig:appendix}.12.\color{black}) is a persistent, low flux source. \color{black} Although it is known to show radio lobe structure and clear jet-ISM interaction \citep{Tetarenko_etal_2020, Saavedra_etal_2022},
during this campaign there are only upper limits with the ATCA. There is also only one detection by the RXTE ASM, suggesting that the source is in the faint, intermediate state (D in Fig. \ref{fig:BHschematic}) consistent with the non-detection in the radio. The SWIFT XRT and BAT results suggest a sharp turn over in flux at high energies.
\color{black}

\begin{figure}[ht]
\hspace{.2in} \includegraphics[width=3.4in]{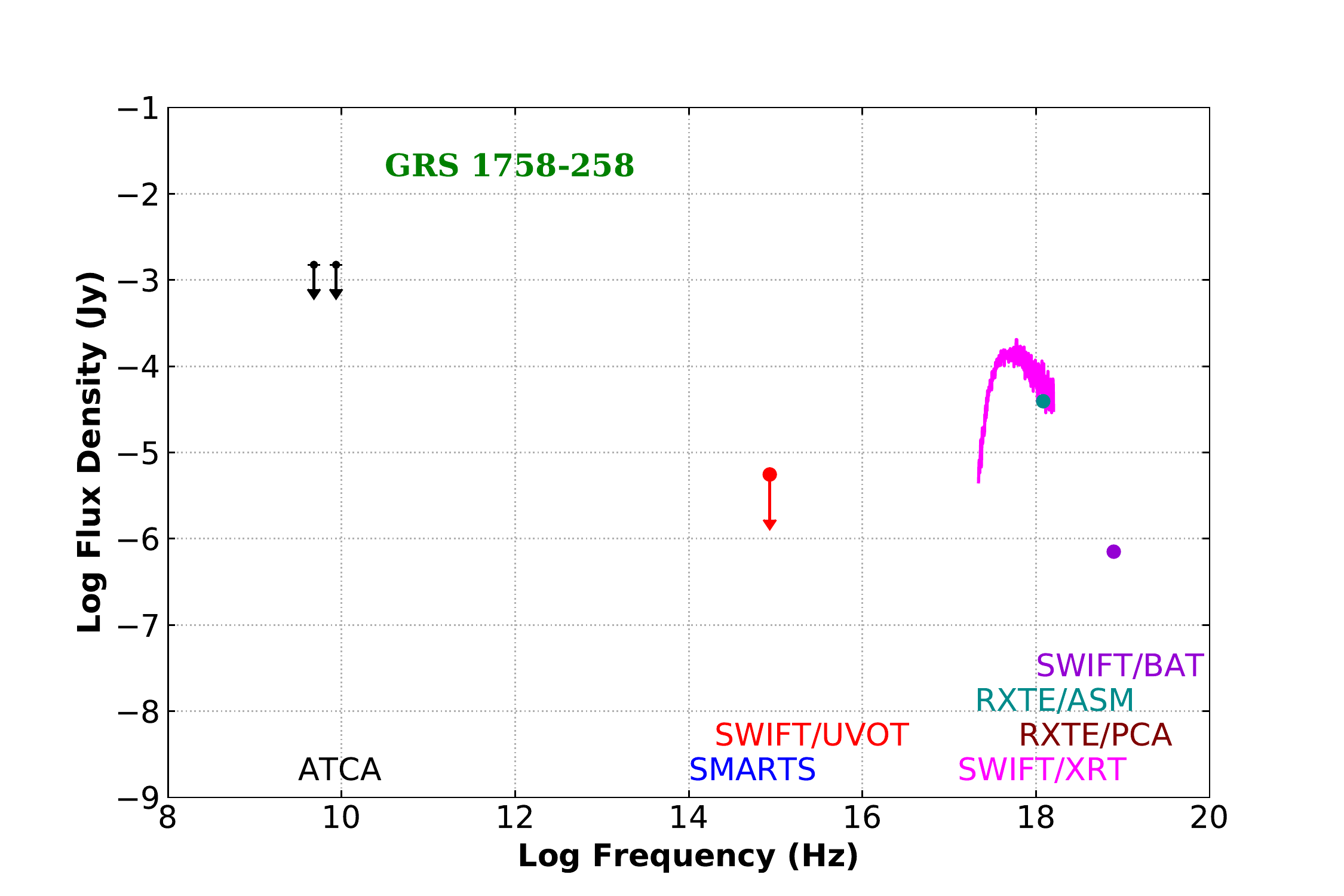}
\hspace{-.5in}
\hspace{.5in} \includegraphics[width=3in]{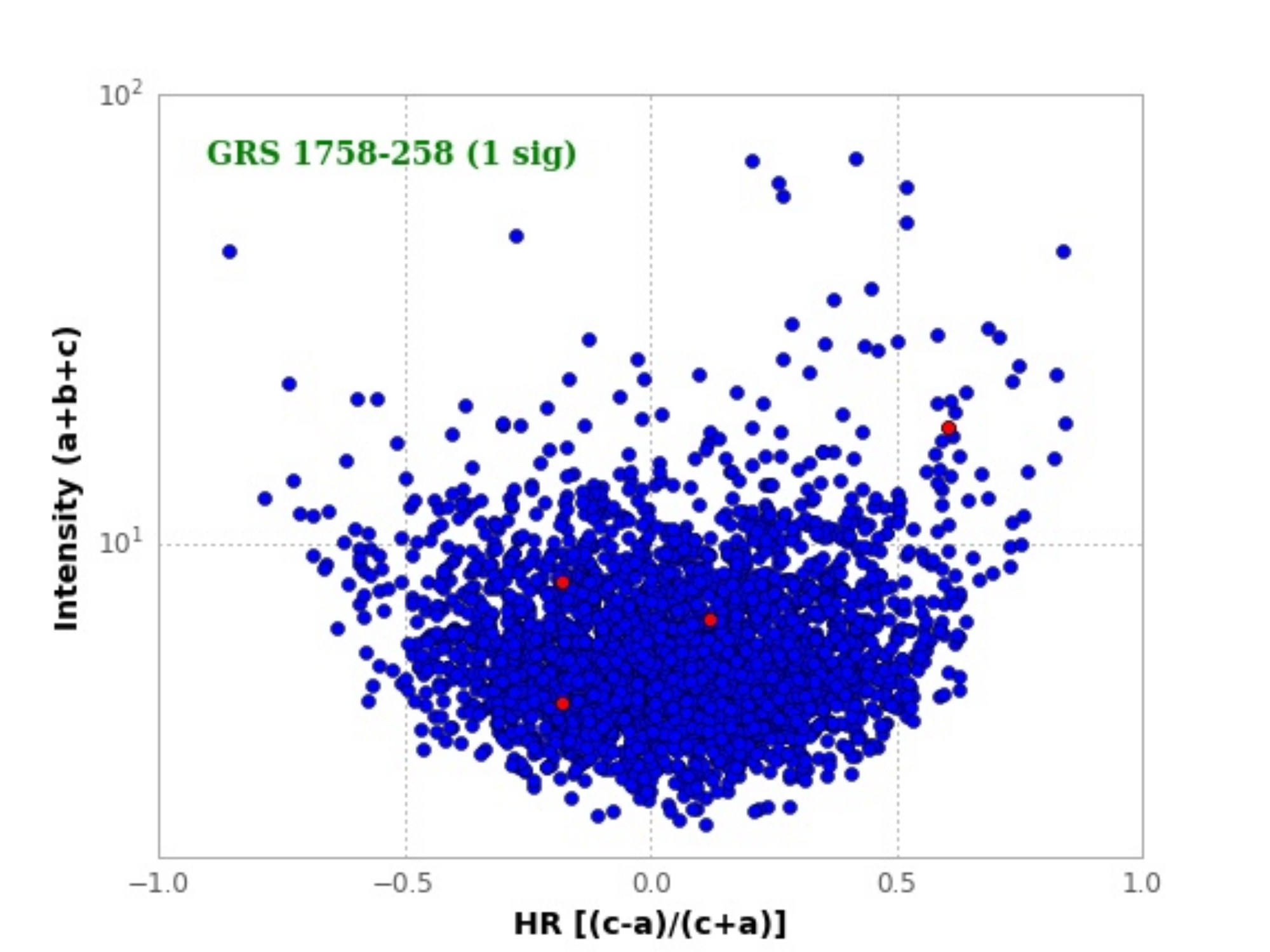}
\caption{Spectral Energy Distribution and color-intensity plots of GRS 1758-258. The
panels and axes are as in figure \ref{fig:SED_1E1740}.
\label{fig:SED_GRS1758}}
\end{figure}

GRS 1758-258 (Fig. \ref{fig:SED_GRS1758}; Fig. \ref{fig:appendix}.13.\color{black}) is a persistent, low-flux source.  It is known to have a clear connection between the radio jet and the ISM \citep{Tetarenko_etal_2020}.
However, during the campaign \color{black} there are only upper limits from the ATCA and Swift UVOT. There are consistent flux densities measured by the Swift XRT and RXTE ASM. It is unclear if the BAT result is high or low: it depends on how the spectrum turns over. The RXTE PCA CI diagram in addition to the non-detection in the radio is consistent with the source being in the intermediate low state (D in Fig. \ref{fig:BHschematic}).

\begin{figure}[ht]
\hspace{.2in} \includegraphics[width=3.4in]{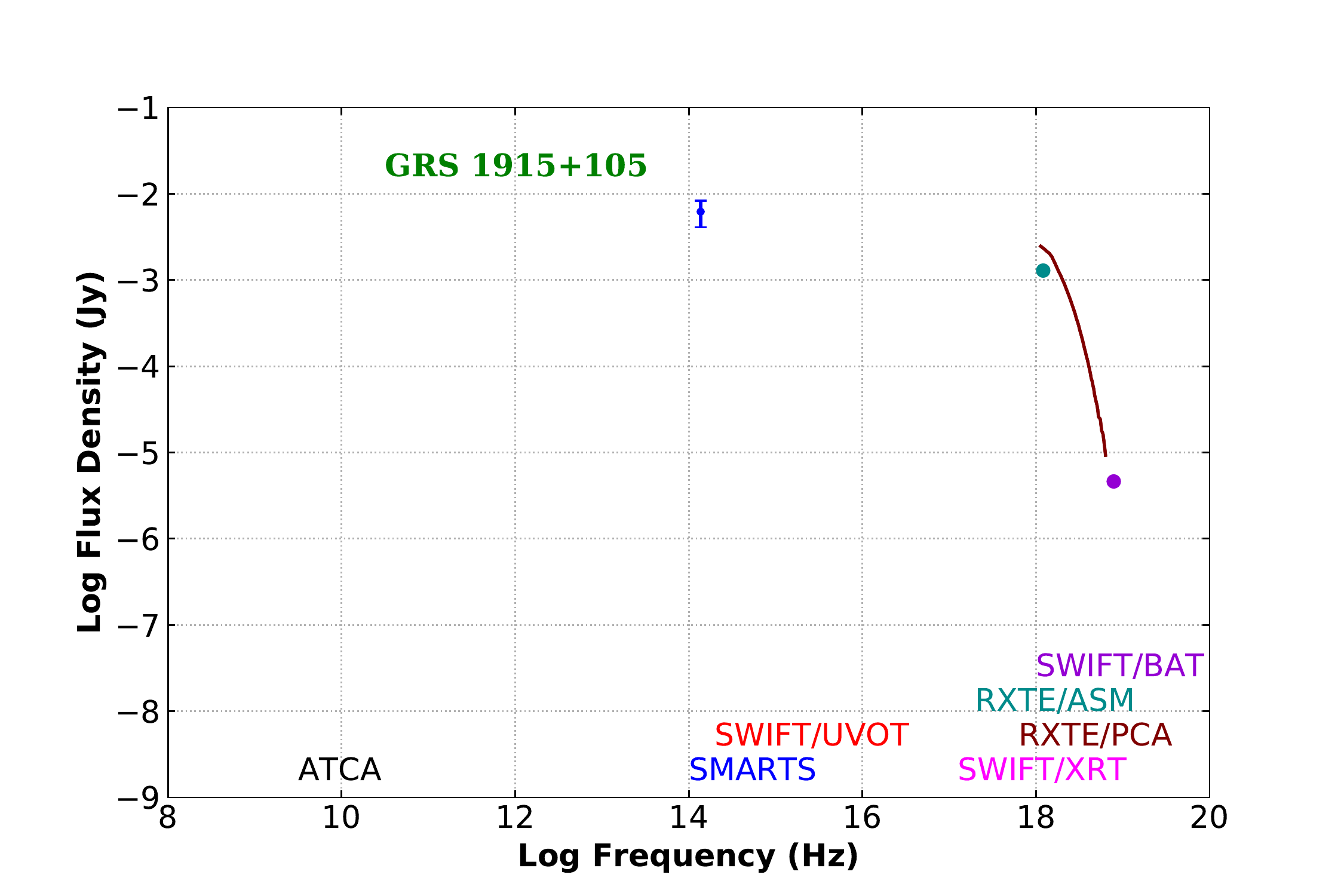}
\hspace{-.5in}
\hspace{.5in} \includegraphics[width=3in]{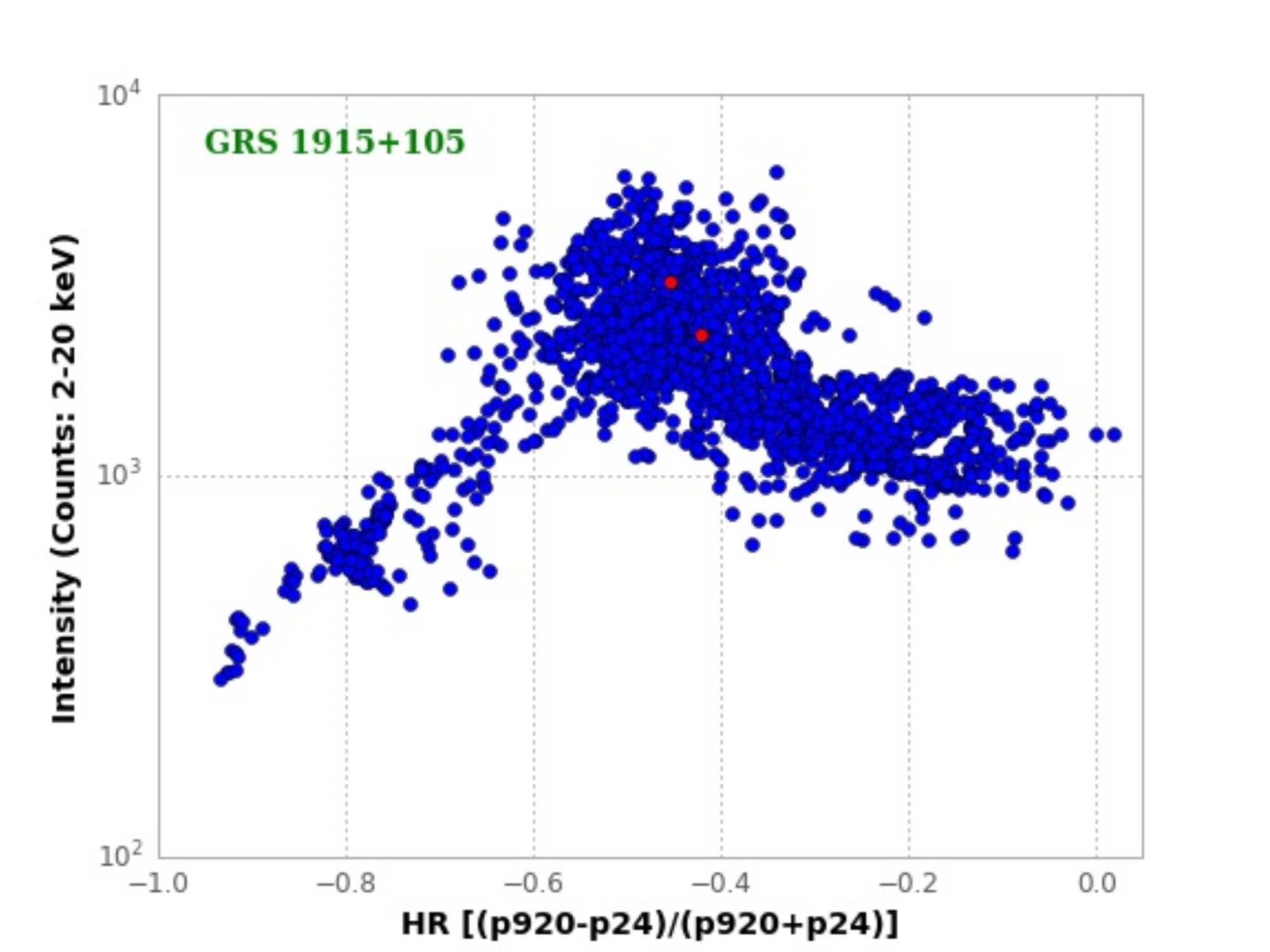}
\caption{Spectral Energy Distribution and color-intensity plots of GRS 1915+105. The panels and axes are as in figure \ref{fig:SED_LMCX3}.
\label{fig:SED_GRS1915}}
\end{figure}

GRS1915+105 (V1487 Aql, Fig. \ref{fig:SED_GRS1915}; Fig. \ref{fig:appendix}.33) \color{black} is a continuously active \color{black}but variable source. It was not observed with the ATCA. It is detected at K-band with SMARTS; the IR flux is likely dominated by the companion star. It was not observed by the Swift XRT. In the ASM data it ranges from 10-170 cts s$^{-1}$. The week of this campaign it was on a steeply rising slope at 80 cts s$^{-1}$. The ASM flux falls below the \color{black}RXTE PCA \color{black}spectrum; the ASM is averaged over the week and the PCA observation was taken later in the week so the two are consistent. The BAT measurement is consistent with the RXTE PCA. The PCA gives the highest flux of all the BHs. During the week of this campaign GRS1915+105 was in the Bright/Intermediate state based on the RXTE PCA CI diagram
\color{black} (B in Fig. \ref{fig:BHschematic}. A resolved radio jet is expected in this state 
\citep{Arur_Maccarone_2022}.\color{black}

\begin{figure}[ht]
\hspace{.2in} \includegraphics[width=3.4in]{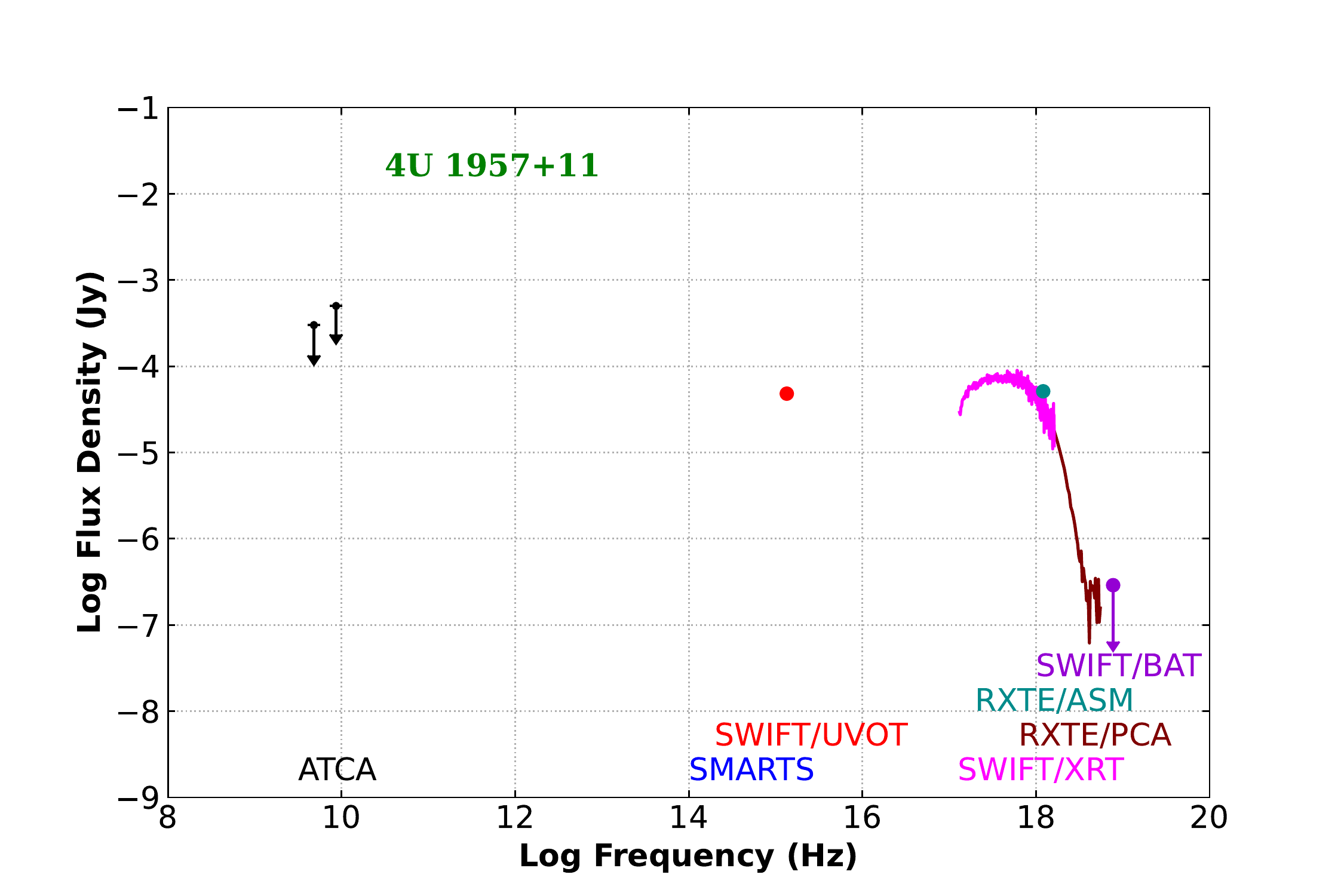}
\hspace{-.5in}
\hspace{.5in} \includegraphics[width=3in]{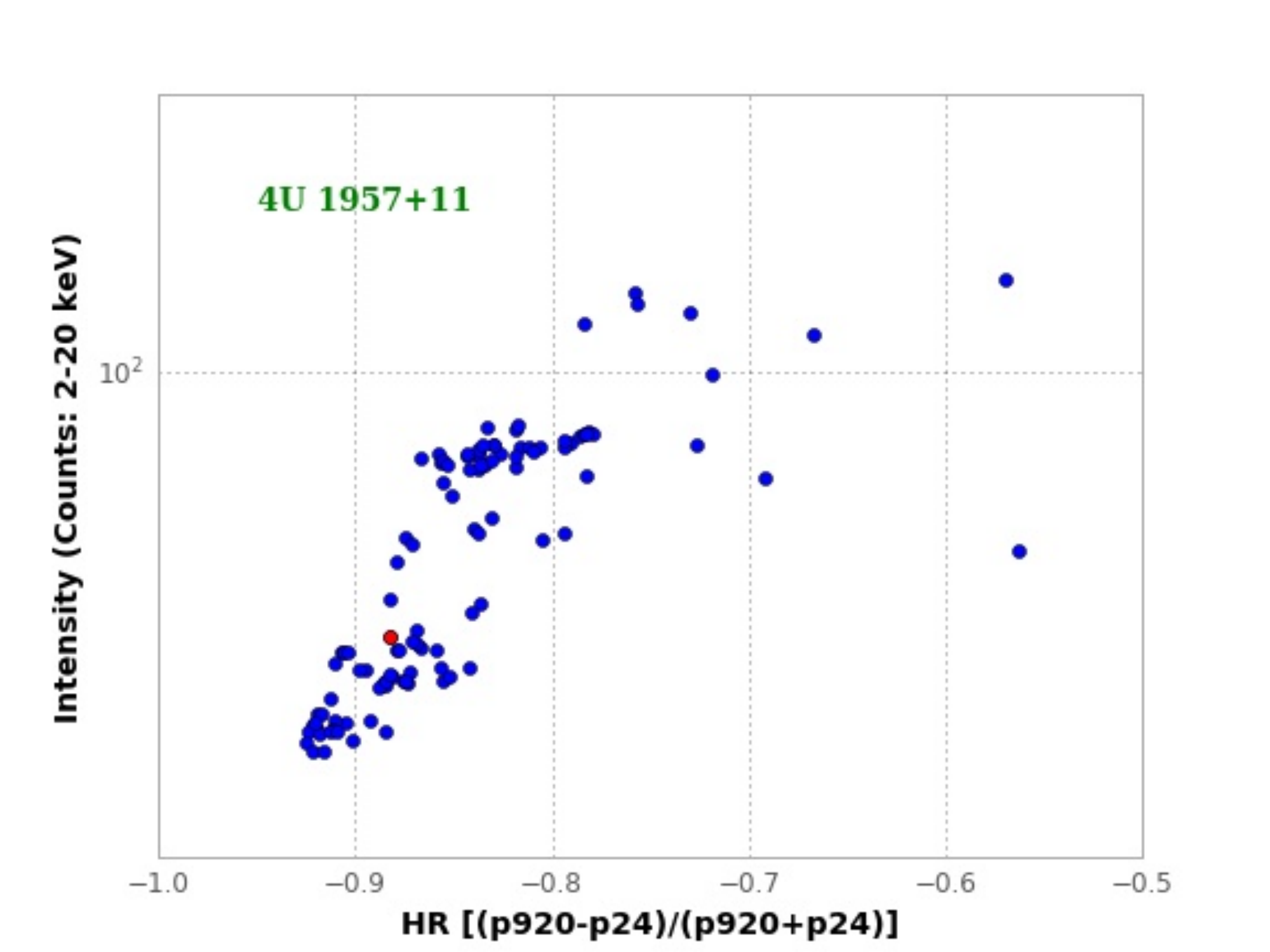}
\caption{Spectral Energy Distribution and color-intensity plots of 4U 1957+11. The panels and axes are as in figure \ref{fig:SED_LMCX3}.
\label{fig:SED_U1957}}
\end{figure}

4U 1957+11 (V1408 Aql, Fig. \ref{fig:SED_U1957}; Fig. \ref{fig:appendix}.16.\color{black}) is a continuously active variable source.  It is one of the fastest spinning BHs 
\citep{Nowak_etal_2011}.
It shows upper limits with the ATCA. The SWIFT/XRT and BAT and the RXTE/PCA and ASM all give consistent flux densities. The PCA CI plot together with the non detection in the radios implies that it is in the A2 state (Fig. \ref{fig:BHschematic}) when the radio jet is quenched.
\color{black}

\subsubsection{X-ray Non-Detections (BHs: 5)}

\color{black} Many of our sources have detections in the infrared, optical, and/or ultraviolet only, but no detection or minimal data in the X-rays. These are generally X-ray transients in quiescence, and only the companion star is visible in this phase. They all show upper limits from the ATCA, consistent with radio flux at the $\mu$Jy level. \color{black} These include LMC X-1 (4U 0540-697), XTE J1550-564 (V381 Nor), 4U 1630-47, GRO J1655-40, and V4641 Sgr. \color{black}

\subsection{Z sources (8 sources; 5 with SEDs)}

\begin{figure}[ht]
\hspace{.2in} \includegraphics[width=3.4in]{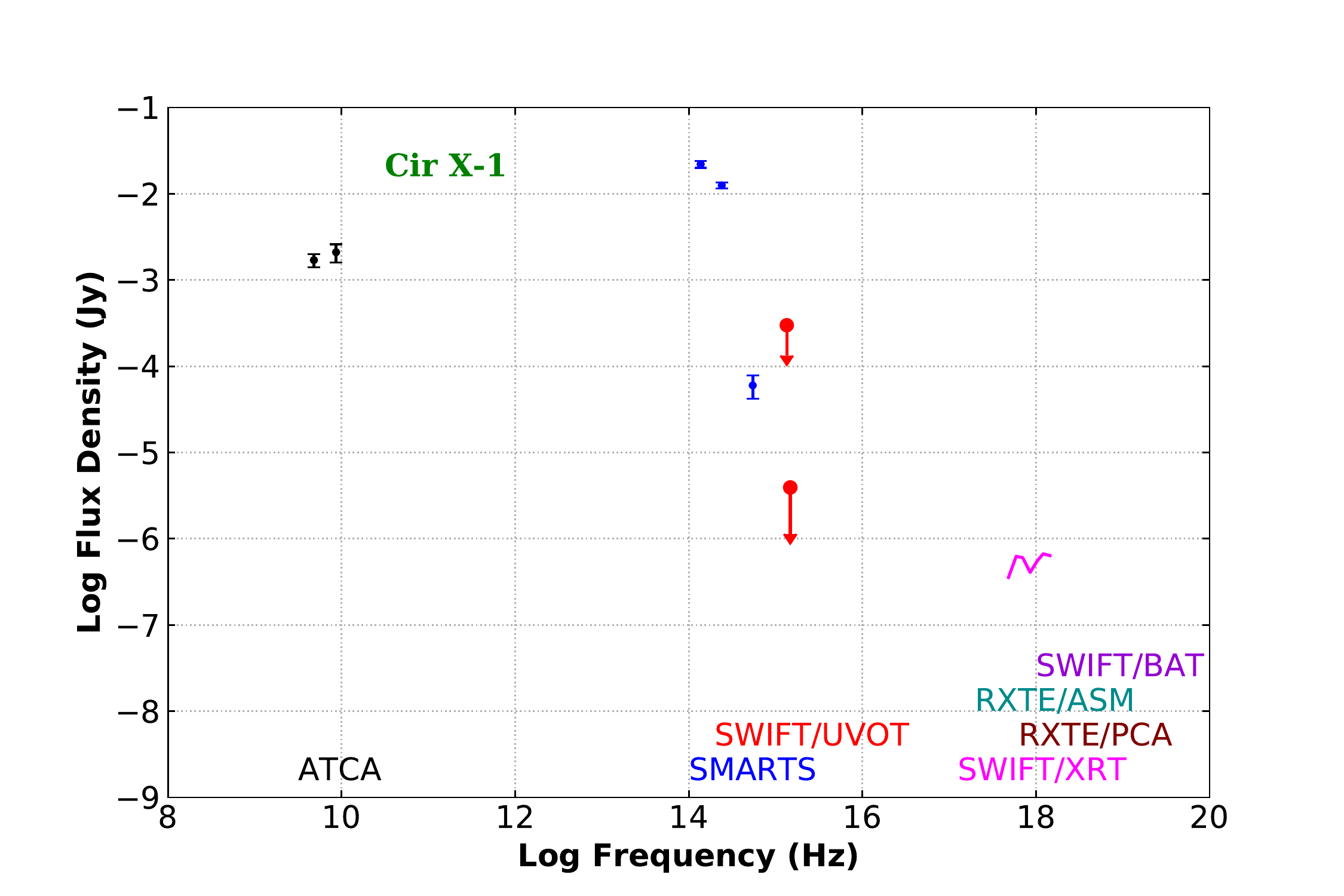}
\hspace{-.5in}
\hspace{.5in} \includegraphics[width=3in]{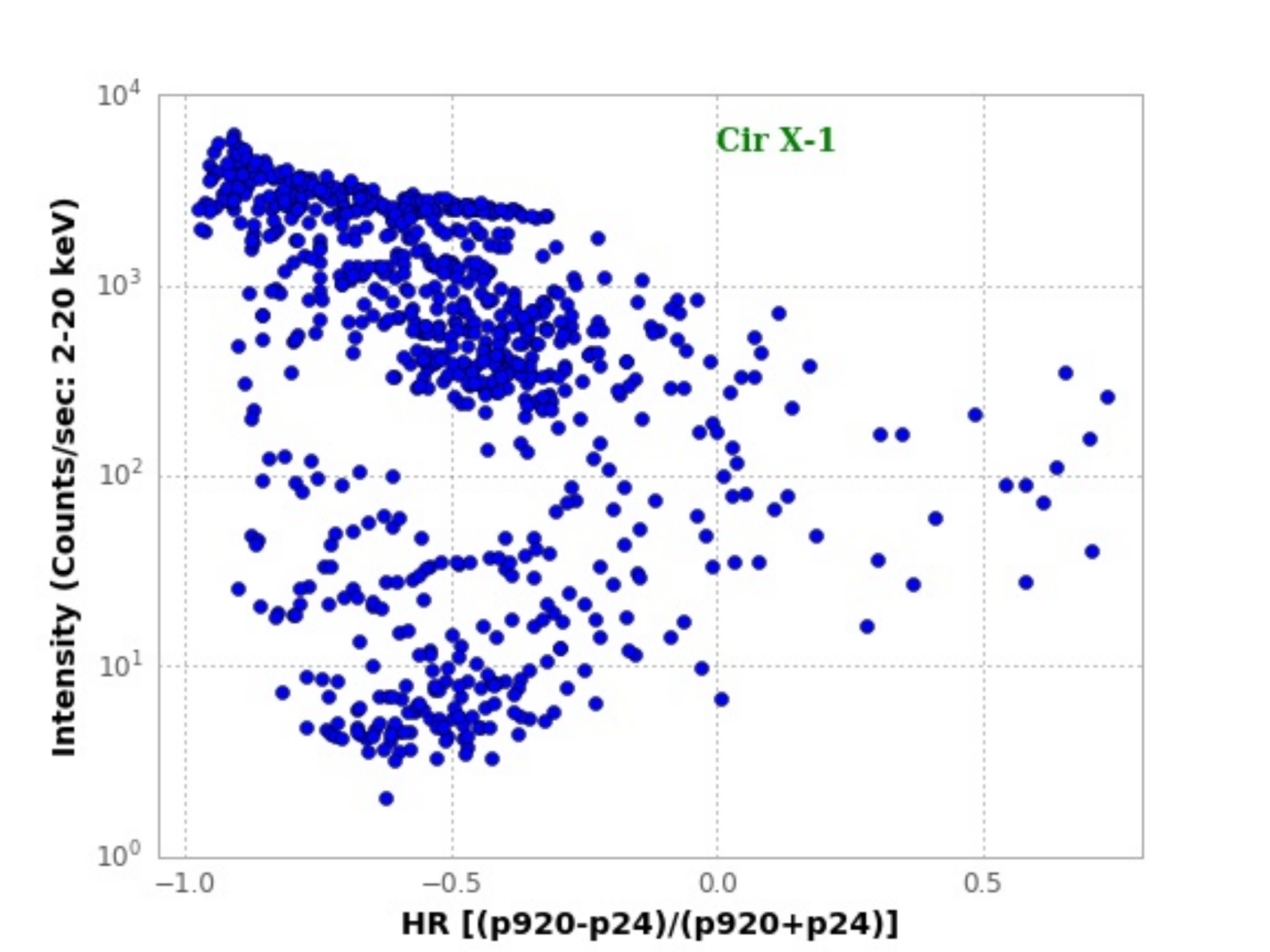}
\caption{Spectral Energy Distribution and color-intensity plots of Cir X-1.
\color{black}
The X-ray spectrum is from Obsid 00030268025 on Table \ref{tab:Swift_xrt} (SWIFT/XRT).
\color{black}
The panels and axes are as in figure \ref{fig:SED_LMCX3}.
\label{fig:SED_CirX1}}
\end{figure}

Cir X-1 (4U 1516-56; BR Cir, Fig. \ref{fig:SED_CirX1}; Fig. \ref{fig:appendix}.32.\color{black}) is a highly variable source ranging from 0-200 ASM cts s$^{-1}$.  During the week of this campaign it was below the detection limit of the ASM.  There are positive detections with the ATCA and SMARTS, upper limits with the Swift UVOT. The XRT spectrum shows a dip at 5.5$\times10^{17}$ Hz that is apparent during very low X-ray flux 
\citep{Schulz_etal_2020}.  While a single blackbody with very small area has been fit at this state 
\citep{Schulz_etal_2020}, the XRT found a blackbody temperature to be unconstrained (25 $\pm$ 100 keV).  There is no significant detection by the Swift BAT during the week of the campaign. It was not observed by the RXTE PCA during the week of the campaign. \color{black}

\begin{figure}[ht]
\hspace{.2in} \includegraphics[width=3.4in]{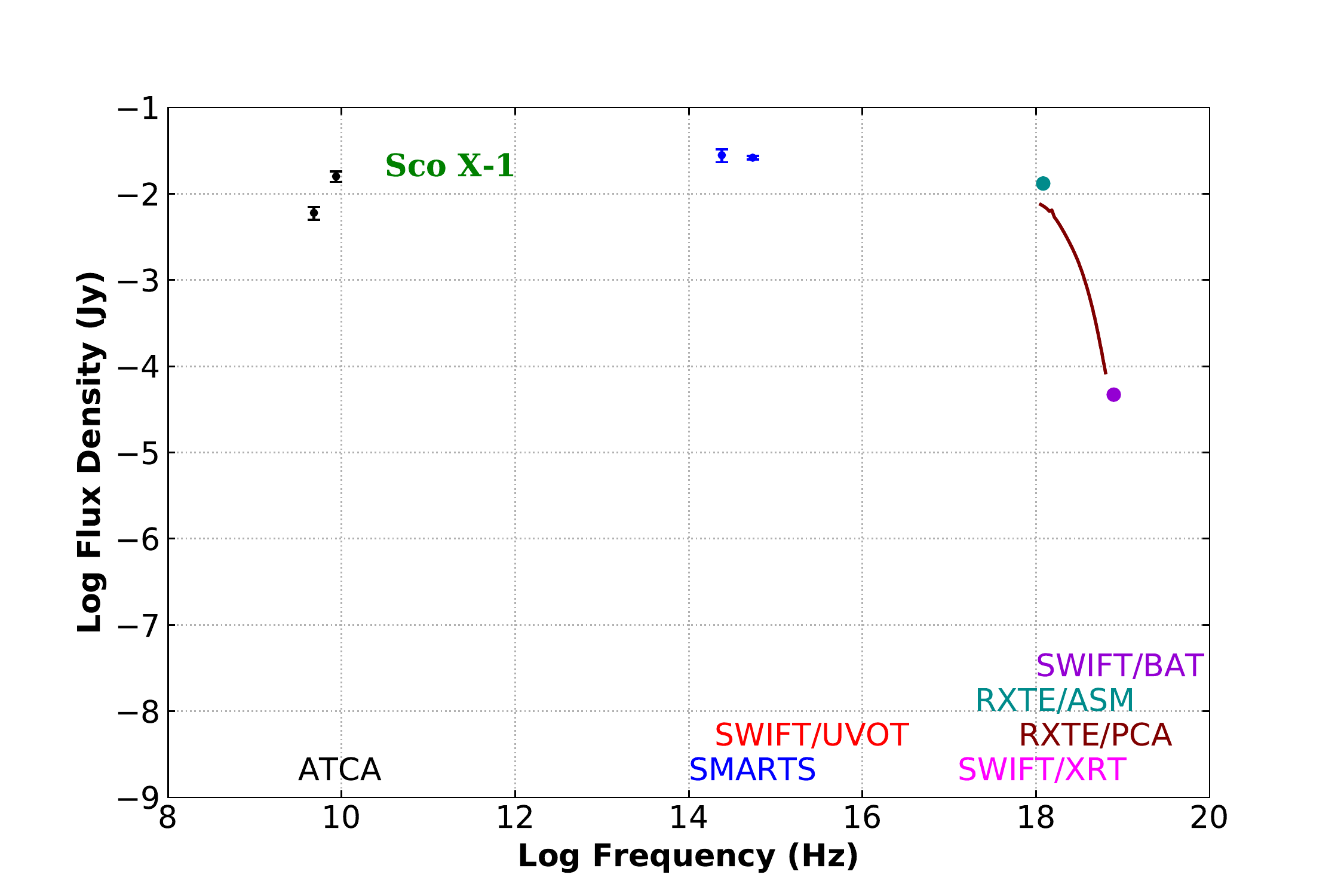}
\hspace{-.5in}
\hspace{.5in} \includegraphics[width=3in]{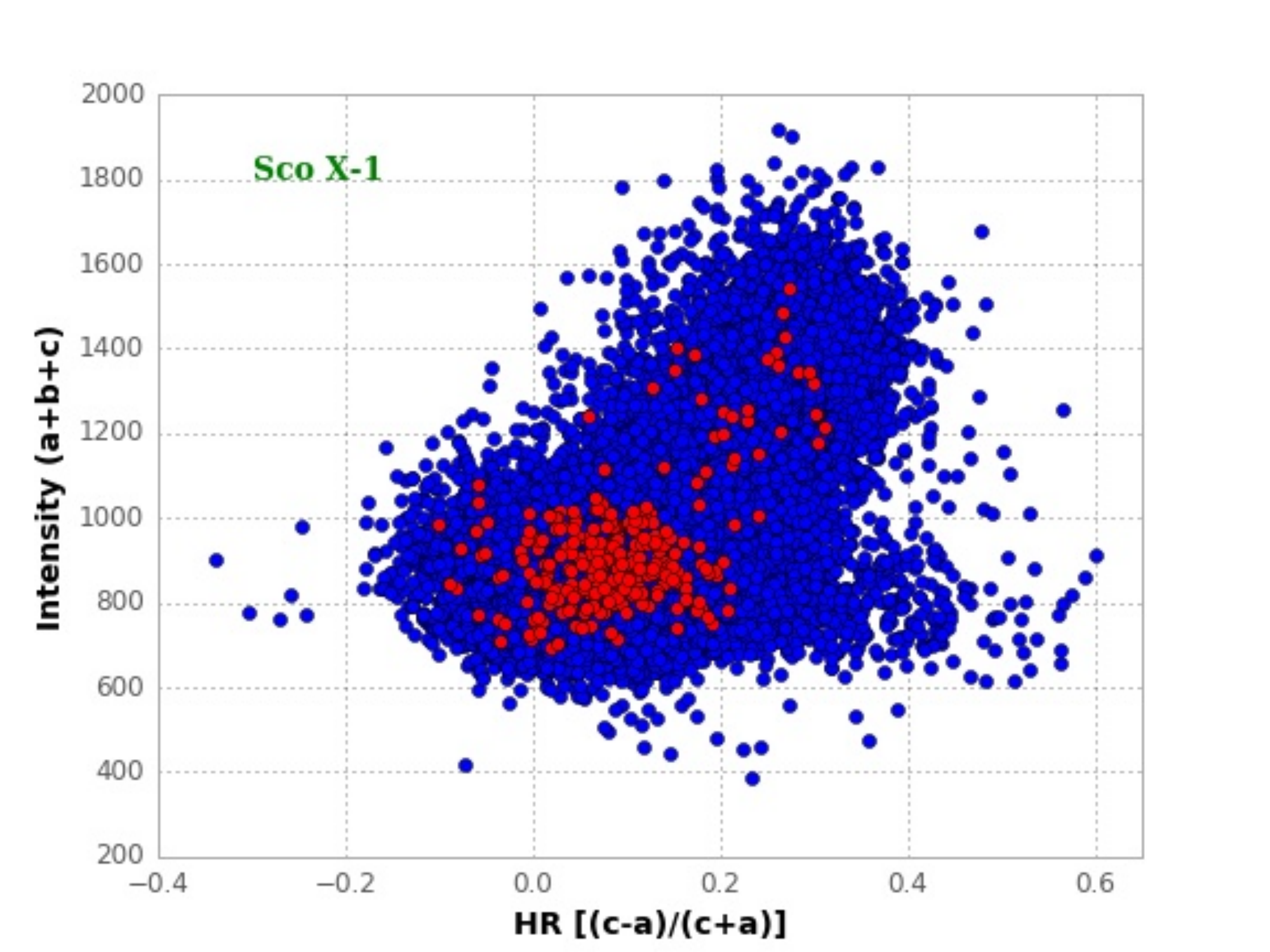}
\caption{Spectral Energy Distribution and color-intensity plots of Sco X-1. The
panels and axes are as in figure \ref{fig:SED_1E1740}.
\label{fig:SED_ScoX1}}
\end{figure}

Sco X-1 (4U 1617-15; V818 Sco, Fig. \ref{fig:SED_ScoX1}; Fig. \ref{fig:appendix}.40.\color{black}) is a very bright, persistent source. We have positive detections with ATCA at the mJy level, and a positive detection with SMARTS. The flux density values are consistent between the
RXTE PCA, and the Swift BAT; the flux density is slightly high in the RXTE ASM.  
The ASM CI plot, using dwell by dwell values (6-8 per day), shows that while Sco X-1 is mostly in the NB there are several excursions into FB. This explains the discrepancy between the ASM and PCA spectra.  The lightcurves plotted with one day averages do not resolve the flares.  This source has the highest radio flux density of all the sources in this campaign.  The rapidly variation seen by the ATCA is consistent with the emergence of a jet as expected for this source in the NB.
\color{black}

\begin{figure}[ht]
\hspace{.2in} \includegraphics[width=3.4in]{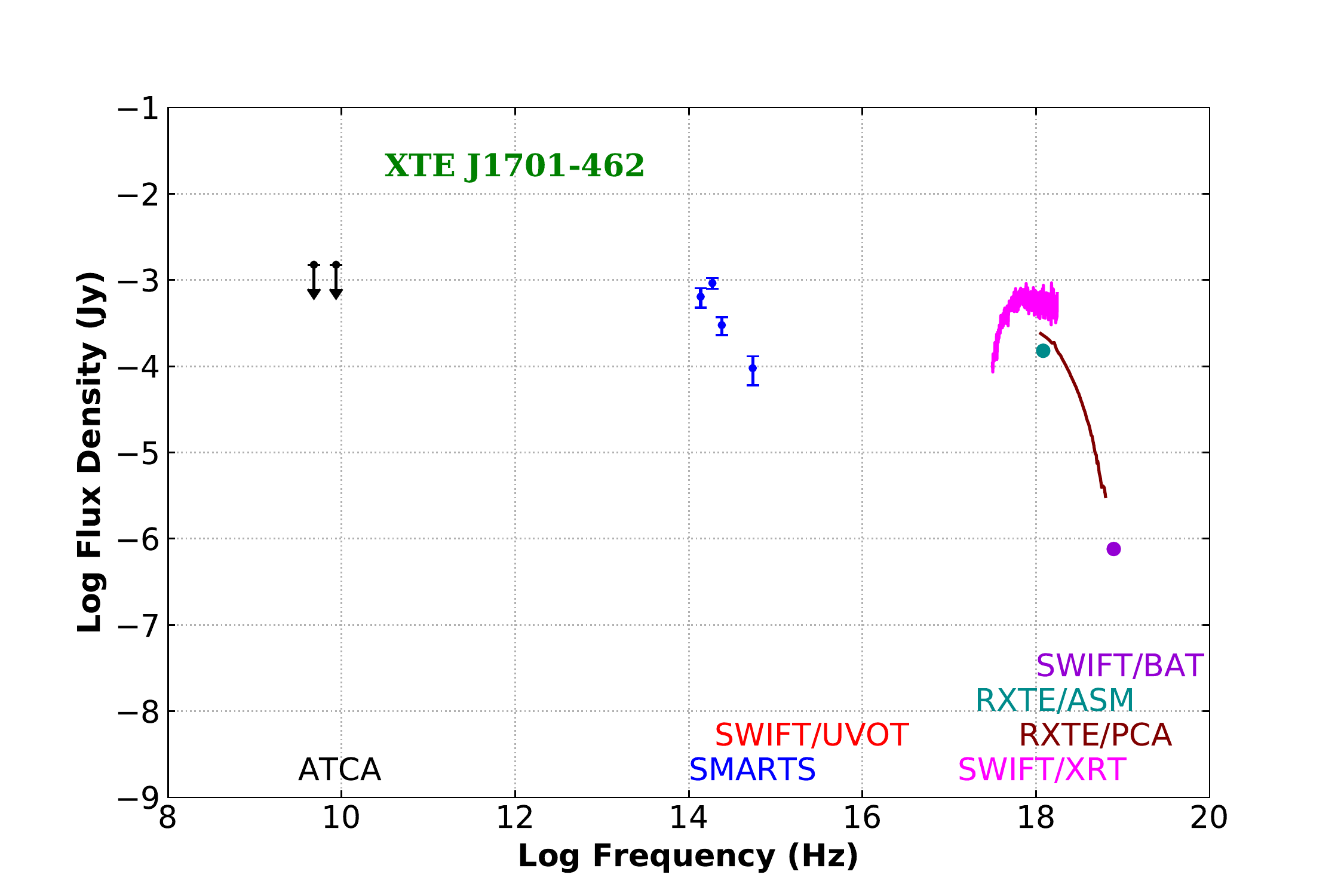}
\hspace{-.5in}
\hspace{.5in} \includegraphics[width=3in]{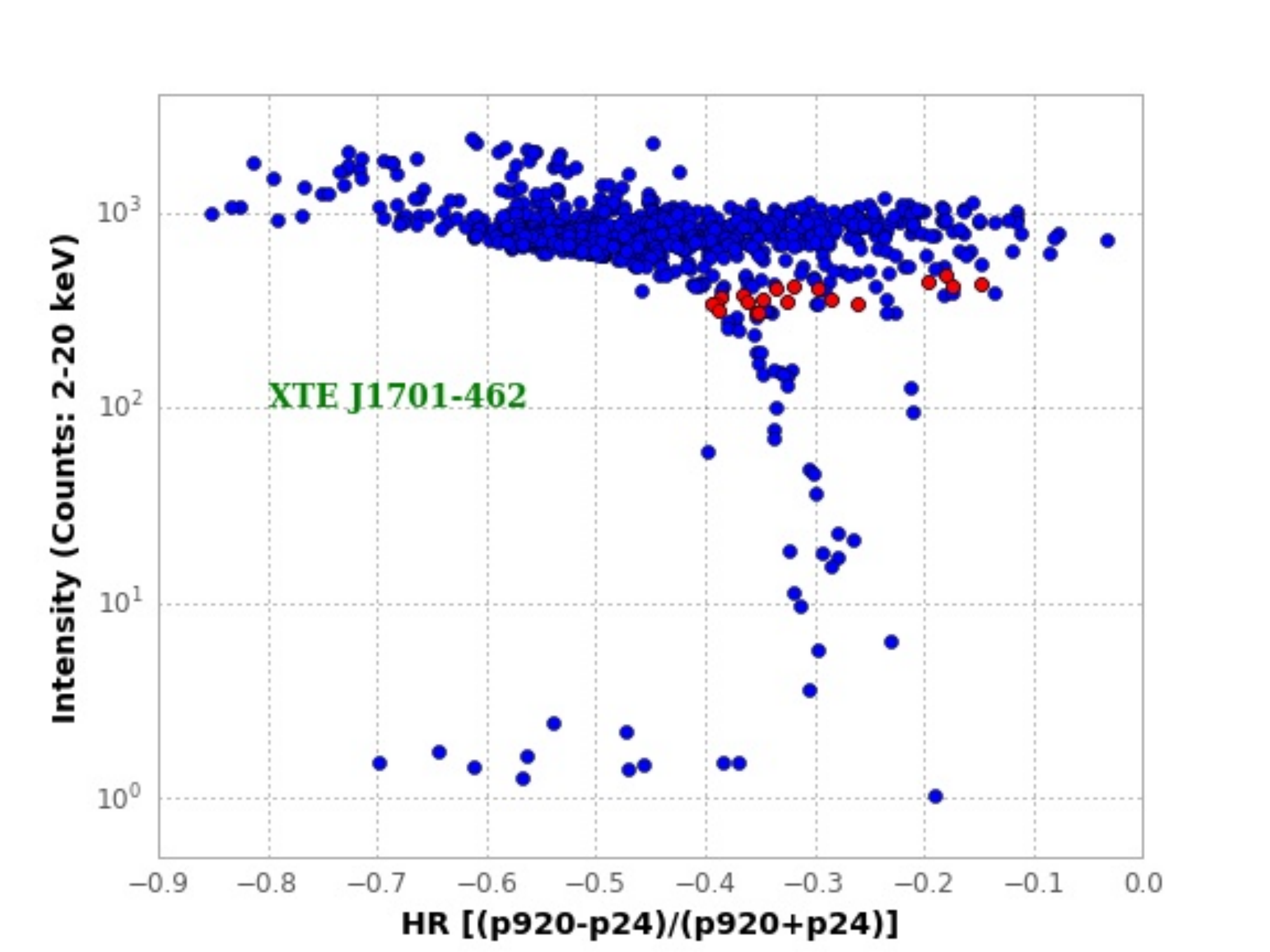}
\caption{Spectral Energy Distribution and color-intensity plots of XTE J1701-462. The panels and axes are as in figure \ref{fig:SED_LMCX3}.
\label{fig:SED_J1701}}
\end{figure}

XTE J1701-462 (Fig. \ref{fig:SED_J1701}; Fig. \ref{fig:appendix}.37.\color{black}) is a transient source on a downward trend of a strong outburst. It shows upper limits with the ATCA. SMARTS has positive detections at all frequencies. The flux density measured with the Swift XRT is high. The PCA data matches the ASM and BAT results. \color{black} As a source that shows both Z and Atoll behavior, it appears to be at a transition between the two based on the RXTE PCA CI plots. 
The lack of a radio detection, consistent with no jet, 
\color{black}
would be consistent with the flaring (Z) branch \citep{Homan_etal_2010}.
\color{black}

\begin{figure}[ht]
\hspace{.2in} \includegraphics[width=3.4in]{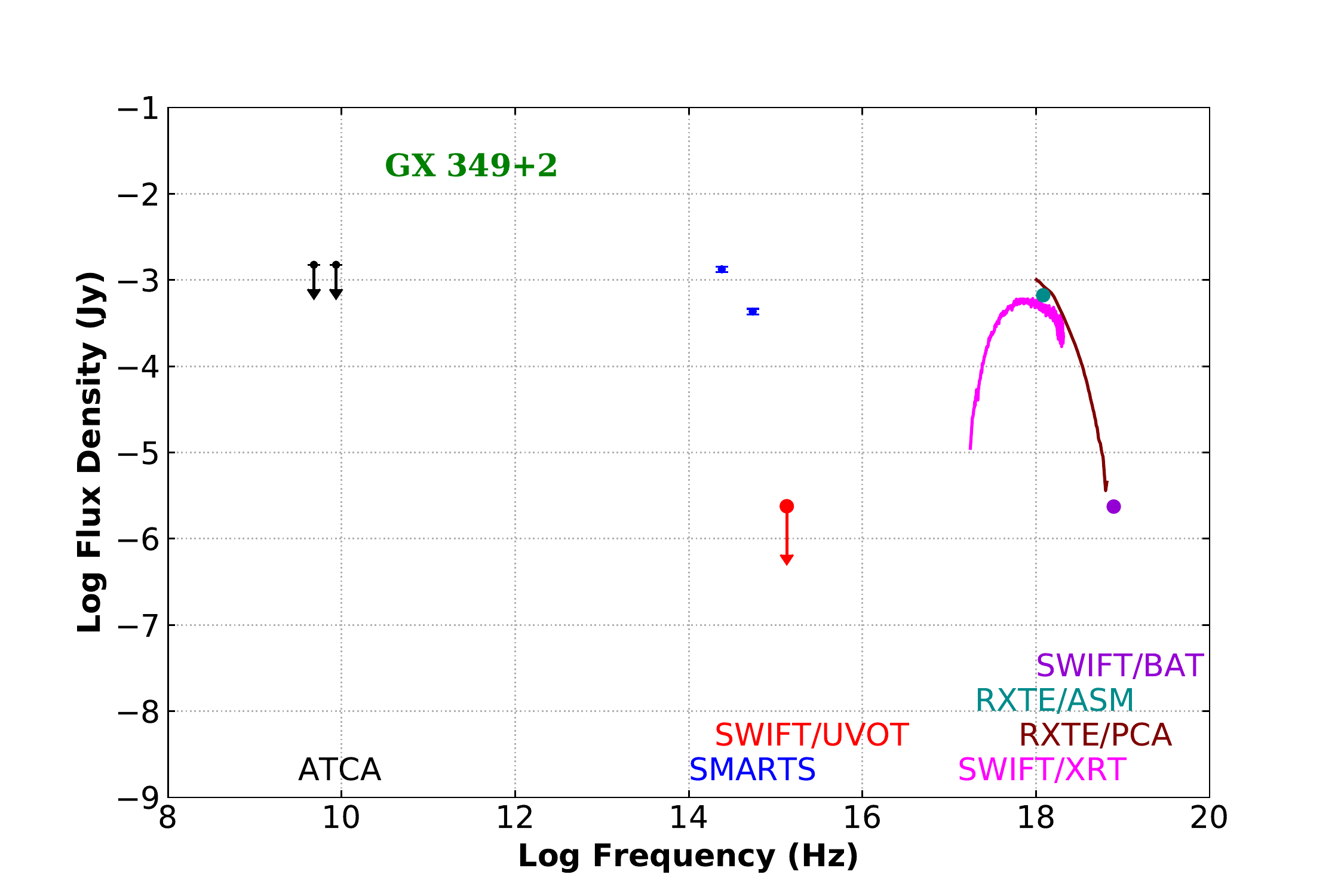}
\hspace{-.5in}
\hspace{.5in} \includegraphics[width=3in]{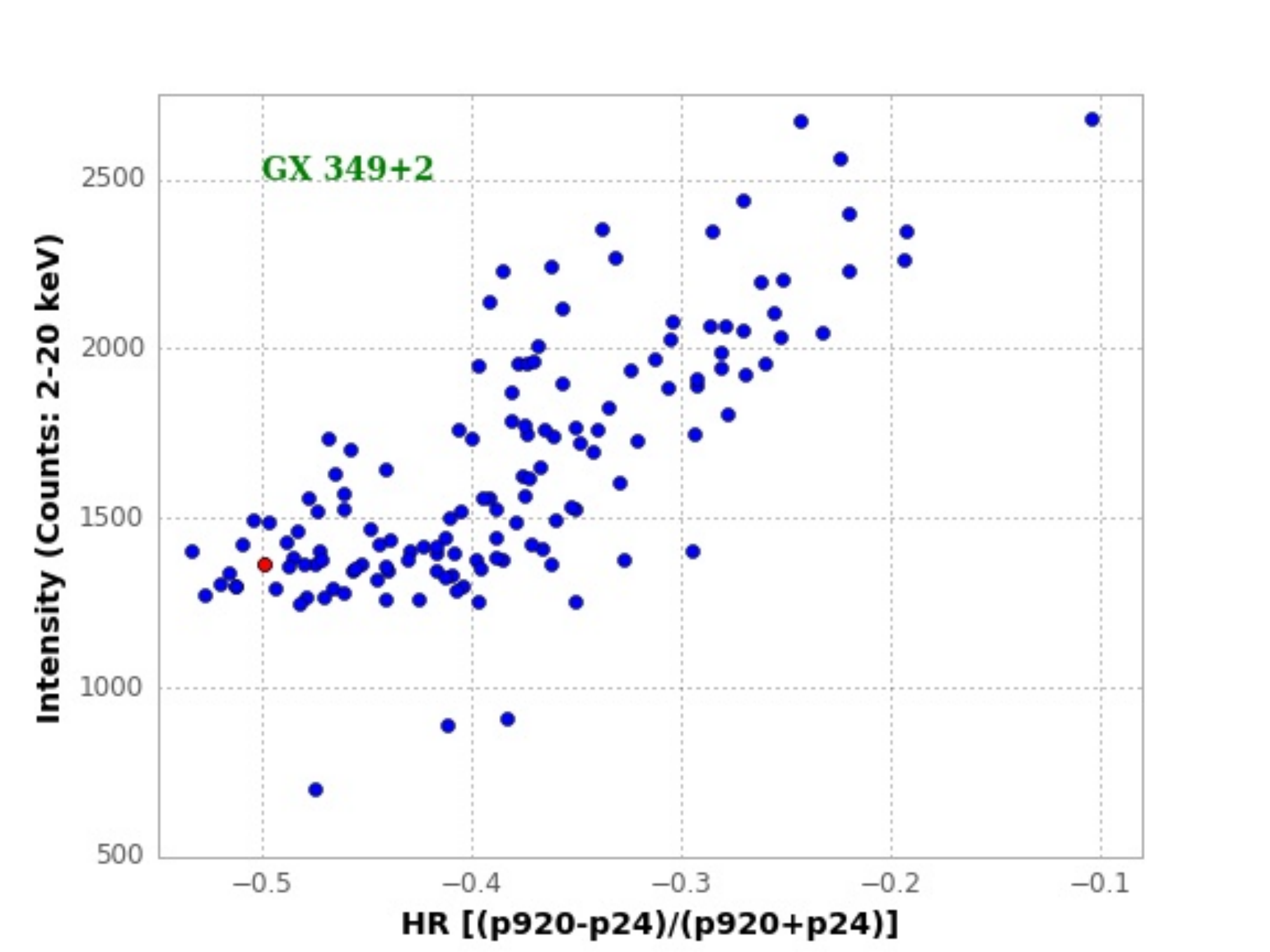}
\caption{Spectral Energy Distribution and color-intensity plots of GX 349+2. The panels and axes are as in figure \ref{fig:SED_LMCX3}.
\label{fig:SED_GX349}}
\end{figure}

GX 349+2 (4U 1702-36; V1101 Sco, Fig. \ref{fig:SED_GX349}; Fig. \ref{fig:appendix}.21.\color{black}) is a bright persistent source that has frequent flares (Coughenour et al. 2018). The ATCA gives flux density upper limits. SMARTS has positive detections. The Swift UVOT has one upper limit. The Swift XRT flux density is low compared with the RXTE PCA, which matches the RXTE ASM and Swift BAT results. During the week of the campaign GX 349+2 was in the cusp between FB and NB based on the RXTE PCA CI diagram.  \color{black} The upper limit in radio favors the FB where the radio is quenched.  The flux discrepancy between XRT and PCA suggests the PCA caught a flare.

\color{black}
\begin{figure}[ht]
\hspace{.2in} \includegraphics[width=3.4in]{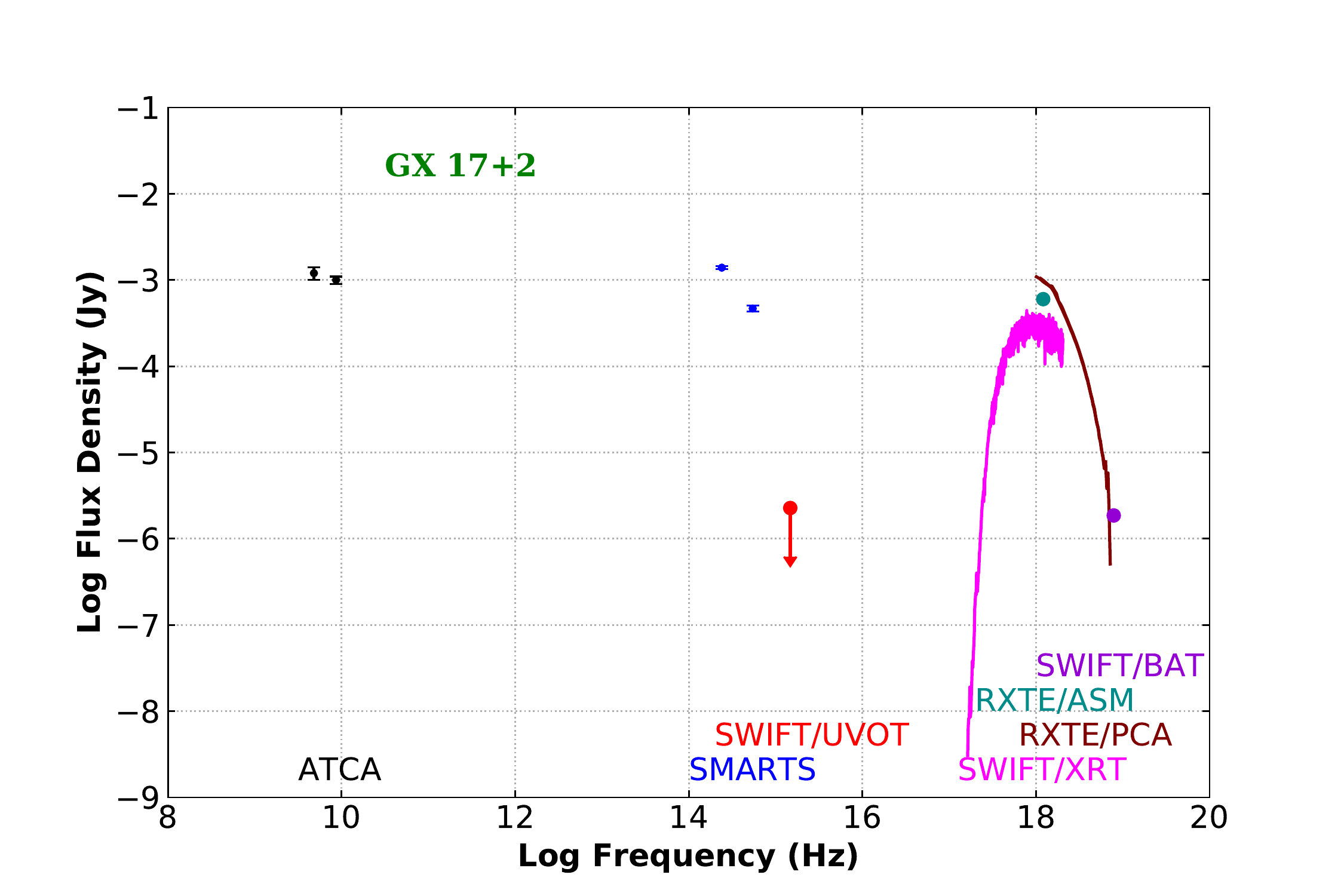}
\hspace{-.5in}
\hspace{.5in} \includegraphics[width=3in]{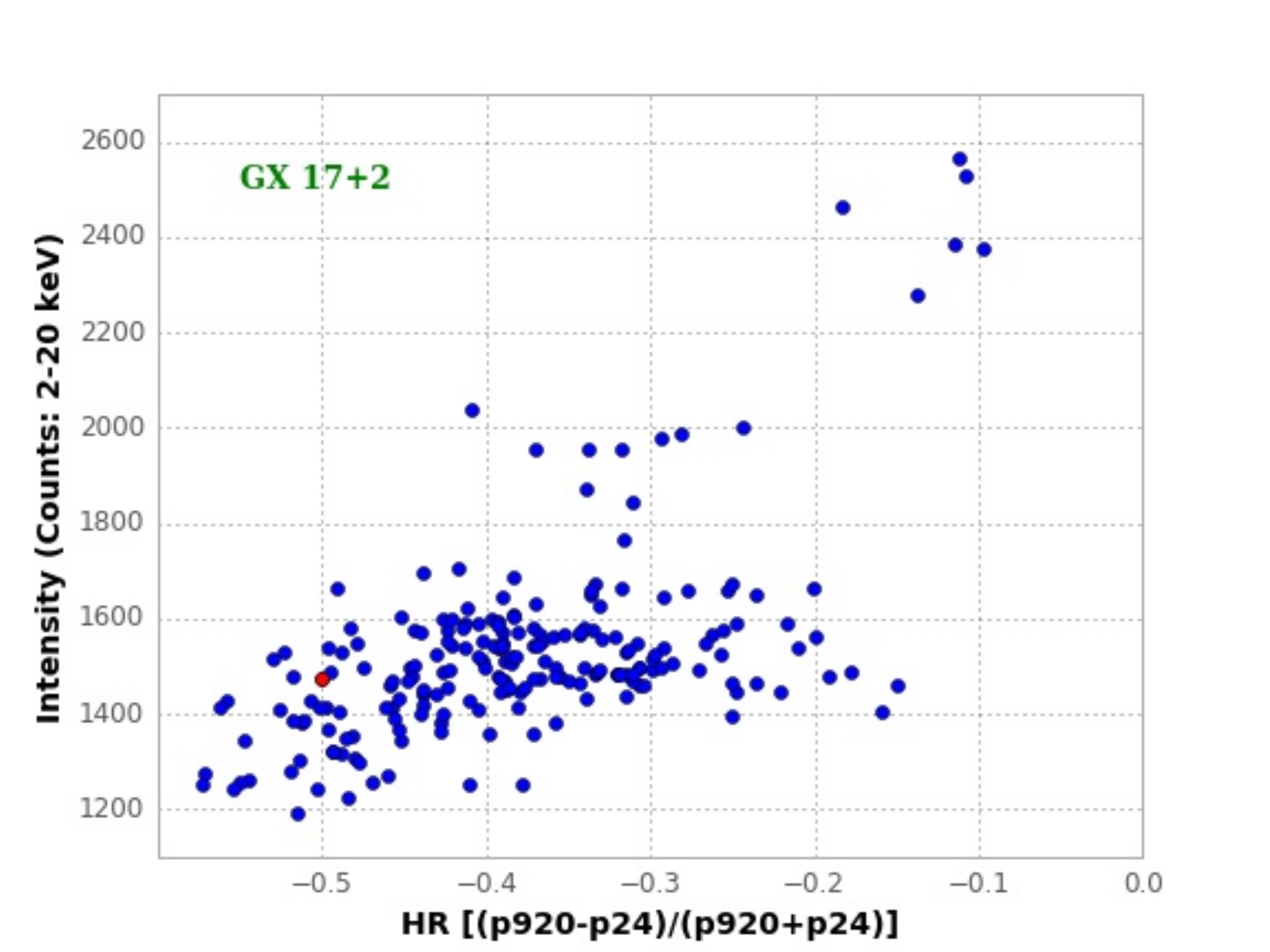}
\caption{Spectral Energy Distribution and color-intensity plots of GX 17+2.
\color{black}
The X-ray spectrum plots both Obsids 00035340006 and 00035340007 on Table \ref{tab:Swift_xrt} (SWIFT/XRT).
\color{black}
The panels and axes are as in figure \ref{fig:SED_LMCX3}.
\label{fig:SED_GX17p2}}
\end{figure}

GX 17+2 (4U 1813-14; NP Ser, Fig. \ref{fig:SED_GX17p2}; Fig. \ref{fig:appendix}.18.\color{black}) is a bright persistent steady source. It shows a detection by the ATCA at the mJy level, and detections by SMARTS. It shows an upper limit with the Swift UVOT. In the campaign week it was in the cusp between FB and NB state based on the RXTE PCA CI diagram; \color{black} the presence of radio emission suggests a jet and hence the NB state.\color{black} 

\subsubsection{X-ray Non-Detections (Z-sources: 3)}

LMC X-2, GX 340+0, and GX 5-1 \color{black} were not detected in either the RXTE-PCA or the SWIFT-XRT. \color{black}

\subsection{Atoll Sources (18 sources; 9 with SEDs)}

\begin{figure}[ht]
\hspace{.2in} \includegraphics[width=3.4in]{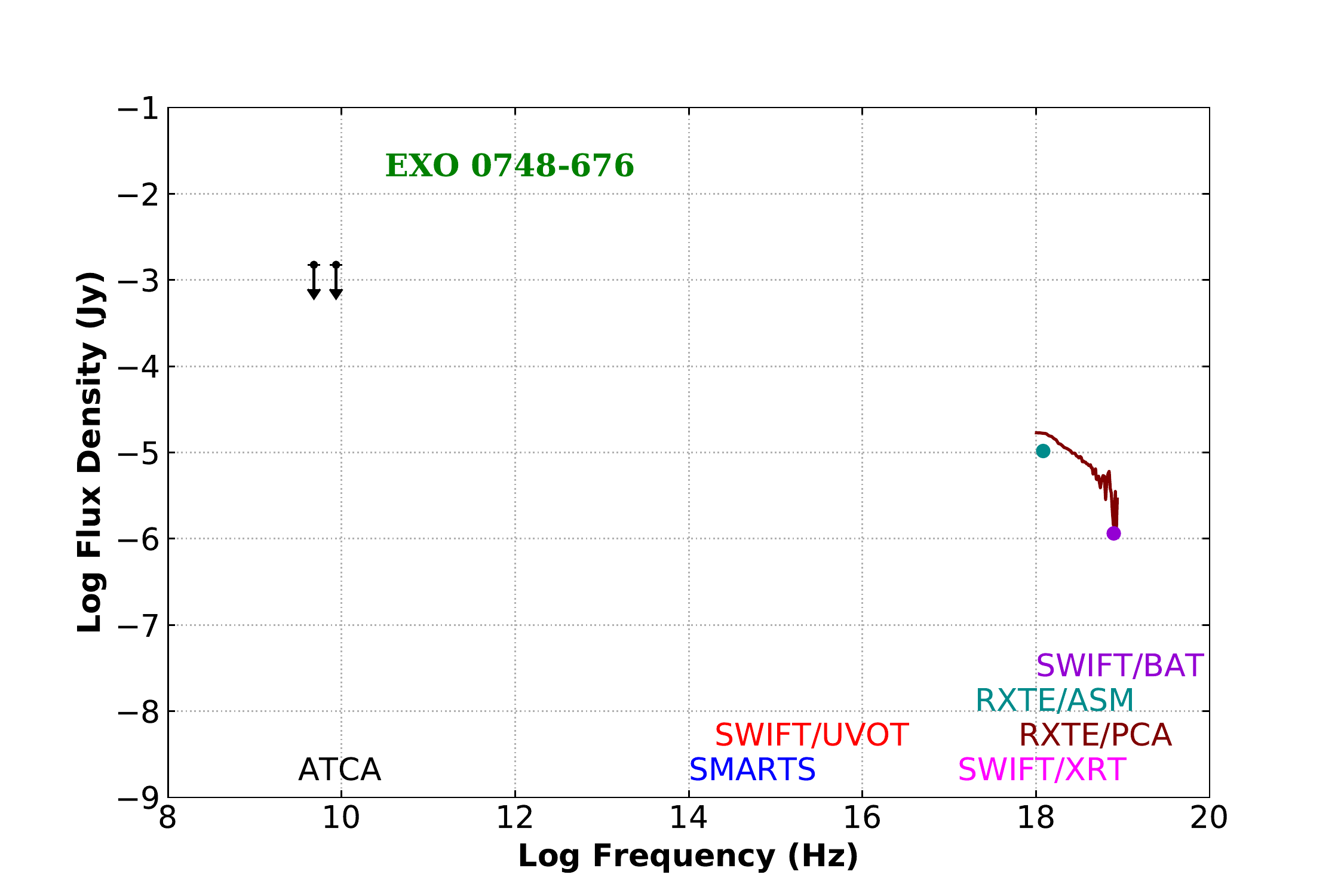}
\hspace{-.5in}
\hspace{.5in} \includegraphics[width=3in]{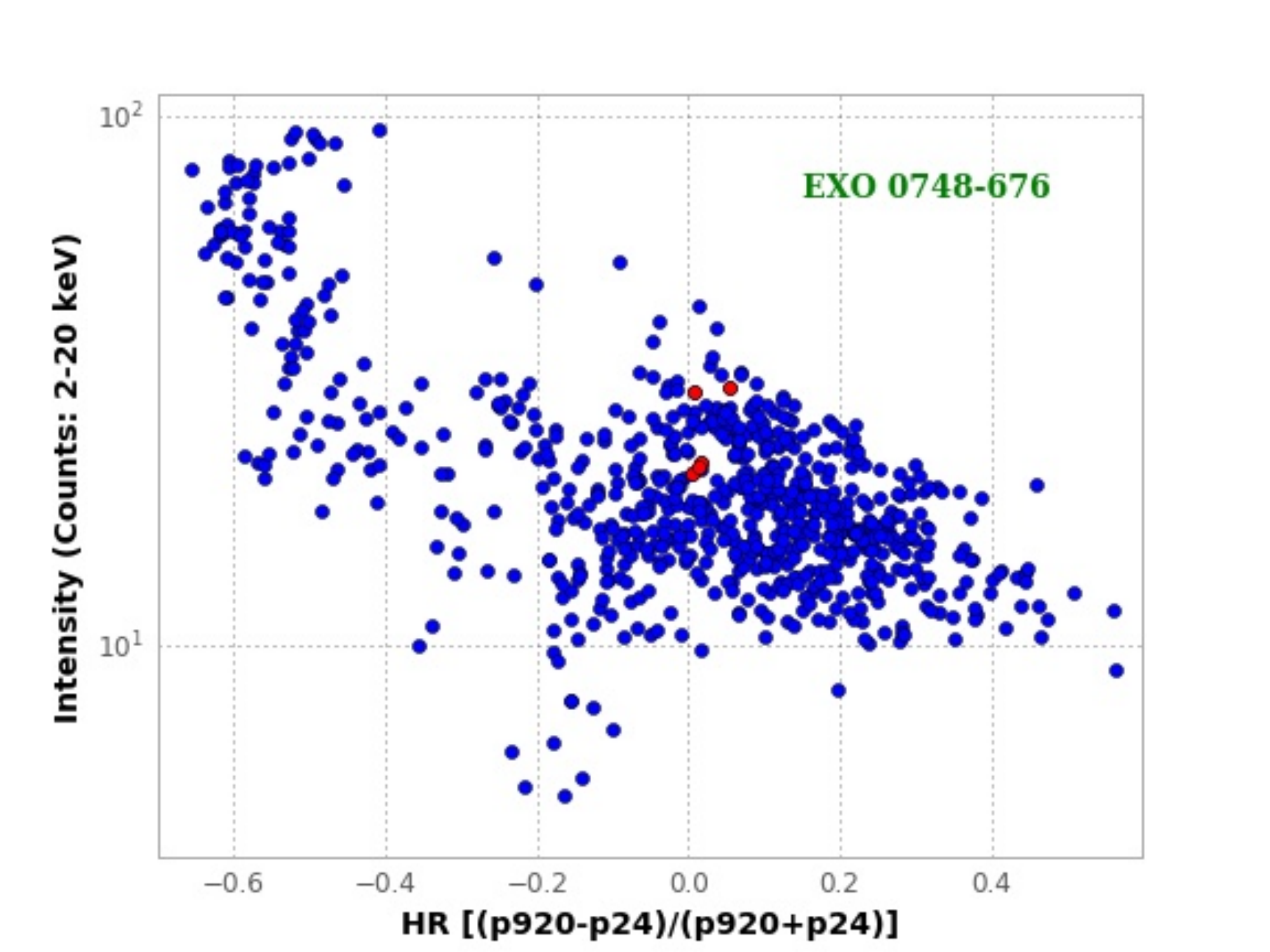}
\caption{Spectral Energy Distribution and color-intensity plots of EXO 0748-676. The panels and axes are as in figure \ref{fig:SED_LMCX3}.
\label{fig:SED_EXO0748}}
\end{figure}

EXO 0748-67 (UY Vol, Fig. \ref{fig:SED_EXO0748}; Fig. \ref{fig:appendix}.2.\color{black}) is a burster. It is mostly at the level of 1 ASM cts s$^{-1}$, with bursts up to 4 ASM cts s$^{-1}$. During the campaign week it was at the 1 ASM cts s$^{-1}$ level. The ATCA gives upper limits. The RXTE PCA flux density is consistent with the Swift BAT upper limit. The RXTE ASM flux density is slightly low. In the campaign week it was in the LB state based on the RXTE PCA CI diagram.

\begin{figure}[ht]
\hspace{.2in} \includegraphics[width=3.4in]{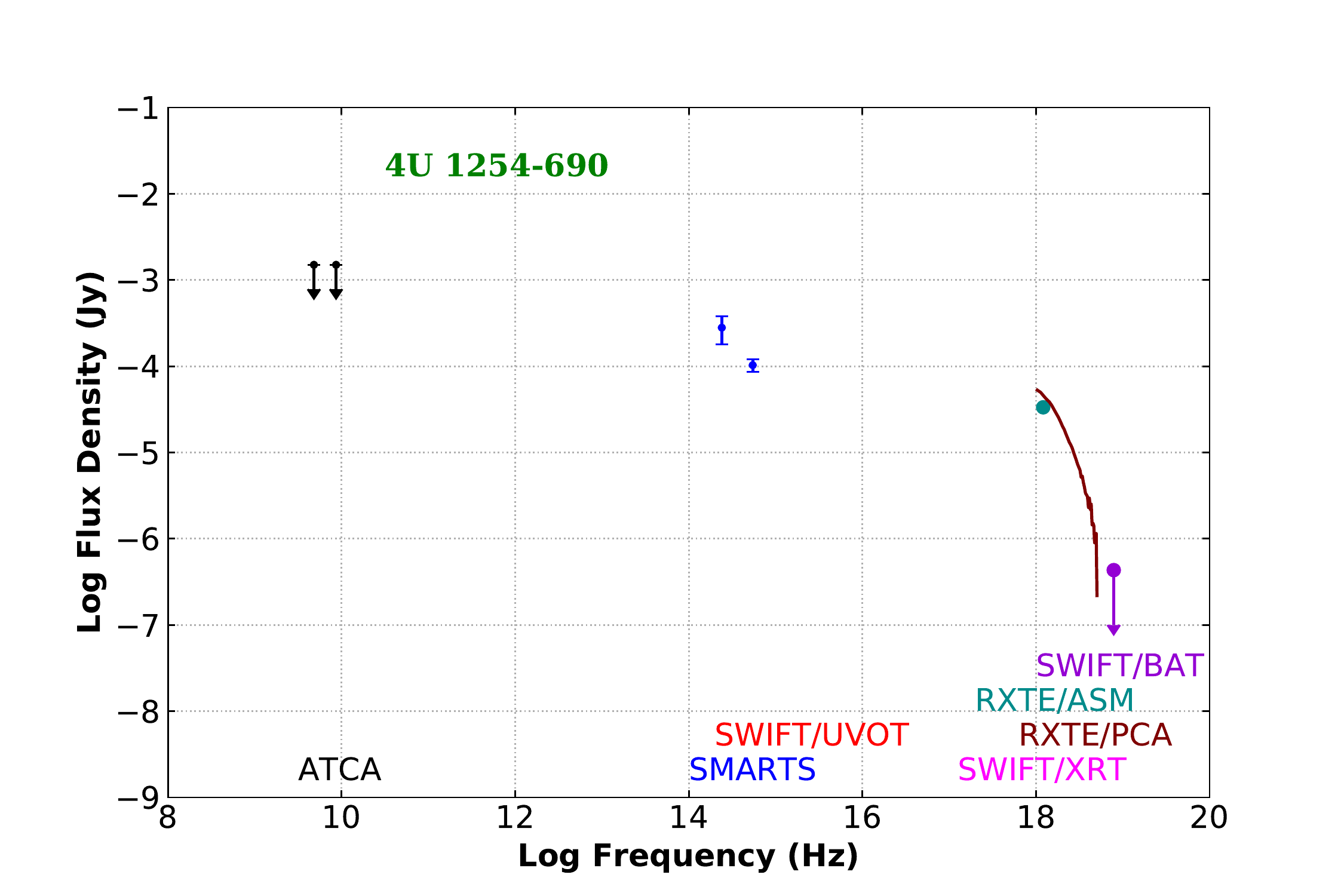}
\hspace{-.5in}
\hspace{.5in} \includegraphics[width=3in]{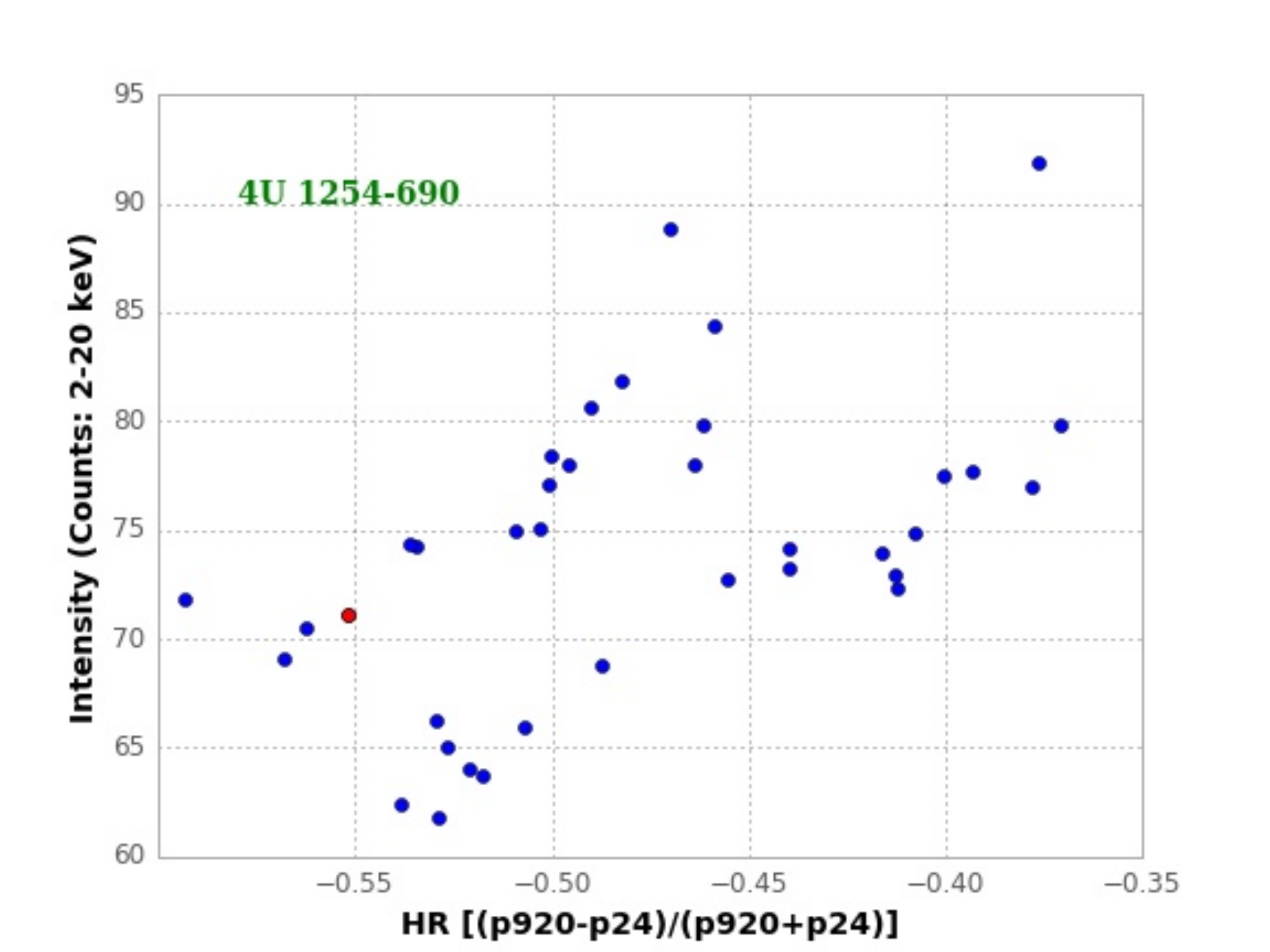}
\caption{Spectral Energy Distribution and color-intensity plots of 4U 1254-690. The panels and axes are as in figure \ref{fig:SED_LMCX3}.
\label{fig:SED_4U1254m69}}
\end{figure}

4U 1254-69 (GR Mus, Fig. \ref{fig:SED_4U1254m69}; Fig. \ref{fig:appendix}.3.\color{black}) is a burster with source persistent at about 3 ASM cts s$^{-1}$.  \color{black} 
It is one of four bursters that have shown a superburst in connection with
the regular burst \citep{int_Zand_etal_2003}. \color{black}
The ATCA gives upper limits. SMARTS gives one upper limit and one detection. The RXTE PCA is consistent with the RXTE ASM. The Swift BAT gives an upper limit. During the campaign week it was on the cusp between the LB and the UB state based on the RXTE PCA CI diagram. 

\begin{figure}[ht]
\hspace{.2in} \includegraphics[width=3.4in]{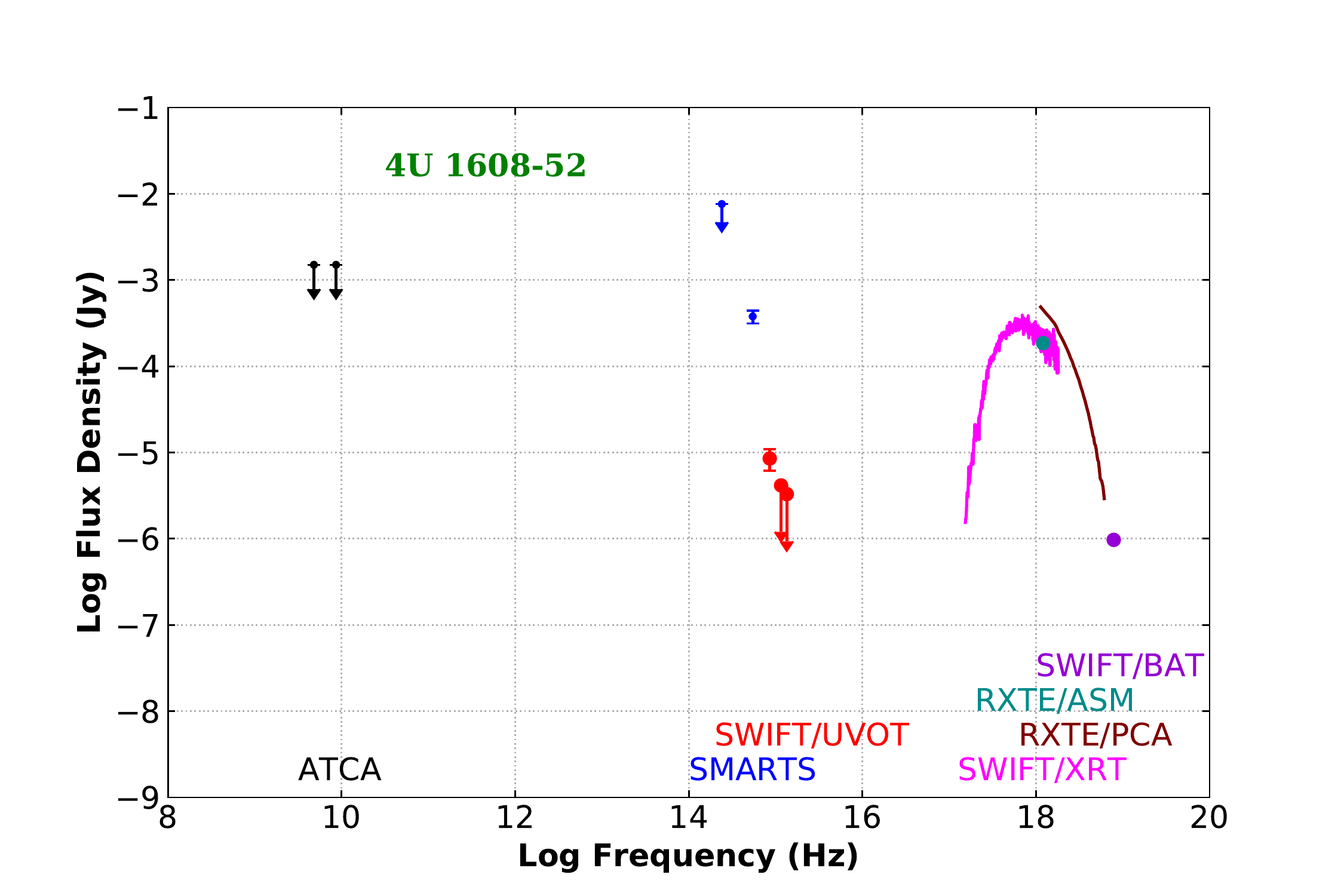}
\hspace{-.5in}
\hspace{.5in} \includegraphics[width=3in]{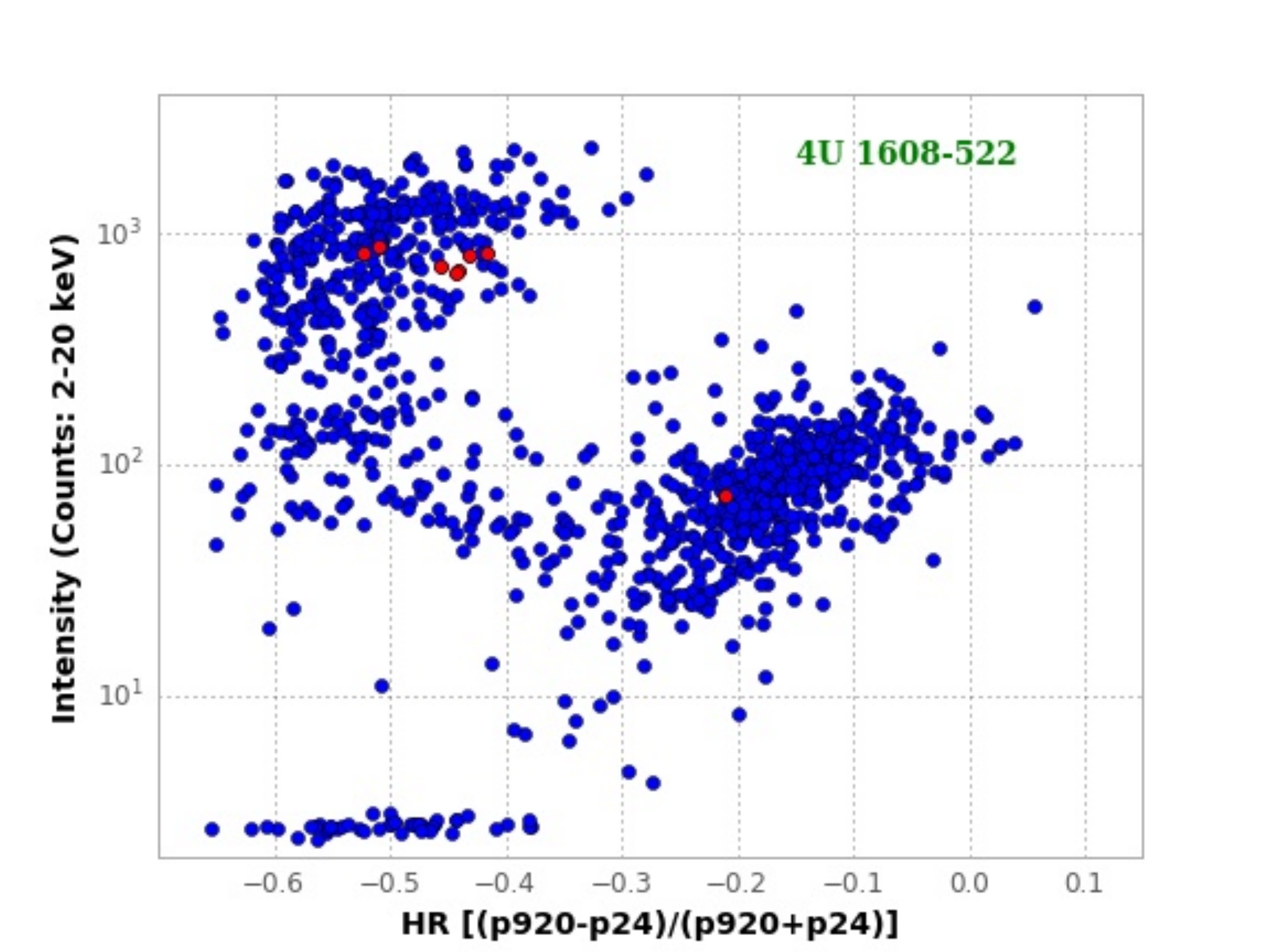}
\caption{Spectral Energy Distribution and color-intensity plots of 4U 1608-52.
\color{black}
The X-ray spectrum plots the average of three Obsids on Table \ref{tab:PCA-1} (RXTE/PCA) and Obsid 00030791018 on Table
\ref{tab:Swift_xrt} (SWIFT/XRT).
\color{black}
The panels and axes are as in figure \ref{fig:SED_LMCX3}.
\label{fig:SED_4U1608m52}}
\end{figure}

4U 1608-522 (QX Nor, Fig. \ref{fig:SED_4U1608m52}; Fig. \ref{fig:appendix}.5.\color{black}) is a transient burster. Most of the time it is not detected by the ASM, but accretion disk outbursts reach up to 65 cts s$^{-1}$. During our campaign week it was at the downturn of an outburst at 20 ASM cts s$^{-1}$.  This is consistent with it being in the propeller state 
\citep{Asai_etal_2013,Matsuoka_Asai_2013}. The derived value of N$_H$ varies by more
than the error bars on Table \ref{tab:Swift_xrt}, col. 8, after MJD 54284.  If this variation is real, it may be linked to the pile-up of accretion disk material expected
during the onset of the propeller phase \citep[e.g.][]{Lii_etal_2014}. 
The PCA data matches the BAT data, but it gives a higher flux density  than the ASM and XRT. Its status during our campaign week is consistent with the UB state based on the RXTE PCA CI diagram.  
\color{black}

\begin{figure}[ht]
\hspace{.2in} \includegraphics[width=3.4in]{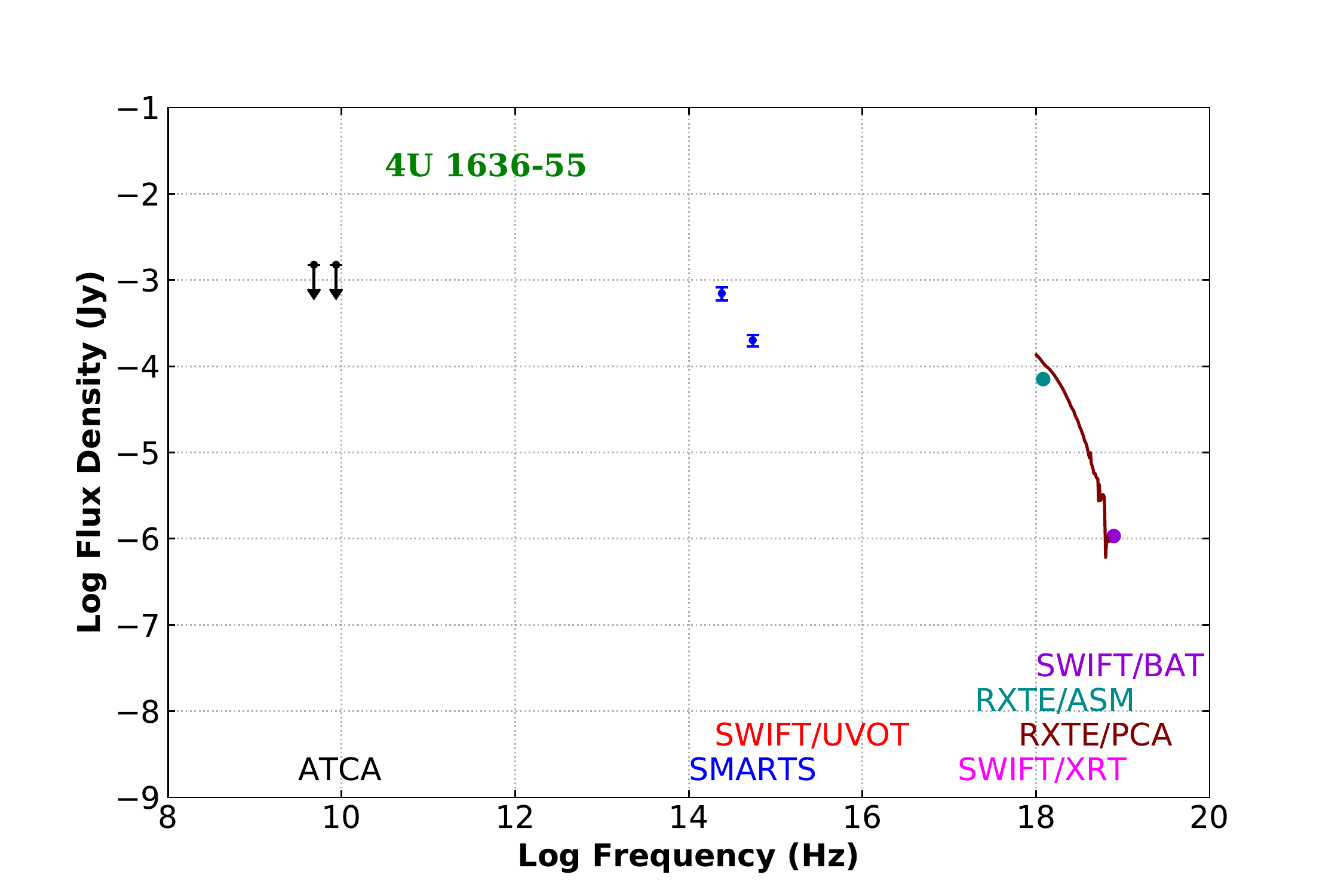}
\hspace{-.5in}
\hspace{.5in} \includegraphics[width=3in]{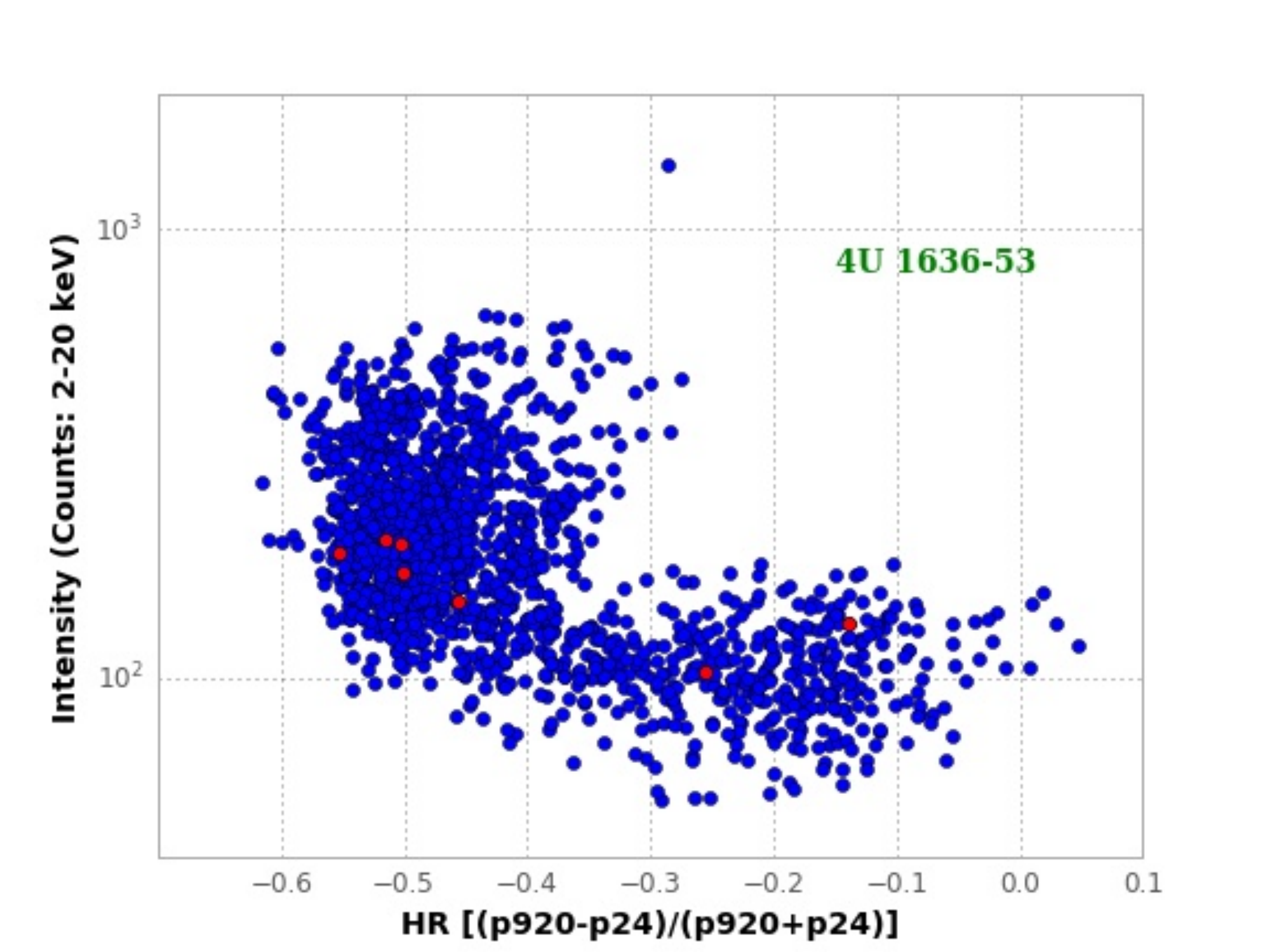}
\caption{Spectral Energy Distribution and color-intensity plots of 4U 1636-53. The panels and axes as in figure \ref{fig:SED_LMCX3}.
\label{fig:SED_4U1636m53}}
\end{figure}

4U 1636-53 (V801 ArA, Fig. \ref{fig:SED_4U1636m53}; Fig. \ref{fig:appendix}.6.\color{black}) is a persistent 
but variable source, with maxima
reaching up to 20 ASM cts s$^{-1}$ lasting from 10-20 days.
It is one of four bursters that have shown a superburst in connection with the regular burst \citep{int_Zand_etal_2003,Strohmayer_Markwardt_2002}. 
 Our campaign week was during the downside of an outburst (10 cts s$^{-1}$). The ATCA shows upper limits. SMARTS has positive detections. The RXTE PCA result matches those of the RXTE ASM and Swift BAT.  During the campaign week it was in the cusp between LB and UB and it shows excursions into the LB state based on the RXTE PCA CI diagram. 
\color{black}

\begin{figure}[ht]
\hspace{.2in} \includegraphics[width=3.4in]{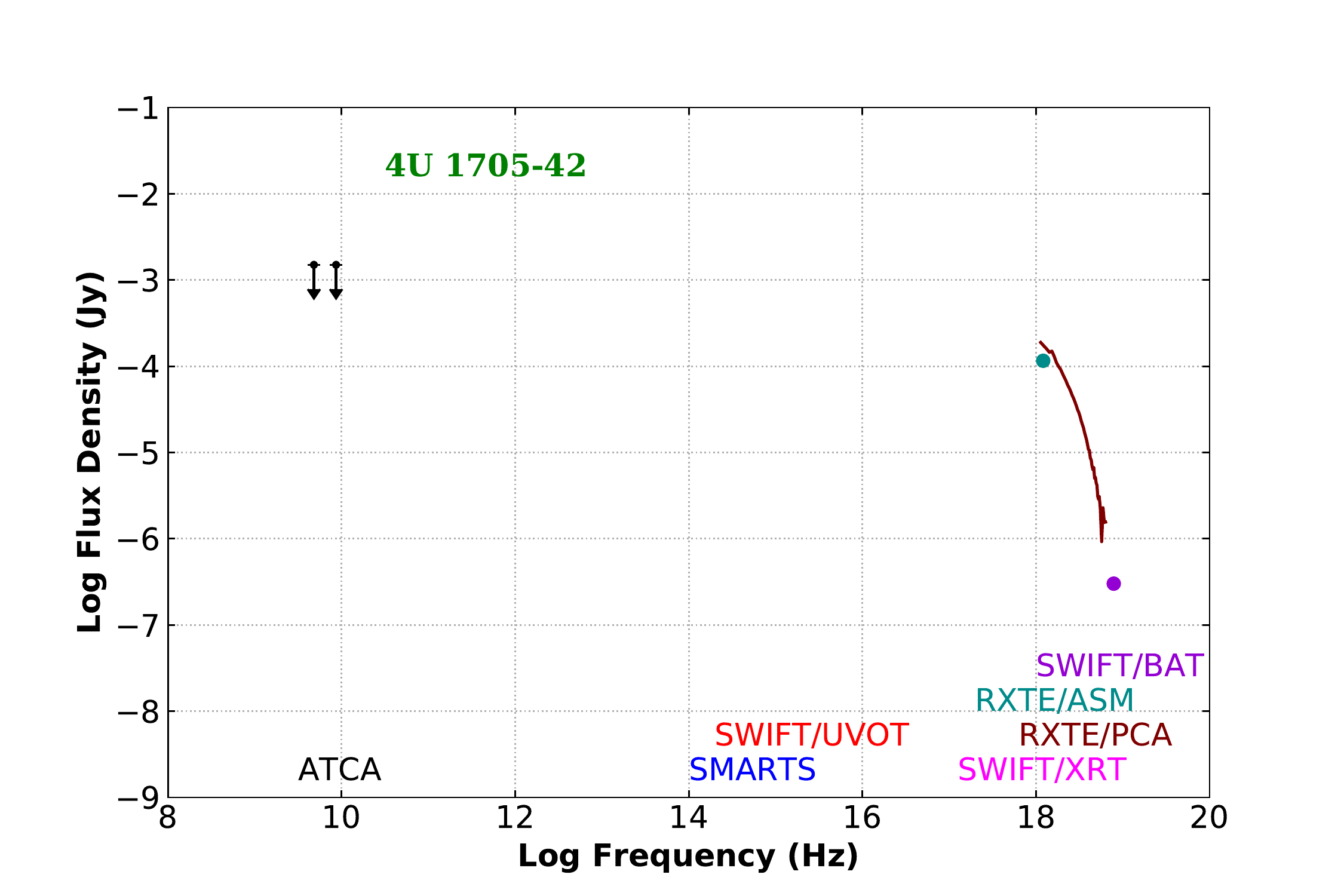}
\hspace{-.5in}
\hspace{.5in} \includegraphics[width=3in]{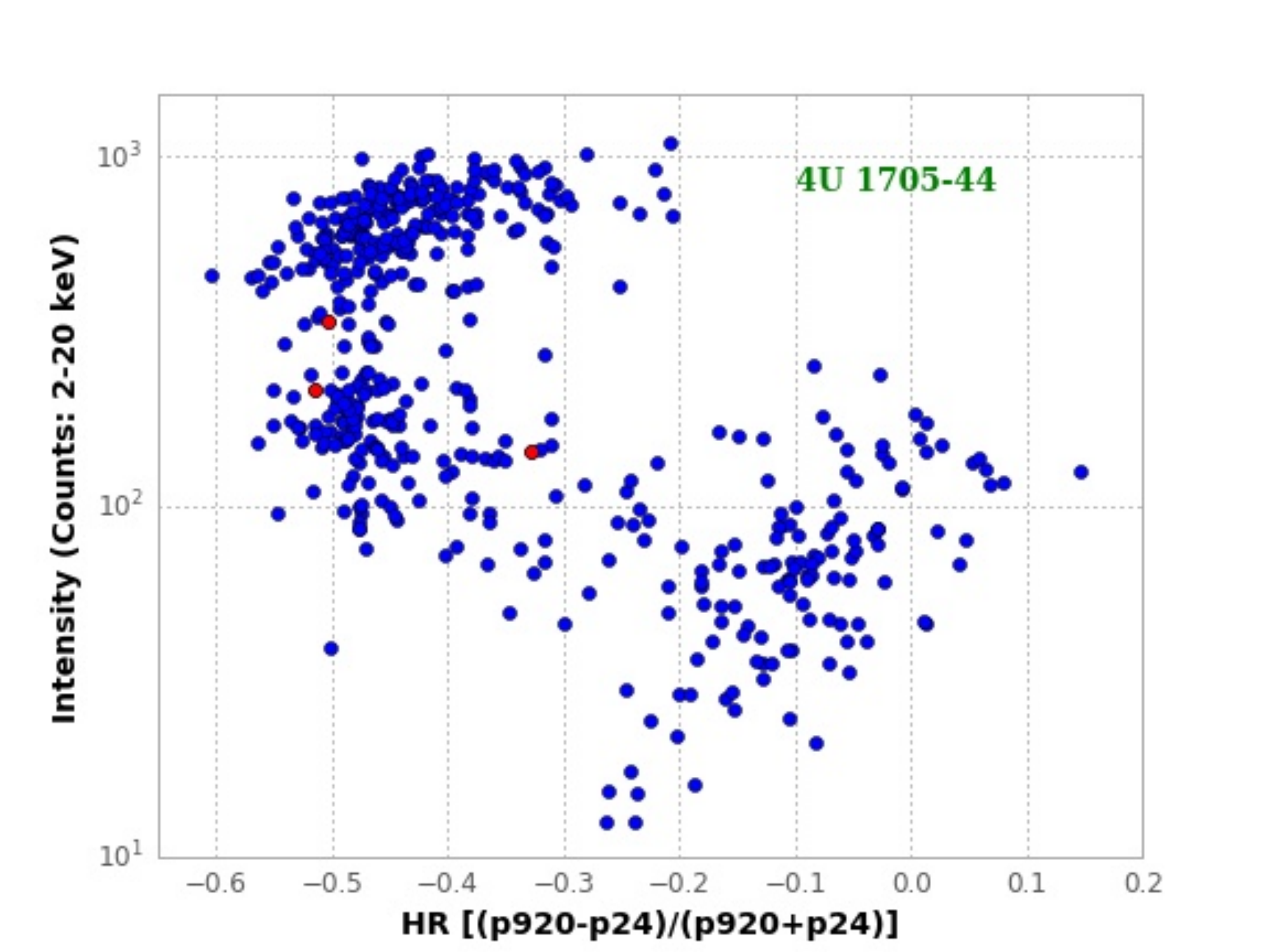}
\caption{Spectral Energy Distribution and color-intensity plots of 4U 1705-44.
\color{black}
The X-ray spectrum plots the average of the two observations on Table \ref{tab:PCA-1} (RXTE/PCA).
\color{black}
The panels and axes as in figure \ref{fig:SED_LMCX3}.
\label{fig:SED_4U1705m44}}
\end{figure}

4U 1705-44 (Fig. \ref{fig:SED_4U1705m44}; Fig. \ref{fig:appendix}.10.\color{black}) is a \color{black}persistent 
but variable source, \color{black}with maxima
reaching up to 30 ASM cts s$^{-1}$ lasting from 20-40 days 
\citep{Phillipson_etal_2018}. During the campaign week was it was on the downside of an outburst (15cts s$^{-1}$). The ATCA gives an upper limit. The RXTE PCA flux density matches the RXTE ASM and Swift BAT results. In the campaign week it was in the UB state based on the RXTE PCA CI diagram. 
\color{black}

\begin{figure}[ht]
\hspace{.2in} \includegraphics[width=3.4in]{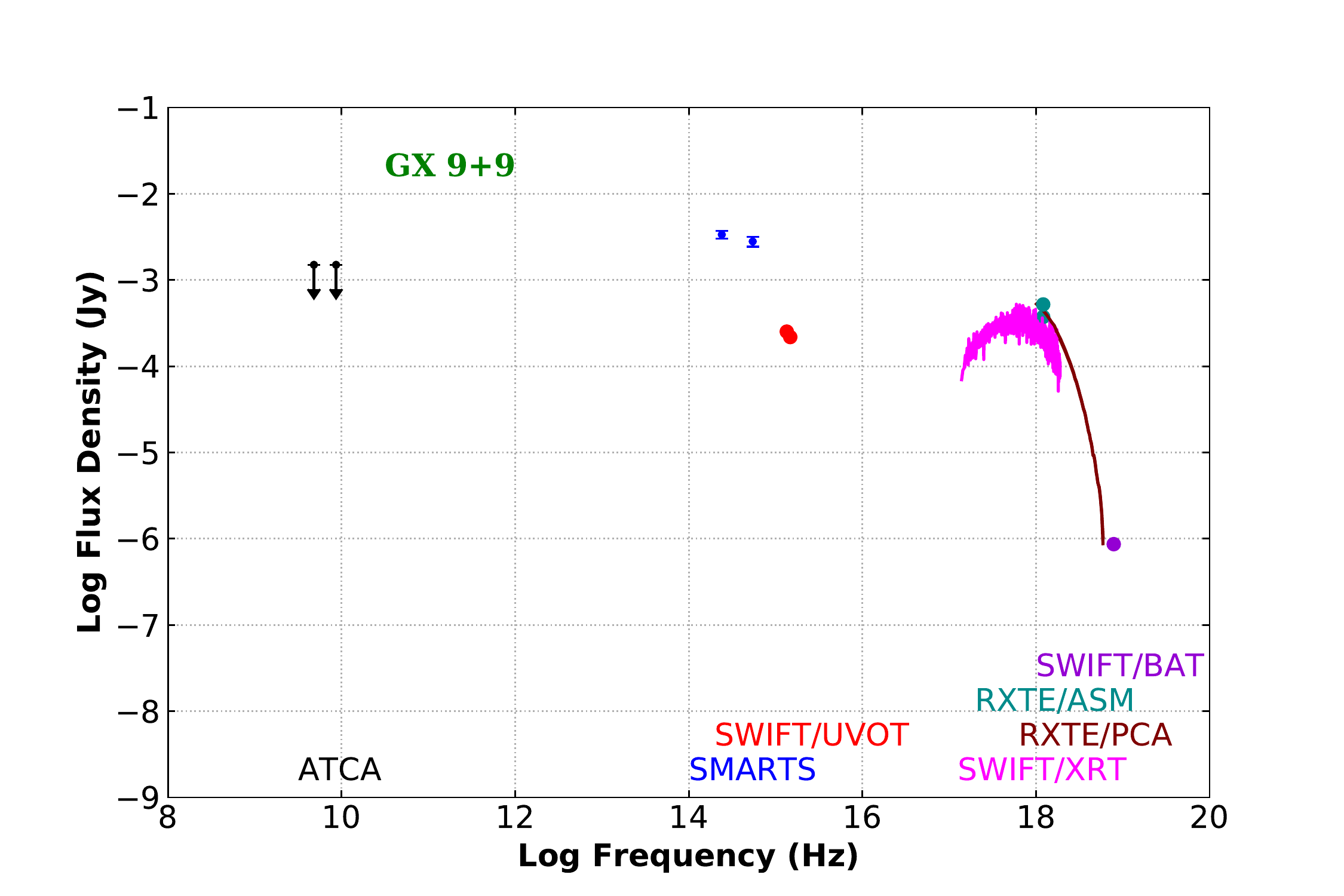}
\hspace{-.5in}
\hspace{.5in} \includegraphics[width=3in]{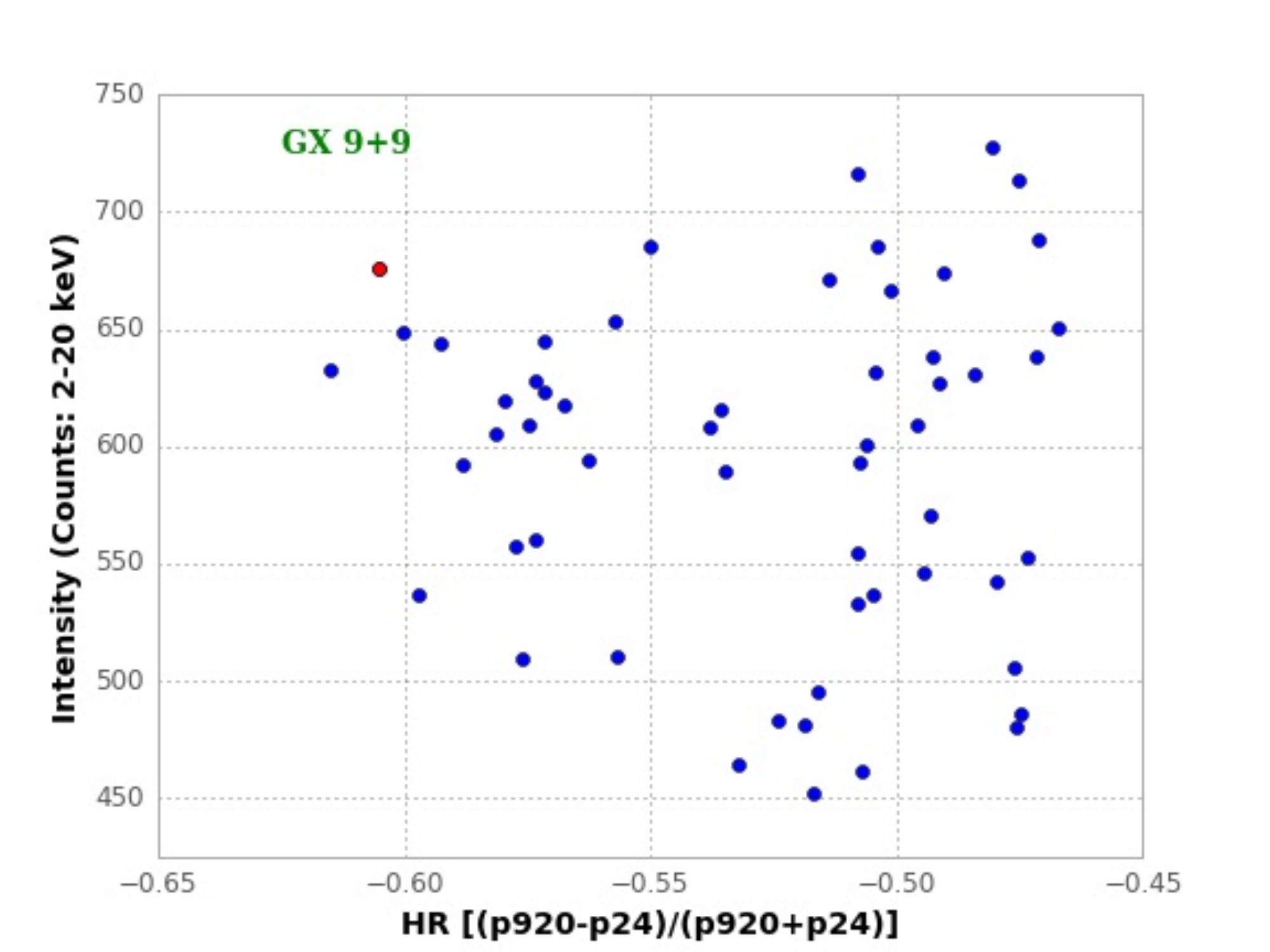}
\caption{Spectral Energy Distribution and color-intensity plots of GX 9+9.
\color{black}
The X-ray spectrum plots all three Obsids on Table \ref{tab:Swift_xrt} (SWIFT/XRT).
\color{black}
The panels and axes as in figure \ref{fig:SED_LMCX3}.
\label{fig:SED_GX9p9}}
\end{figure}

GX 9+9 (4U 1728-16; V2216 Oph, Fig. \ref{fig:SED_GX9p9}; Fig. \ref{fig:appendix}.37.\color{black})  is a bright, persistent, smoothly varying source, typically giving 15 to 25 ASM cts s$^{-1}$; during our campaign it was at a level of 25 cts s$^{-1}$. The ATCA gives upper limits; SMARTS and the UVOT give positive detections. Flux densities from the Swift XRT, RXTE PCA, ASM, and BAT are all consistent. During the campaign week it was in the UB state based on the RXTE PCA CI diagram. 

\begin{figure}[ht]
\hspace{.2in} \includegraphics[width=3.4in]{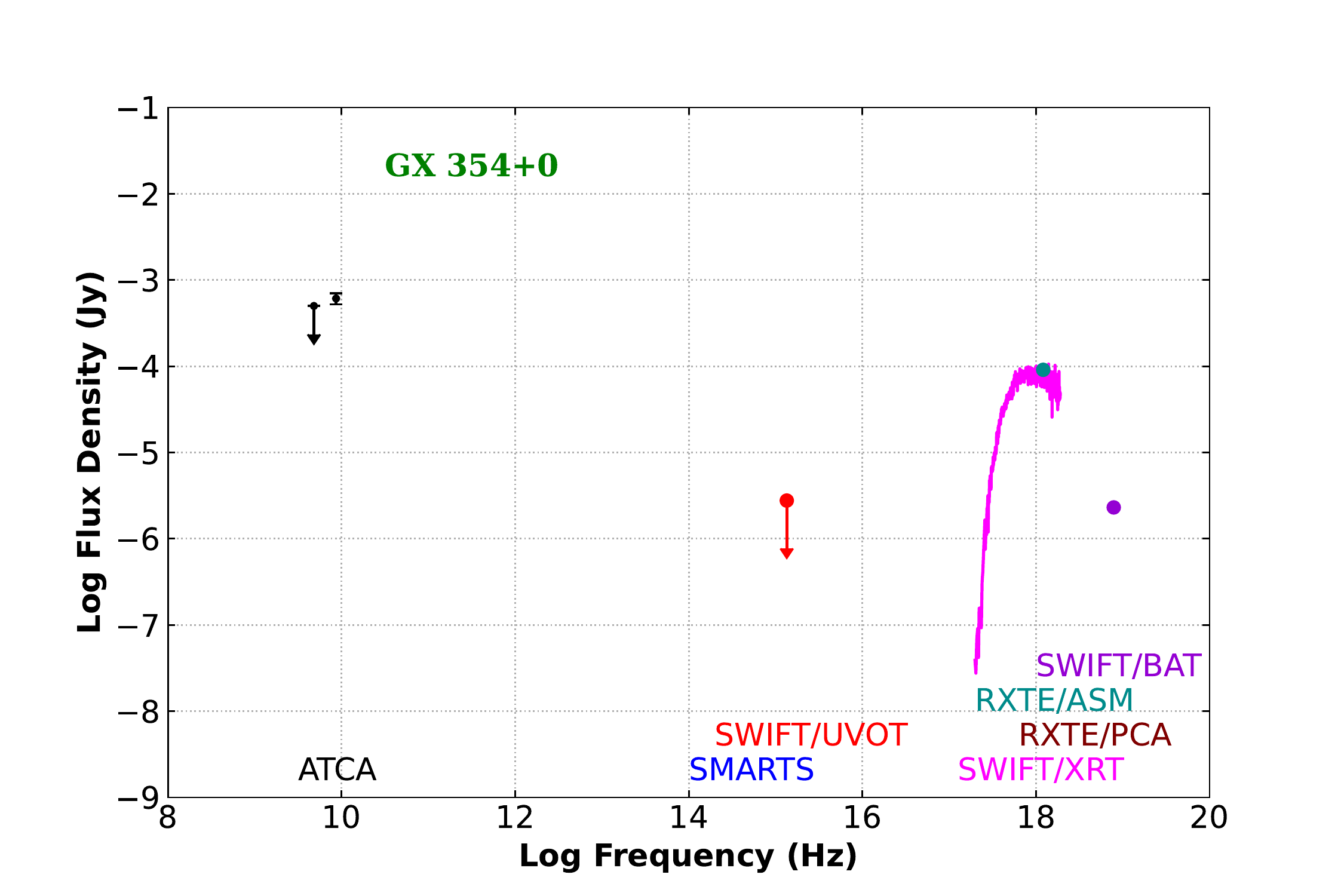}
\hspace{-.5in}
\hspace{.5in} \includegraphics[width=3in]{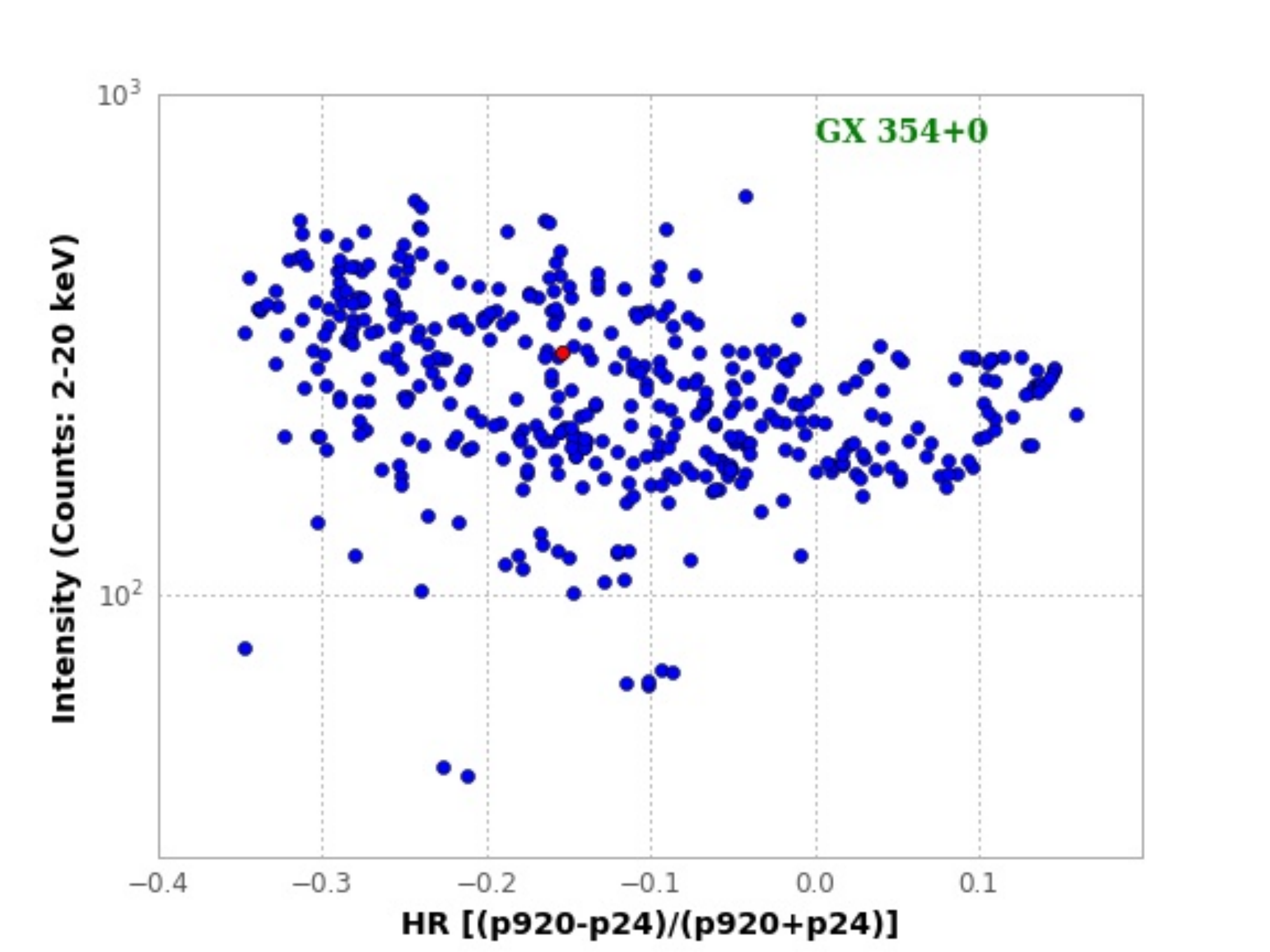}
\caption{Spectral Energy Distribution and color-intensity plots of GX 354+0. The panels and axes as in figure \ref{fig:SED_LMCX3}.
\label{fig:SED_GX354m0}}
\end{figure}

GX 354+0 (4U 1728-337, Fig. \ref{fig:SED_GX354m0}; Fig. \ref{fig:appendix}.22.\color{black}) is a persistent burster.  It typically ranges from 5-20 ASM cts s$^{-1}$. During our week it was at 7 cts s$^{-1}$. 
 \color{black}
It is unusual for an Atoll source in that radio detections consistent with a jet have been reported \citep{Migliari_etal_2003}.
 \color{black}
The ATCA and Swift UVOT give upper limits. The Swift XRT flux density is consistent with  the RXTE ASM. It is unclear if the BAT is high or low: that depends on how the spectrum turns over. During our campaign week it was in the IS based on the RXTE PCA CI diagram. 

\begin{figure}[ht]
\hspace{.2in} \includegraphics[width=3.4in]{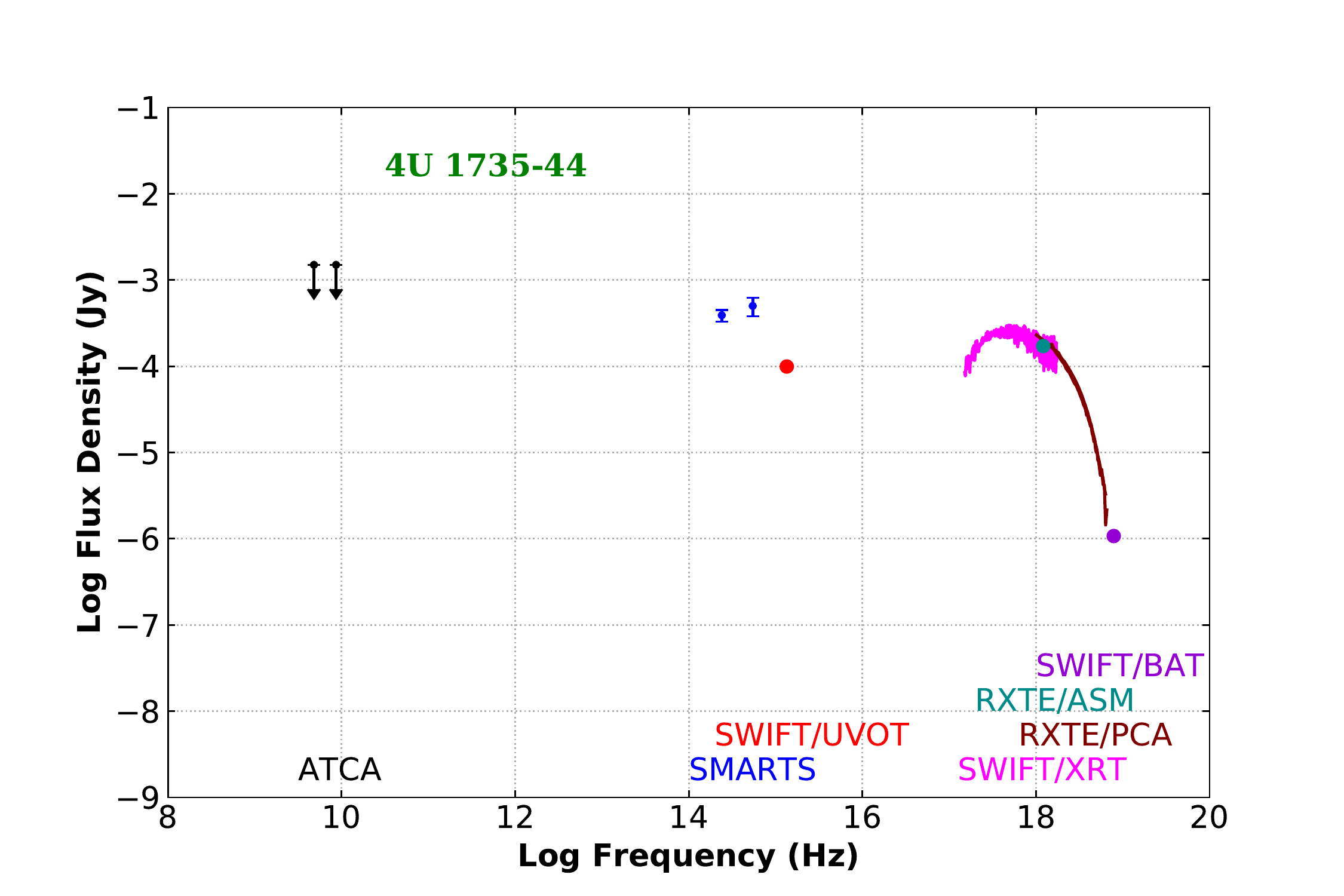}
\hspace{-.5in}
\hspace{.5in} \includegraphics[width=3in]{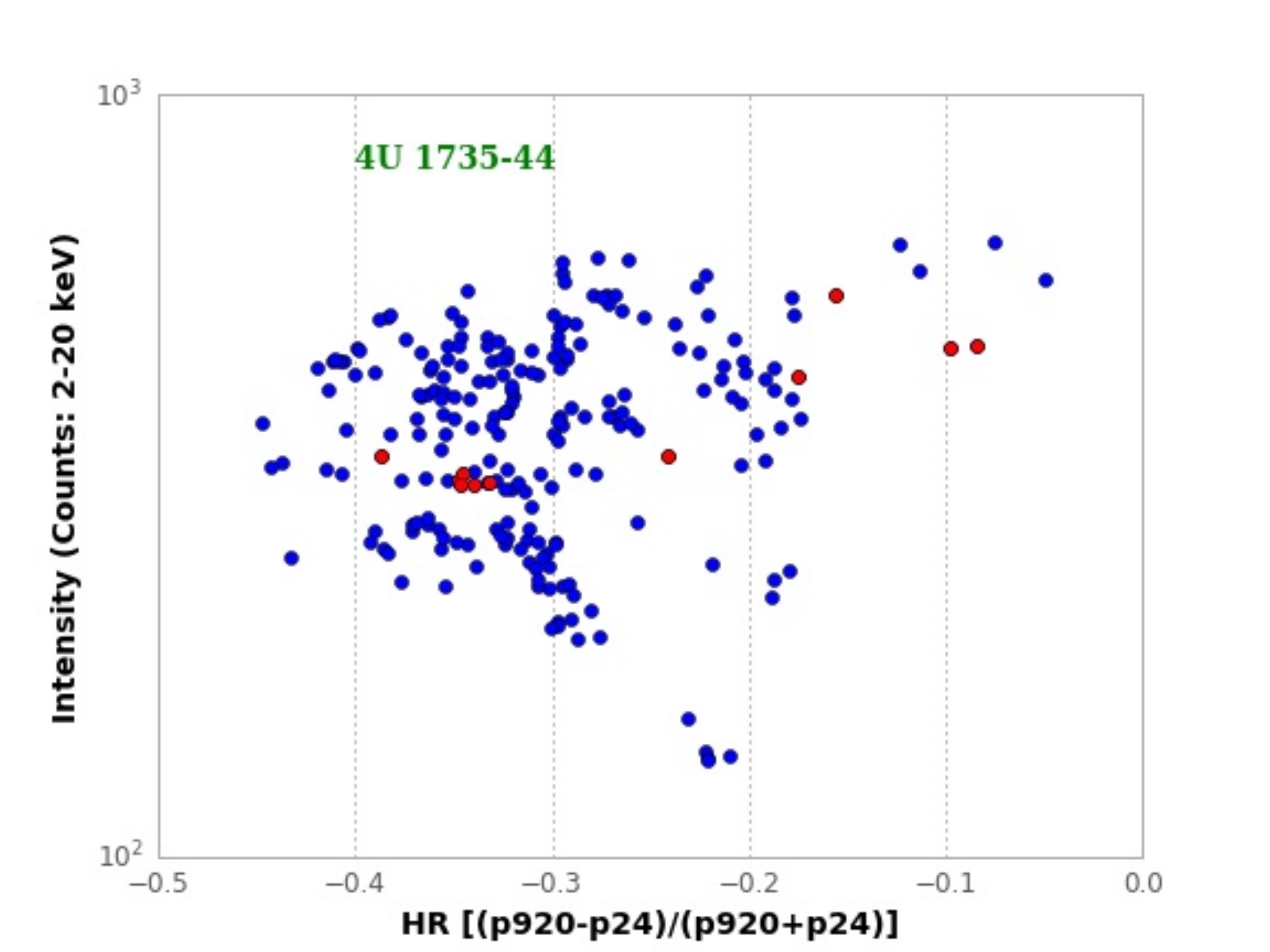}
\caption{Spectral Energy Distribution and color-intensity plots of 4U 1735-44. The panels and axes as in figure \ref{fig:SED_LMCX3}.
\label{fig:SED_4U1735m44}}
\end{figure}

4U 1735-44 (V926 Sco, Fig. \ref{fig:SED_4U1735m44}; Fig. \ref{fig:appendix}.11.\color{black}) is a bright, persistent, smoothly varying source; it typically ranges from 5 to 30  ASM cts s$^{-1}$. During our week it was between 12 and 15 cts s$^{-1}$. The ATCA gives upper limits. SMARTS and the Swift UVOT give positive detections. Results from the RXTE PCA, RXTE ASM, Swift XRT, and Swift BAT all match. It was in UB and IS based on the RXTE PCA CI diagram. 

\begin{figure}[ht]
\hspace{.2in} \includegraphics[width=3.4in]{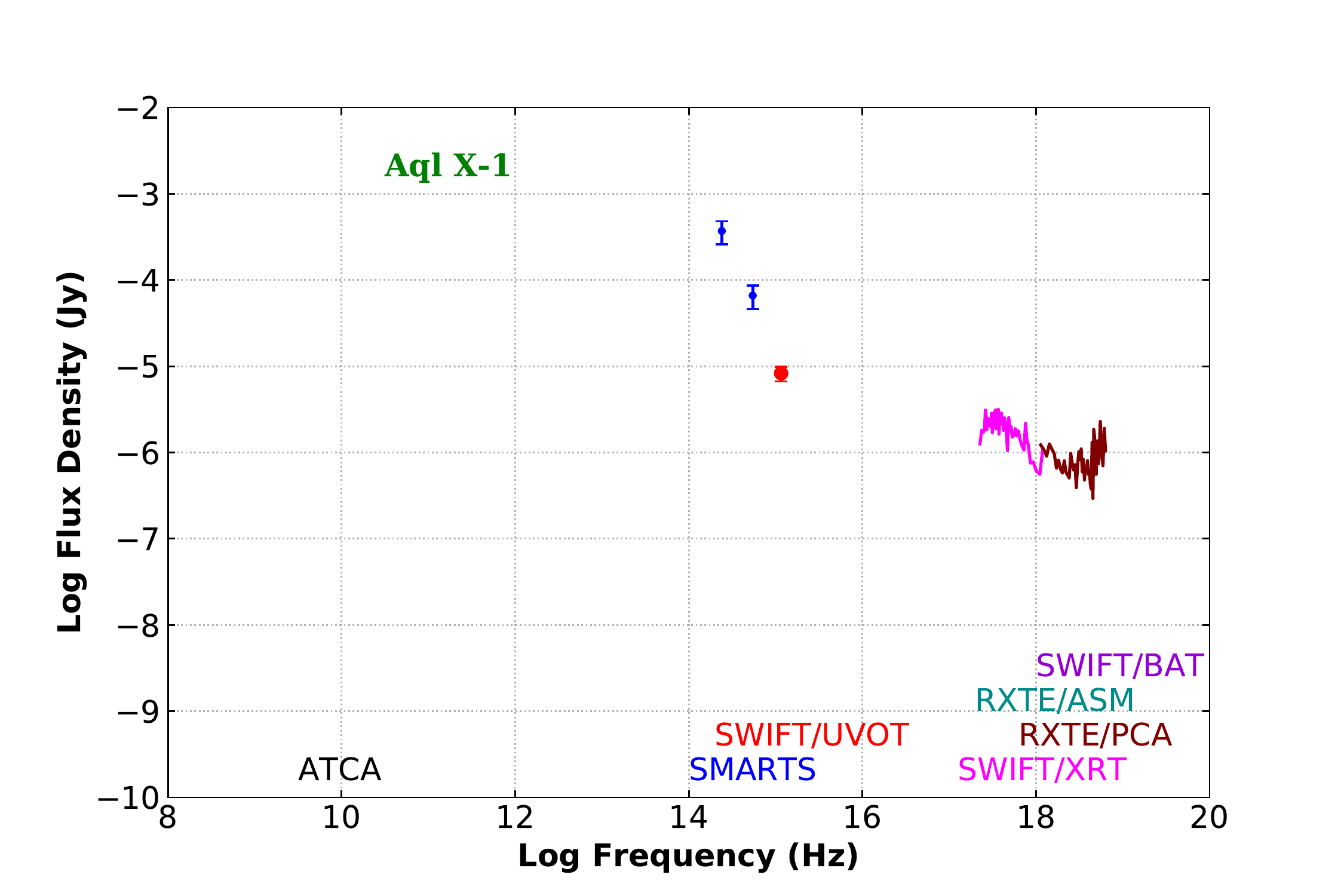}
\hspace{-.5in}
\hspace{.5in} \includegraphics[width=3in]{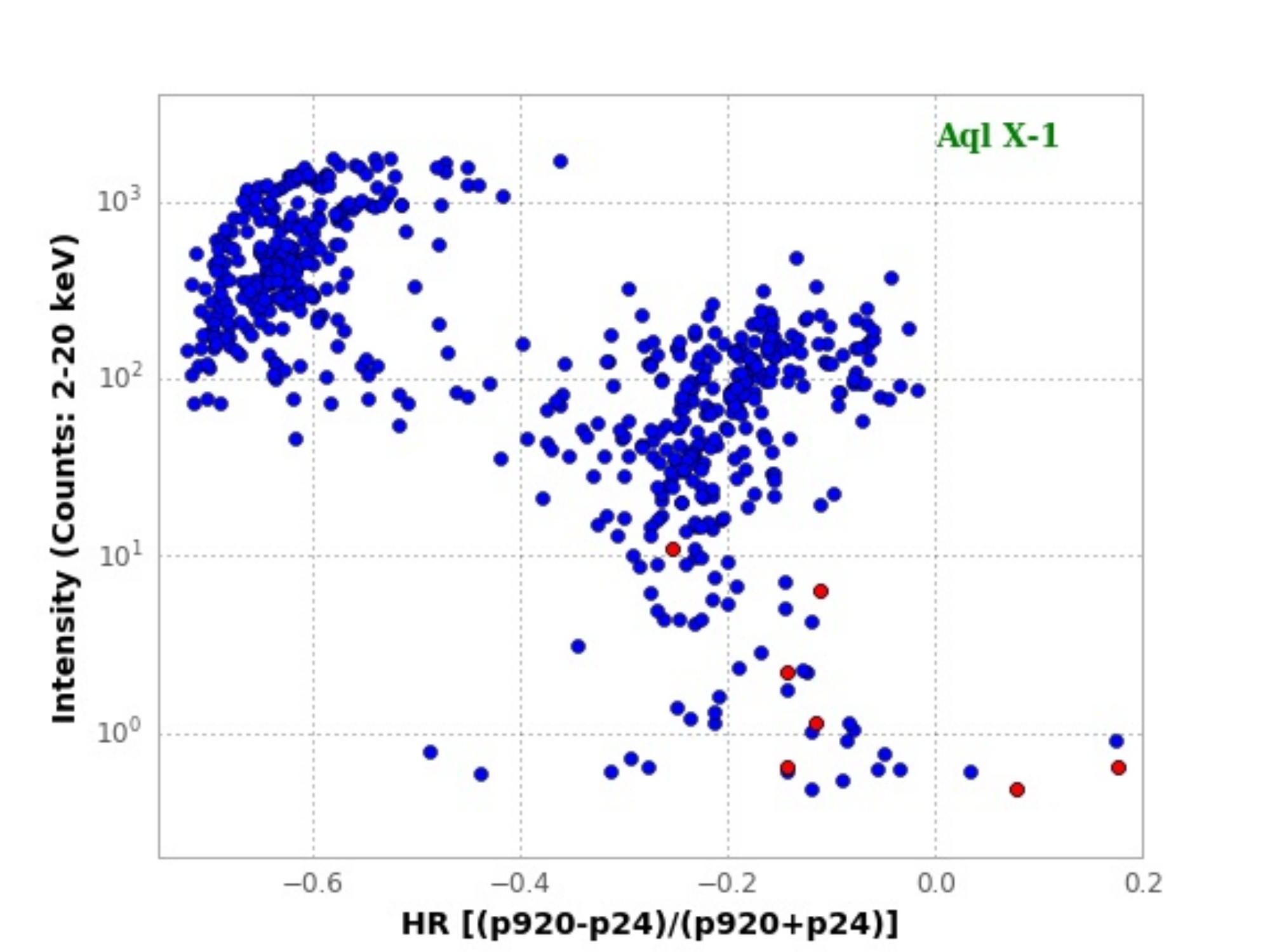}
\caption{Spectral Energy Distribution and color-intensity plots of Aql X-1.
\color{black}
The X-ray spectrum plots Obsid 30796022 on Table \ref{tab:Swift_xrt} (SWIFT/XRT).
\color{black}
The panels and axes as in figure \ref{fig:SED_LMCX3}.
\label{fig:SED_AqlX1}}
\end{figure}

Aql X-1 (4U 1908+00, Fig. \ref{fig:SED_AqlX1}; Fig. \ref{fig:appendix}.22.\color{black}) is a transient burster; it ranges from 0-50 ASM cts s$^{-1}$ec. It was not observed by  the ATCA. It shows positive detections with SMARTS and the Swift UVOT. Flux densities from the Swift XRT and RXTE PCA are consistent. There was no significant detection with the RXTE ASM during the week of the campaign, however the lightcurve shows that it was caught shortly after a small outburst.  It was in the LB state based on the RXTE PCA CI diagram. 

\subsubsection{X-ray Non-Detections (Atolls: 9)}

\color{black}
2S 0921-630 (V395 Car) and Cen X-4 (4U 1456-32; V822 Cen) give only upper limits with the ATCA and the RXTE  ASM. Cen X-4 is an SXT and a burster. It gives positive detections in SMARTS. \color{black} 4U 1543-624 (QU TrA), 4U 1556-60 (LU TrA), MXB 1658-298 (4U 1704-30, V2134 Oph), 4U 1702-429 (Ara X-1), GX 3+1 (4U 1744-26, Sgr X-1), GX 9+1 (4U 1758-20, Sgr X-3), and GX13+1 (Sgr X-2, V5512 Sgr) also show radio non-detections and minimal X-ray data.  
\color{black}

\subsection{Pulsars (11 sources; 3 with seds)}

\begin{figure}[ht]
\hspace{.2in} \includegraphics[width=3.4in]{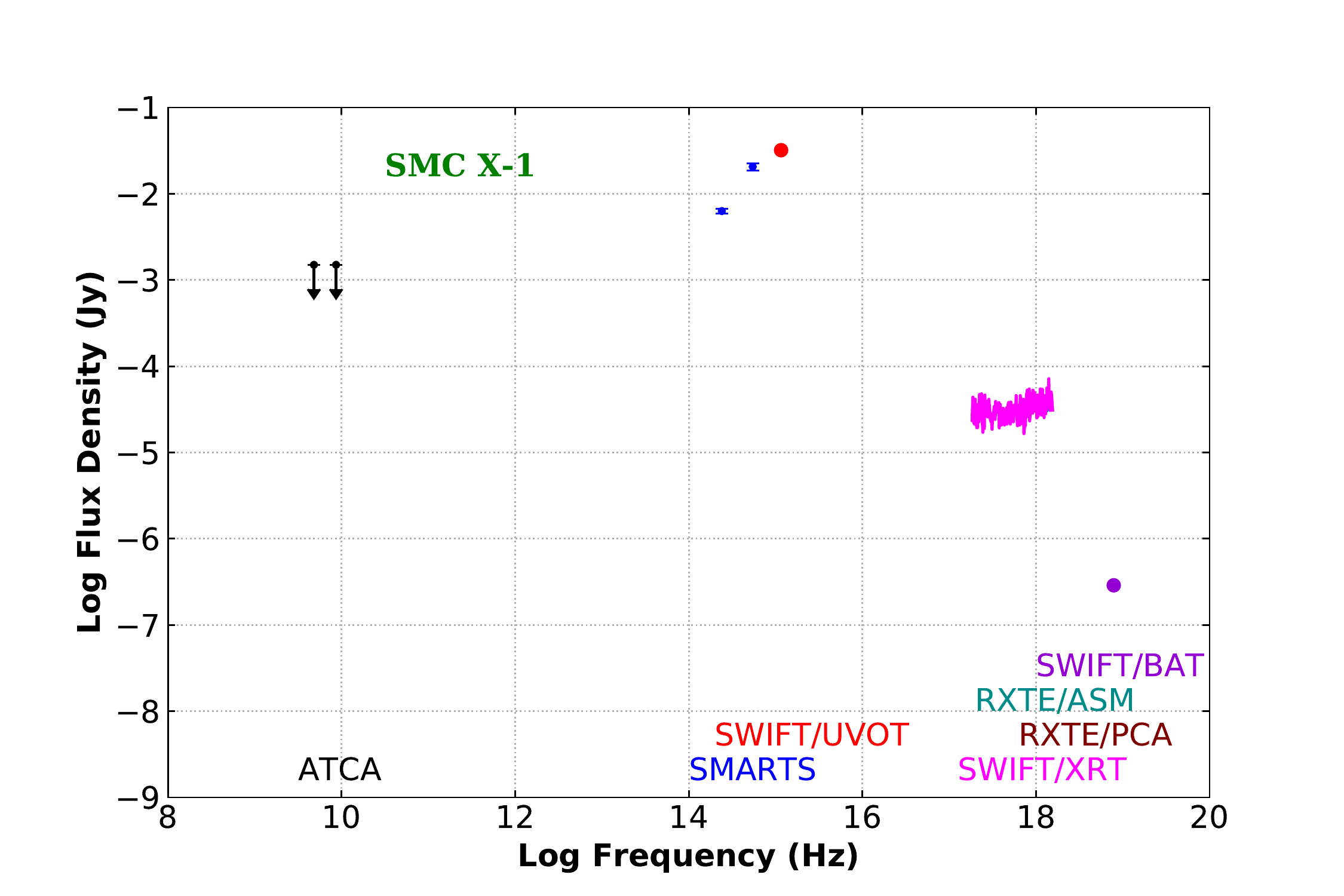}
\hspace{-.5in}
\hspace{.5in} \includegraphics[width=3in]{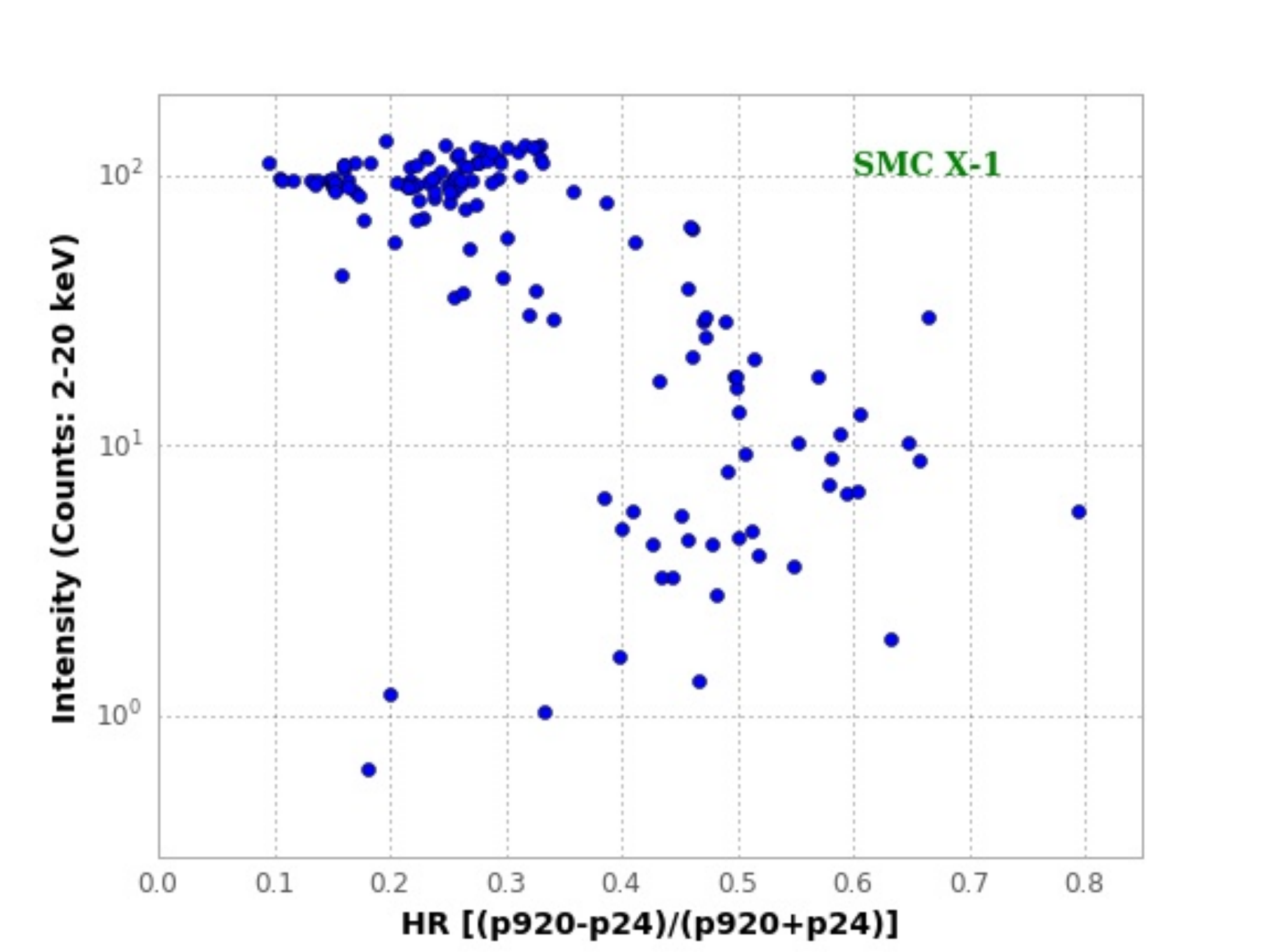}
\color{black}
\caption{Spectral Energy Distributions and color-intensity plots of SMC X-1. The panels and axes as in figure \ref{fig:SED_LMCX3}.
\label{fig:SED_SMCX1}}
\end{figure}

\begin{figure}[ht]
\hspace{.2in} \includegraphics[width=3.4in]{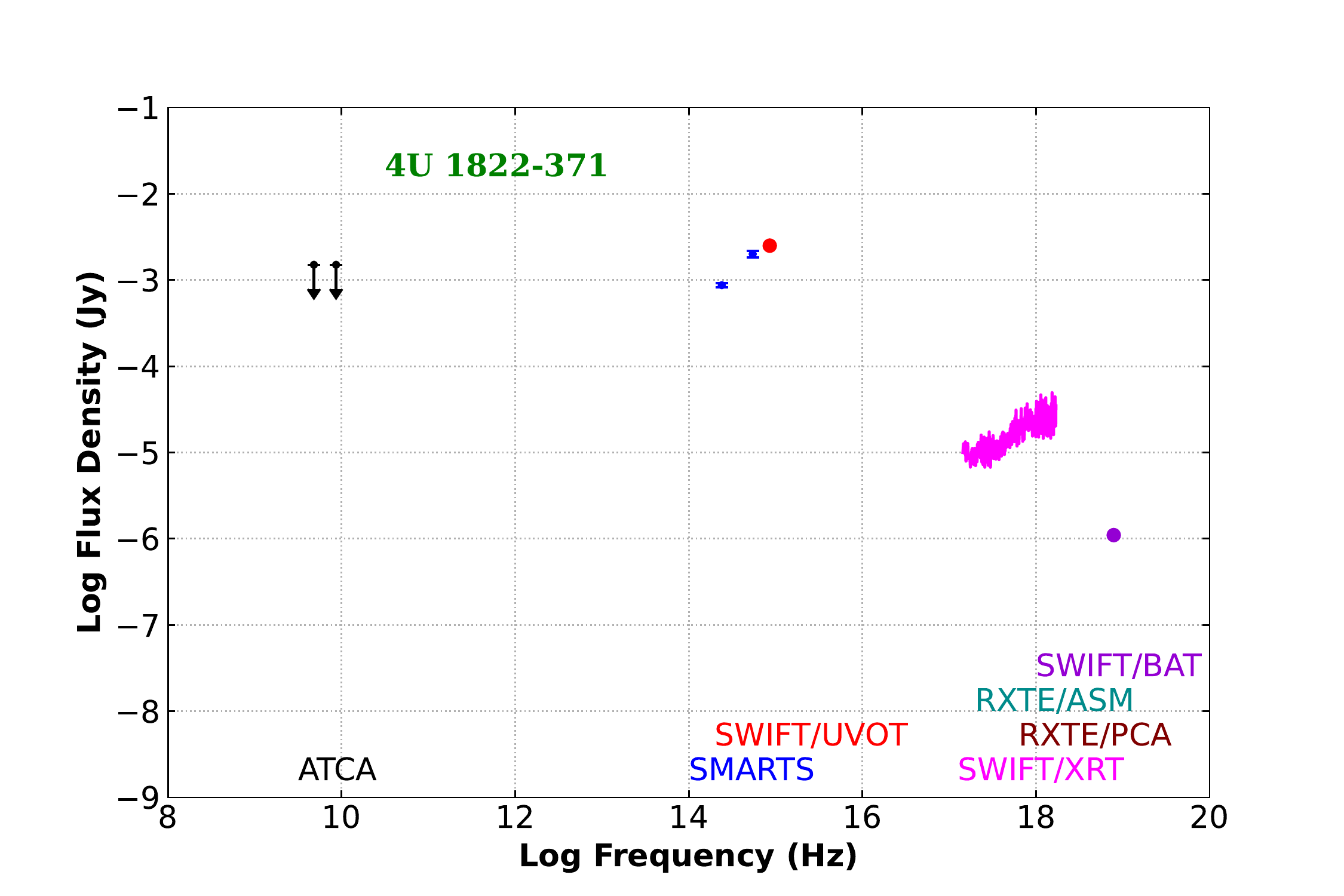}

\vspace{-2.7in}
\hspace{3in} \includegraphics[width=4.4in]{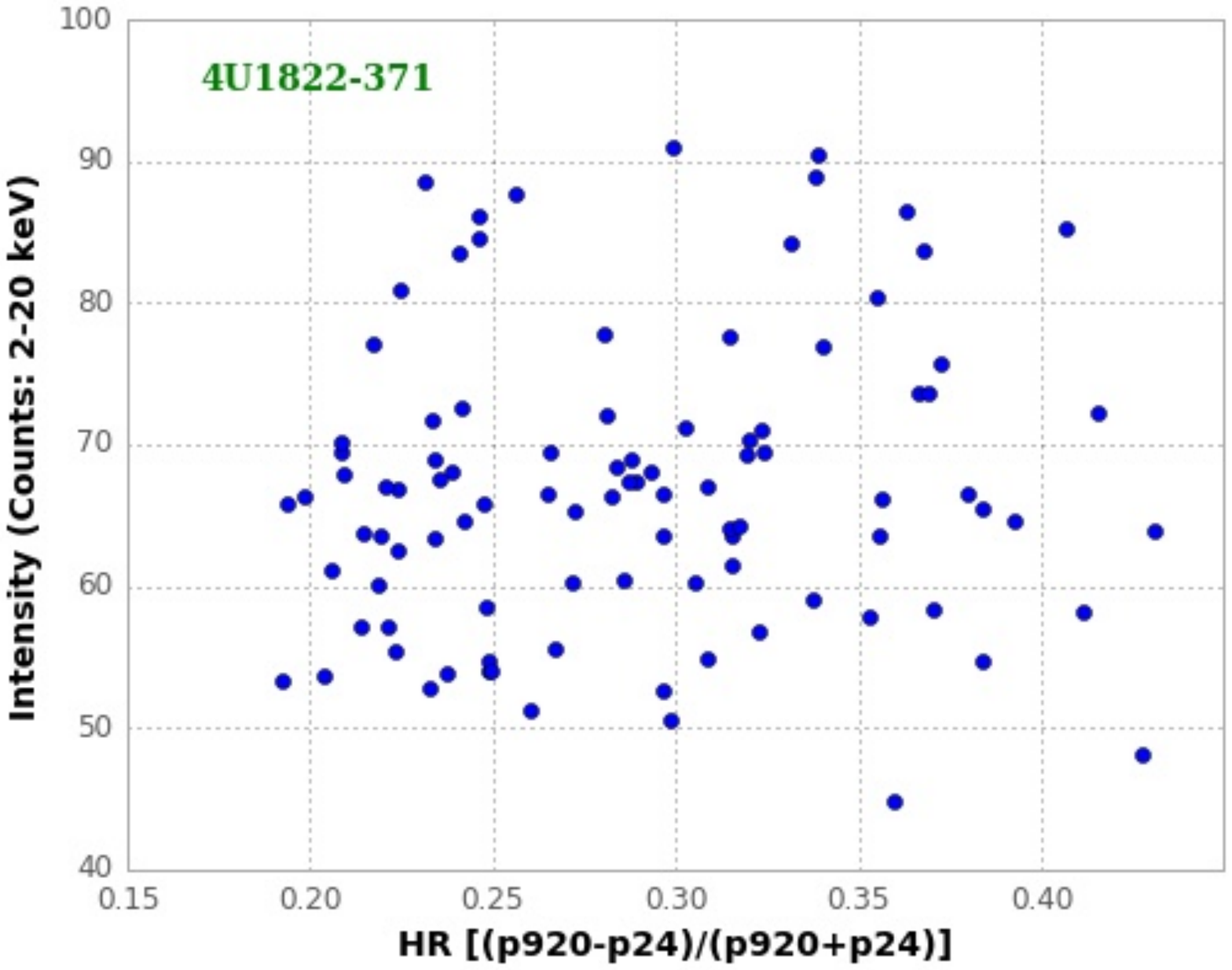}

\vspace{-.3in}
\caption{\color{black} Spectral Energy Distribution and color-intensity plots of 4U 1822-371. The panels and axes as in figure \ref{fig:SED_LMCX3}.  
\label{fig:SED_4U1822}}
\end{figure}

SMC X-1 (4U 0115-73; SK 160, Fig. \ref{fig:SED_SMCX1}; Fig. \ref{fig:appendix}.41.\color{black}) is a pulsar with a superorbital period varying from 40-60 days 
\citep{Pradhan_etal_2020}.  There is some evidence of a correlation between spin and superorbital
periods 
\citep{Hu_etal_2023}.  During the week of the campaign it was in the uprise of the superorbital period.  It gives upper limits with the ATCA, and detections in SMARTS, the UVOT,  and the SWIFT XRT and BAT. Its SED is very similar to that of 4U 1822-371 (Fig. \ref{fig:SED_4U1822}). 

4U 1822-371 (V691 CrA, Fig. \ref{fig:SED_4U1822}; Fig. \ref{fig:appendix}.14.\color{black} ) is one of the few low mass XRBs to show pulsations.  During the campaign week it was below the detection limit of the ASM and PCA.  It shows upper limits with the ATCA, there are positive detections with SMARTS, the UVOT, the SWIFT XRT and the SWIFT BAT. The SEDs for SMC X-1 (Fig. \ref{fig:SED_SMCX1}) and 4U 1822-371 are remarkably similar, both showing rising spectra near 10 Hz (4 keV) with the SWIFT XRT, but a rapid drop in flux density at  the higher energies of the BAT (30 keV). 
\color{black}

\begin{figure}[ht]
\hspace{.2in} \includegraphics[width=3.4in]{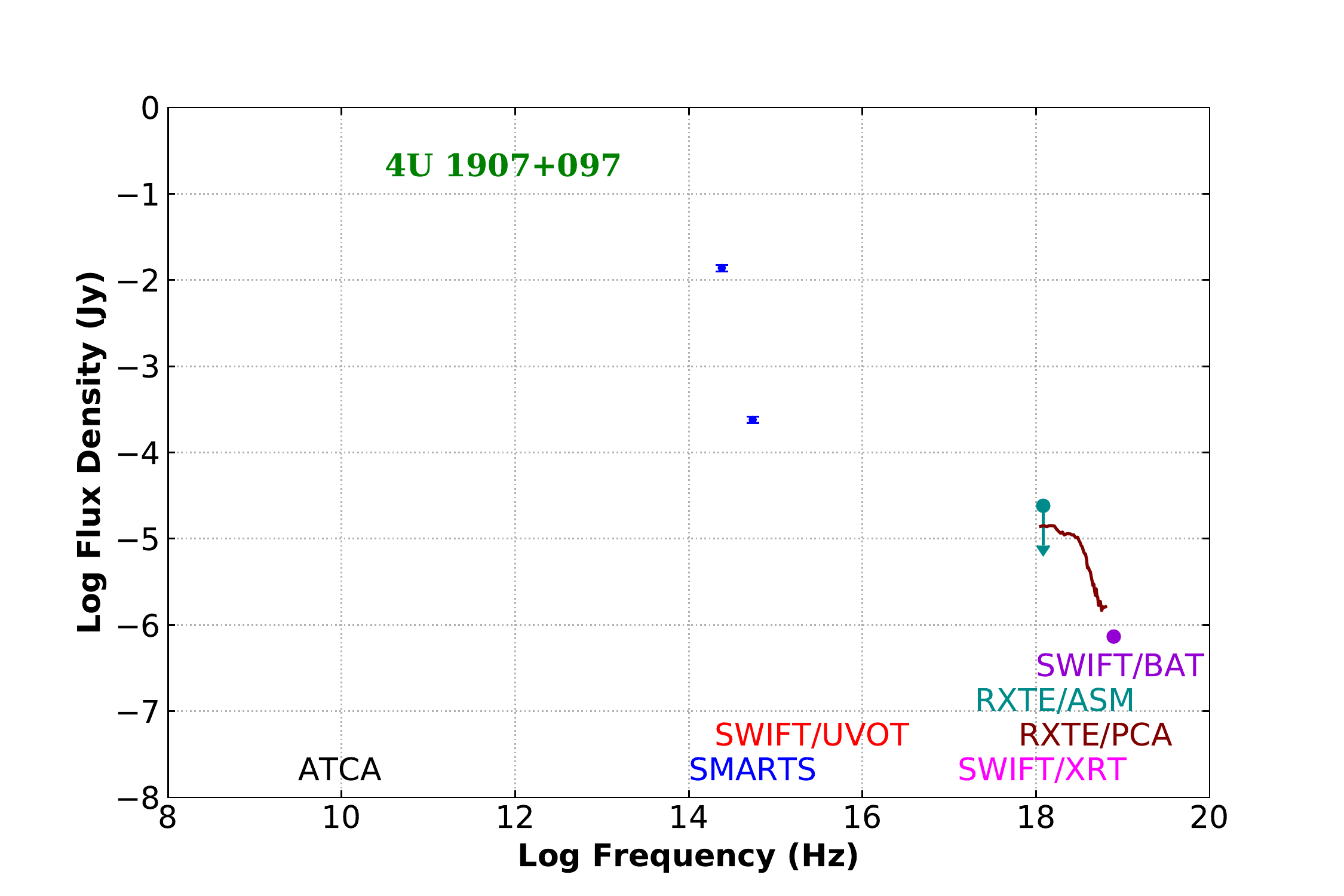}
\hspace{-.5in}
\hspace{.5in} \includegraphics[width=3in]{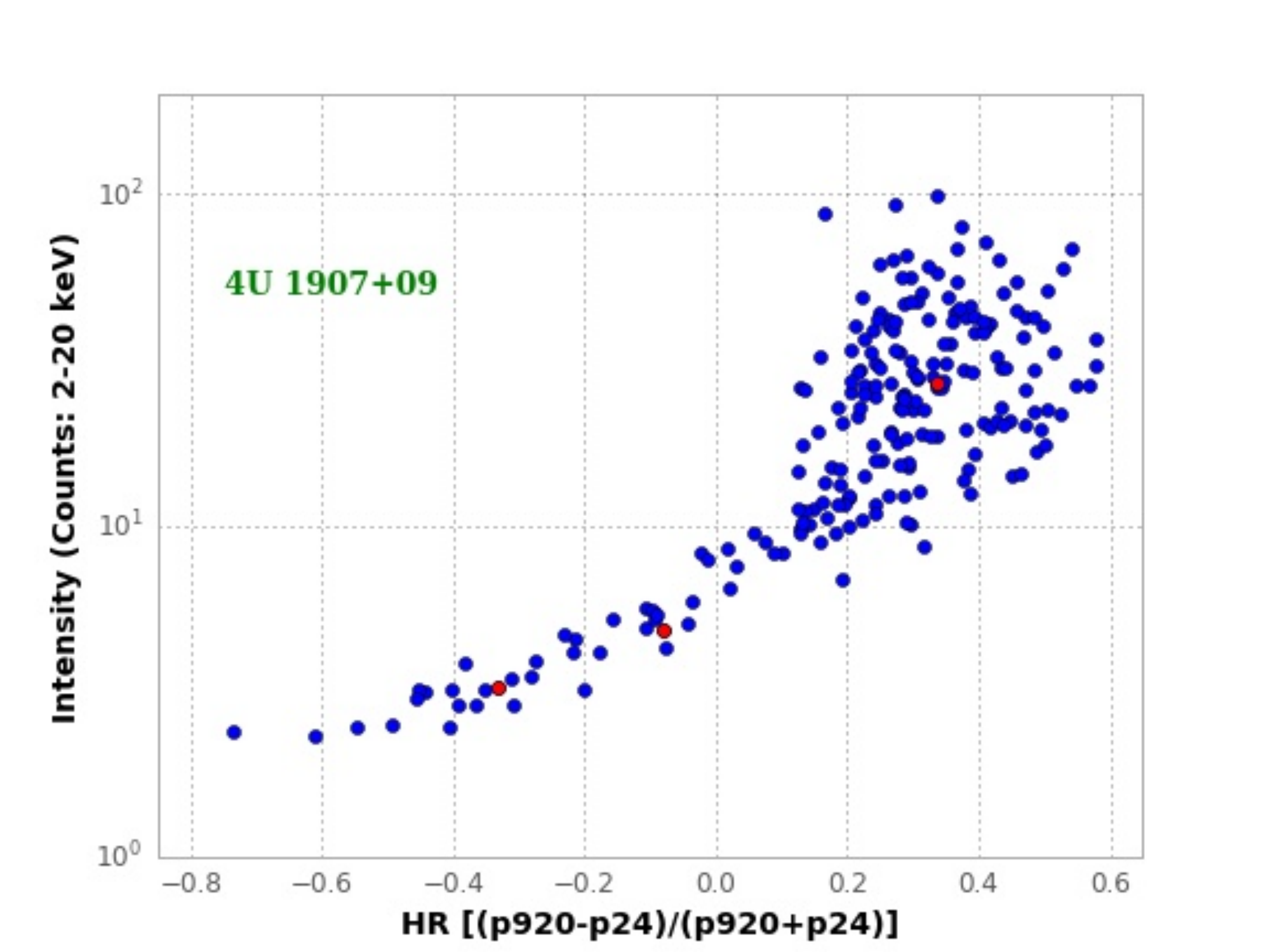}
\caption{Spectral Energy Distribution and color-intensity plots of 4U 1907+097.
\color{black}
The X-ray spectrum plots the observation at MJD 54282.398 on Table \ref{tab:PCA-1} (RXTE/PCA).
\color{black}
The panels and axes as in figure \ref{fig:SED_LMCX3}.
\label{fig:SED_U1907}}
\end{figure}

 4U 1907+09 (Fig. \ref{fig:SED_U1907}; Fig. \ref{fig:appendix}.28.\color{black}) is relatively weak in X-rays. \color{black} It was not observed with the ATCA. It shows positive detections in SMARTS. There is no significant detection by the RXTE ASM during the week of the campaign. The Swift XRT shows a sharp drop off consistent with the Swift BAT detection. It covers the full area typical for  pulsars in the RXTE PCA CI diagram. 

\subsubsection{X-ray Non-Detections (8)}

\color{black} SMC X-3 (Lin 198), GX 1+4 (4U 1728-24, V2216 Oph (sometimes classed as a symbiotic star), and Swift J1756.9- 2508, show only upper limits in the radio and X-rays. \color{black}SMC X-1 (4U 0115-73, SK 160)  Vela X-1 (4U 0900-40, GP Vel), Cen  X-3 (4U 1119-603, V779 Cen), GX 301-2 (4U 1223-62, BP Cru), 4U 1538-522 (QV Nor), and 4U 1700-37 (HD 153919) all show only upper limits in the radio and have little or no X-ray data.
\color{black}

\section{Conclusions}\label{sec:conclusions}

Multi-wavelength observations of black hole and neutron star XRBs show trends that demonstrate important connections between the accretion disk, the nature of the compact object and jet production.  Jet theories invoke a poloidal magnetic field configuration arising from the accretion disk that collimates the jet flow \citep{Ferreira_2008}.  \citet{Fender_etal_2004} use magnetic field strengths to rank XRBs from most likely to least likely to produce jets: BHs, Z-types, Atoll-types, and Pulsars. \citet{Massi_Bernado_2008} quantify this progression noting that high magnetic fields hinder the production of jets and high accretion rates enhance the production of jets.  Our observations are consistent with these interpretations: we detect no significant radio emission from pulsars or atolls, but we do detect radio emission from Z sources in the normal or horizontal branch, and from BHs in the soft/high, hard/low and quiescent states. 

Ten out of 11 BH systems were observed by the ATCA, four of these (LMC X-3, GX~339-4, 1E 1740.7-2942, and GRS~1758-258) have simultaneous X-ray spectra from Swift/XRT and/or RXTE/PCA.  LMC X-3 is in the faint intermediate state with radio upper limits consistent with no jet; LMC X-1 is in the intermediate/low state (inferred from the RXTE/ASM CI) with upper limits in radio emission consistent with no jet;  GX~339-4 is in a hard/low state with radio flux of a few mJy consistent with a continuously active,
flat-spectrum, compact, resolved jet; 1E 1740.7-2942 is relatively faint, with a radio flux upper limit consistent with a $\mu$Jy level compact radio jet; GRS~1758-258 is in a faint intermediate state with radio upper limits consistent with no jet; XTE J1550-564, 4U~1630-47, and GRO J1655-40 are all below the detection limit in the ASM with undetected radio flux, consistent with the expected levels at quiescence.  V4641 Sgr was not observed in X-rays; GRS~1915+105 is in the bright/intermediate state but was not observed in the radio.  4U~1957+11 is in the soft/low state with upper limits from the ATCA consistent with no jet. 

Five of eight Z sources observed by the ATCA have simultaneous X-ray spectra from Swift XRT and/or RXTE PCA (Cir X-1, Sco X-1, XTE J1701-462, GX~349+2, and GX~17+2); for the other three (LMC X-2, GX~340+0, and GX~5-1) we are able to infer the X-ray state from the RXTE ASM data.  LMC X-2 is in a mini-flare with only upper limits from the ATCA consistent with no radio emission; Cir X-1 is a peculiar source, at very low X-ray luminosity with radio emission at the mJy level consistent with the HB, but no RXTE ASM detection during the week of the campaign.  Sco X-1, known to be a resolved jet source, was mostly in the NB with excursions into FB, having radio emission of tens of mJy consistent with the presence of a jet. GX~340+0, GX~5-1, and GX~17+2 are all in the NB of the Z and have radio detections at the mJy level consistent with a transient jet; J1701-462 is a transient source in outburst, with upper limits from the ATCA. The HB state inferred from the RXTE PCA CI is consistent with no jets at this X-ray state.

The 18 Atoll sources yielded only upper limits from the ATCA. This is consistent with Atoll sources having at most an unresolved radio source with flux density at the microJy level. Of nine pulsars none were detected by the ATCA, consistent with expectations.

\begin{acknowledgements}

{We are grateful to Alberto Pasten for running the observations at CTIO.

The Australia Telescope Compact Array is part of the Australia Telescope National Facility which is funded by the Australian Government for operation as a National Facility managed by CSIRO.

This publication makes use of data products from the Two Micron All Sky Survey, which is a joint project of the University of Massachusetts and the Infrared Processing and Analysis Center/California Institute of Technology, funded by the National Aeronautics and Space Administration and the National Science Foundation. \color{black} This research also used data from the AAVSO Photometric All Sky Survey (APASS) DR9 \citep{APASS}.\color{black}
This research has also made use of NASA's Astrophysics Data System as well as the VizieR catalogue access tool, Centre de Donn{`e}es astronomiques de Strasbourg, Strasbourg, France.
}

The authors are grateful to an anonymous referee for helpful criticism and suggestions.

\software{
This research made use of SciPy \citep{SciPy2020}, NumPy \citep{van2011numpy}, Matplotlib, a Python library for publication quality graphics \citep{Hunter_2007}, Astropy (\url{http://www.astropy.org}), a community-developed core Python package for Astronomy \citep{astropy:2013}, MIRIAD \citep{Sault_etal_1995}, HEAsoft \citep{HEASARC_2014}, KARMA \citep{Gooch_1996}, and XSPEC \citep{Arnaud_1996}. 
}

\facilities{Australia Telescope Compact Array (ATCA), Two Micron All Sky Survey (2MASS), Small and Moderate Aperture Research Telescope System (SMARTS),
Swift Burst Alert Telescope (BAT), Swift Ultarviolet and Optical Telescope (UVOT), Swift X-Ray Telescope (XRT), RXTE All Sky Monitor (ASM), Rossi X-ray Timing Explorer Proportional Counter Array (RXTE-PCA)}
\end{acknowledgements}

 \newcommand{\noop}[1]{}

\appendix  \color{black}
\section{ASM 14-year Lightcurves \label{sec:appendix}}

\begin{figure}[ht]
\hspace{-.2in} \includegraphics[width=3.7in]{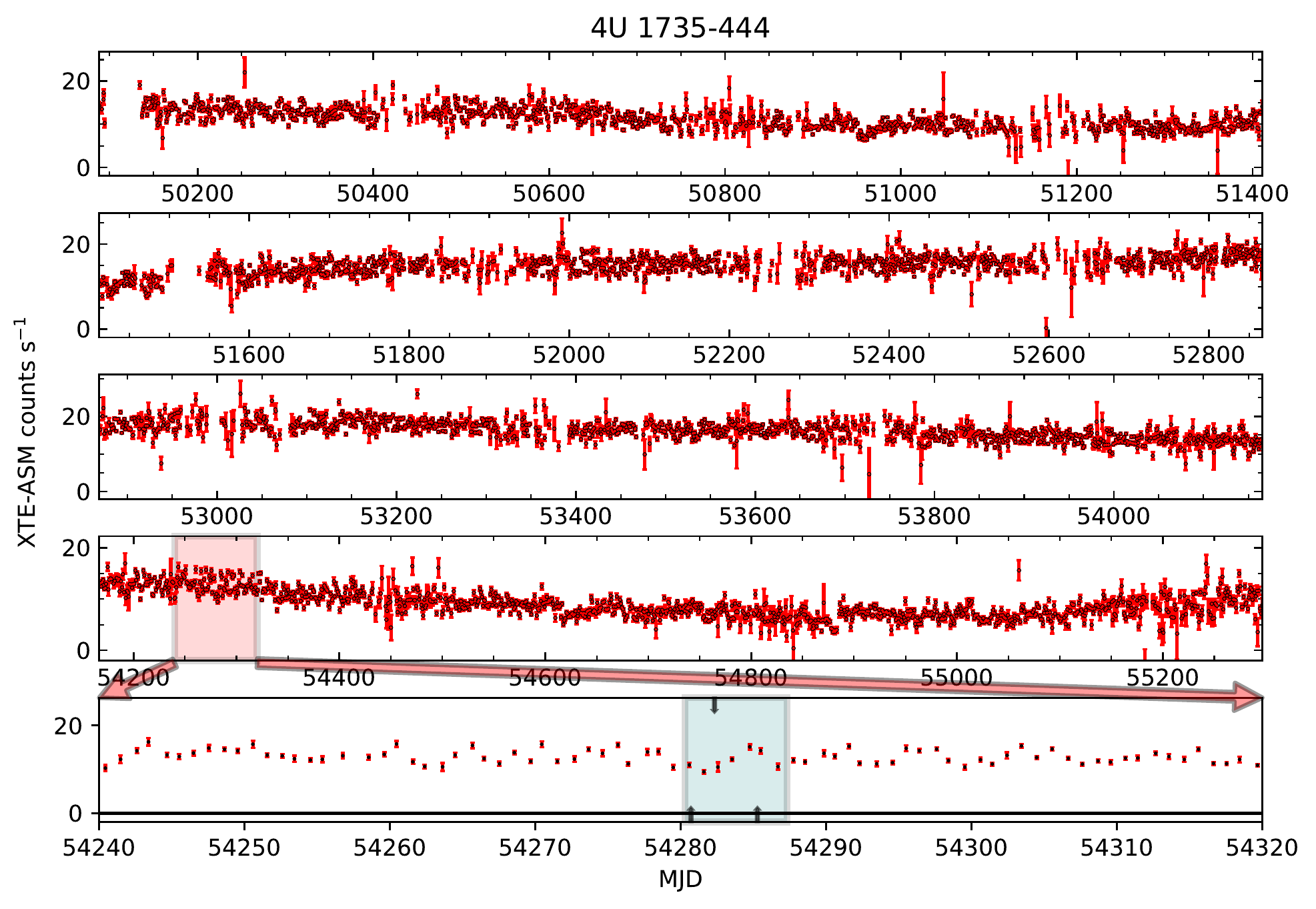}
\hspace{-.1in} \includegraphics[width=3.7in]{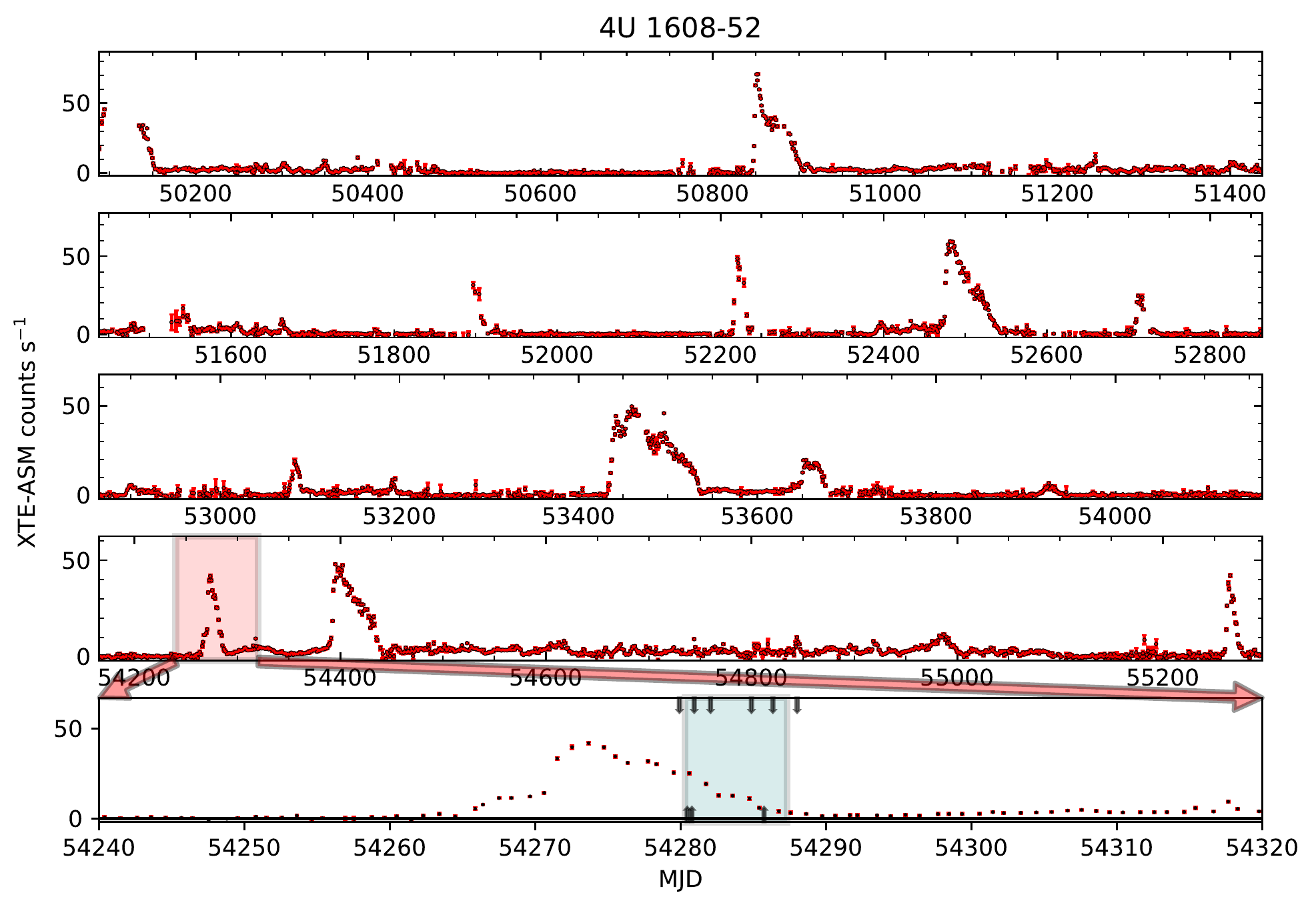}
\caption{X-ray light curves for two sample XRBs with data from the RXTE ASM.
The upper four rows show the X-ray flux in counts per second for most of
the lifetime of the instrument (50100 $<$ MJD $<$ 55300,
i.e. January 1996 through April 2010).  The bottom row is an expanded scale
view of the 80 days surrounding our observing campaign, highlighed in pink
on the fourth row above.  The seven days of the campaign are highlighted
in blue.  
\color{black}
Markers on the top scale of the bottom panel indicate the times of observations with the Swift XRT (Table \ref{tab:Swift_xrt}).  Markers on the bottom scale indicate times of observations with the RXTE PCA (Tables \ref{tab:PCA-2} and \ref{tab:PCA-1}), or the end points of the range in cases where all observations were averaged together.
\color{black}
Error bars on the daily averages are drawn 
in red.  The complete figure set (42 images) is available in the online journal. \label{fig:appendix}}
\end{figure}

\figsetstart
\figsetnum{\ref{fig:appendix}}
\figsettitle{RXTE-ASM Lightcurves}

\figsetgrpstart
\figsetgrpnum{\ref{fig:appendix}.1}
\figsetgrptitle{4U 1543-624}
\figsetplot{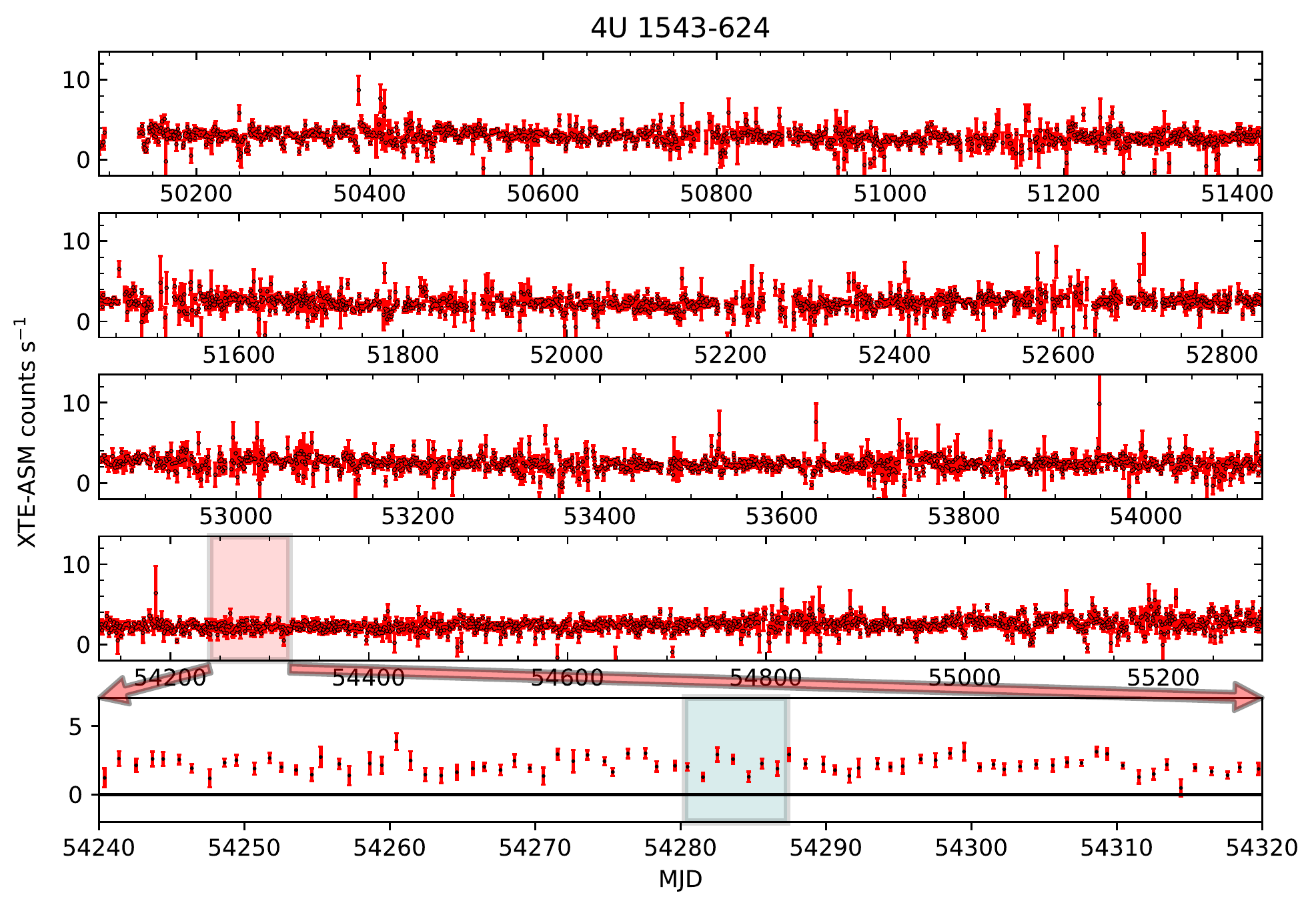}
\figsetgrpnote{X-ray lightcurve from RXTE-ASM.  The bottom panel shows the 80 days
around our campaign with an expanded scale.  The campaign dates are highlighted in blue.}
\figsetgrpend

\figsetgrpstart
\figsetgrpnum{\ref{fig:appendix}.2}
\figsetgrptitle{EXO 0748-676}
\figsetplot{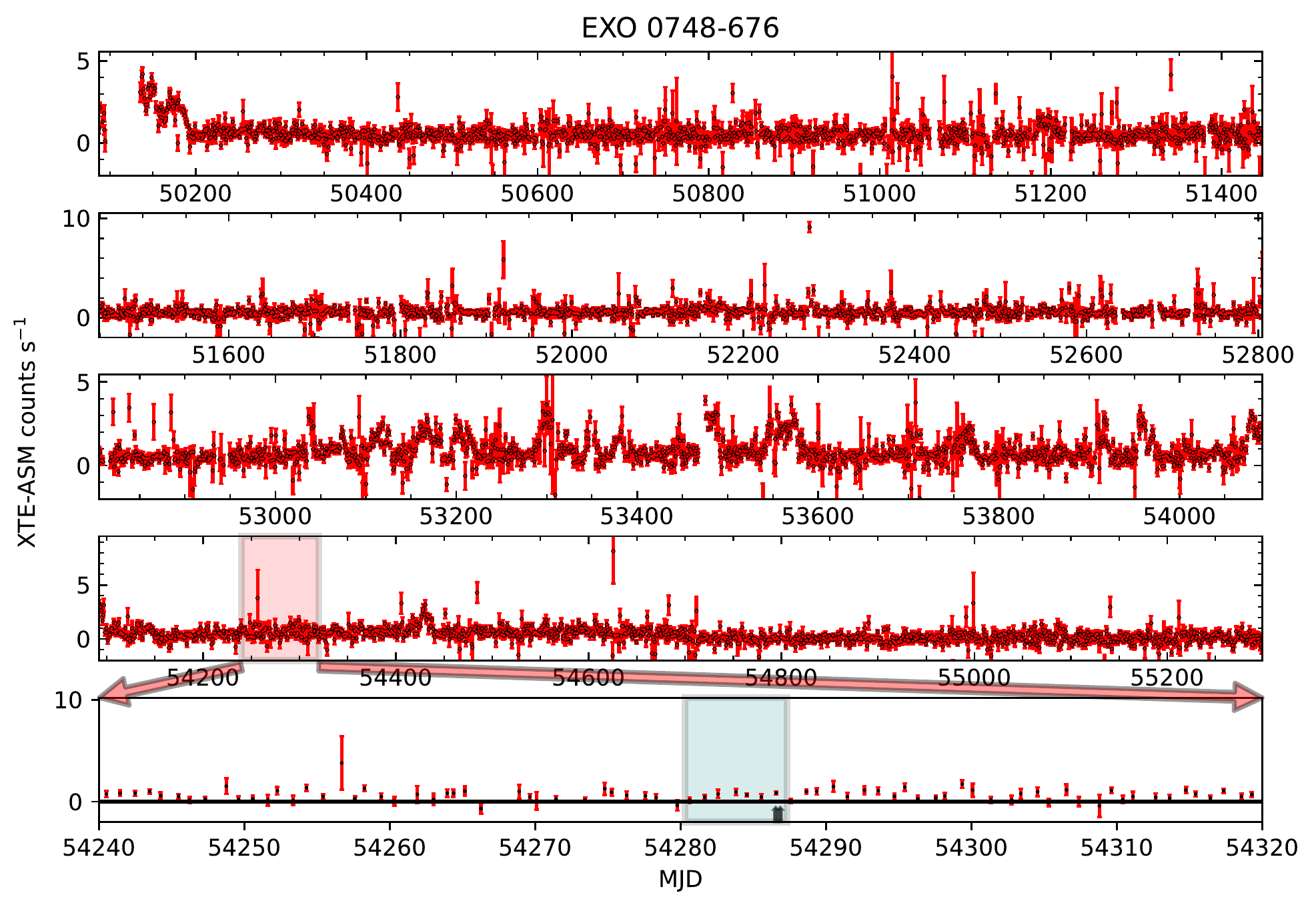}
\figsetgrpnote{X-ray lightcurve from RXTE-ASM.  The bottom panel shows the 80 days
around our campaign with an expanded scale.  The campaign dates are highlighted in blue.}
\figsetgrpend

\figsetgrpstart
\figsetgrpnum{\ref{fig:appendix}.3}
\figsetgrptitle{4U 1254-690}
\figsetplot{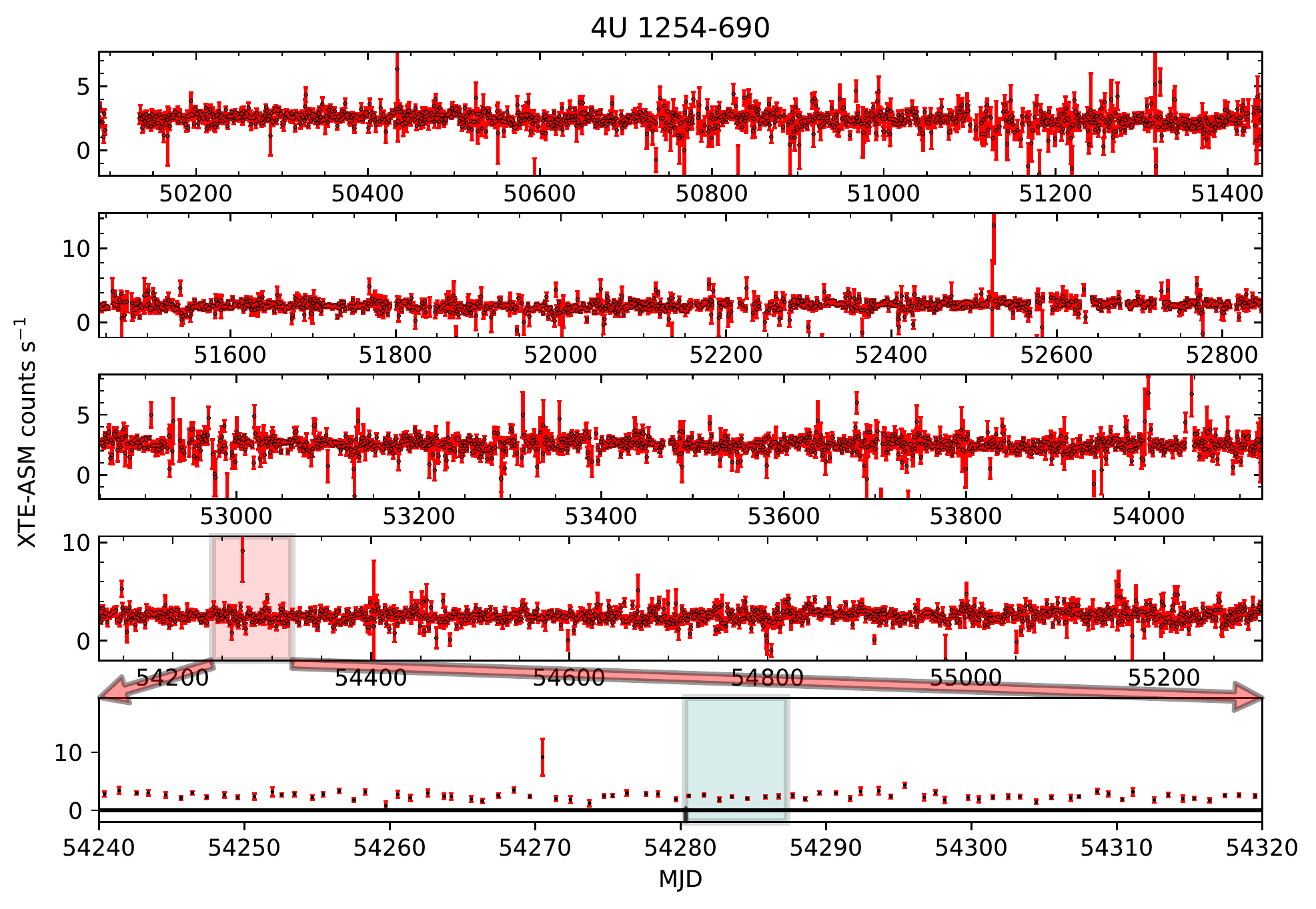}
\figsetgrpnote{X-ray lightcurve from RXTE-ASM.  The bottom panel shows the 80 days
around our campaign with an expanded scale.  The campaign dates are highlighted in blue.}
\figsetgrpend

\figsetgrpstart
\figsetgrpnum{\ref{fig:appendix}.4}
\figsetgrptitle{4U 1538-522}
\figsetplot{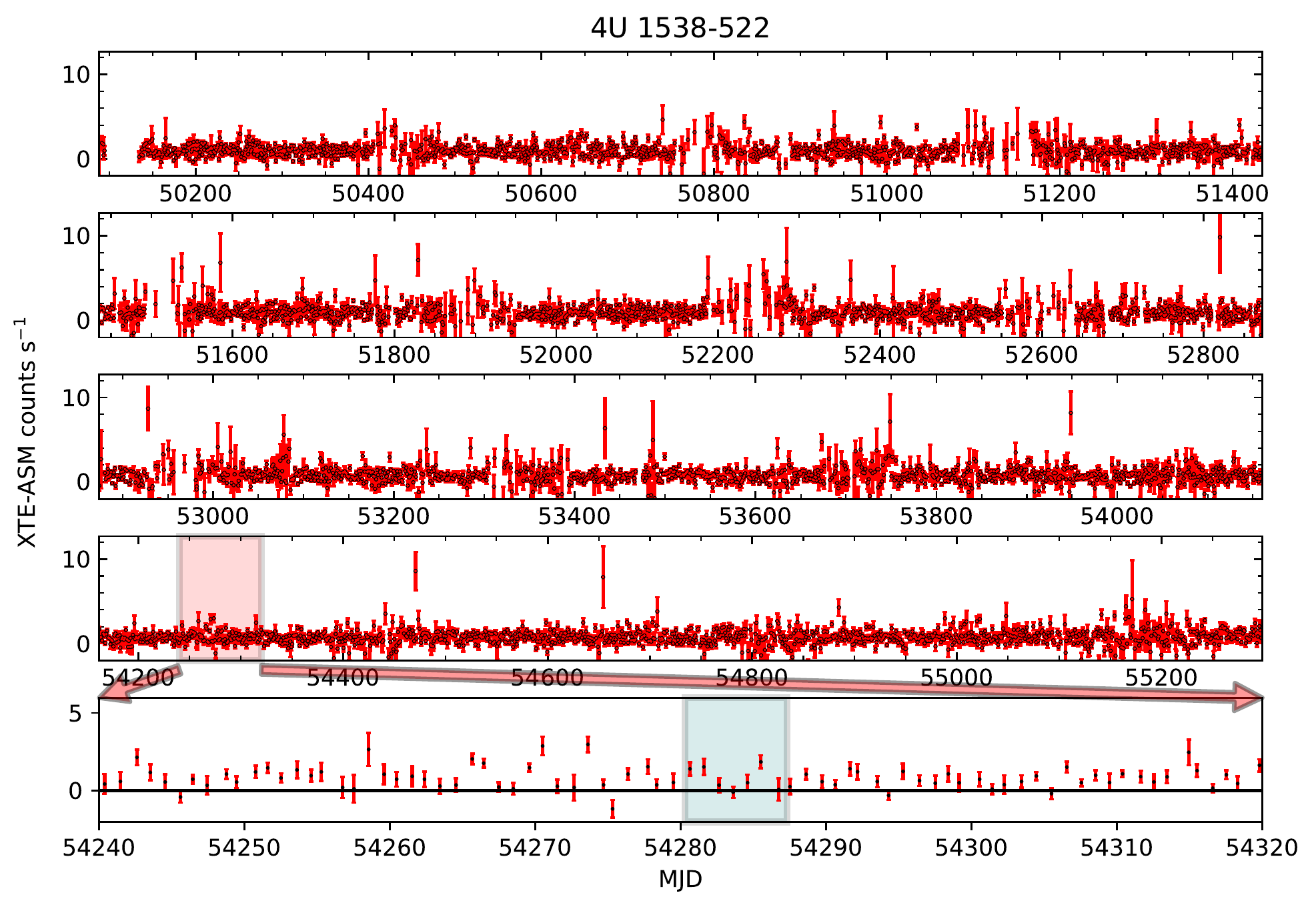}
\figsetgrpnote{X-ray lightcurve from RXTE-ASM.  The bottom panel shows the 80 days
around our campaign with an expanded scale.  The campaign dates are highlighted in blue.}
\figsetgrpend

\figsetgrpstart
\figsetgrpnum{\ref{fig:appendix}.5}
\figsetgrptitle{4U 1608-52}
\figsetplot{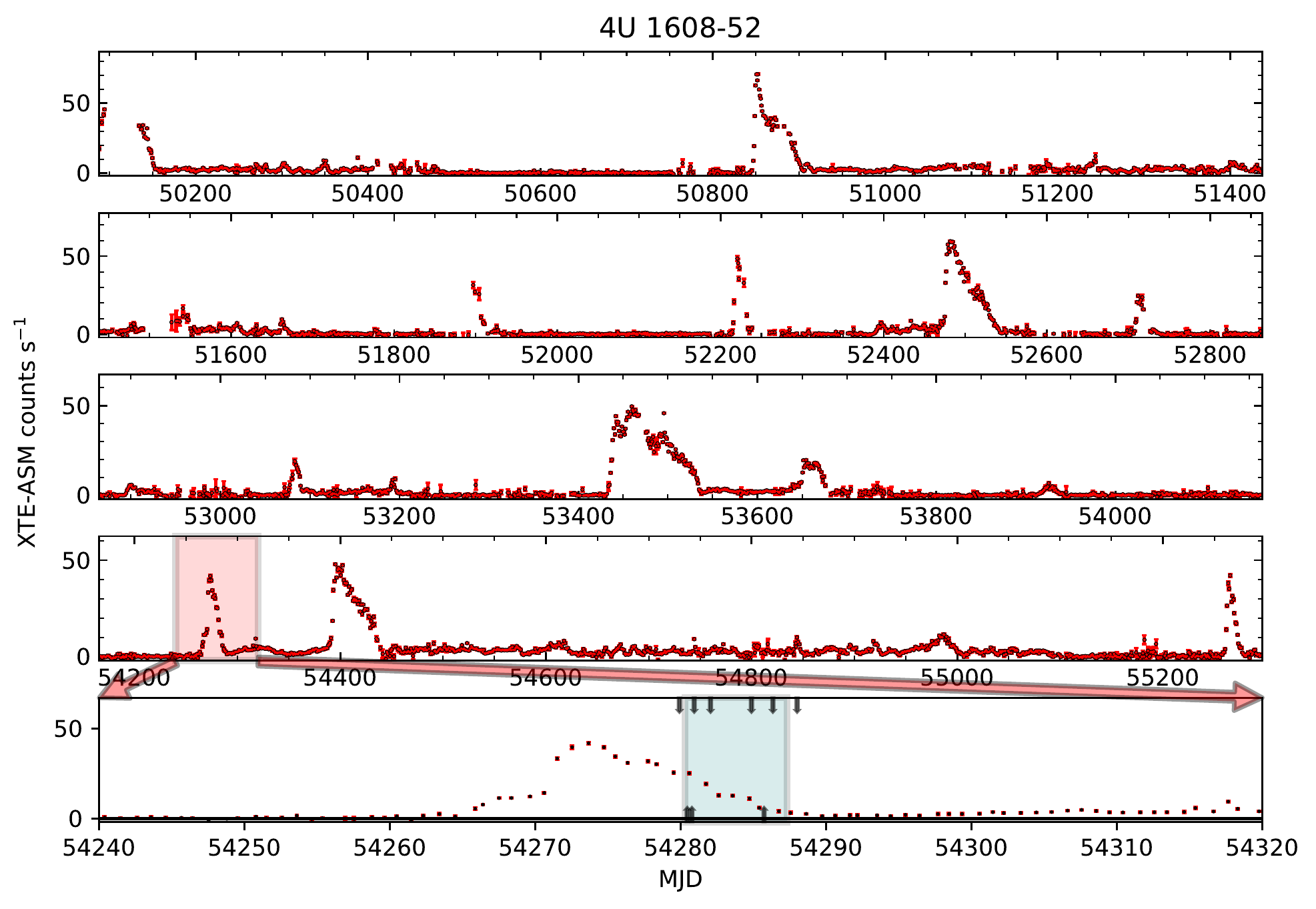}
\figsetgrpnote{X-ray lightcurve from RXTE-ASM.  The bottom panel shows the 80 days
around our campaign with an expanded scale.  The campaign dates are highlighted in blue.}
\figsetgrpend

\figsetgrpstart
\figsetgrpnum{\ref{fig:appendix}.6}
\figsetgrptitle{4U 1636-536}
\figsetplot{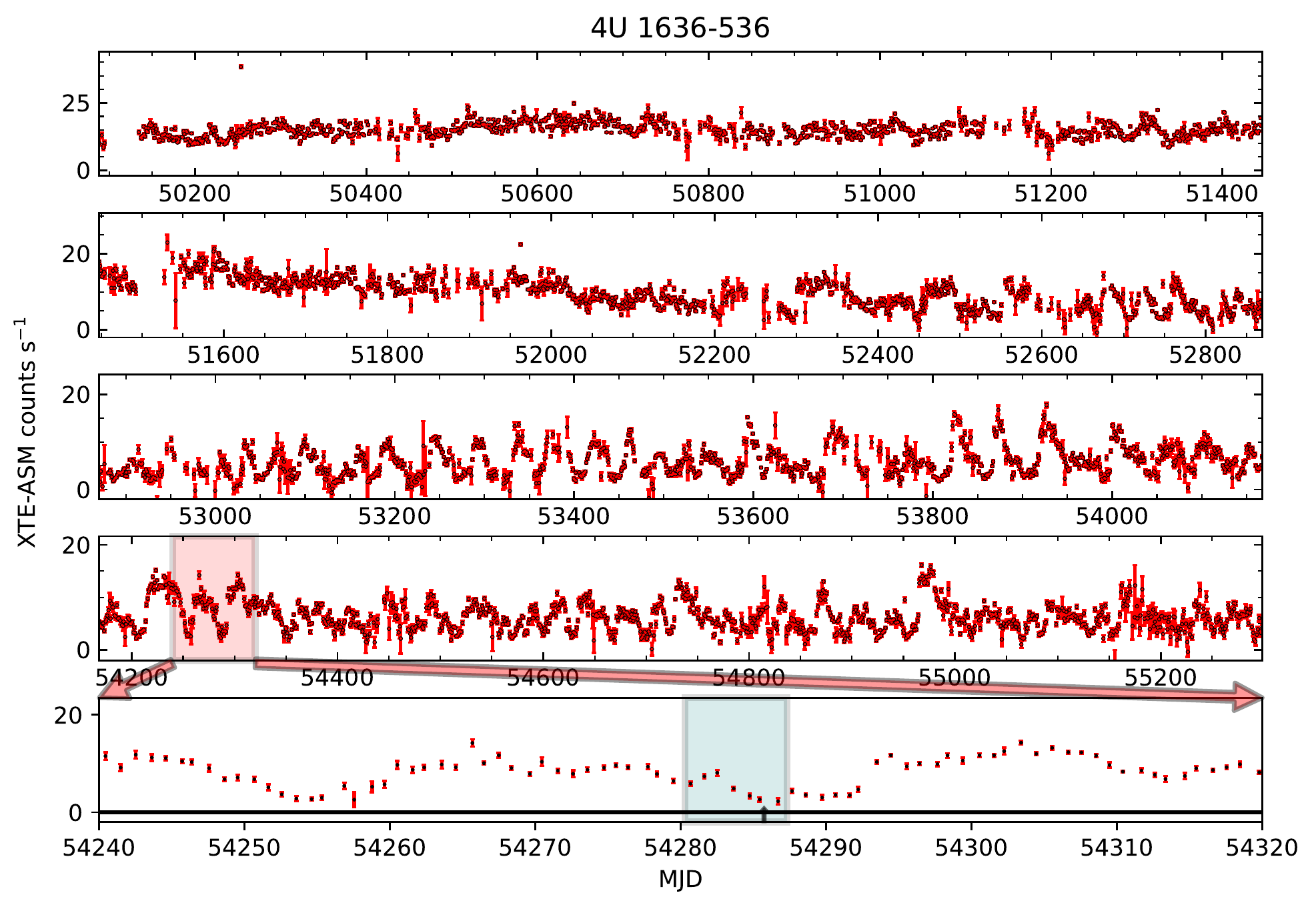}
\figsetgrpnote{X-ray lightcurve from RXTE-ASM.  The bottom panel shows the 80 days
around our campaign with an expanded scale.  The campaign dates are highlighted in blue.}
\figsetgrpend

\figsetgrpstart
\figsetgrpnum{\ref{fig:appendix}.7}
\figsetgrptitle{MXB 1658-298}
\figsetplot{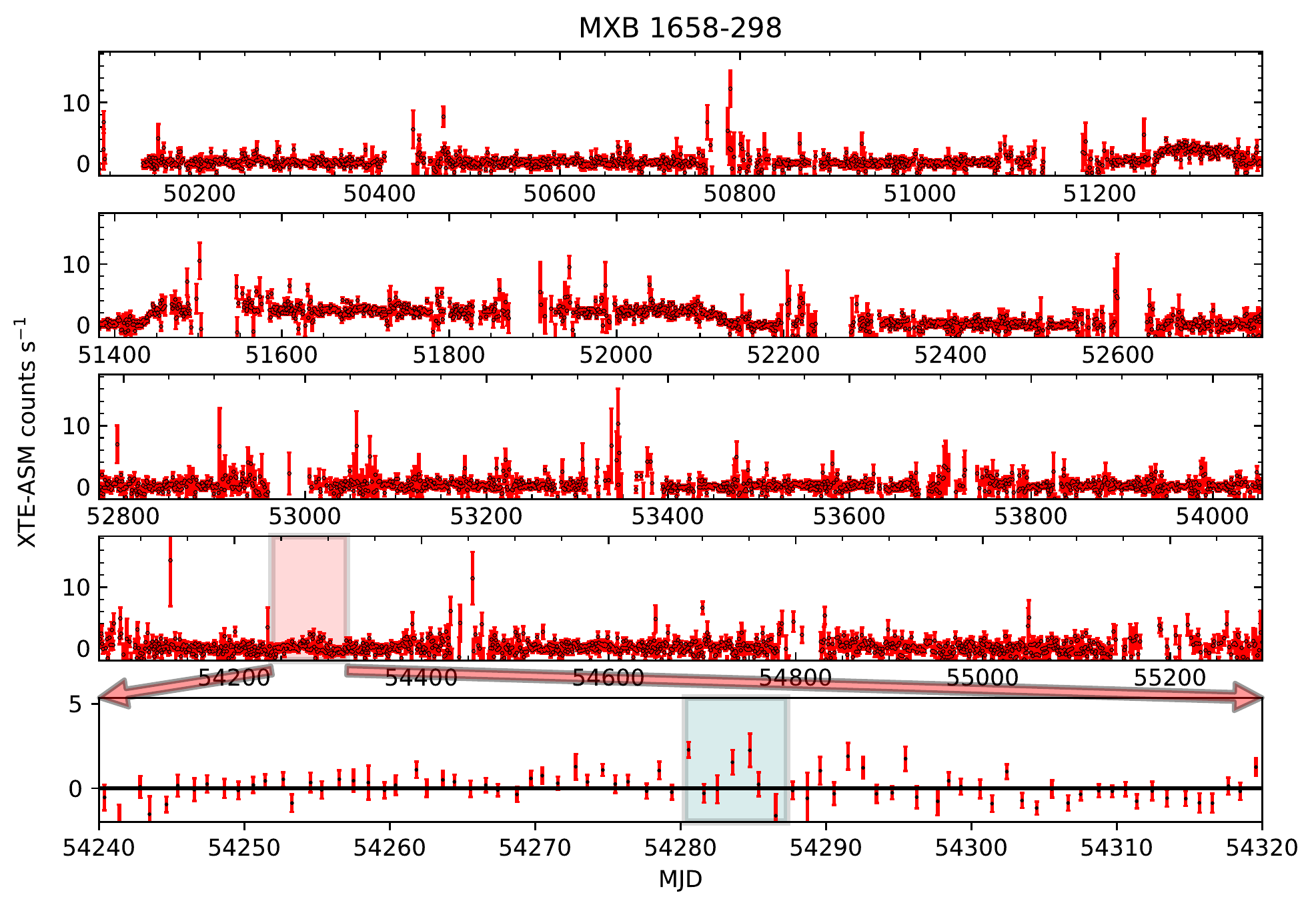}
\figsetgrpnote{X-ray lightcurve from RXTE-ASM.  The bottom panel shows the 80 days
around our campaign with an expanded scale.  The campaign dates are highlighted in blue.}
\figsetgrpend

\figsetgrpstart
\figsetgrpnum{\ref{fig:appendix}.8}
\figsetgrptitle{4U 1700-37}
\figsetplot{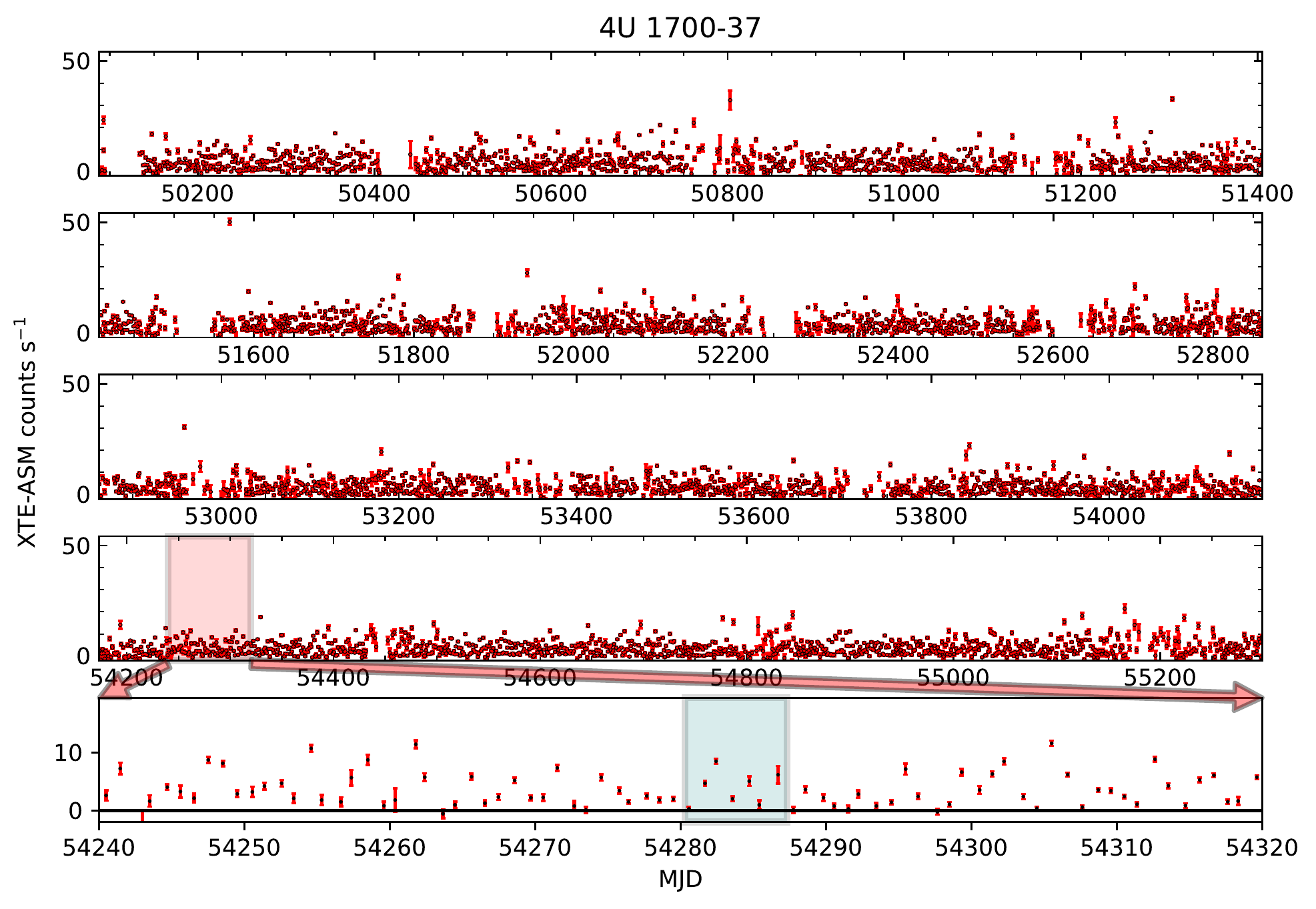}
\figsetgrpnote{X-ray lightcurve from RXTE-ASM.  The bottom panel shows the 80 days
around our campaign with an expanded scale.  The campaign dates are highlighted in blue.}
\figsetgrpend

\figsetgrpstart
\figsetgrpnum{\ref{fig:appendix}.9}
\figsetgrptitle{4U 1702-429}
\figsetplot{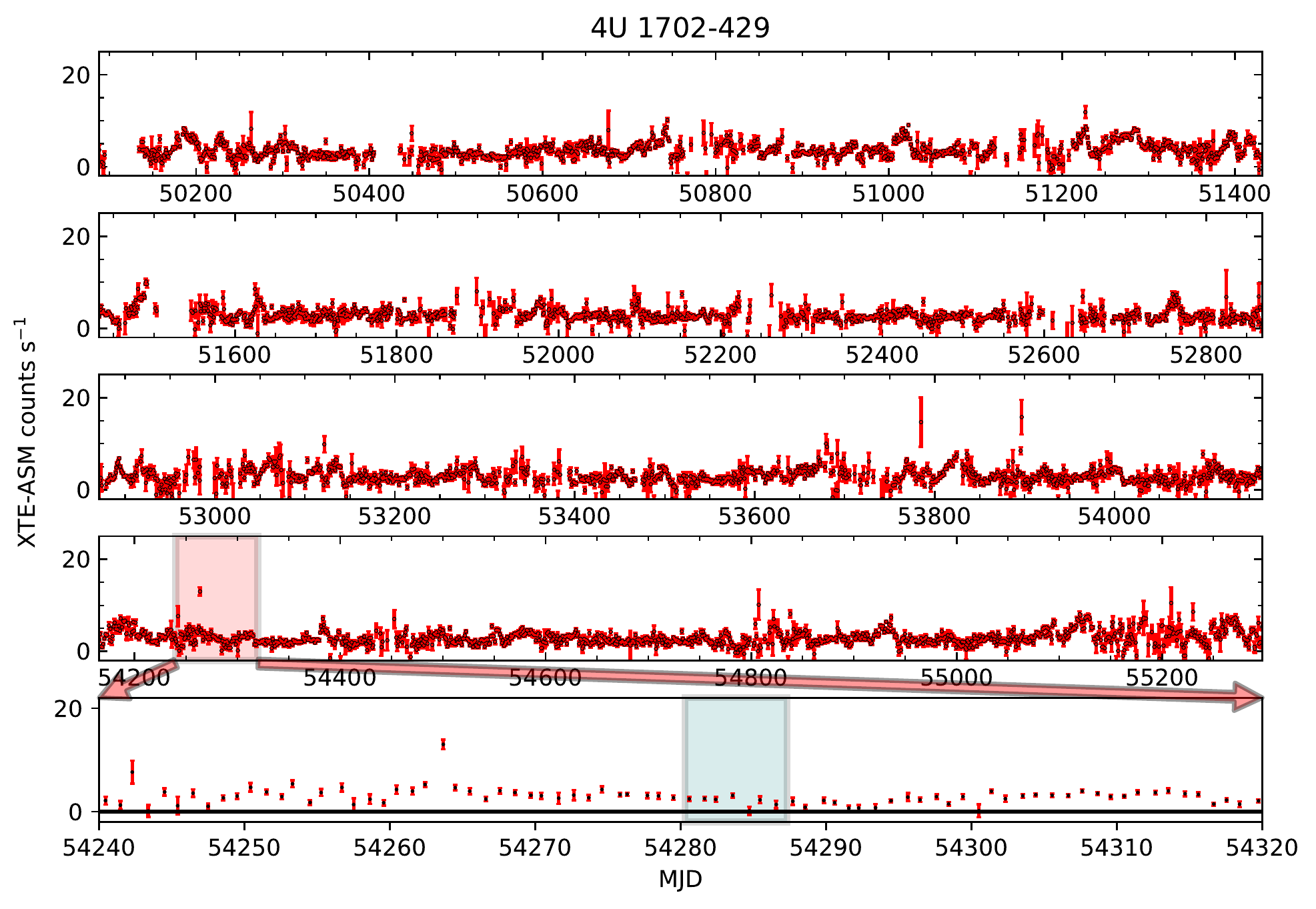}
\figsetgrpnote{X-ray lightcurve from RXTE-ASM.  The bottom panel shows the 80 days
around our campaign with an expanded scale.  The campaign dates are highlighted in blue.}
\figsetgrpend

\figsetgrpstart
\figsetgrpnum{\ref{fig:appendix}.10}
\figsetgrptitle{4U 1705-44}
\figsetplot{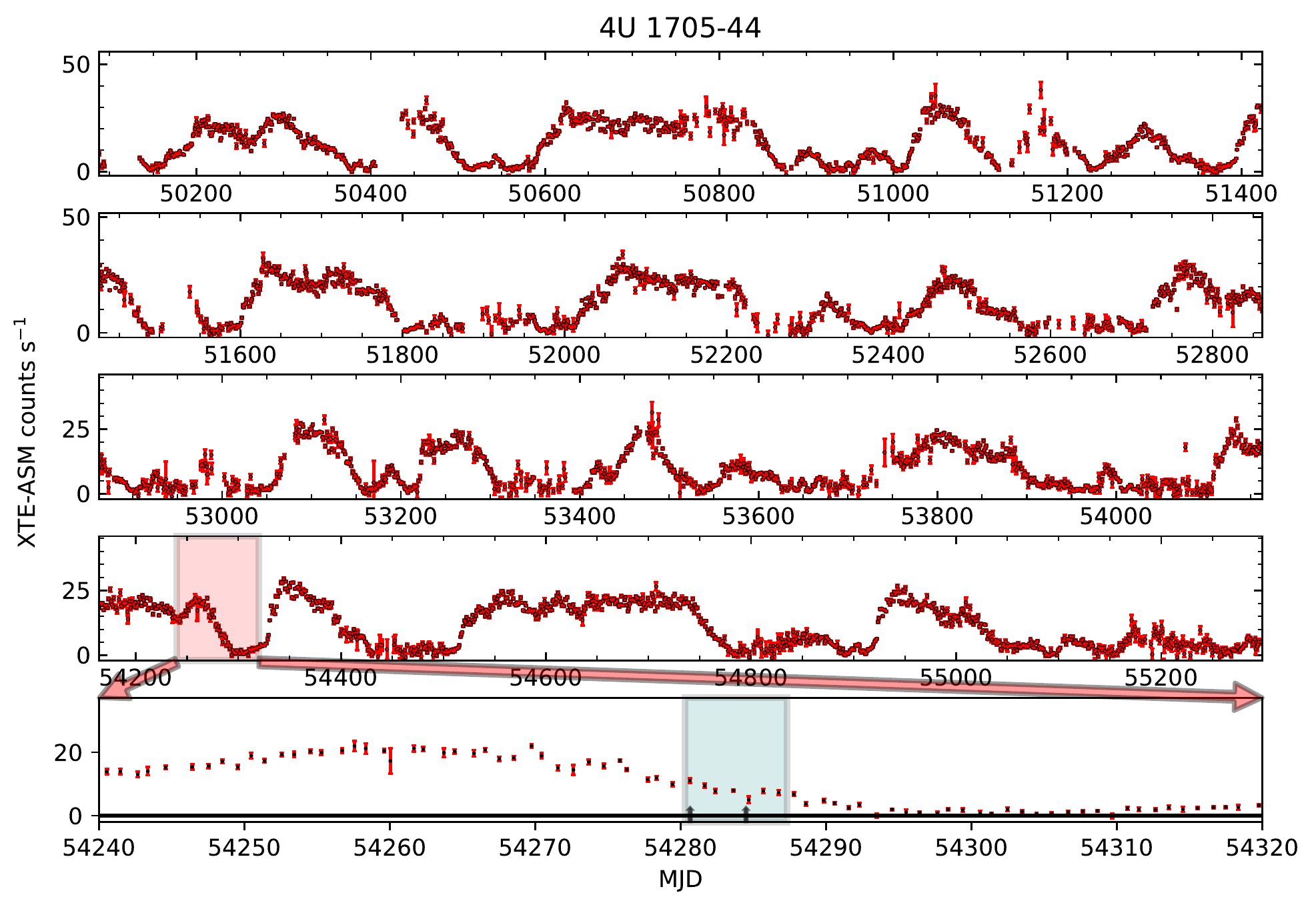}
\figsetgrpnote{X-ray lightcurve from RXTE-ASM.  The bottom panel shows the 80 days
around our campaign with an expanded scale.  The campaign dates are highlighted in blue.}
\figsetgrpend

\figsetgrpstart
\figsetgrpnum{\ref{fig:appendix}.11}
\figsetgrptitle{4U 1735-444}
\figsetplot{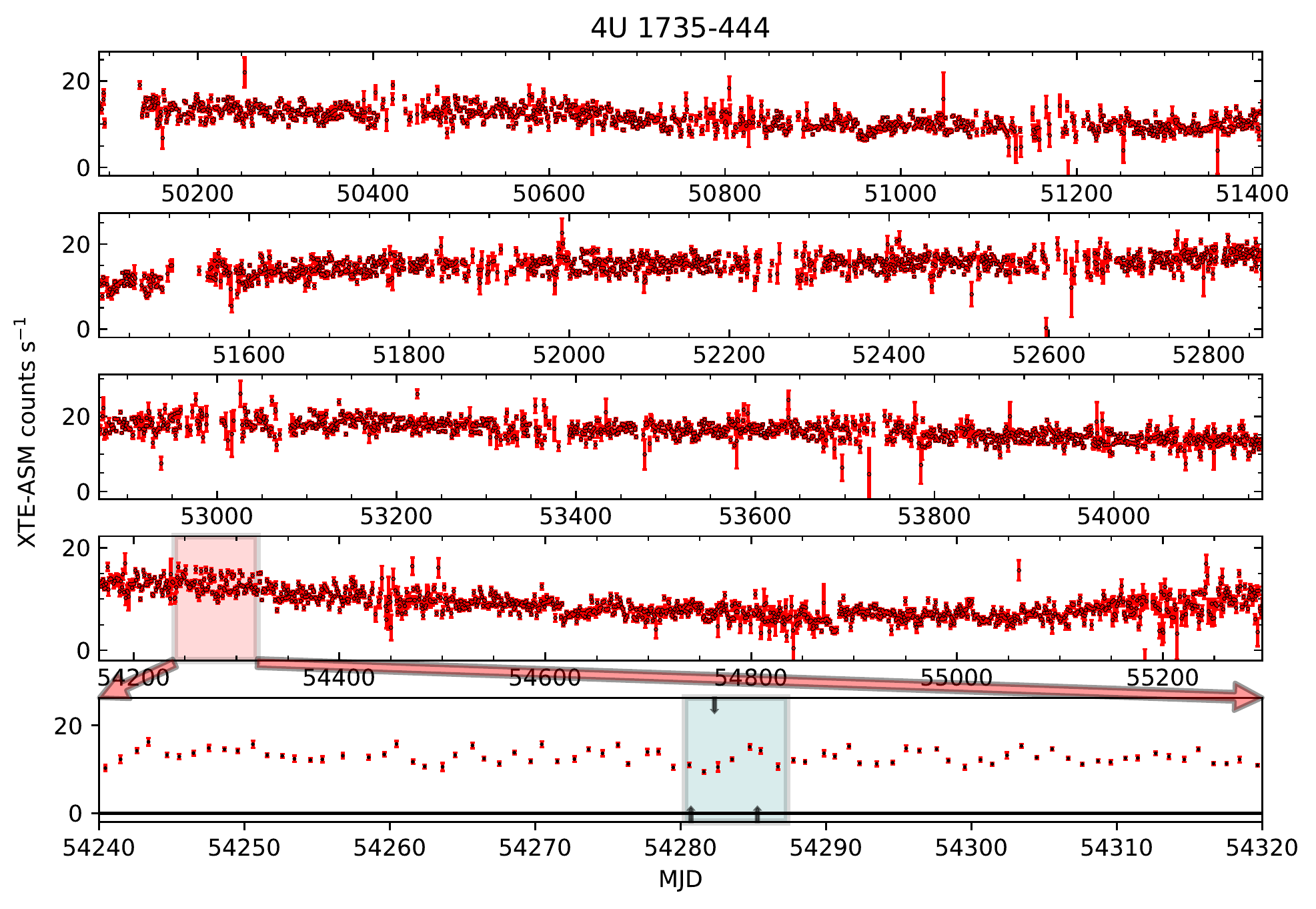}
\figsetgrpnote{X-ray lightcurve from RXTE-ASM.  The bottom panel shows the 80 days
around our campaign with an expanded scale.  The campaign dates are highlighted in blue.}
\figsetgrpend

\figsetgrpstart
\figsetgrpnum{\ref{fig:appendix}.12}
\figsetgrptitle{1E 1740.7-2942}
\figsetplot{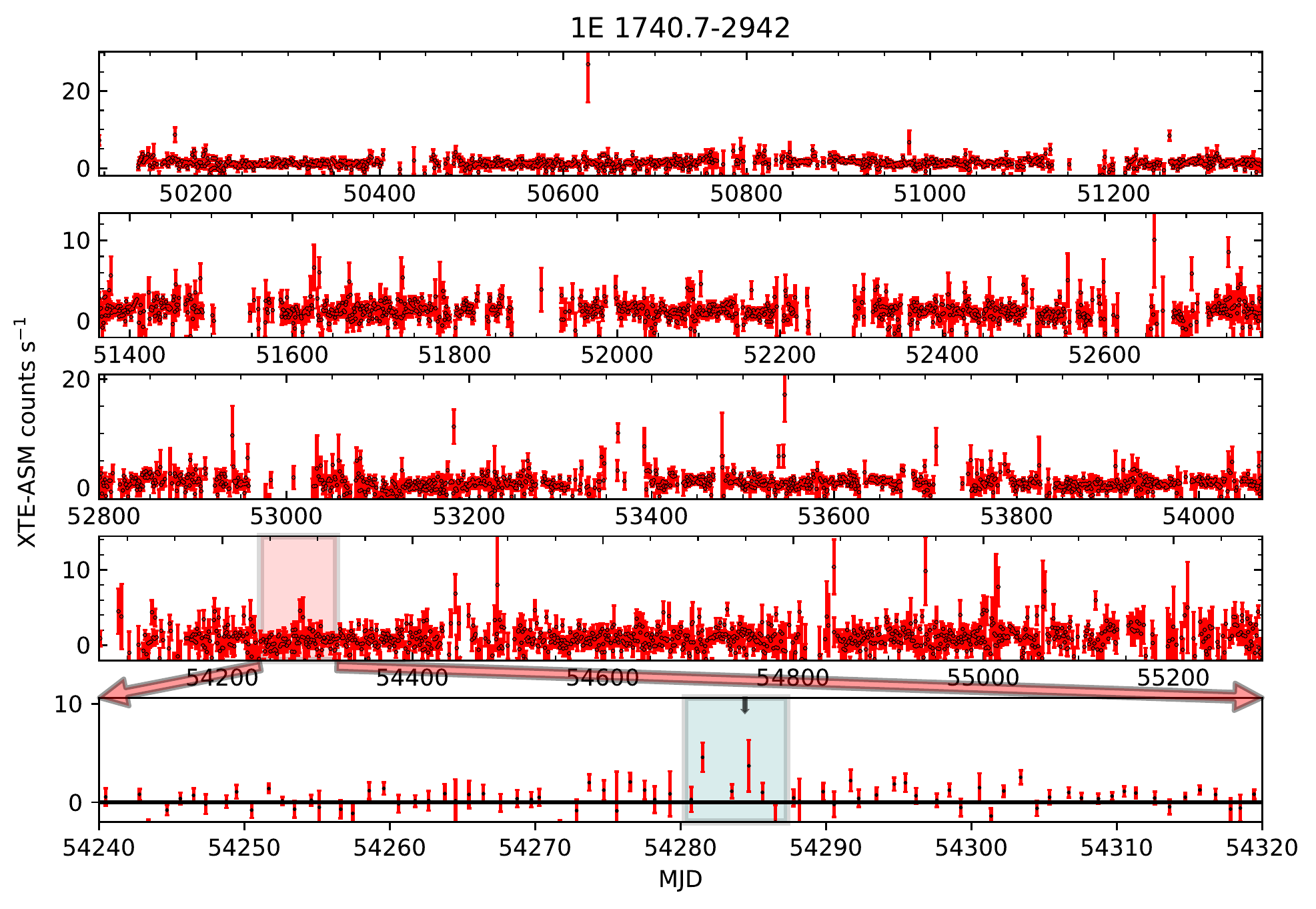}
\figsetgrpnote{X-ray lightcurve from RXTE-ASM.  The bottom panel shows the 80 days
around our campaign with an expanded scale.  The campaign dates are highlighted in blue.}
\figsetgrpend

\figsetgrpstart
\figsetgrpnum{\ref{fig:appendix}.13}
\figsetgrptitle{4U 1758-258}
\figsetplot{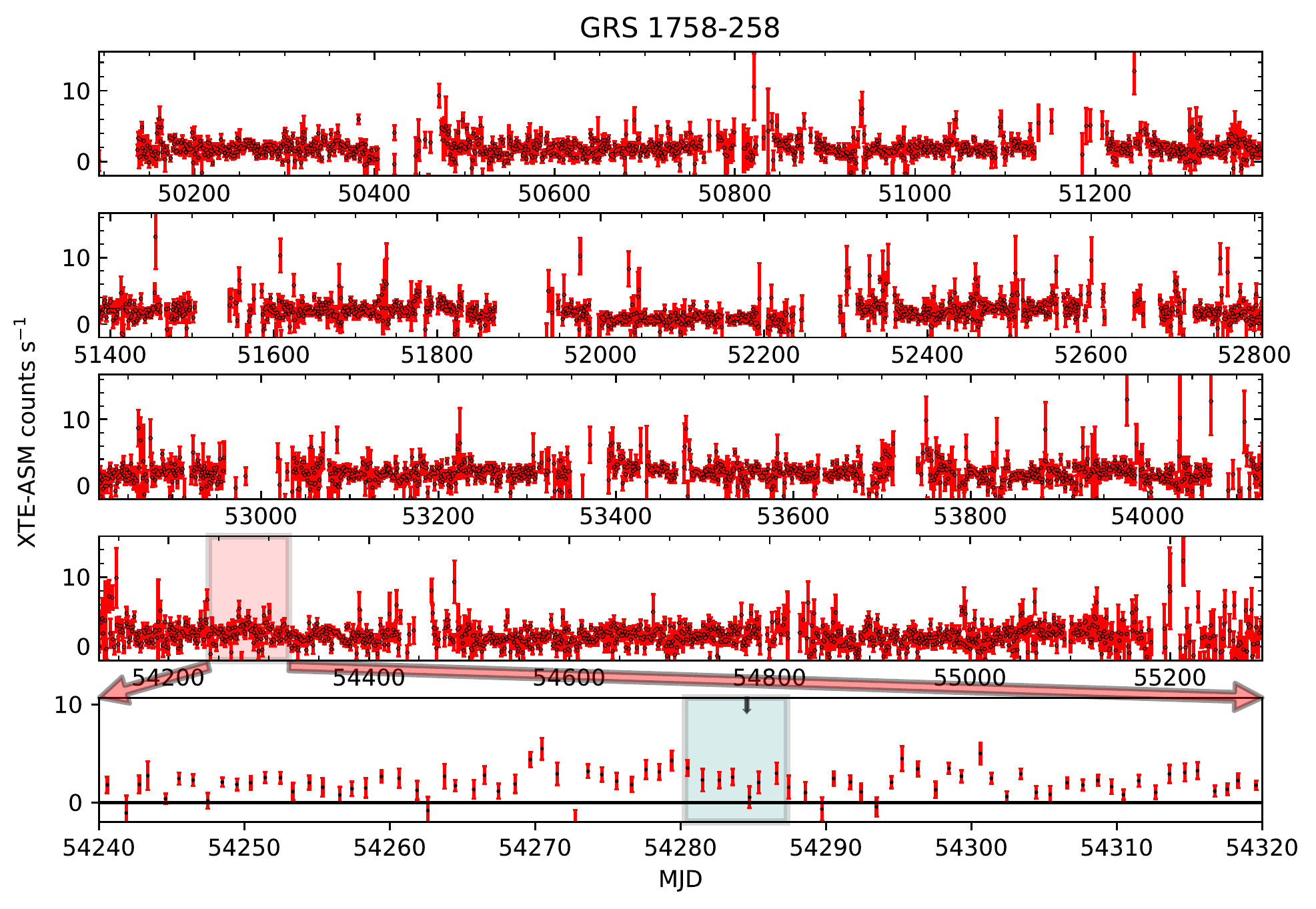}
\figsetgrpnote{X-ray lightcurve from RXTE-ASM.  The bottom panel shows the 80 days
around our campaign with an expanded scale.  The campaign dates are highlighted in blue.}
\figsetgrpend

\figsetgrpstart
\figsetgrpnum{\ref{fig:appendix}.14}
\figsetgrptitle{4U 1822-371}
\figsetplot{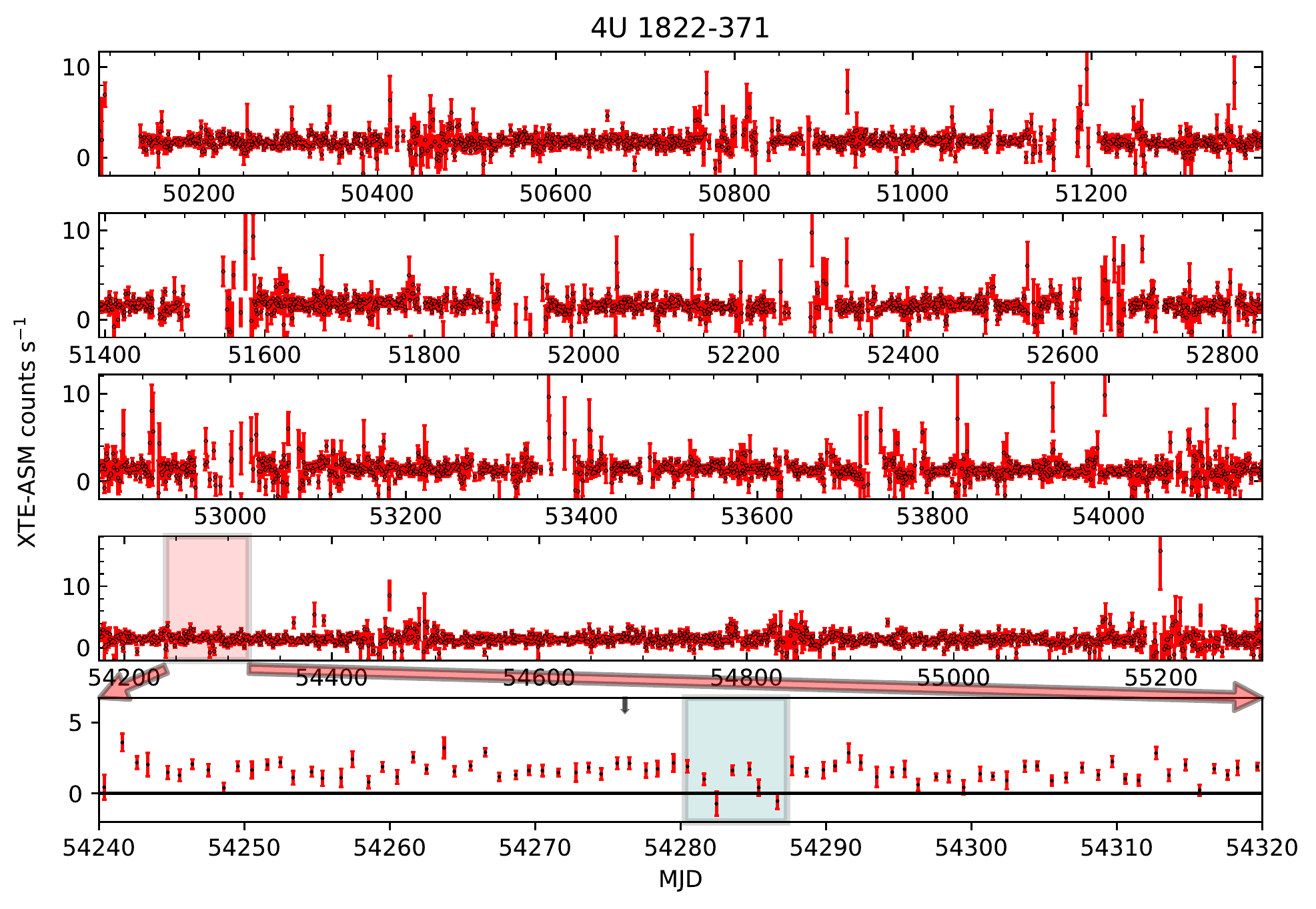}
\figsetgrpnote{X-ray lightcurve from RXTE-ASM.  The bottom panel shows the 80 days
around our campaign with an expanded scale.  The campaign dates are highlighted in blue.}
\figsetgrpend

\figsetgrpstart
\figsetgrpnum{\ref{fig:appendix}.15}
\figsetgrptitle{4U 1907+097}
\figsetplot{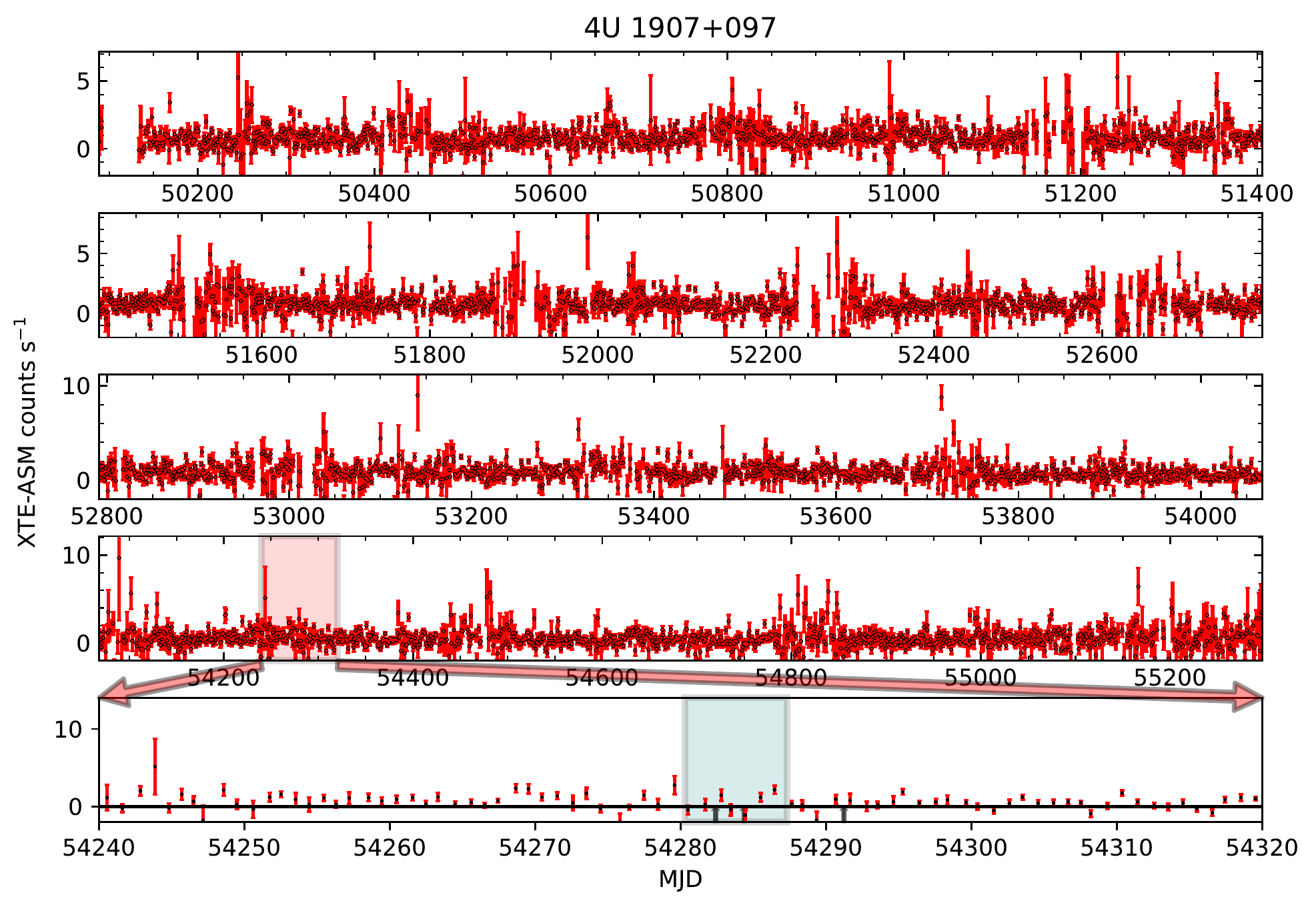}
\figsetgrpnote{X-ray lightcurve from RXTE-ASM.  The bottom panel shows the 80 days
around our campaign with an expanded scale.  The campaign dates are highlighted in blue.}
\figsetgrpend

\figsetgrpstart
\figsetgrpnum{\ref{fig:appendix}.16}
\figsetgrptitle{4U 1957+11}
\figsetplot{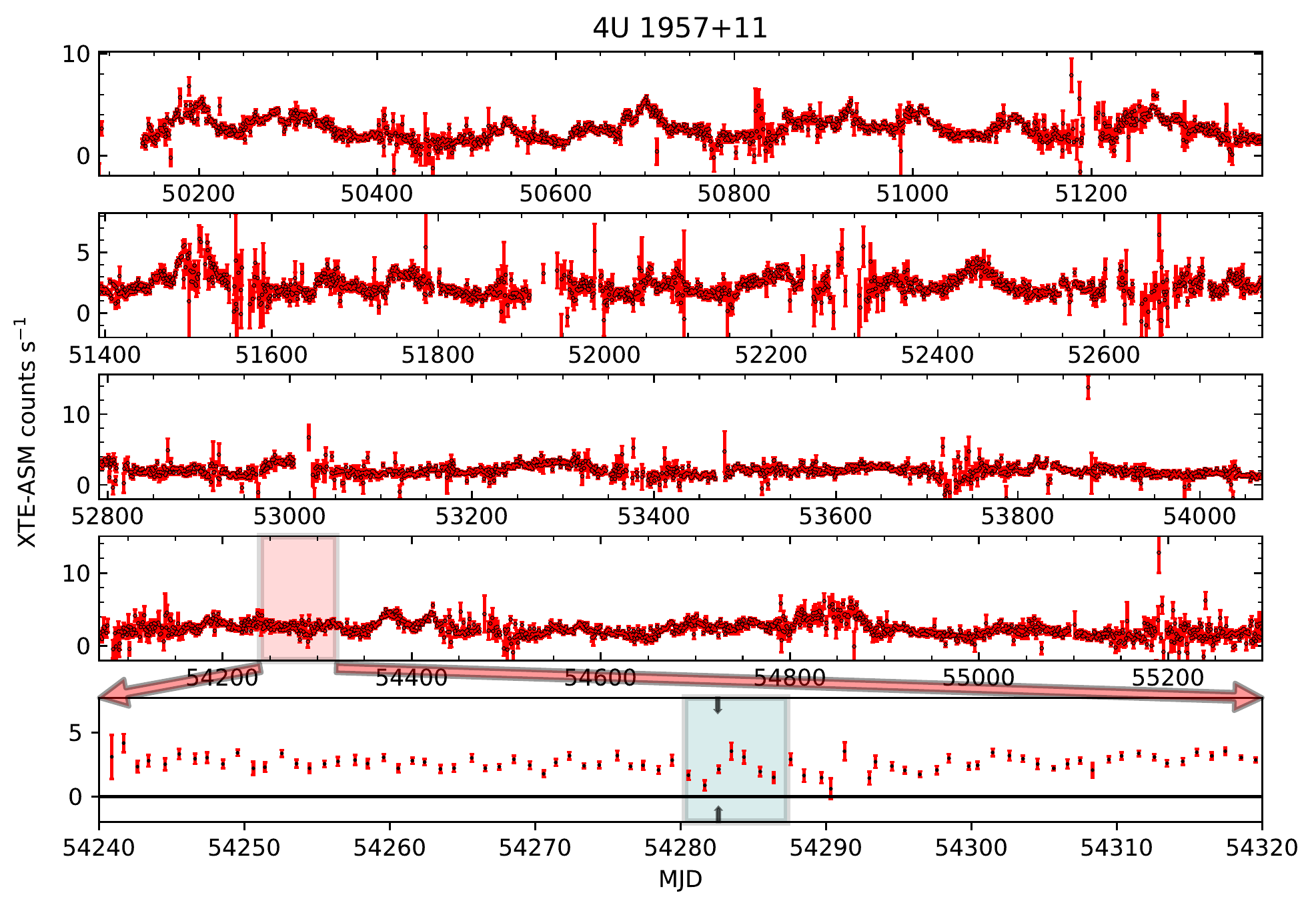}
\figsetgrpnote{X-ray lightcurve from RXTE-ASM.  The bottom panel shows the 80 days
around our campaign with an expanded scale.  The campaign dates are highlighted in blue.}
\figsetgrpend

\figsetgrpstart
\figsetgrpnum{\ref{fig:appendix}.17}
\figsetgrptitle{2S 0921-630}
\figsetplot{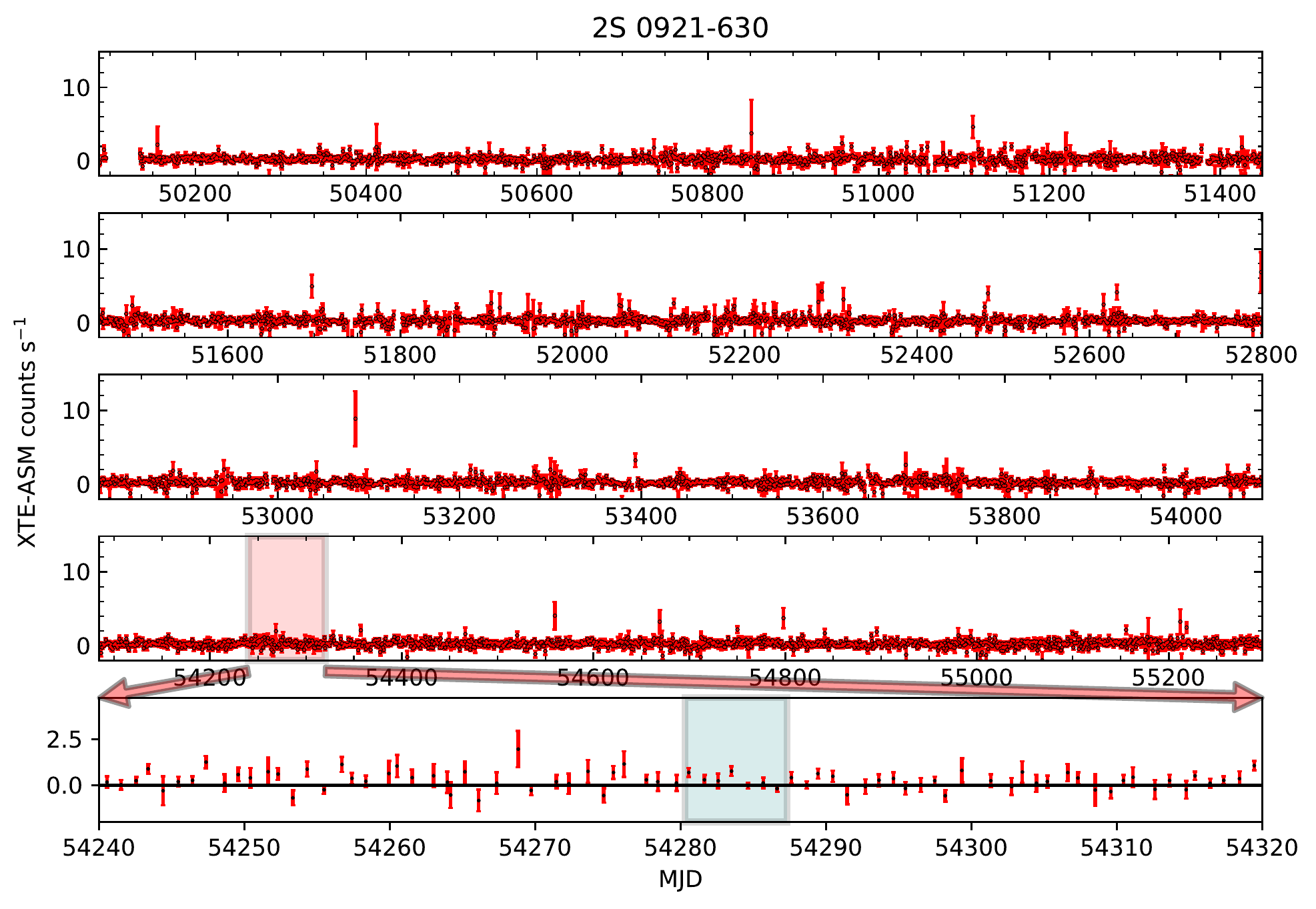}
\figsetgrpnote{X-ray lightcurve from RXTE-ASM.  The bottom panel shows the 80 days
around our campaign with an expanded scale.  The campaign dates are highlighted in blue.}
\figsetgrpend

\figsetgrpstart
\figsetgrpnum{\ref{fig:appendix}.18}
\figsetgrptitle{GX 17+2}
\figsetplot{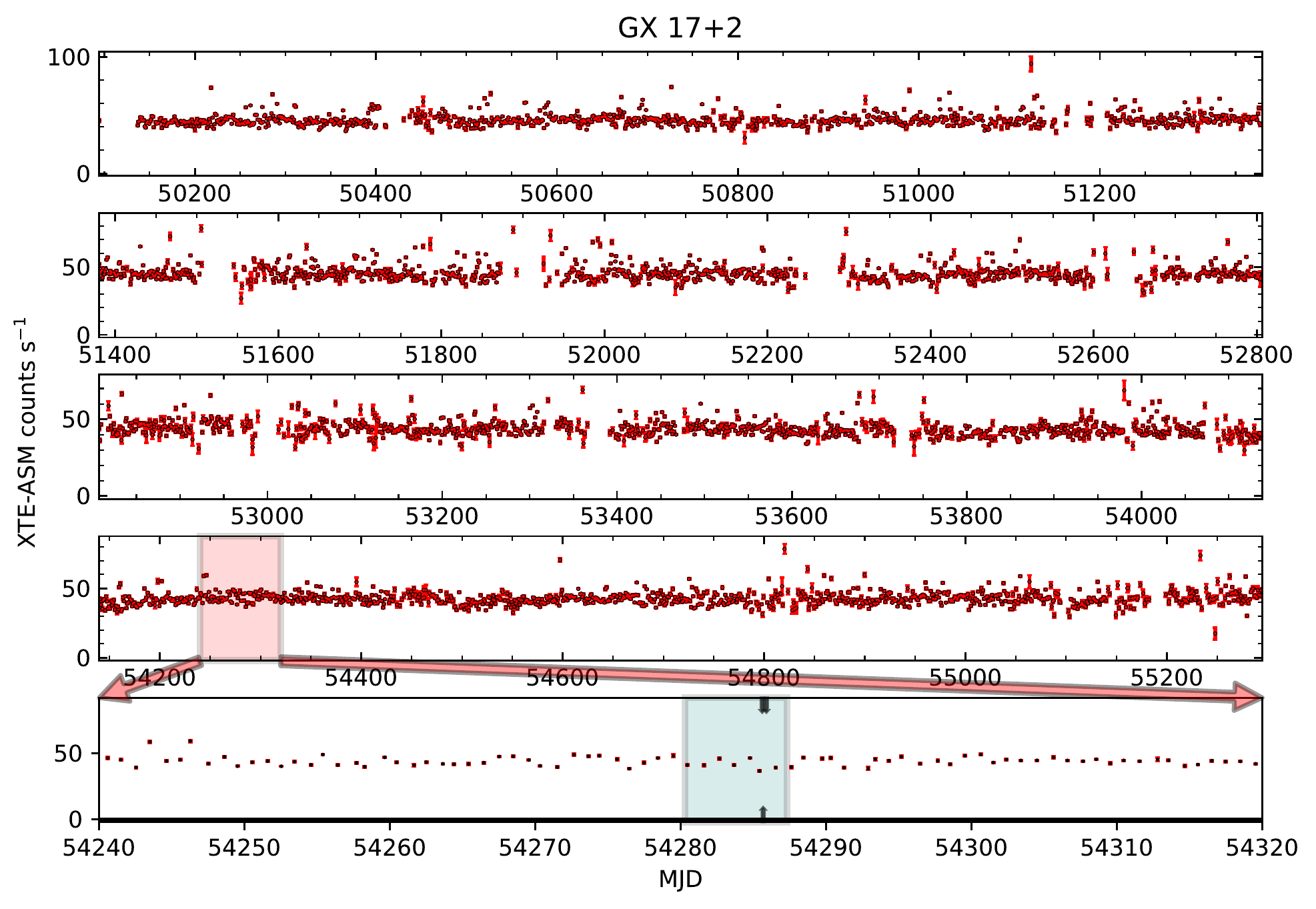}
\figsetgrpnote{X-ray lightcurve from RXTE-ASM.  The bottom panel shows the 80 days
around our campaign with an expanded scale.  The campaign dates are highlighted in blue.}
\figsetgrpend

\figsetgrpstart
\figsetgrpnum{\ref{fig:appendix}.19}
\figsetgrptitle{GX 339-4}
\figsetplot{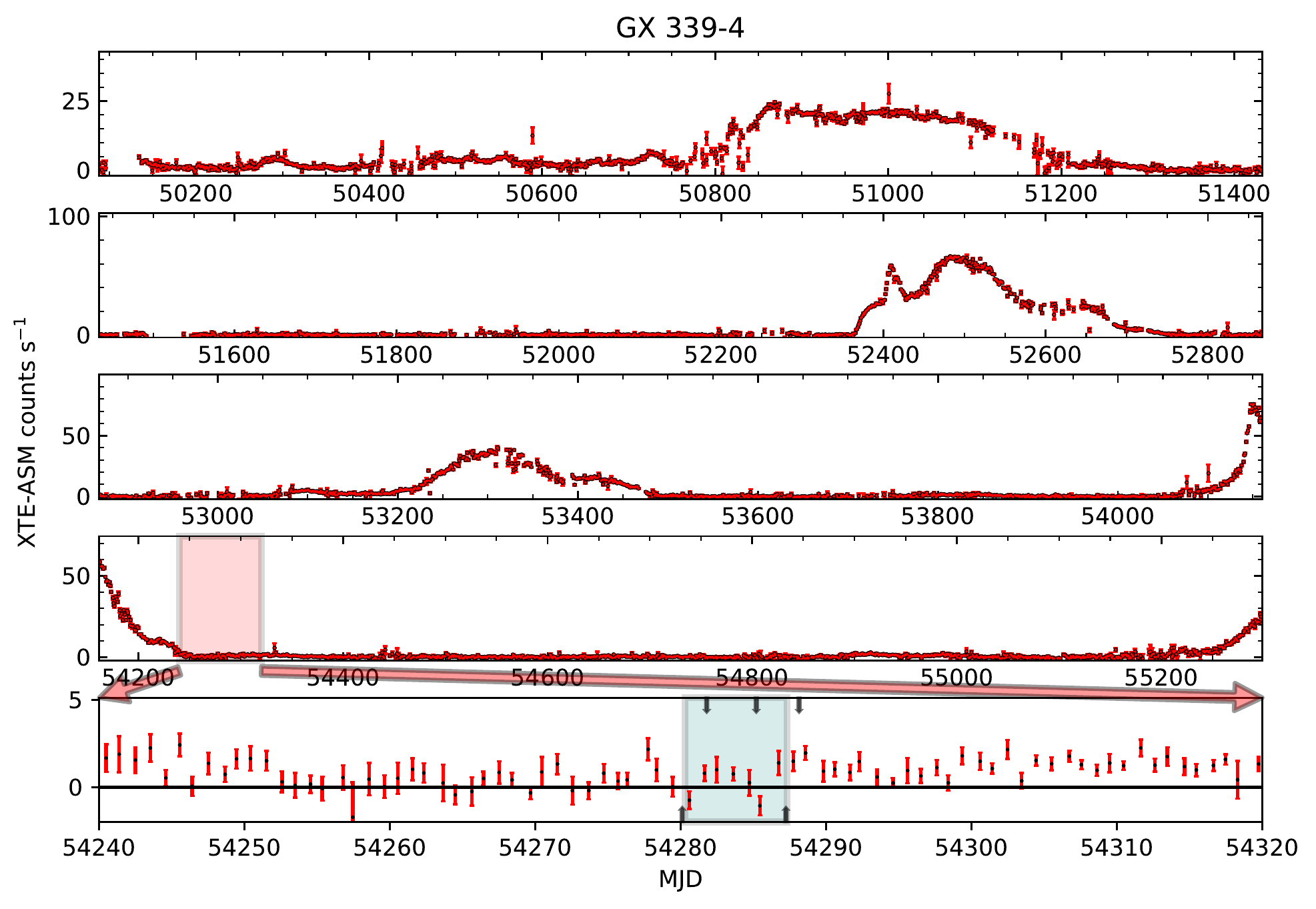}
\figsetgrpnote{X-ray lightcurve from RXTE-ASM.  The bottom panel shows the 80 days
around our campaign with an expanded scale.  The campaign dates are highlighted in blue.}
\figsetgrpend

\figsetgrpstart
\figsetgrpnum{\ref{fig:appendix}.20}
\figsetgrptitle{GX 340+0}
\figsetplot{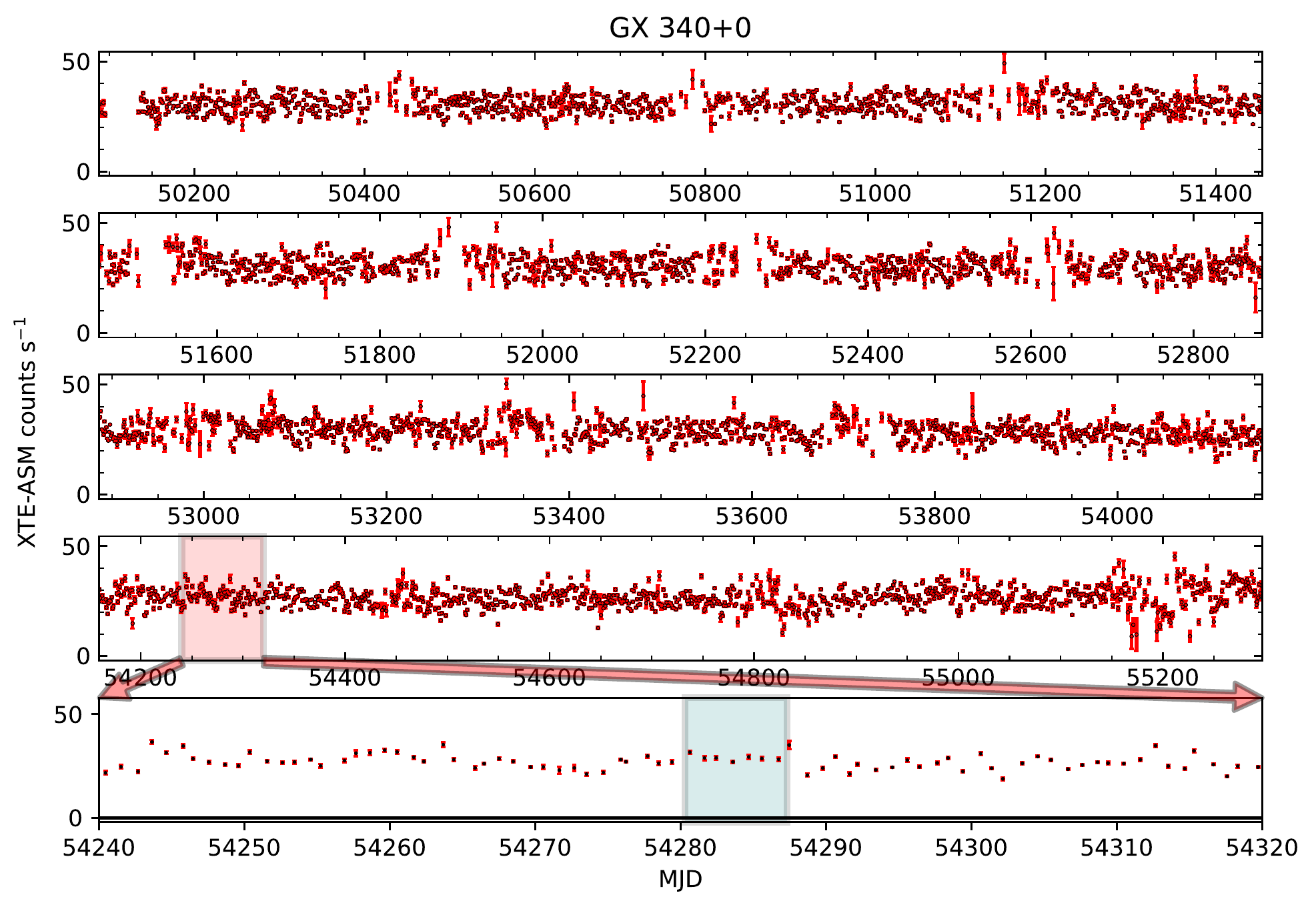}
\figsetgrpnote{X-ray lightcurve from RXTE-ASM.  The bottom panel shows the 80 days
around our campaign with an expanded scale.  The campaign dates are highlighted in blue.}
\figsetgrpend

\figsetgrpstart
\figsetgrpnum{\ref{fig:appendix}.21}
\figsetgrptitle{GX 349+2}
\figsetplot{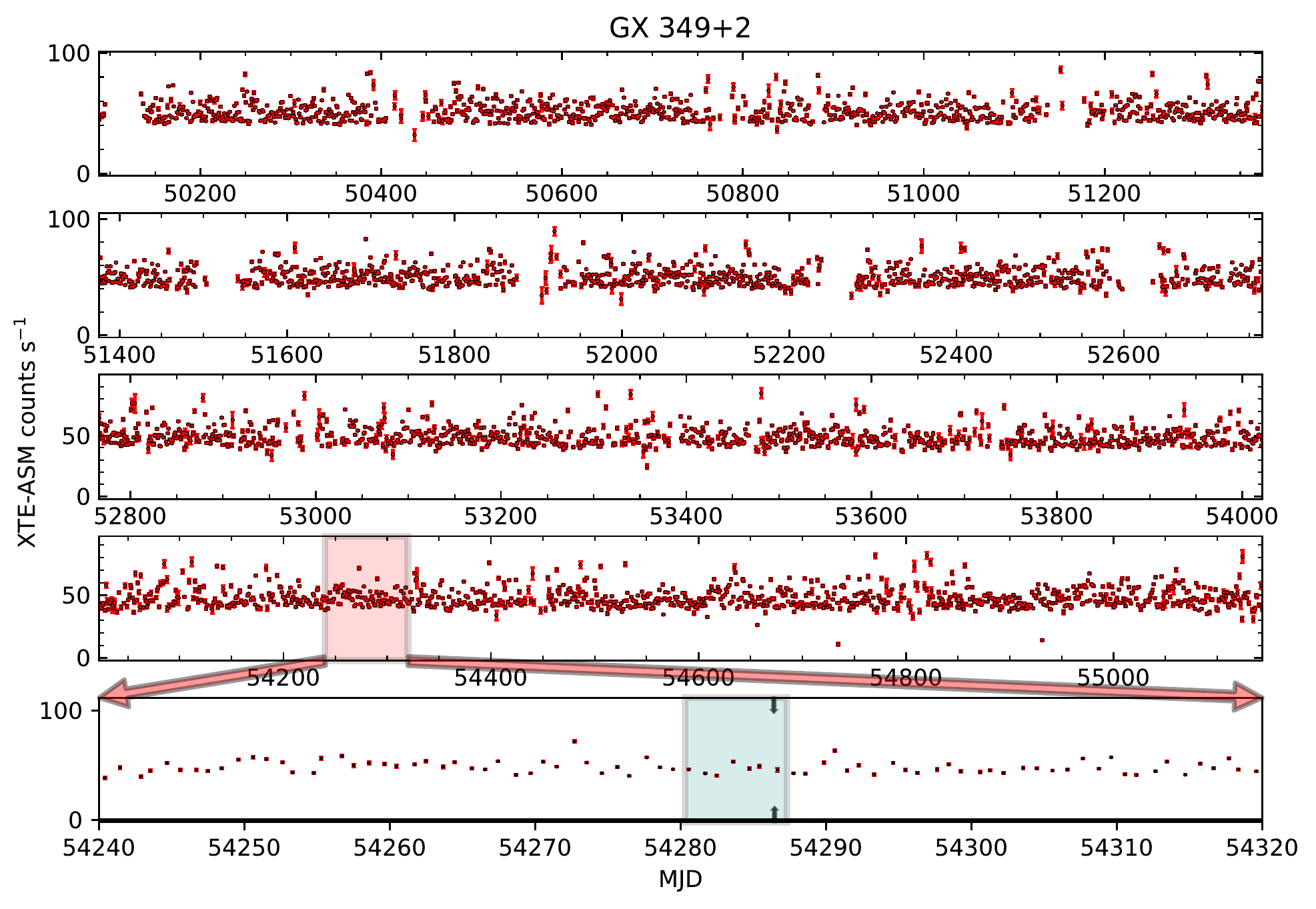}
\figsetgrpnote{X-ray lightcurve from RXTE-ASM.  The bottom panel shows the 80 days
around our campaign with an expanded scale.  The campaign dates are highlighted in blue.}
\figsetgrpend

\figsetgrpstart
\figsetgrpnum{\ref{fig:appendix}.22}
\figsetgrptitle{GX 354-0}
\figsetplot{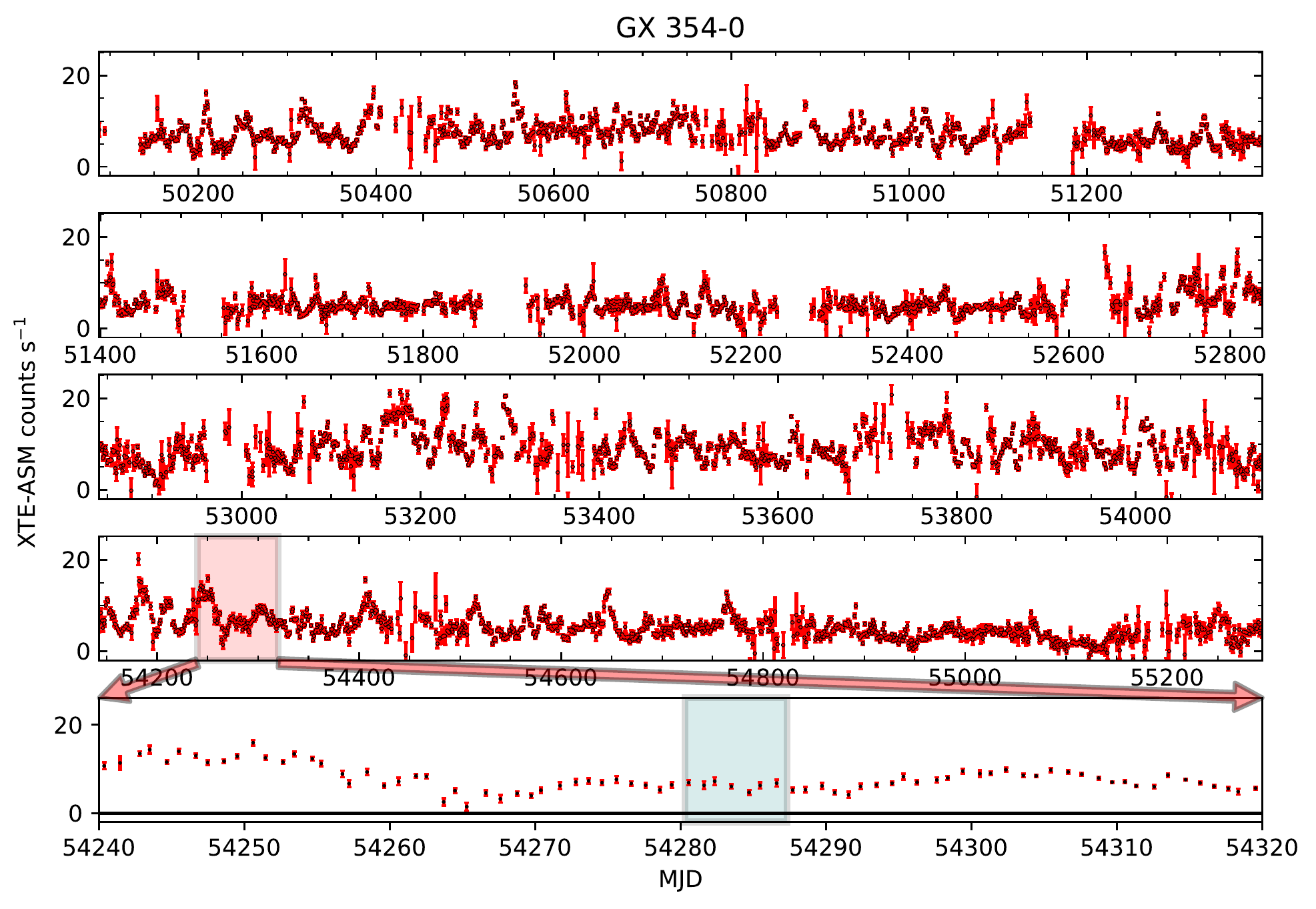}
\figsetgrpnote{X-ray lightcurve from RXTE-ASM.  The bottom panel shows the 80 days
around our campaign with an expanded scale.  The campaign dates are highlighted in blue.}
\figsetgrpend

\figsetgrpstart
\figsetgrpnum{\ref{fig:appendix}.23}
\figsetgrptitle{GX 3+1}
\figsetplot{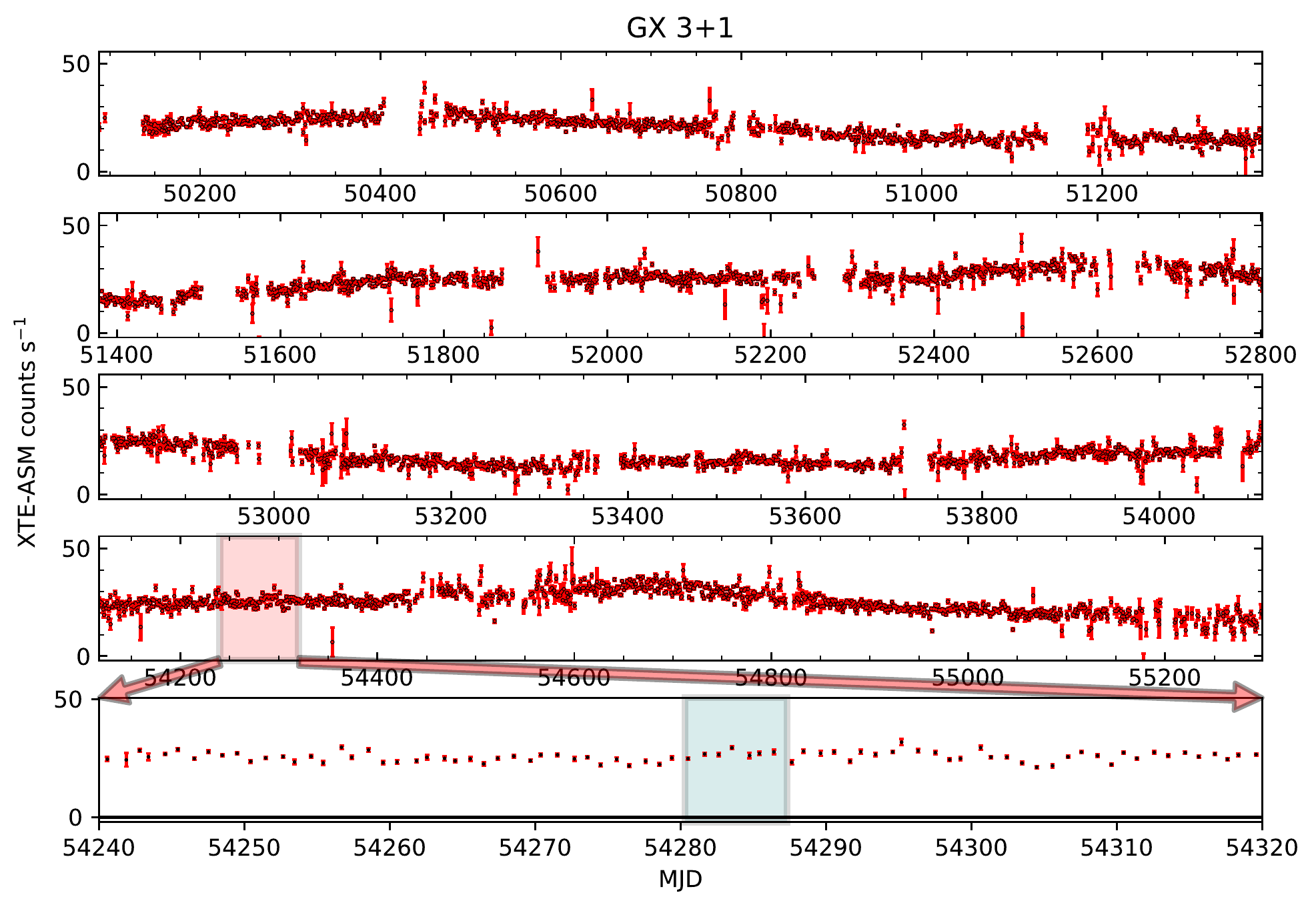}
\figsetgrpnote{X-ray lightcurve from RXTE-ASM.  The bottom panel shows the 80 days
around our campaign with an expanded scale.  The campaign dates are highlighted in blue.}
\figsetgrpend

\figsetgrpstart
\figsetgrpnum{\ref{fig:appendix}.24}
\figsetgrptitle{GX 5-1}
\figsetplot{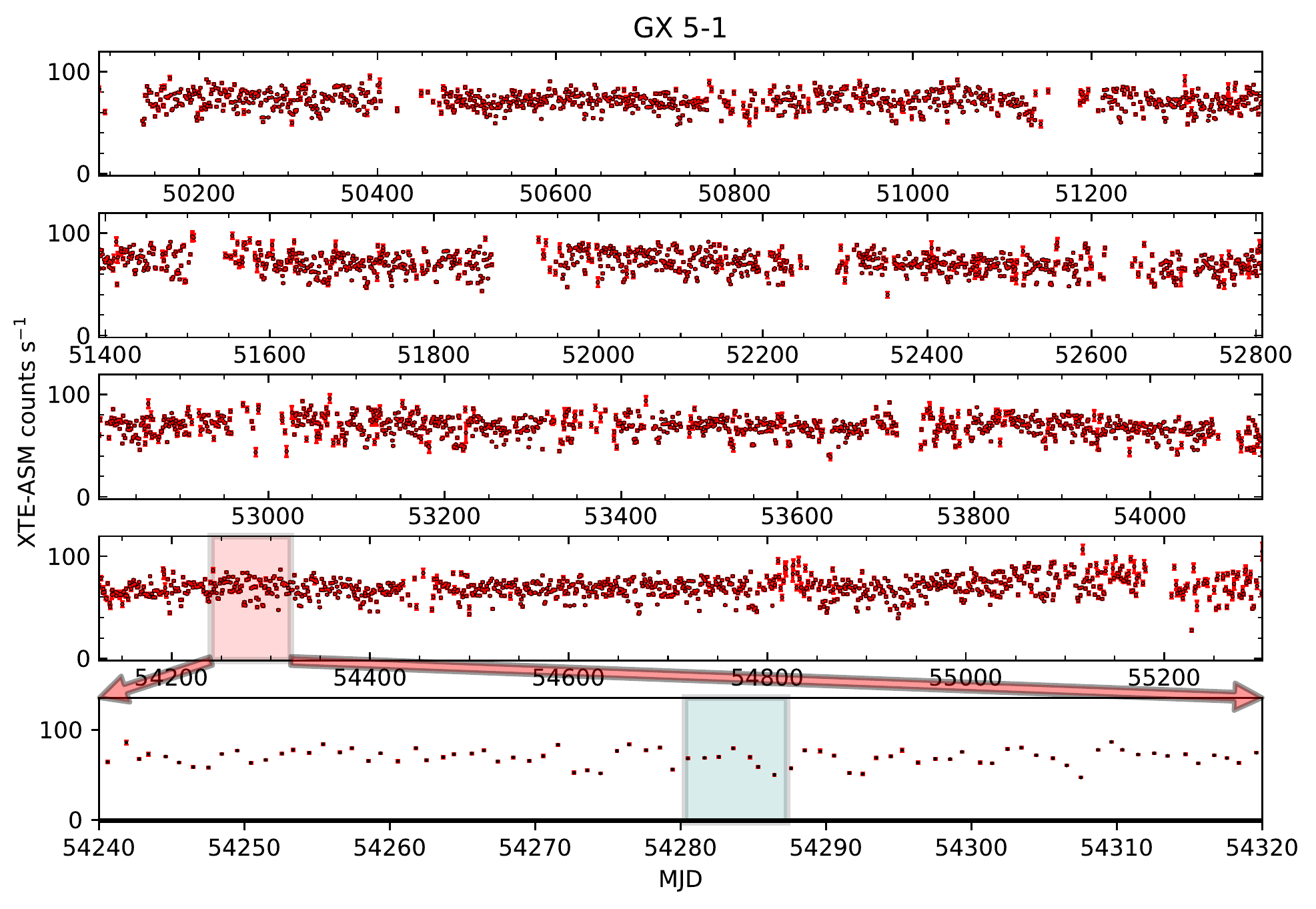}
\figsetgrpnote{X-ray lightcurve from RXTE-ASM.  The bottom panel shows the 80 days
around our campaign with an expanded scale.  The campaign dates are highlighted in blue.}
\figsetgrpend

\figsetgrpstart
\figsetgrpnum{\ref{fig:appendix}.25}
\figsetgrptitle{XTE J1550-564}
\figsetplot{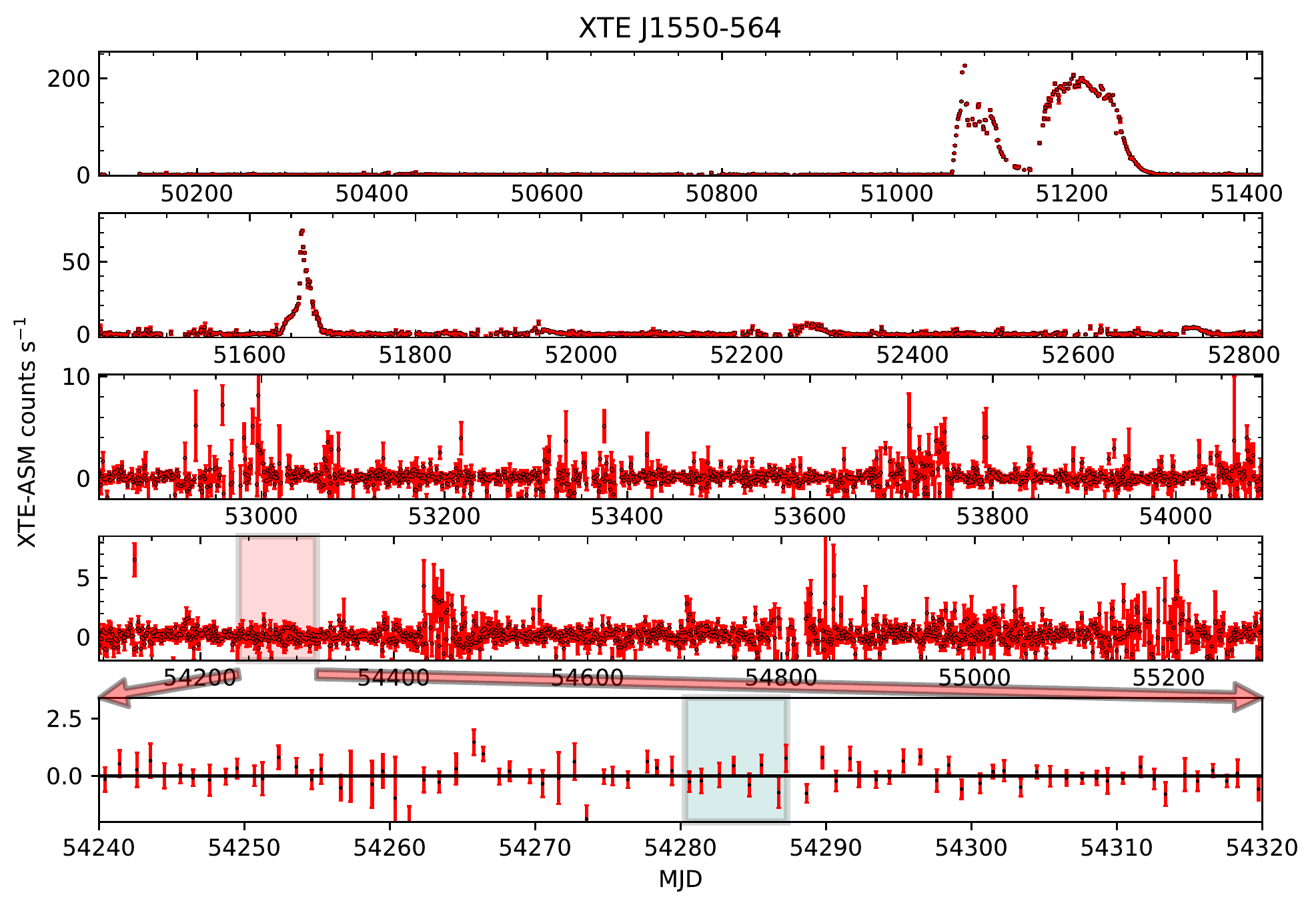}
\figsetgrpnote{X-ray lightcurve from RXTE-ASM.  The bottom panel shows the 80 days
around our campaign with an expanded scale.  The campaign dates are highlighted in blue.}
\figsetgrpend

\figsetgrpstart
\figsetgrpnum{\ref{fig:appendix}.26}
\figsetgrptitle{LMC X-1}
\figsetplot{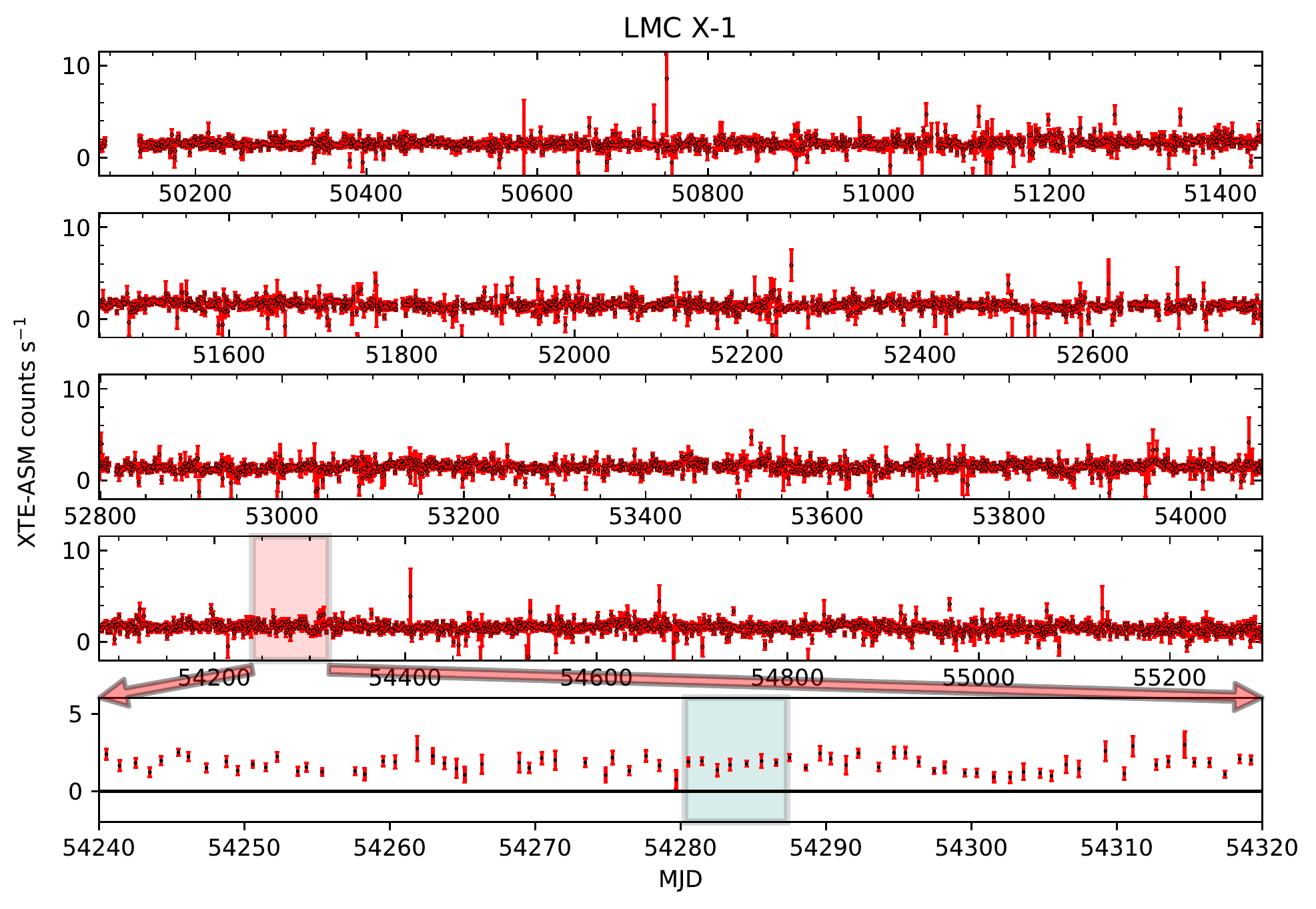}
\figsetgrpnote{X-ray lightcurve from RXTE-ASM.  The bottom panel shows the 80 days
around our campaign with an expanded scale.  The campaign dates are highlighted in blue.}
\figsetgrpend

\figsetgrpstart
\figsetgrpnum{\ref{fig:appendix}.27}
\figsetgrptitle{LMC X-2}
\figsetplot{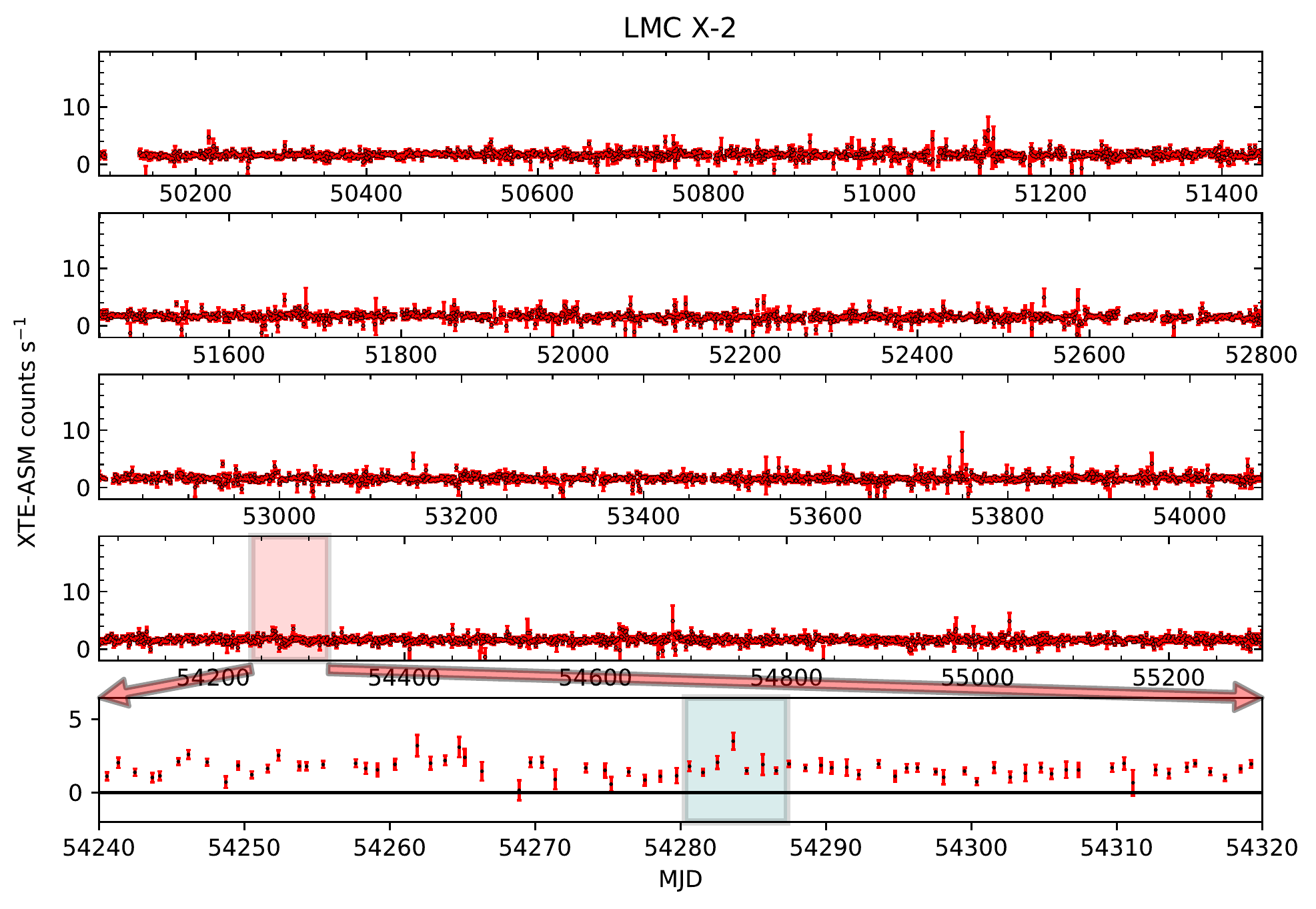}
\figsetgrpnote{X-ray lightcurve from RXTE-ASM.  The bottom panel shows the 80 days
around our campaign with an expanded scale.  The campaign dates are highlighted in blue.}
\figsetgrpend

\figsetgrpstart
\figsetgrpnum{\ref{fig:appendix}.28}
\figsetgrptitle{LMC X-3}
\figsetplot{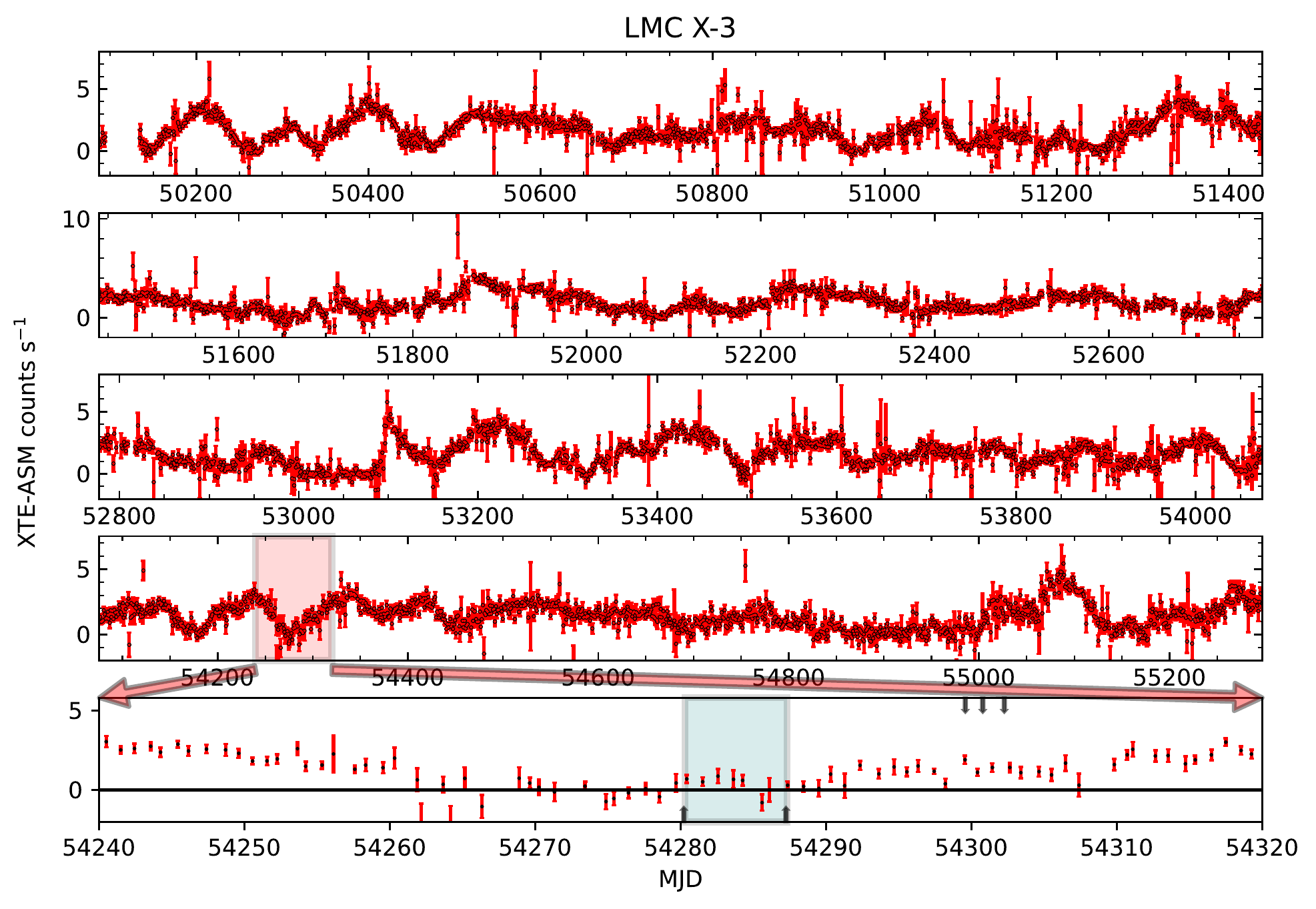}
\figsetgrpnote{X-ray lightcurve from RXTE-ASM.  The bottom panel shows the 80 days
around our campaign with an expanded scale.  The campaign dates are highlighted in blue.}
\figsetgrpend

\figsetgrpstart
\figsetgrpnum{\ref{fig:appendix}.29}
\figsetgrptitle{SMC X-3}
\figsetplot{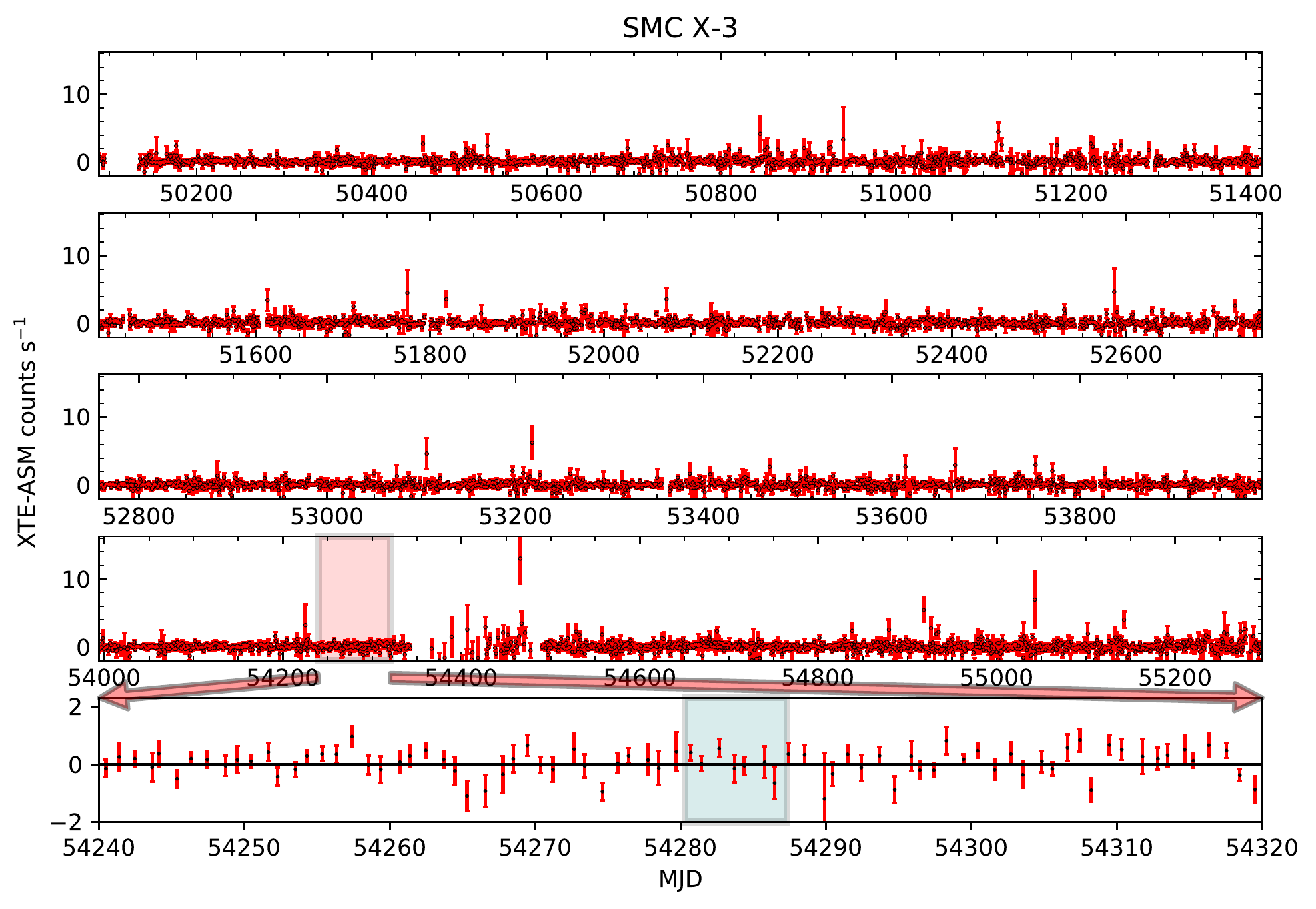}
\figsetgrpnote{X-ray lightcurve from RXTE-ASM.  The bottom panel shows the 80 days
around our campaign with an expanded scale.  The campaign dates are highlighted in blue.}
\figsetgrpend

\figsetgrpstart
\figsetgrpnum{\ref{fig:appendix}.30}
\figsetgrptitle{Aql X-1}
\figsetplot{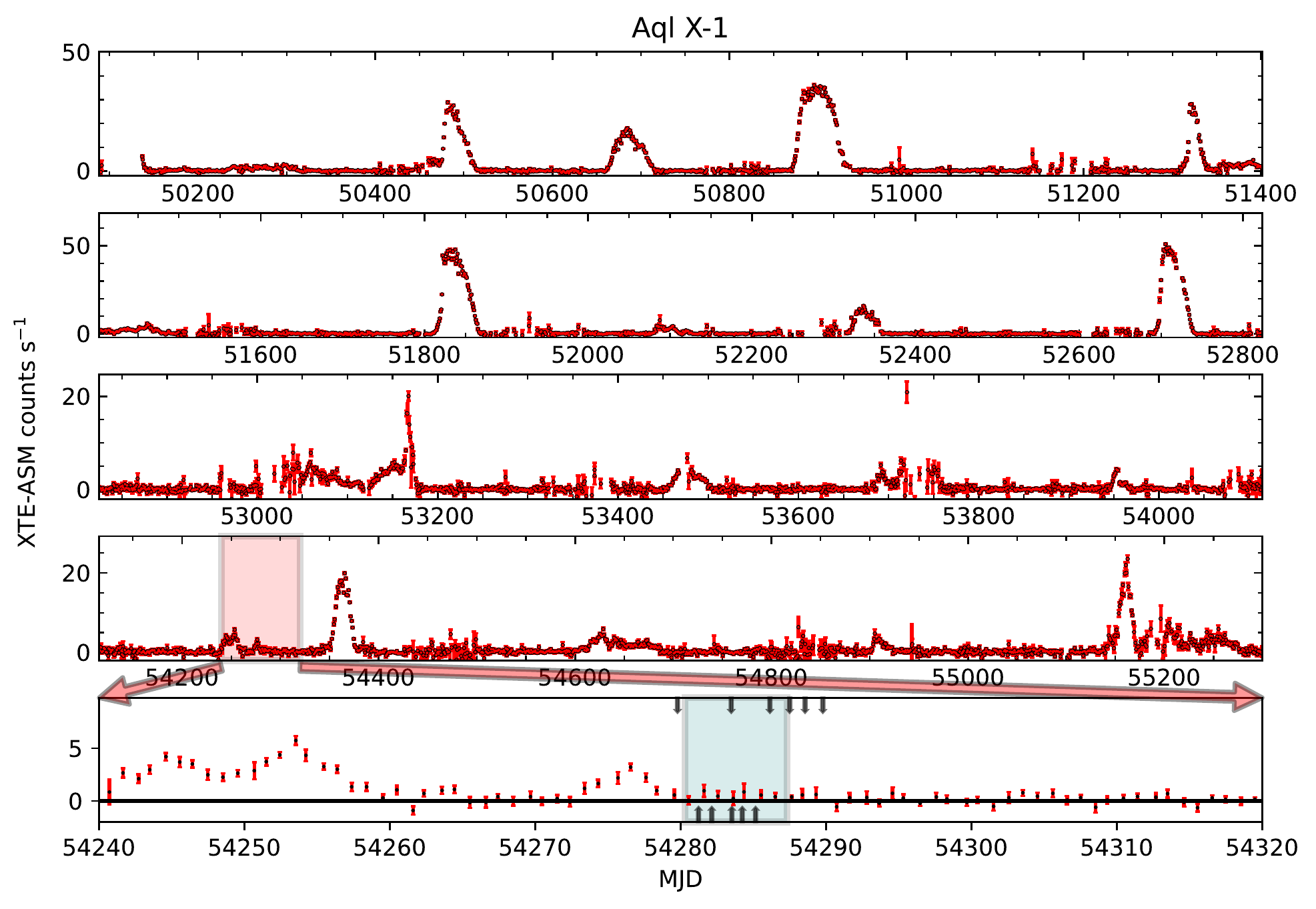}
\figsetgrpnote{X-ray lightcurve from RXTE-ASM.  The bottom panel shows the 80 days
around our campaign with an expanded scale.  The campaign dates are highlighted in blue.}
\figsetgrpend

\figsetgrpstart
\figsetgrpnum{\ref{fig:appendix}.31}
\figsetgrptitle{Cen X-3}
\figsetplot{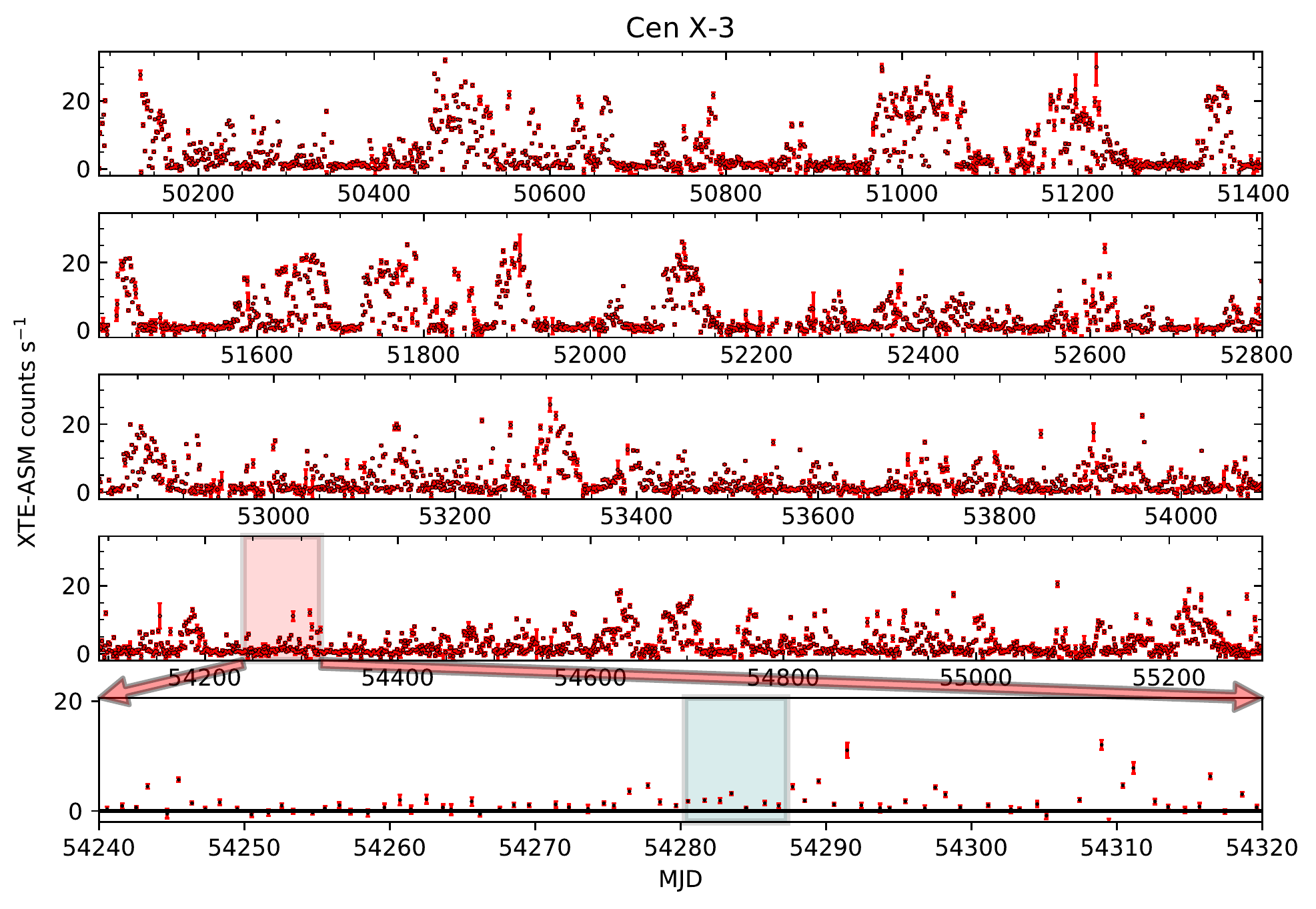}
\figsetgrpnote{X-ray lightcurve from RXTE-ASM.  The bottom panel shows the 80 days
around our campaign with an expanded scale.  The campaign dates are highlighted in blue.}
\figsetgrpend

\figsetgrpstart
\figsetgrpnum{\ref{fig:appendix}.32}
\figsetgrptitle{Cir X-1}
\figsetplot{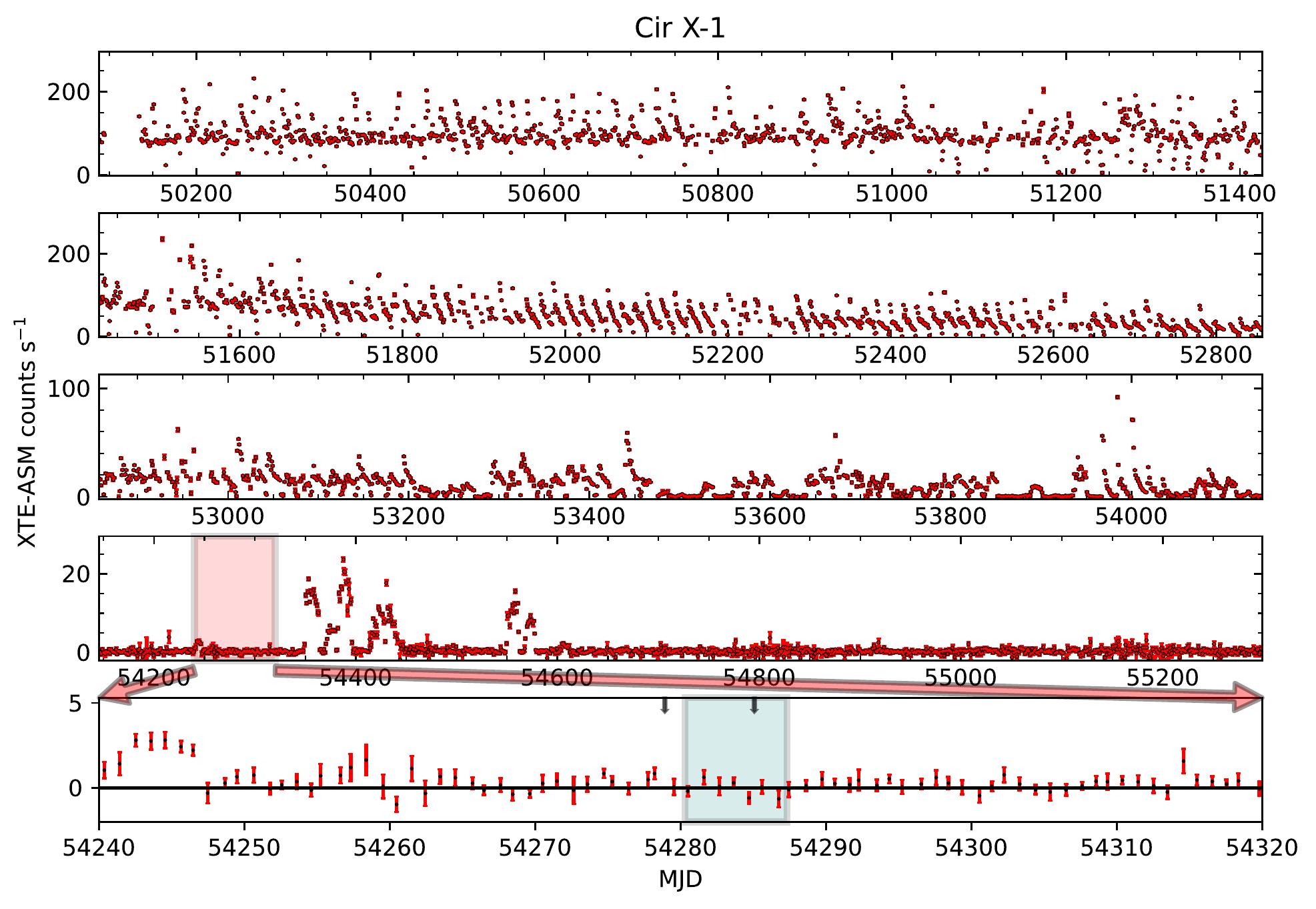}
\figsetgrpnote{X-ray lightcurve from RXTE-ASM.  The bottom panel shows the 80 days
around our campaign with an expanded scale.  The campaign dates are highlighted in blue.}
\figsetgrpend

\figsetgrpstart
\figsetgrpnum{\ref{fig:appendix}.33}
\figsetgrptitle{GRS 1915+105}
\figsetplot{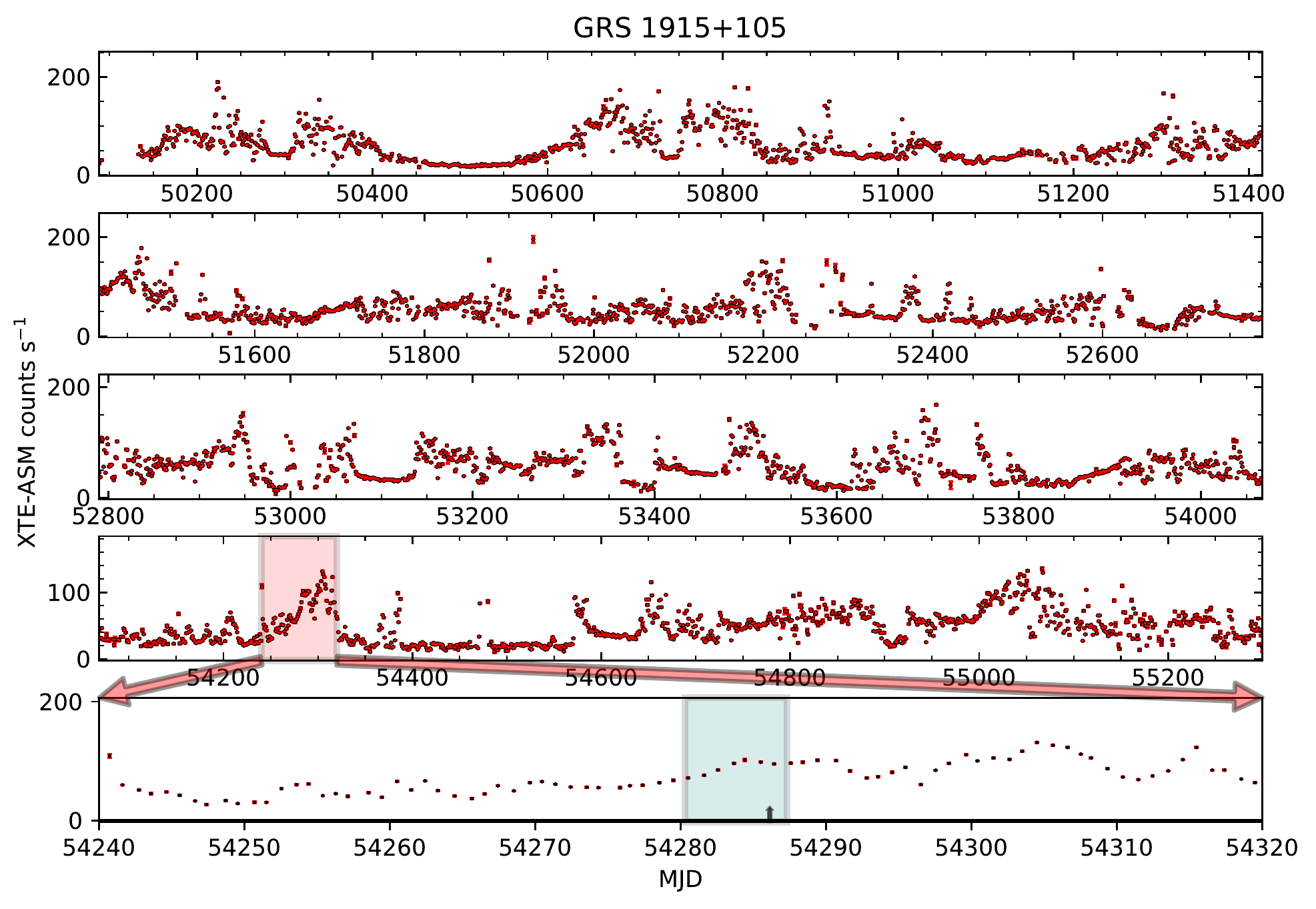}
\figsetgrpnote{X-ray lightcurve from RXTE-ASM.  The bottom panel shows the 80 days
around our campaign with an expanded scale.  The campaign dates are highlighted in blue.}
\figsetgrpend

\figsetgrpstart
\figsetgrpnum{\ref{fig:appendix}.34}
\figsetgrptitle{GX 13+1}
\figsetplot{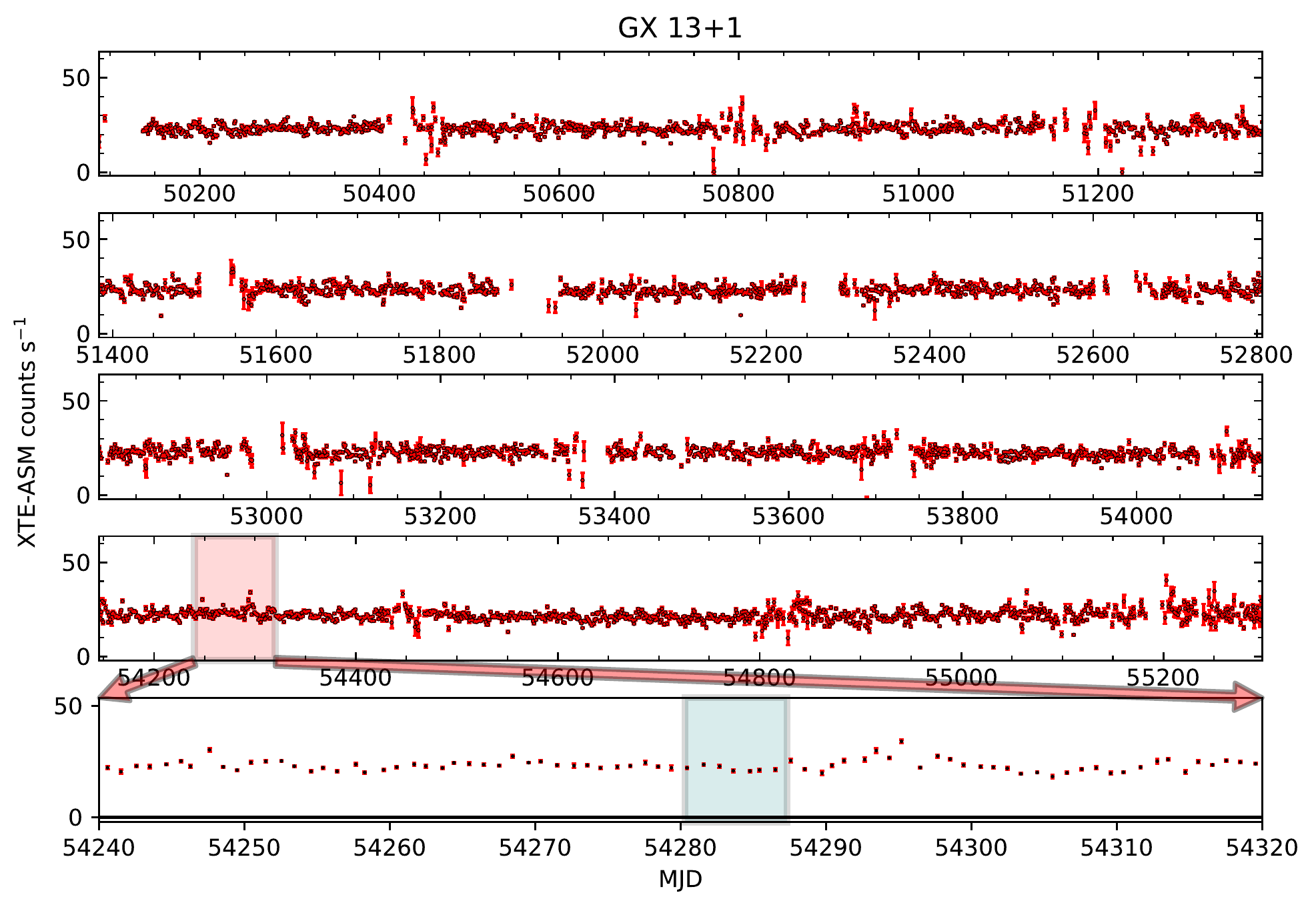}
\figsetgrpnote{X-ray lightcurve from RXTE-ASM.  The bottom panel shows the 80 days
around our campaign with an expanded scale.  The campaign dates are highlighted in blue.}
\figsetgrpend

\figsetgrpstart
\figsetgrpnum{\ref{fig:appendix}.35}
\figsetgrptitle{GX 301-2}
\figsetplot{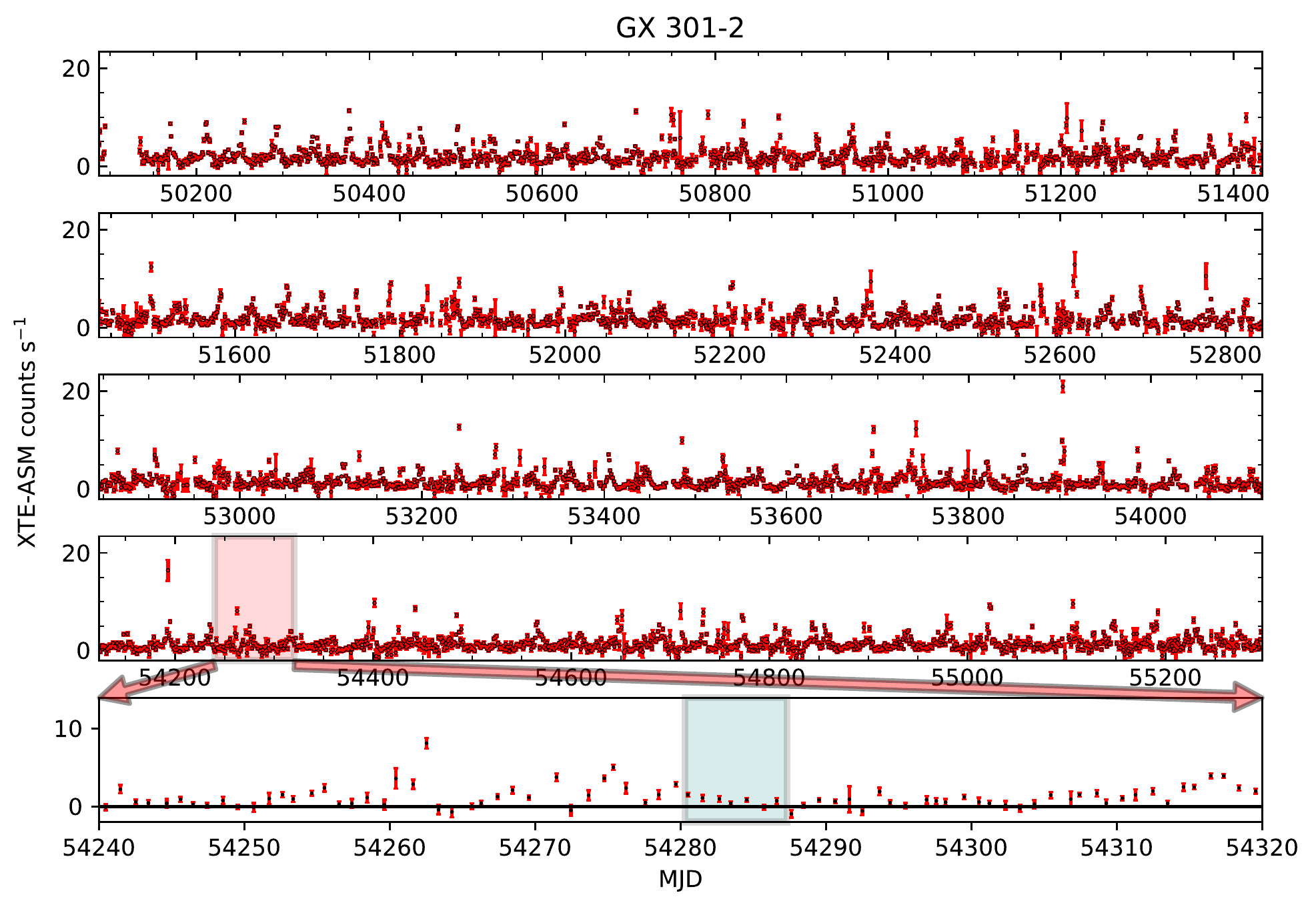}
\figsetgrpnote{X-ray lightcurve from RXTE-ASM.  The bottom panel shows the 80 days
around our campaign with an expanded scale.  The campaign dates are highlighted in blue.}
\figsetgrpend

\figsetgrpstart
\figsetgrpnum{\ref{fig:appendix}.36}
\figsetgrptitle{GX 9+1}
\figsetplot{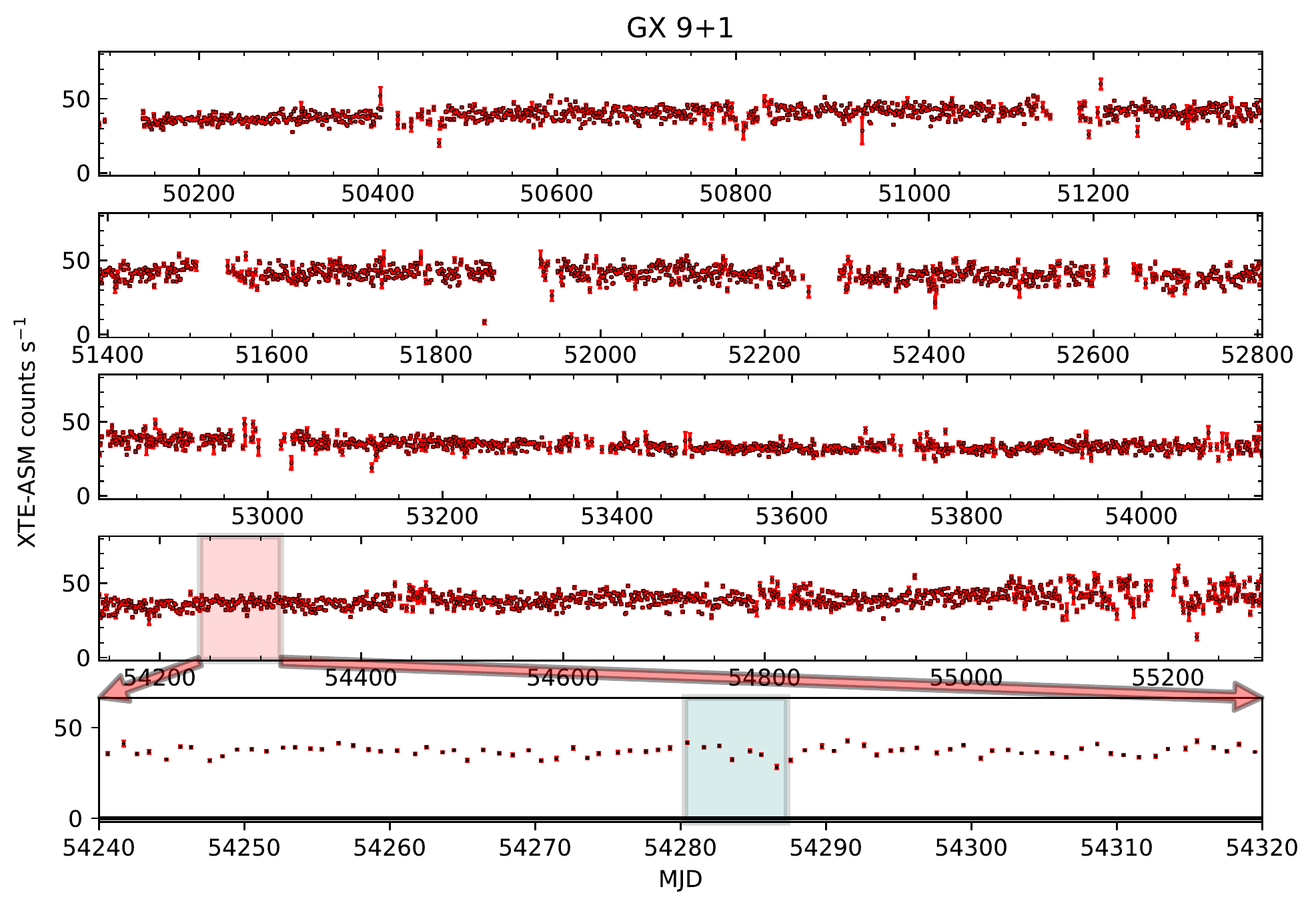}
\figsetgrpnote{X-ray lightcurve from RXTE-ASM.  The bottom panel shows the 80 days
around our campaign with an expanded scale.  The campaign dates are highlighted in blue.}
\figsetgrpend

\figsetgrpstart
\figsetgrpnum{\ref{fig:appendix}.37}
\figsetgrptitle{GX 9+9}
\figsetplot{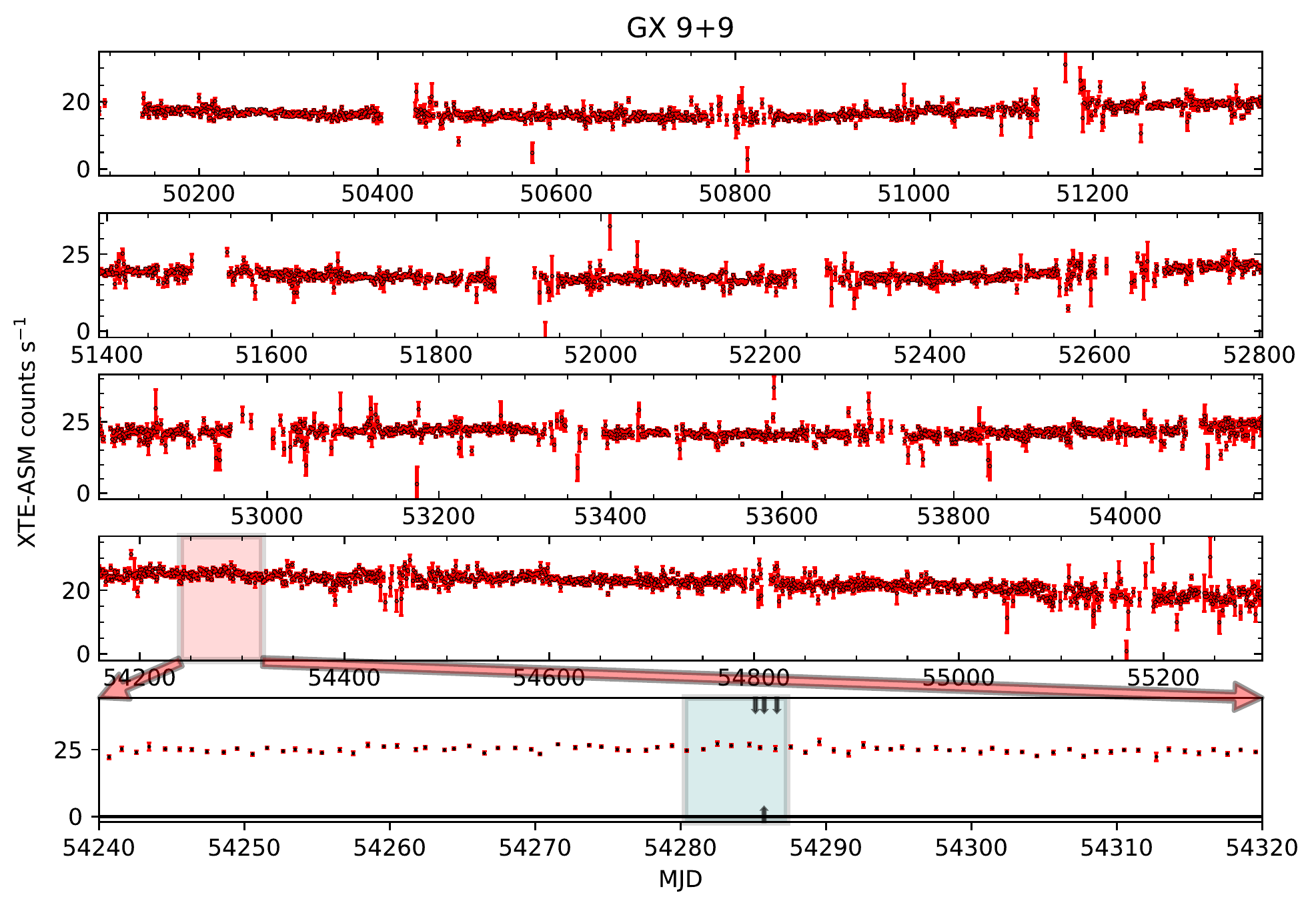}
\figsetgrpnote{X-ray lightcurve from RXTE-ASM.  The bottom panel shows the 80 days
around our campaign with an expanded scale.  The campaign dates are highlighted in blue.}
\figsetgrpend

\figsetgrpstart
\figsetgrpnum{\ref{fig:appendix}.38}
\figsetgrptitle{4U 1556-60}
\figsetplot{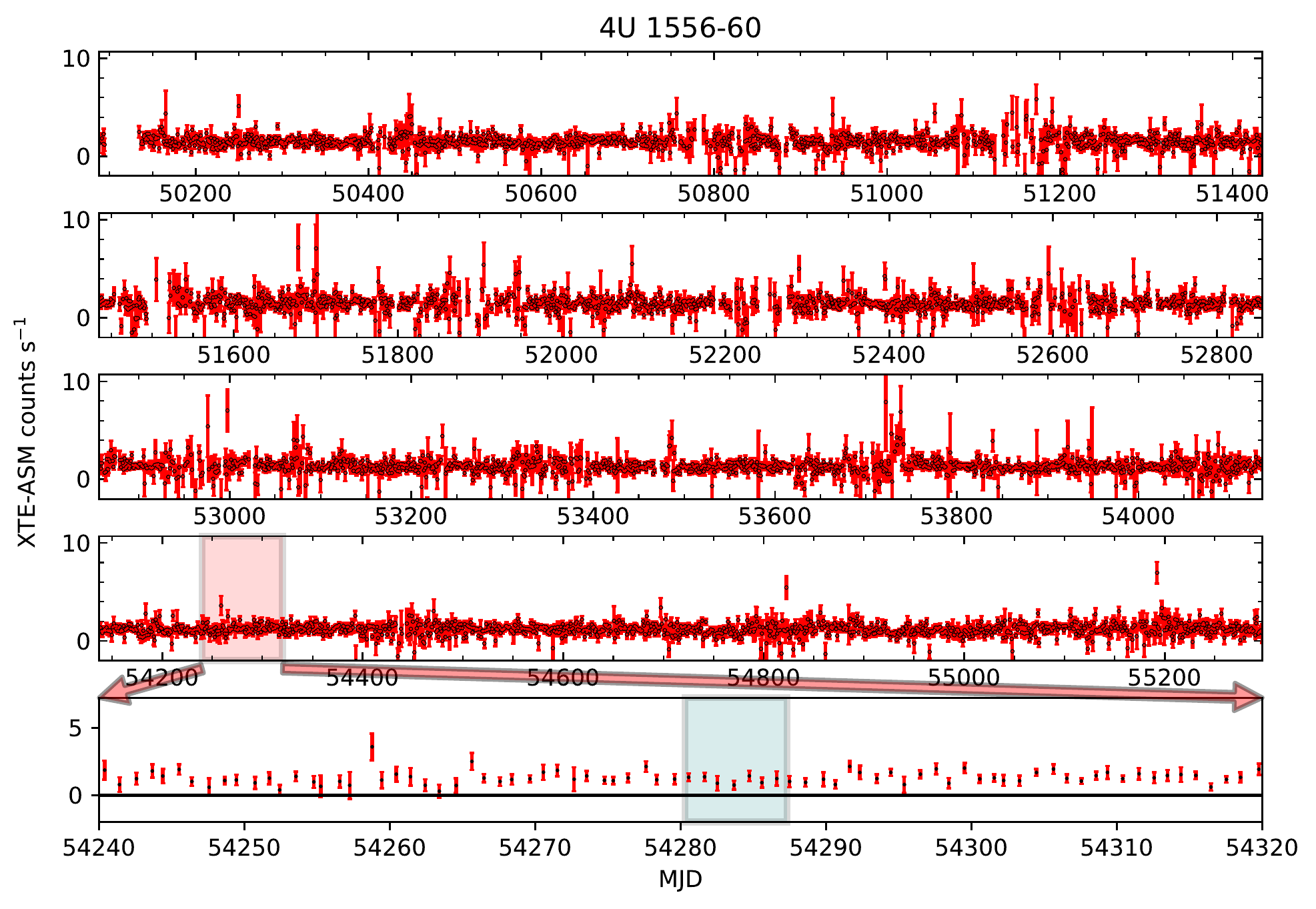}
\figsetgrpnote{X-ray lightcurve from RXTE-ASM.  The bottom panel shows the 80 days
around our campaign with an expanded scale.  The campaign dates are highlighted in blue.}
\figsetgrpend

\figsetgrpstart
\figsetgrpnum{\ref{fig:appendix}.39}
\figsetgrptitle{XTE J1701-462}
\figsetplot{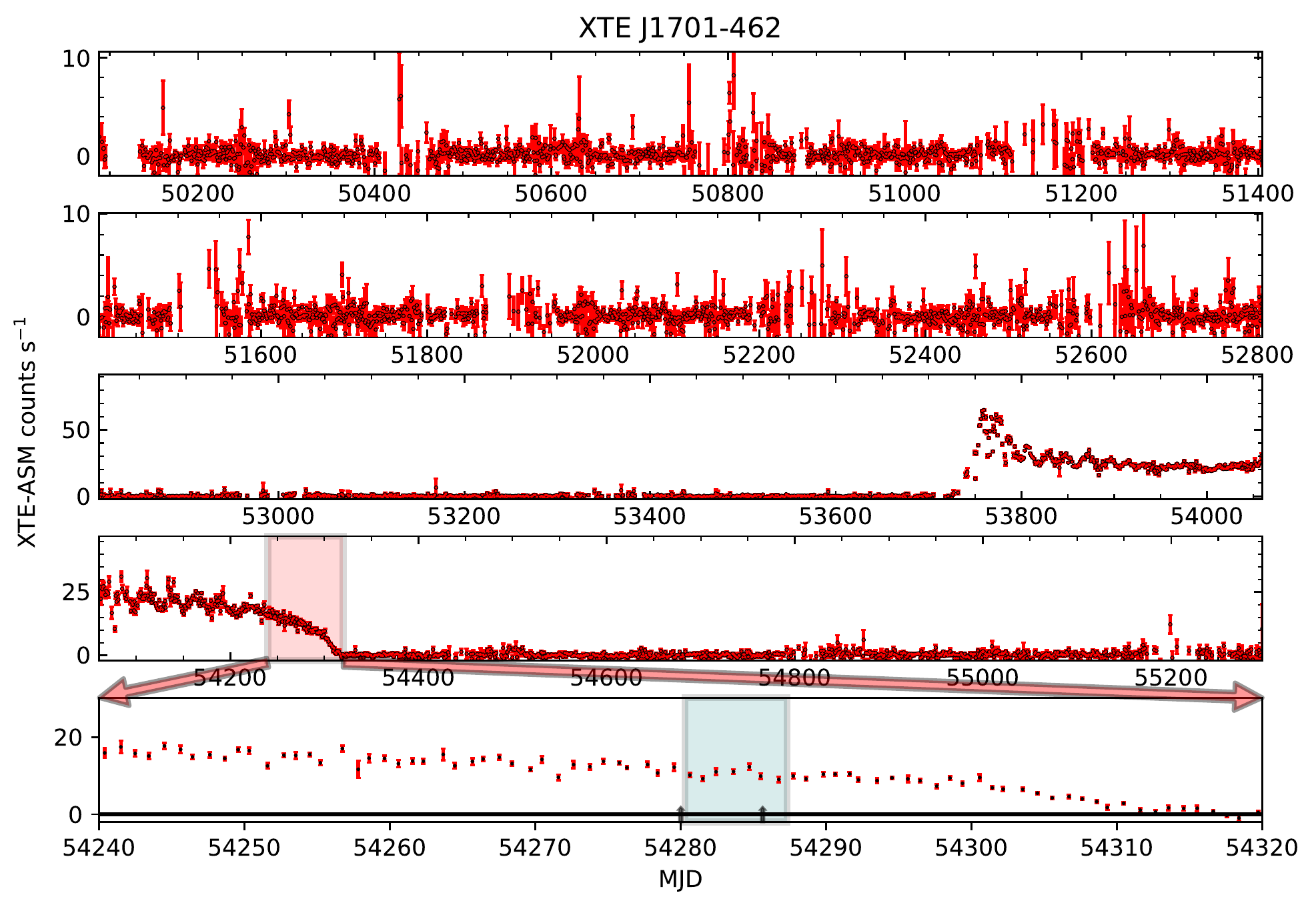}
\figsetgrpnote{X-ray lightcurve from RXTE-ASM.  The bottom panel shows the 80 days
around our campaign with an expanded scale.  The campaign dates are highlighted in blue.}
\figsetgrpend

\figsetgrpstart
\figsetgrpnum{\ref{fig:appendix}.40}
\figsetgrptitle{Sco X-1}
\figsetplot{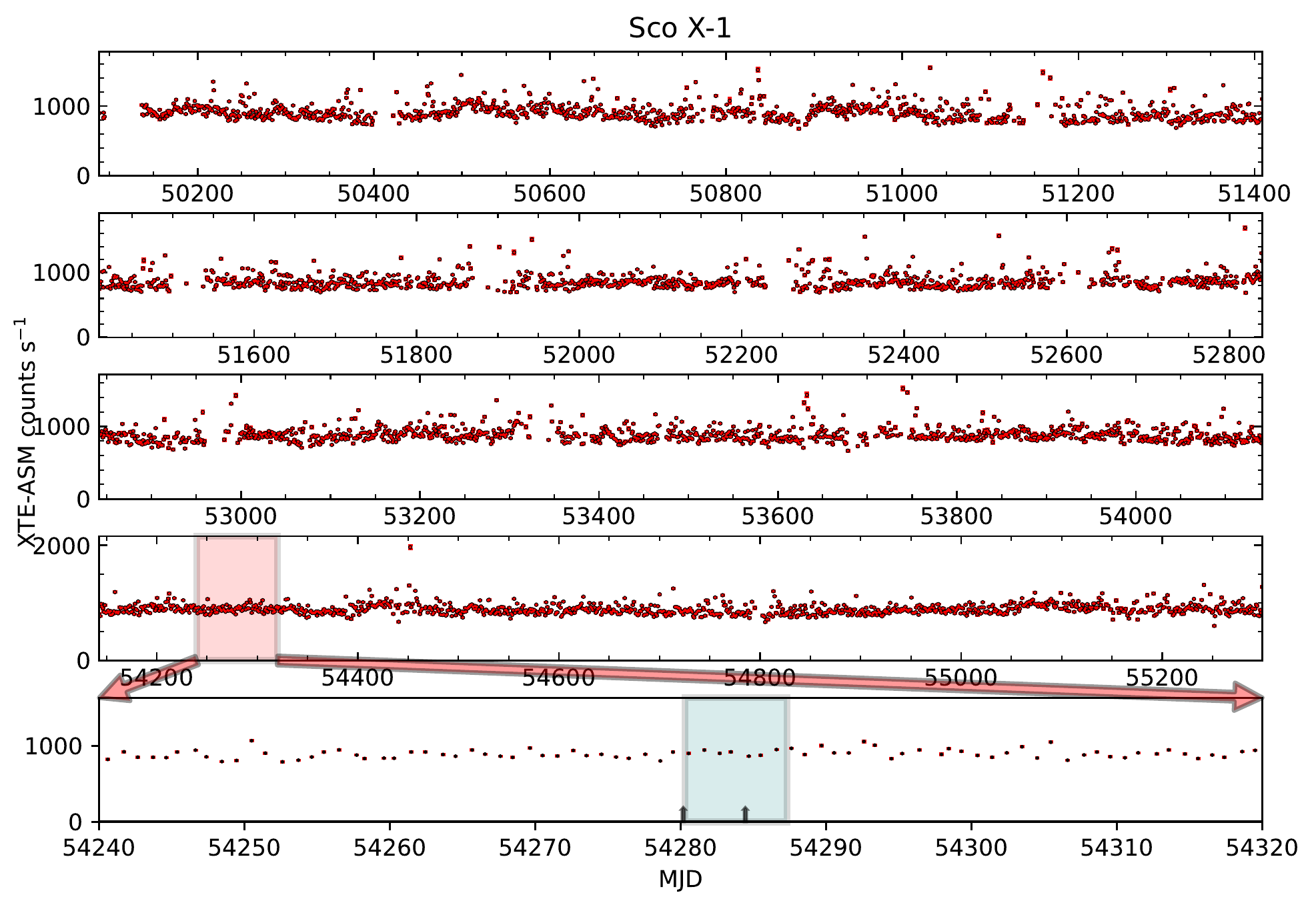}
\figsetgrpnote{X-ray lightcurve from RXTE-ASM.  The bottom panel shows the 80 days
around our campaign with an expanded scale.  The campaign dates are highlighted in blue.}
\figsetgrpend

\figsetgrpstart
\figsetgrpnum{\ref{fig:appendix}.41}
\figsetgrptitle{SMC X-1}
\figsetplot{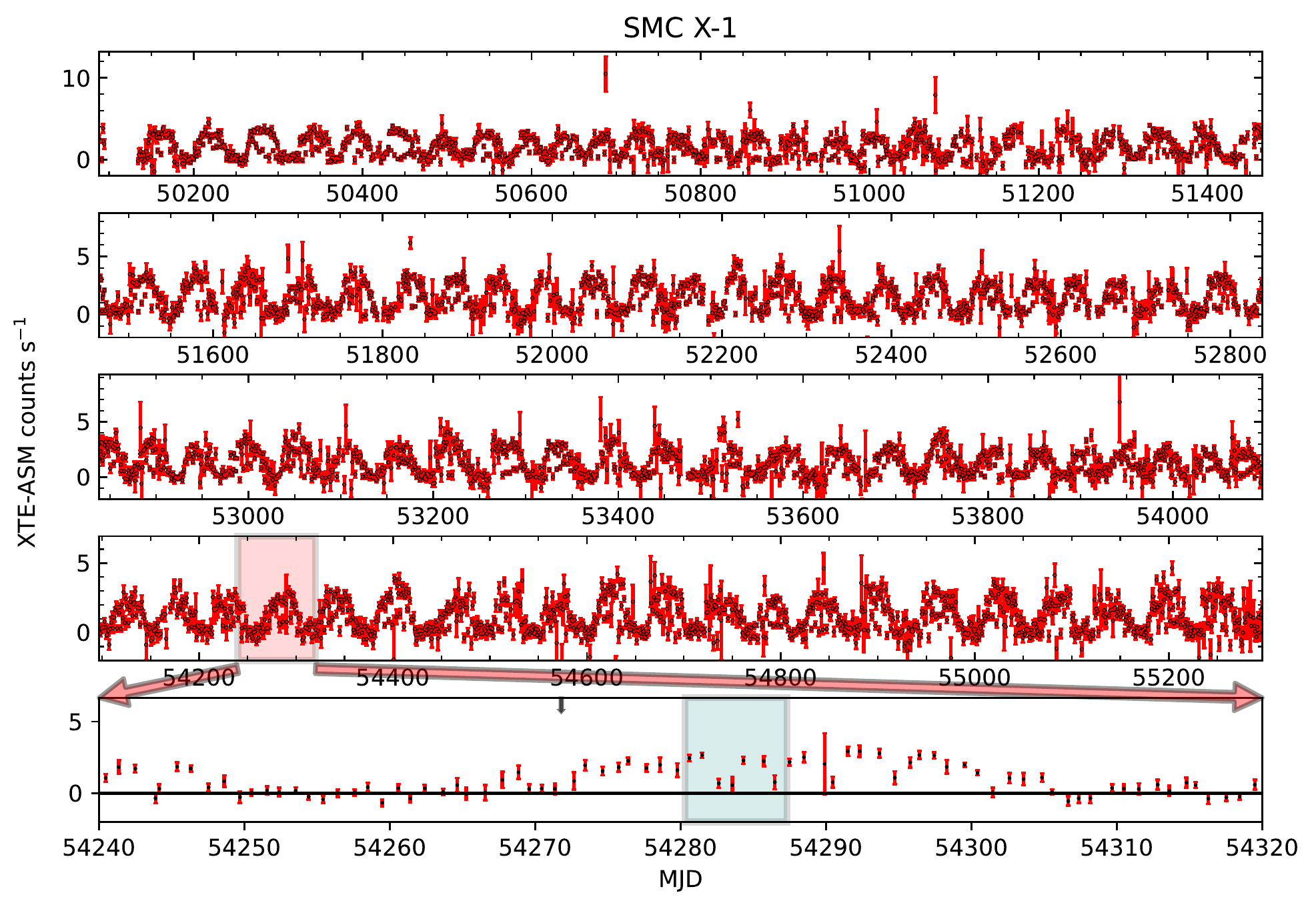}
\figsetgrpnote{X-ray lightcurve from RXTE-ASM.  The bottom panel shows the 80 days
around our campaign with an expanded scale.  The campaign dates are highlighted in blue.}
\figsetgrpend

\figsetgrpstart
\figsetgrpnum{\ref{fig:appendix}.42}
\figsetgrptitle{Vela X-1}
\figsetplot{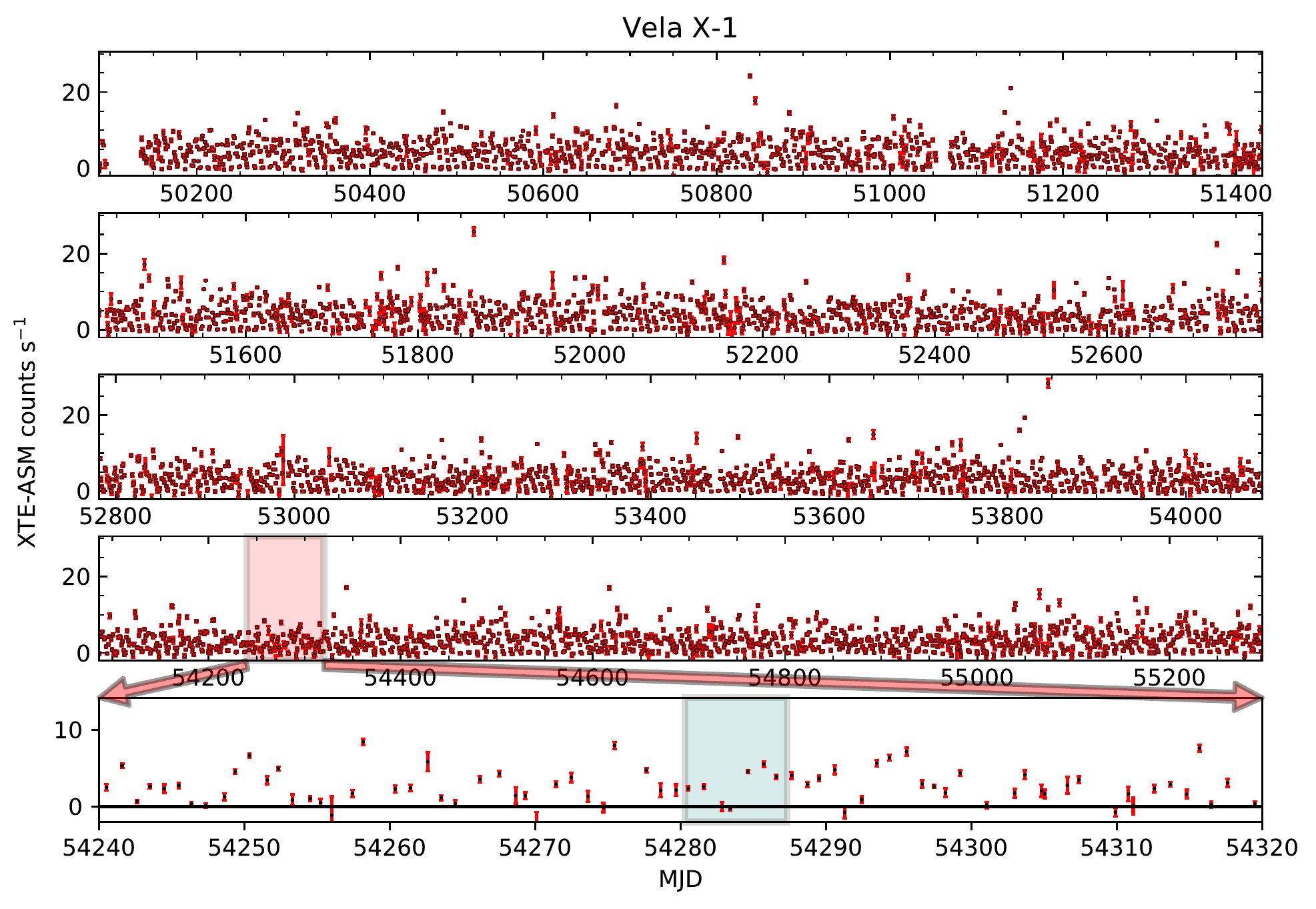}
\figsetgrpnote{X-ray lightcurve from RXTE-ASM.  The bottom panel shows the 80 days
around our campaign with an expanded scale.  The campaign dates are highlighted in blue.}
\figsetgrpend

\figsetend

Fig. \ref{fig:appendix} and the associated figure set shows RXTE All Sky Monitor lightcurves of 42 of the sources in this study for most of the fifteen years that the RXTE was active.  They illustrate the variability type of the object.  For example: 4U 1735-44 is a persistent source and 4U1608-52 is a transient source with accretion disk outbursts.  The 80 day interval around the week of the 2007 campaign is expanded in the bottom panel to emphasize the context of the source luminosity variation during our observations.  For example, 4U1608-52 was in the downward side of an accretion disk outburst whereas 4U 1735-44 was varying around its typical flux density during our campaign.  The y axis is flux density in counts per second summed over the three ASM bands 1.3 - 12 keV.  As discussed by \citet{Vrtilek_Boroson_2013}, the last two years of
ASM operation showed gradually more frequent gain variations that were not present in the earlier 14 years.  Thus we cut off the lightcurves at about MJD=55300, nearly three years after our observing campaign.
\color{black}

\end{document}